\newcommand{\xBc}{\langle}
\newcommand{\xBe}{\rangle}

\newcommand{\xbD}{\Delta}
\newcommand{\xbF}{\Phi}

\newcommand{\xbP}{\Pi}

\newcommand{\xbS}{\Sigma}

\newcommand{\xba}{\alpha}
\newcommand{\xbb}{\beta}

\newcommand{\xbd}{\delta}
\newcommand{\xbe}{\in}
\newcommand{\xbf}{\phi}
\newcommand{\xbg}{\gamma}

\newcommand{\xbm}{\mu}

\newcommand{\xbo}{\omega}
\newcommand{\xbp}{\pi}
\newcommand{\xbq}{\psi}
\newcommand{\xbr}{\rho}
\newcommand{\xbs}{\sigma}
\newcommand{\xbt}{\tau}

\newcommand{\xCK}{\times}
\newcommand{\xCL}{\pm}

\newcommand{\xCN}{\neg}
\newcommand{\xCO}{ }
\newcommand{\xCQ}{\emptyset}

\newcommand{\xCc}{<}

\newcommand{\xCe}{>}
\newcommand{\xCf}{\hspace{0.1em}}

\newcommand{\xcA}{\forall}
\newcommand{\xcB}{\subsetneqq}
\newcommand{\xcC}{\not\subseteq}
\newcommand{\xcD}{\not\supseteq}
\newcommand{\xcE}{\exists}

\newcommand{\xcN}{\hspace{0.2em}\not\sim\hspace{-0.9em}\mid\hspace{0.8em}}
\newcommand{\xcO}{\bigvee}
\newcommand{\xcP}{\not\rightarrow}

\newcommand{\xcS}{\bigcap}
\newcommand{\xcT}{\bot}
\newcommand{\xcU}{\bigwedge}
\newcommand{\xcV}{\bigcup}

\newcommand{\xcb}{\subset}
\newcommand{\xcc}{\subseteq}
\newcommand{\xcd}{\supseteq}
\newcommand{\xce}{\not\in}

\newcommand{\xcg}{\geq}
\newcommand{\xch}{\Rightarrow}
\newcommand{\xci}{\Leftarrow}
\newcommand{\xcj}{\Leftrightarrow}
\newcommand{\xck}{\leq}

\newcommand{\xcm}{\models}
\newcommand{\xcn}{\hspace{0.2em}\sim\hspace{-0.9em}\mid\hspace{0.58em}}

\newcommand{\xco}{\vee}
\newcommand{\xcp}{\rightarrow}

\newcommand{\xcr}{\leftrightarrow}
\newcommand{\xcs}{\cap}
\newcommand{\xct}{\top}
\newcommand{\xcu}{\wedge}
\newcommand{\xcv}{\cup}

\newcommand{\xcx}{\Diamond}

\newcommand{\xcz}{\Box}

\newcommand{\xDB}{\\[2ex]}

\newcommand{\xDH}{\item }

\newcommand{\xDK}{\otimes}

\newcommand{\xDN}{\ominus}

\newcommand{\xdN}{\mbox{\boldmath$N$}}

\newcommand{\xdR}{\Re}

\newcommand{\xdZ}{\mbox{\boldmath$Z$}}

\newcommand{\xda}{{\cal A}}
\newcommand{\xdb}{{\cal B}}
\newcommand{\xdc}{{\cal C}}
\newcommand{\xdd}{{\cal D}}

\newcommand{\xdf}{{\cal F}}
\newcommand{\xdg}{{\cal G}}

\newcommand{\xdi}{{\cal I}}

\newcommand{\xdl}{{\cal L}}
\newcommand{\xdm}{{\cal M}}

\newcommand{\xdo}{{\cal O}}
\newcommand{\xdp}{{\cal P}}

\newcommand{\xdx}{{\cal X}}

\newcommand{\xEI}{\begin{itemize}}
\newcommand{\xEJ}{\end{itemize}}

\newcommand{\xET}{\%}

\newcommand{\xEc}{\not<}
\newcommand{\xEd}{\neq}

\newcommand{\xEh}{\begin{enumerate}}

\newcommand{\xEj}{\end{enumerate}}

\newcommand{\xEn}{\begin{description}}
\newcommand{\xEp}{\end{description}}

\newcommand{\xeA}{\nabla}

\newcommand{\xea}{\heartsuit}
\newcommand{\xeb}{\prec}

\newcommand{\xer}{\sqsubset}

\newcommand{\xew}{\sqcap}
\newcommand{\xex}{\upharpoonright}

\newcommand{\xfA}{\mid}
\newcommand{\xfB}{\uparrow}

\newcommand{\xfI}{\mbox{I}}

\newcommand{\xfK}{\not\uparrow}

\newcommand{\xfw}{\sqcup}

\newcommand{\Xl}{\ldots}

\newcommand{\ol}{\overline}

\newcommand{\bl}{\begin{lemma} \rm}
\newcommand{\el}{\end{lemma}}
\newcommand{\br}{\begin{remark} \rm}
\newcommand{\er}{\end{remark}}
\newcommand{\be}{\begin{example} \rm}
\newcommand{\ee}{\end{example}}
\newcommand{\bco}{\begin{corollary} \rm}
\newcommand{\eco}{\end{corollary}}
\newcommand{\bc}{\begin{claim} \rm}
\newcommand{\ec}{\end{claim}}
\newcommand{\bfa}{\begin{fact} \rm}
\newcommand{\efa}{\end{fact}}
\newcommand{\bp}{\begin{proposition} \rm}
\newcommand{\ep}{\end{proposition}}
\newcommand{\bd}{\begin{definition} \rm}
\newcommand{\ed}{\end{definition}}
\newcommand{\bcs}{\begin{construction} \rm}
\newcommand{\ecs}{\end{construction}}
\newcommand{\bcd}{\begin{condition} \rm}
\newcommand{\ecd}{\end{condition}}
\newcommand{\bt}{\begin{theorem} \rm}
\newcommand{\et}{\end{theorem}}
\newcommand{\bn}{\begin{notation} \rm}
\newcommand{\en}{\end{notation}}
\newcommand{\bfi}{\begin{bild} \rm}
\newcommand{\efi}{\end{bild}}
\newcommand{\bsta}{\begin{statement} \rm}
\newcommand{\esta}{\end{statement}}
\newcommand{\bcom}{\begin{comment} \rm}
\newcommand{\ecom}{\end{comment}}
\newcommand{\bdia}{\begin{diagram} \rm}
\newcommand{\edia}{\end{diagram}}

\newcommand{\bfc}{\begin{figure}[htb] \begin{center}}
\newcommand{\efc}{\end{center} \end{figure}}

\makeindex

\documentclass{book}

\usepackage{amssymb,latexsym,epic,eepic,rotating}

\title{Truth and Knowledge
\thanks{File: Tak, arXiv 2104.13573
}
\thanks{Preliminary Version, Final Version to appear in College Publications, Studies in Logic and
Argumentation, ISBN 978-1-84890-403-3
}
}

\author{Karl Schlechta
\thanks{
schcsg@gmail.com - https://sites.google.com/site/schlechtakarl/ -
Koppeweg 24, D-97833 Frammersbach, Germany}
\thanks{
Retired, formerly: Aix-Marseille Universit\'{e}, CNRS, LIF UMR 7279, F-13000
Marseille, France
}
}

\begin{document}

\newtheorem{lemma}{Lemma}[section]
\newtheorem{theorem}[lemma]{Theorem}
\newtheorem{proposition}[lemma]{Proposition}
\newtheorem{corollary}[lemma]{Corollary}
\newtheorem{claim}[lemma]{Claim}
\newtheorem{fact}[lemma]{Fact}
\newtheorem{remark}[lemma]{Remark}
\newtheorem{definition}{Definition}[section]
\newtheorem{construction}{Construction}[section]
\newtheorem{condition}{Condition}[section]
\newtheorem{example}{Example}[section]
\newtheorem{notation}{Notation}[section]
\newtheorem{bild}{Figure}[section]
\newtheorem{comment}{Comment}[section]
\newtheorem{statement}{Statement}[section]
\newtheorem{diagram}{Diagram}[section]

\renewcommand{\labelenumi}
  {(\arabic{enumi})}
\renewcommand{\labelenumii}
  {(\arabic{enumi}.\arabic{enumii})}
\renewcommand{\labelenumiii}
  {(\arabic{enumi}.\arabic{enumii}.\arabic{enumiii})}
\renewcommand{\labelenumiv}
  {(\arabic{enumi}.\arabic{enumii}.\arabic{enumiii}.\arabic{enumiv})}

\maketitle

\setcounter{secnumdepth}{3}

\setcounter{tocdepth}{3}

\tableofcontents
\clearpage

%
%
%
\markboth{\centerline{\scriptsize Truth and Knowledge}}
{\centerline{\scriptsize Truth and Knowledge}}

$ \xCO $
\chapter{
Introduction
}

\label{Chapter INT}

$ \xCO $
\section{
Overview
}

Remark: This is a preliminary version, the final text will appear
in College Publications, Rickmansworth, UK,
Series Studies in Logic and Argumentation, ISBN 978-1-84890-403-3
\subsection{
Background
}

We adhere to ``normal'' scepticism, i.e. we are certainly aware
of various fallacies of perception etc., but assume that reality
exists, that we are conscious, and so is the reader, etc.

Modern science, like physics, medical sciences, etc. are good
examples of deep and efficient knowledge about (aspects of) the world.

From these assumptions and ``gold standards'' of systematic knowledge,
we try to investigate other sets of knowledge.
 \xEh
 \xDH
First, we are aware that different areas may have different fallacies
and interferences in the observations. E.g., the placebo/nocebo effects
are important in medicine, but not in physics.
 \xDH
In some areas, it seems impossible to have direct access to phenomena,
$ \xfI $ have no access to your consciousness, and vice versa, still, we
can communicate about our experiences. It is unclear how such questions
can
be approached in problems about consciousness of animals. We would have to
extrapolate. (The author thinks that we should not take an easy way out,
in the style of ``everything is a bit conscious  \Xl.''.)

There is nothing mysterious about such situations.
If the ideal (gold standard, ``absolute truth'') cannot be or is too
difficult to
achieve, we have to be pragmatic and not give up.
If we have no freeway, we have to take back country roads.
 \xDH
Physics works with well structured knowledge, we may add, multiply,
compare
reals etc. Sometimes, our knowledge is less complete and structured, we
may
have only a partial order, but still would like to do probability theory,
so we need suitable approximations. This is e.g. the case in legal
reasoning.
 \xDH
If we have no more than perhaps contradictory data, we can still try to
come to a reasonable conclusion, based on majorities and past
reliability. If a source of data was often wrong in the past, we should
be more sceptical than for data from a more reliable source.
 \xDH
It is human that researchers tend to try to confirm their own theories,
so we have to confirm/disprove results independently - as far as
possible. (In questions about consciousness, this is often impossible.)

 \xDH
Finally, our knowledge should be consistent, and have a possible
solution.
 \xEj

In (partial) summary:
 \xEI
 \xDH
different areas have different traps and fallacies (e.g. placebo effects),
to
identify them is part of the game.
 \xDH
if we cannot be as good as in physics, this no reason to
give up, we have to do the best we can do, without direct access to
data (consciousness problems), partial knowledge, etc.
 \xEJ

We also have to be clear about our aims. A successful philosophical
analysis
of a notion need not mean that we actually think as described in the
analysis.

The present text addresses some of above issues, and others.

\br

$\hspace{0.01em}$


\label{Remark WT}

To the author's knowlege, there is no systematic overview of the issues,
techniques, and solutions, in epistemology and
philosophy of science for different areas,
as hinted at above.

It might be a book that should be written.
\subsection{
Details
}

\er

This text looks at problems of truth and knowledge from different angles.
The subject of truth and knowledge binds the chapters together,
otherwise, they are mostly independent from each other.
Their choice is due to the author's interests, and his limited competence.

 \xEI
 \xDH
We might have insufficient knowledge for certain operations (e.g.
comparisons) -
how can we approximate sufficient knowledge by a best guess?
(Chapter \ref{Chapter PAR} (page \pageref{Chapter PAR})).

 \xDH
Are changes from one case to another relevant for a certain question?
(Chapter \ref{Chapter REL} (page \pageref{Chapter REL})).

 \xDH
Can we discern honest arguments from propaganda?
(Chapter \ref{Chapter TRU} (page \pageref{Chapter TRU})).

 \xDH
To which extent corresponds a formal philosophical analysis to actual
thinking about a problem? (This is a meta-question, applicable in many
situations.)
(Chapter \ref{Chapter CFC} (page \pageref{Chapter CFC})).

 \xDH
And, on a more specific level, we use well known approaches, e.g. to the
analysis
of counterfactuals, to an analysis of analogical reasoning
(Chapter \ref{Chapter ANA} (page \pageref{Chapter ANA})),
and, finally,
we look at Yablo's paradox, analyse his construction, and generalize it
to arbitrary formulas of the type $ \xcO \xcU \xbf_{i,j}.$ The latter is
an attempt to
to come closer to a characterisation of Yablo-like constructions
(Chapter \ref{Chapter YAB} (page \pageref{Chapter YAB})).

 \xEJ

Thus, in other words, the main subjects of this text are:

 \xEh
 \xDH Generalization of concepts and operations, like distance and size,
to situations where they are not definable in the usual way.
 \xDH A pragmatic theory of handling information (and
contradictions) using reliability of the
information sources.
 \xDH Relation of formal semantics to brain processes.
 \xDH Remarks on Yablo's coding of the liar paradox in infinite acyclic
graphs.
 \xEj

From another perspective, we treat
 \xEh
 \xDH aspects of human reasoning and their formal analoga, and their
discrepancies
(Chapter \ref{Chapter ANA} (page \pageref{Chapter ANA})) and
(Chapter \ref{Chapter CFC} (page \pageref{Chapter CFC})),
 \xDH ``softening'' of formal constructions to better fit some requirements
of various situations
(Chapter \ref{Chapter PAR} (page \pageref{Chapter PAR})),
 \xDH meta-properties of formal reasoning
(Chapter \ref{Chapter REL} (page \pageref{Chapter REL})),
 \xDH assessment of information using reliability of its sources
(Chapter \ref{Chapter TRU} (page \pageref{Chapter TRU})),
and
 \xDH comments on the Yablo paradox
(Chapter \ref{Chapter YAB} (page \pageref{Chapter YAB})).
 \xEj

In more detail:

 \xEh

 \xDH
Chapter \ref{Chapter PAR} (page \pageref{Chapter PAR})
generalizes usual operations to structures with weaker properties.
In Section \ref{Section Boole} (page \pageref{Section Boole})
and
in Section \ref{Section Sign} (page \pageref{Section Sign})
we generalize set operations to subsets of the powerset which are not
closed
under those operations.
In Section \ref{Section Height} (page \pageref{Section Height})
we use the height of a element in a partial order to determine
the size of that element, and apply our ideas
in Section 
\ref{Section Mean} (page 
\pageref{Section Mean})  to the problems seen in
Chapter \ref{Chapter TRU} (page \pageref{Chapter TRU}).

Section \ref{Section ETH} (page \pageref{Section ETH})
gives, among other things, our motivation to discuss the
generalizations of the present chapter.

 \xDH

Chapter \ref{Chapter CFC} (page \pageref{Chapter CFC})
discusses the (very probable) difference between the simple beauty
of the Stalnaker/Lewis semantics for counterfactuals and what
happens in our brain, when working with counterfactuals.

Section 
\ref{Section NEU} (page 
\pageref{Section NEU})  gives a short and very simplified picture
of the structure of, and processes in the brain.

 \xDH

Chapter \ref{Chapter ANA} (page \pageref{Chapter ANA})
presents a highly abstract approach to
analogical reasoning, in the spirit of (generalized) distance.
This is in the spirit of e.g. the philosophical analysis
of counterfactual conditionals, as done by Stalnaker and Lewis.

 \xDH

The main contribution of
Chapter \ref{Chapter REL} (page \pageref{Chapter REL})
is a detailed examination of the size relation between sets
based on filters and ideals - and thus on nonmonotonic
logics - of different strengths.
Such size relations are used in
Chapter \ref{Chapter PAR} (page \pageref{Chapter PAR}).
(The author has discussed other aspects of this problem in other
books.)

 \xDH

Chapter 
\ref{Chapter TRU} (page 
\pageref{Chapter TRU})  presents a theory of truth, describing how we
can
solve conflicts between contradictory information by assigning
a dynamic reliability to information sources. We need here generalized
operations discussed in
Section \ref{Section Mean} (page \pageref{Section Mean})  of
Chapter \ref{Chapter PAR} (page \pageref{Chapter PAR}).

 \xDH

Finally,
Chapter \ref{Chapter YAB} (page \pageref{Chapter YAB})
presents formal comments and ideas
about a solution of the representation problem
for Yablo-like structures.

 \xEj

In particular,
Chapter \ref{Chapter PAR} (page \pageref{Chapter PAR})  and
Chapter \ref{Chapter TRU} (page \pageref{Chapter TRU})  may be
seen as examples of how to try to find truth in less than perfect
situations.
In other such situations, we may need different approaches and techniques.
We should see ourselves as detectives who will use all clues at hand to
find truth.

The formal material in
Chapter \ref{Chapter PAR} (page \pageref{Chapter PAR})  through
Chapter \ref{Chapter TRU} (page \pageref{Chapter TRU})
is largely elementary.
\section{
Acknowledgements
}

The author would like to thank Andre Fuhrmann, Dov Gabbay, and
David Makinson for many very valuable discussions.
\clearpage

$ \xCO $

$ \xCO $
\markboth{\centerline{\scriptsize Partial Orders}}
{\centerline{\scriptsize Partial Orders}}

$ \xCO $
\chapter{
Operations on Partial Orders
}

\label{Section PAR}

\label{Chapter PAR}

$ \xCO $
\section{
Introduction
}

\label{Section Intro}
\subsection{
Motivation
}

In reasoning about complicated situations, e.g. in
legal reasoning, see for instance  \cite{Haa14},
the chapter on legal probabilism,
classical probability theory is often criticised for
imposing comparisons which seem arbitrary.
Our approach tries to counter such criticism by a more
flexible approach.

We do not have ``the best solution'', we rather present some suggestions,
first, how to work within one partial order, then, how to associate
to an element in a partial order in a reasonable way
a (rational) number, often in
the interval $[0,1],$ so comparisons over different partial orders
are possible, as well as operations between partial orders, like
multiplication, etc.

The ideas, as well as the formal results, are elementary, and only meant
as suggestions.
\subsection{
Overview
}

 \xEh

 \xDH
Boolean operators:

We discuss in Section \ref{Section Boole} (page \pageref{Section Boole})  and
Section \ref{Section Sign} (page \pageref{Section Sign})
possibilities to approximate the result
of the usual operations of sup, inf, etc. in partial
orders which are not complete under these operations.

 \xEh
 \xDH
In Section 
\ref{Section Boole} (page 
\pageref{Section Boole}), we first give the basic definitions, see
Definition \ref{Definition Elements} (page \pageref{Definition Elements}),
they are quite standard, but due to incompleteness,
the results may be sets of several elements, and not single elements (or
singletons). This forces us to consider operators on sets of elements,
which sometimes complicates the picture, see
Definition \ref{Definition Sets} (page \pageref{Definition Sets}).

We then discuss basic properties of our definitions in
Fact \ref{Fact Rules} (page \pageref{Fact Rules}).

 \xDH
An alternative definition for sets is given in
Definition 
\ref{Definition Alternative} (page 
\pageref{Definition Alternative}), but
Fact \ref{Fact Alternative} (page \pageref{Fact Alternative})  shows
why we will not use this definition.

 \xDH
In Section \ref{Section Sign} (page \pageref{Section Sign}),
we discuss in preliminary outline
a (new, to our knowledge) approach, by adding supplementary information to
the
results of
the operations, which may help further processing.
The operators now do not only
work on elements or sets of elements, but also the additional information,
e.g., instead of considering $X \xew Y,$ we consider $inf(X) \xew inf(Y),$
$sup(X) \xew sup(Y),$ etc., where
``inf'' and ``sup'' is the supplementary information.

 \xEj

 \xDH Height, size, and probability:

Section 
\ref{Section Height} (page 
\pageref{Section Height})  discusses ways to associate size with
elements
in partial orders, so we can compare them, calculate probabilities
of such elements, etc. There are different ways to do this,
the ``right'' way probably depends on the context.
This section is related to
Section \ref{Section FormalProp} (page \pageref{Section FormalProp})  in
Chapter \ref{Chapter REL} (page \pageref{Chapter REL}), where
we discussed size comparison in a
non-monotonical setting.
The approach here is more abstract, the relation is supposed to be
given.

 \xEh

 \xDH
In Section \ref{Section Height-Basic} (page \pageref{Section Height-Basic}),
we introduce the ``height'' of an element, as the
maximal length of a chain from $ \xcT $ to that element, see
Definition 
\ref{Definition Height} (page 
\pageref{Definition Height}). We also define
relative height, a value between 0 and 1.

 \xDH
In Section 
\ref{Section Sequences} (page 
\pageref{Section Sequences}), we argue that the situation for
sequences of partial
orders may be more complicated than their product - basically as
a warning about perhaps unexpected problems.

 \xDH
Section 
\ref{Section Proba} (page 
\pageref{Section Proba})  introduces two notions of size for sets of
elements,
one, Definition \ref{Definition P} (page \pageref{Definition P}),
by the maximal height of its elements, the other, in
Remark 
\ref{Remark MoreStandard} (page 
\pageref{Remark MoreStandard}), as the sum of
their heights. We also discuss some basic properties of these
definitions.

 \xEj

 \xDH
In Section 
\ref{Section Mean} (page 
\pageref{Section Mean}), we discuss the operations needed in
Chapter \ref{Chapter TRU} (page \pageref{Chapter TRU}).

 \xDH
Section 
\ref{Section PAR-Appendix} (page 
\pageref{Section PAR-Appendix})  presents some other remarks on
abstract operations.

 \xDH
Finally, we discuss our initial motivation for the present
chapter in criticism by S. Haack on the use of probabilities
in legal reasoning. This is done in
Section \ref{Section ETH} (page \pageref{Section ETH}),
where we present some general remarks on questions of ethics, too.

 \xEj
\clearpage
\section{
Boolean Operations in Partial Orders
}

\label{Section Boole}

We define here Boolean operations on not necessarily complete partial
orders,
and then probability measures on such orders.

In a way, this is a continuation of work in
 \cite{Leh96} and  \cite{DR15}.

First, a general remark:

\br

$\hspace{0.01em}$


\label{Remark Adeq}

We are not perfectly happy with our generalizations of the usual
operations
of $ \xew,$ $ \xfw,$ and $ \xDN $ to not necessarily complete partial
orders. We looked at a
few alternative definitions, but none is fully satisfactory.

There are a number of possible considerations when working on a new
definition, here a generalization of a standard definition:
 \xEI
 \xDH
Do we have a clear intuition?
 \xDH
Is there a desired behaviour?
 \xDH
Are there undesirable properties, like trivialisation in certain cases?
 \xDH
Can we describe it as an approximation to some ideal? Perhaps with some
natural distance?
 \xDH
How does the new definition behave for the original situation, here
complete
partial orders, etc.?
 \xEJ
\subsection{
Framework
}

\er

Assume a finite partial order $(\xdx,<)$ with TOP, $ \xct,$ and BOTTOM,
$ \xcT,$ and
$ \xcT < \xct,$ i.e. $ \xdx $ has at least two elements. $<$ is assumed
transitive.
We do not assume that the order is complete.

We will not always detail the order, so if we do not explicitly say that
$x<y$ or $y<x,$ we will assume that they are incomparable - with the
exception
$ \xcT <x< \xct $ for any $x,$ and transitivity is always assumed to hold.
\subsection{
Basic Definitions
}

\bd

$\hspace{0.01em}$


\label{Definition MinMax}

 \xEh
 \xDH
For $x,y \xbe \xdx,$ set $x \xfB y$ iff $a \xck x$ and $a \xck y$ implies
$a= \xcT.$
 \xDH
For $X \xcc \xdx,$ define

$min(X)$ $:=$ $\{x \xbe X:$ $ \xCN \xcE x' \xbe X.x' <x\}$

$max(X)$ $:=$ $\{x \xbe X:$ $ \xCN \xcE x' \xbe X.x' >x\}$

 \xDH
For $Y \xcc \xdx,$ define

$sup(Y)$ $:=$ $min(\{y':$ $ \xcA y \xbe Y.y \xck y' \})$

$inf(Y)$ $:=$ $max(\{y':$ $ \xcA y \xbe Y.y \xcg y' \})$

If $sup(Y)$ (or $inf(Y))$ is a singleton, we also write $SUP(Y)$ (or
$INF(Y)).$

 \xEj

\ed

\bfa

$\hspace{0.01em}$


\label{Fact Up}

$x \xfB y,$ $x' \xck x$ $ \xcp $ $x' \xfB y.$

(Trivial by transitivity.)

\efa

We define

\bd

$\hspace{0.01em}$


\label{Definition Set-Smaller}

 \xEh
 \xDH
$X^{y}:=\{x \xbe X.x \xck y\}$
 \xDH
$X \xck Y$ iff $ \xcA x \xbe X \xcE y \xbe Y.x \xck y$
 \xDH
$X<Y$ iff $X \xck Y$ and $ \xcE y \xbe Y \xcA x \xbe X^{y}.x<y$
 \xEj

\ed

\br

$\hspace{0.01em}$


\label{Remark Set-Smaller}

 \xEh
 \xDH
$X \xck \{ \xct \}$ (trivial).
 \xDH
$X \xcc Y$ $ \xch $ $X \xck Y$ (trivial).
 \xDH
The alternative definition:

$X \xck_{1}Y$ iff $ \xcA y \xbe Y \xcE x \xbe X.x \xck y$

does not seem right, as the example $X:=\{a, \xct \},$ $Y:=\{b\},$ and
$a<b$ shows, as then
$X \xck_{1}Y.$
 \xEj

\er

We want to define analogues of the usual boolean operators, written
here $ \xew,$ $ \xfw,$ $ \xDN.$

We will see below that the result of a simple operation will not always
give a
simple result, i.e. an element (or a singleton), but a set with
several elements as result.
Consequently, we will, in the general case,
have to define operations on sets of elements, not only on single
elements.
Note that we will often not distinguish between singletons and their
element,
what is meant will be clear from the context.
\subsection{
Definitions of the Operators $\xew, \xfw, \xDN$
}

\bd

$\hspace{0.01em}$


\label{Definition Elements}

 \xEh
 \xDH
Let $x,y \xbe \xdx.$ The ususal $x \xew y$ might not exist, as the order
is not necessarily
complete. So, instead of a single ``best'' element, we might have only a set
of ``good'' elements.

Define
 \xEh
 \xDH
$x \xew y$ $:=$ $\{a \xbe \xdx:$ $a \xck x$ and $a \xck y\}$

This is not empty, as $ \xcT \xbe x \xew y.$

If $X \xcc \xdx $ is a set, we define

$ \xew X$ $:=$ $\{a \xbe \xdx:$ $a \xck x$ for all $x \xbe X\}.$

In particular, $x \xew y \xew z$ $:=$ $\{a \xbe \xdx:$ $a \xck x,$ $a
\xck y,$ $a \xck z\}.$
 \xDH
We may refine, and consider

$x \xew' y$ $:=$ $max(x \xew y)$

Usually, also $x \xew' y$ will contain more than one element.

We will consider in the next section a subset $x \xew'' y$ of $x \xew'
y,$
but $x \xew'' y$ may still contain several elements.
 \xEj

 \xDH
Consider now $ \xfw.$ The same remark as for $ \xew $ applies here, too.

Define
 \xEh
 \xDH
$x \xfw y$ $:=$ $\{a \xbe \xdx:$ $a \xcg x$ and $a \xcg y\}.$ Note that $
\xct \xbe x \xfw y.$

If $X \xcc \xdx $ is a set, we define

$ \xfw X$ $:=$ $\{a \xbe \xdx:$ $a \xcg x$ for all $x \xbe X\}.$

In particular, $x \xfw y \xfw z$ $:=$ $\{a \xbe \xdx:$ $a \xcg x,$ $a
\xcg y,$ $a \xcg z\}.$
 \xDH
Next, we define

$x \xfw' y$ $:=$ $min(x \xfw y)$

Again, we will also define some $x \xfw'' y \xcc x \xfw' y$ later.
 \xEj

 \xDH
Consider now $ \xDN.$

Define
 \xEh
 \xDH Unary $ \xDN $
 \xEh
 \xDH
$ \xDN x$ $:=$ $\{a \xbe \xdx:$ $a \xfB x\},$ note that $ \xcT \xbe \xDN
x.$

If $X \xcc \xdx $ is a set, we define

$ \xDN X$ $:=$ $\{a \xbe \xdx:$ $a \xfB x$ for all $x \xbe X\}$
 \xDH
Define

$ \xDN' x$ $:=$ $max(\xDN x)$

$ \xDN' X$ $:=$ $max(\xDN X)$

Again, we will also define some $ \xDN'' x \xcc \xDN' x$ later.
 \xEj

It is not really surprising that the seemingly intuitively correct
definition
for the set variant of $ \xDN $ behaves differently from that for $ \xew $
and $ \xfw,$
negation often does this. We will, however, discuss an alternative
definition in Definition 
\ref{Definition Alternative} (page 
\pageref{Definition Alternative}), (3),
and will show in Fact 
\ref{Fact Alternative} (page 
\pageref{Fact Alternative}), (3),
that it seems inadequate.

 \xDH Binary $ \xDN $
We may define $x-y$ either by $x \xew (\xDN y)$ or directly:
 \xEh
 \xDH
$x \xDN y$ $:=$ $\{a \xbe \xdx:$ $a \xck x$ and $a \xfB y\},$ note again
that $ \xcT \xbe x \xDN y,$

and
 \xDH
$x \xDN' y$ $:=$ $max(x \xDN y)$

 \xEj

For a comparison between direct and indirect definition, see
Fact \ref{Fact Rules} (page \pageref{Fact Rules}), (4.4).
 \xEj
 \xEj

\ed

We turn to the set operations, so assume $X,Y \xcc \xdx $ are sets of
elements, and we
define $X \xew Y,$ $X \xfw Y.$

One idea is to consider all pairs $(x,y),$ $x \xbe X,$ $y \xbe Y$
so we define (in contrast to above
Definition \ref{Definition Elements} (page \pageref{Definition Elements}))
for $ \xew $ and $ \xfw $:

\bd

$\hspace{0.01em}$


\label{Definition Sets}

We define the set operators:
 \xEh
 \xDH $ \xew $
 \xEh
 \xDH $ \xew $

$X \xew Y$ $:=$ $ \xcV \{x \xew y:$ $x \xbe X,$ $y \xbe Y\}$

 \xDH $ \xew' $

$X \xew' Y$ $:=$ $max(X \xew Y)$
 \xEj
 \xDH $ \xfw $
 \xEh
 \xDH $ \xfw $

$X \xfw Y$ $:=$ $ \xcV \{x \xfw y:$ $x \xbe X,$ $y \xbe Y\}$

 \xDH $ \xfw' $

$X \xfw' Y$ $:=$ $min(X \xfw Y)$
 \xEj
 \xDH $ \xDN $

$ \xDN X$ and $ \xDN' X$ were already defined. We do not define $X \xDN
Y,$ but see it as an
abbreviation for $X \xew (\xDN Y).$
 \xEj

\ed

See Definition 
\ref{Definition Alternative} (page 
\pageref{Definition Alternative})  and
Fact \ref{Fact Alternative} (page \pageref{Fact Alternative})
for an alternative definition for sets, and its discussion.
\subsection{
Properties of the Operators $\xew, \xfw, \xDN$
}

We now look at a list of properties,
for the element and the set versions.

\bfa

$\hspace{0.01em}$


\label{Fact Versions-1}

Consider $ \xdx:=\{ \xcT,a,b, \xct \}$ with $a \xfB b.$
We compare $ \xew $ with $ \xew',$ $ \xfw $ with $ \xfw',$ and $ \xDN
$ with $ \xDN'.$
 \xEh
 \xDH
$ \xct \xew a=\{x:x \xck a\}=\{ \xcT,a\},$ so $ \xct \xew a \xEd \{a\},$
but ``almost'', and $ \xct \xew' a=max(\xct \xew a)=\{a\}.$
 \xDH
$ \xcT \xfw a=\{x:x \xcg a\}=\{ \xct,a\},$ so $ \xcT \xfw a \xEd \{a\},$
but ``almost'', and $ \xcT \xfw' a=min(\xcT \xfw a)=\{a\}.$
 \xDH
$ \xDN a=\{ \xcT,b\},$ $ \xDN' a=\{b\},$ and by
Definition 
\ref{Definition Elements} (page 
\pageref{Definition Elements}), (3.1.1),
$ \xDN \xDN a=\{ \xcT,a\},$
and $ \xDN' \xDN' a=\{a\}.$

 \xDH

Consider $X=\{a,b\} \xcc \xdx.$

Then by Definition \ref{Definition Sets} (page \pageref{Definition Sets}),
(1), $ \xct \xew X=\{a, \xcT \} \xcv \{b, \xcT \}=\{a,b, \xcT \},$ and
$ \xct \xew' X=max(\xct \xew X)=X.$

 \xDH

Consider again $X=\{a,b\} \xcc \xdx.$

Then by Definition \ref{Definition Sets} (page \pageref{Definition Sets}), (2),
$ \xcT \xfw X=\{a, \xct \} \xcv \{b, \xct \}=\{a,b, \xct \},$ and
$ \xcT \xfw' X=min(\xcT \xfw X)=X.$
 \xEj
Thus, $ \xew',$ $ \xfw',$ $ \xDN' $ seem the better variants.

\efa

We first show some simple facts about the $ \xck $ relation for elements
and sets
(as defined in
Definition 
\ref{Definition Set-Smaller} (page 
\pageref{Definition Set-Smaller})),
and the operators $ \xew,$ $ \xfw,$ $ \xDN.$

\bfa

$\hspace{0.01em}$


\label{Fact Set-Smaller}

 \xEh
 \xDH $ \xcc $

$X \xcc X' $ $ \xch $ $X \xck X' $
 \xDH $ \xew $
 \xEh
 \xDH
$x \xck x' $ $ \xch $ $x \xew y \xcc x' \xew y$
 \xDH
$x \xck x' $ $ \xch $ $x \xew y \xck x' \xew y$
 \xDH
$X \xcc X' $ $ \xch $ $X \xew Y \xck X' \xew Y$
 \xDH
$X \xck X' $ $ \xch $ $X \xew Y \xck X' \xew Y$
 \xEj

 \xDH $ \xfw $
 \xEh
 \xDH
$x \xck x' $ $ \xch $ $x' \xfw y \xck x \xfw y$
 \xDH
$X \xcc X' $ $ \xch $ $X \xfw Y \xck X' \xfw Y$
 \xDH
Neither

$X \xck X' $ $ \xch $ $X \xfw Y \xck X' \xfw Y$

nor

$X \xck X' $ $ \xch $ $X' \xfw Y \xck X \xfw Y$

holds

 \xEj

 \xDH $ \xDN $
 \xEh
 \xDH
$x \xck x' $ $ \xch $ $ \xDN x' \xck \xDN x$
 \xDH
$X \xcc X' $ $ \xch $ $ \xDN X' \xck \xDN X$
 \xDH
$X \xck X' $ $ \xch $ $ \xDN X' \xck \xDN X$
 \xEj

 \xEj

\efa

\subparagraph{
Proof
}

$\hspace{0.01em}$


 \xEh
 \xDH $ \xcc $

By definition of $ \xck $

 \xDH $ \xew $
 \xEh
 \xDH
$x \xew y$ $=$ $\{a:$ $a \xck x$ $ \xcu $ $a \xck y\}$ $ \xcc $ $\{a:$ $a
\xck x' $ $ \xcu $ $a \xck y\}$ $=$ $x' \xew y.$
 \xDH
By (1) and (2.1)
 \xDH
$X \xew Y$ $=$ $ \xcV \{x \xew y:x \xbe X,y \xbe Y\}.$

$X \xcc X' $ $ \xch $ $X \xew Y \xcc X' \xew Y$ $ \xch $ $X \xew Y \xck X'
\xew Y.$

 \xDH
$X \xck X' $ $ \xch $ $ \xcA x \xbe X \xcE x' \xbe X'.x \xck x'.$

$X \xew Y$ $=$ $ \xcV \{x \xew y:x \xbe X,y \xbe Y\}.$

Let $x \xew y \xbe X \xew Y,$ then there is $x' \xbe X'.x \xck x',$ and
$x' \xew y \xbe X' \xew Y,$ but
$x \xew y \xcc x' \xew y,$ so $X \xew Y \xcc X' \xew Y,$ so $X \xew Y \xck
X' \xew Y,.$
 \xEj

 \xDH $ \xfw $

 \xEh
 \xDH

By $x \xck x' $ $x' \xfw y$ $ \xcc $ $\{a:a \xcg x \xcu a \xcg y\}$ $=$ $x
\xfw y.$

 \xDH

Analogous to (2.3).

 \xDH

Consider $ \xdx =(\xcT,a,b,c, \xct \},$ $b<a,$ $X:=\{b\},$ $X'
:=\{a,c\},$ so $X \xck X',$ and $Y:=\{b,c\}.$

Then $X \xfw Y$ $=$ $(b \xfw b) \xcv (b \xfw c)$ $=$ $\{b,a, \xct \},$ and
$X' \xfw Y$ $=$ $(a \xfw b) \xcv (a \xfw c) \xcv (c \xfw b) \xcv (c \xfw
c)$ $=$ $\{a, \xct, \xct, \xct,c, \xct \}$ $=$ $\{a,c, \xct \}.$

 \xEj

 \xDH $ \xDN $

 \xEh
 \xDH

$x \xck x',$ so by Fact \ref{Fact Up} (page \pageref{Fact Up}),
$a \xfB x' $ $ \xch $ $a \xfB x.$

$ \xDN x' $ $=$ $\{a:a \xfB x' \}$ $ \xcc $ $\{a:a \xfB x\}$ $=$ $ \xDN
x.$

 \xDH
$X \xcc X',$ $a \xfB x$ for all $x \xbe X',$ so $a \xfB x$ for all $x
\xbe X.$

Thus $ \xDN X' $ $=$ $\{a:$ $a \xfB x$ for all $x \xbe X' \}$ $ \xcc $
$\{a:$ $a \xfB x$ for all $x \xbe X\}=$ $ \xDN X.$

 \xDH

$a \xbe \xDN X' $ $ \xch $ $a \xfB x' $ for all $x' \xbe X'.$ Let $x \xbe
X,$ then there is $x' \xbe X'.x \xck x',$ as
$a \xfB x',$ and $x \xck x' $ $a \xfB x.$ Thus $a \xbe \xDN X,$ so $ \xDN
X' \xcc \xDN X.$

 \xEj
 \xEj

We now examine the properties of $ \xew,$ $ \xfw,$ and $ \xDN.$

\bfa

$\hspace{0.01em}$


\label{Fact Rules}

Commutativity of $ \xew $ and $ \xfw $ is trivial. We check simple cases
like
$ \xct \xew x,$ $ \xcT \xfw x,$ show that associativity holds, but
distributivity
fails. Concerning $ \xDN,$ we see that $ \xDN \xDN x$ is not
well-behaved, and neither
is the combination of $ \xDN $ with $ \xfw.$

 \xEh
 \xDH $ \xew $ and $ \xew' $
 \xEh
 \xDH $ \xct \xew x=x$?

$ \xct \xew x$ $=$ $\{a \xbe \xdx:$ $a \xck x\}.$

$ \xct \xew' x$ $=$ $\{x\}.$
 \xDH $ \xct \xew X=X$?

$ \xct \xew X$ $=$ $\{a:$ $a \xck x$ for some $x \xbe X\}$

$ \xct \xew' X$ $=$ $max(X)$ - which is not necessarily $X$ (if there are
$a,a' \xbe X$ with
$a<a').$

 \xDH $x \xew x=x$?

$x \xew x$ $=$ $\{a \xbe \xdx:$ $a \xck x\}$

$x \xew' x$ $=$ $\{x\}.$
 \xDH $X \xew X=X$?

$X \xew X$ $=$ $ \xcV \{x \xew y:$ $x,y \xbe X\}.$

Note that $x \xew y \xcc x \xew x$ for all $x,y \xbe \xdx,$ thus
$X \xew X$ $=$ $ \xcV \{x \xew x:x \xbe X\}$ $=$ $\{a \xbe \xdx:$ $a \xck
x$ for some $x \xbe X\}.$

$X \xew' X$ $=$ $max(X)$ - which is not necessarily $X.$

 \xDH $x \xew y \xew z$ $=$ $x \xew (y \xew z)$?

Let $A:=$ $x \xew y \xew z$ $=$ $\{a:$ $a \xck x,$ $a \xck y,$ $a \xck
z\}$

Set $B:=$ $y \xew z$ $=$ $\{b:$ $b \xck y,$ $b \xck z\}$
 \xEh
 \xDH $ \xew $

We have to show $A=x \xew B.$

$x \xew B$ $=$ $ \xcV \{x \xew b:$ $b \xbe B\}$ by
Definition \ref{Definition Sets} (page \pageref{Definition Sets}), (1).

If $a \xbe A,$ then $a \xbe B,$ moreover $a \xck x,$ so $a \xbe x \xew a
\xcc x \xew B.$

Let $a \xbe x \xew B,$ then there is $b \xbe B,$ $a \xbe x \xew b.$
As $b \xbe B,$ $b \xck y,$ $b \xck z,$ so $a \xck x,$ $a \xck b \xck y,$
$a \xck b \xck z,$ so $a \xbe A$
by transitivity.

 \xDH $ \xew' $

(This just due to the fact that $max(max(A) \xcv max(B))=max(A \xcv B).)$

Set $A':=max(A),$ $B':=max(B),$ so $x \xew' B' $ $=$ $max(\xcV \{x
\xew' b:$ $b \xbe B' \}).$

Let $a \xbe A' \xcc A \xcc B,$ so there is $b' \xcg a,$ $b' \xbe B',$ and
by $a \xbe A,$ $a \xck x,$ so $a \xbe x \xew b'.$

Suppose there is $a' \xbe \xcV \{x \xew' b:$ $b \xbe B' \}$ $ \xcc $ $x
\xew B,$ $a' >a.$
Then by (1.5.1) $a' \xbe A,$ contradicting maximality of a.

Conversely,
let $a \xbe max(\xcV \{x \xew' b:$ $b \xbe B' \})$ $ \xcc $ $x \xew B,$
then $a \xbe A$ by (1.5.1). Suppose there is
$a' >a,$ $a' \xbe x \xew' y \xew' z,$ so we may assume $a' \xbe A',$
then
$a' \xbe max(\xcV \{x \xew' b:$ $b \xbe B' \}),$ as we just saw,
contradiction.

Thus, it works for $ \xew',$ too.

 \xEj
 \xEj

 \xDH $ \xfw $ and $ \xfw' $
 \xEh
 \xDH $ \xcT \xfw x=x$?

$ \xcT \xfw x$ $=$ $\{a \xbe \xdx:$ $a \xcg x\}.$

$ \xcT \xfw' x$ $=$ $\{x\}.$

 \xDH $ \xcT \xfw X=X$?

$ \xcT \xfw X$ $=$ $\{a:$ $a \xcg x$ for some $x \xbe X\}$

$ \xcT \xfw' X$ $=$ $min(X)$ - which is not necessarily $X.$

 \xDH $x \xfw x=x$?

$x \xfw x$ $=$ $\{a \xbe \xdx:$ $a \xcg x\}.$

$x \xfw' x$ $=$ $\{x\}.$

 \xDH $X \xfw X=X$?

$X \xfw X$ $=$ $ \xcV \{x \xfw y:$ $x,y \xbe X\}.$

Note that $x \xfw y \xcc x \xfw x$ for all $x,y \xbe \xdx,$ thus
$X \xfw X$ $=$ $ \xcV \{x \xfw x:x \xbe X\}$ $=$ $\{a \xbe \xdx:$ $a \xcg
x$ for some $x \xbe X\}.$

$X \xfw' X$ $=$ $min(X)$ - which is not necessarily $X.$

 \xDH $x \xfw y \xfw z$ $=$ $x \xfw (y \xfw z)$?

Let $A:=$ $x \xfw y \xfw z$ $=$ $\{a:$ $a \xcg x,$ $a \xcg y,$ $a \xcg
z\}$

Set $B:=$ $y \xfw z$ $=$ $\{b:$ $b \xcg y,$ $b \xcg z\}$
 \xEh
 \xDH $ \xfw $

We have to show $A=x \xfw B.$

$x \xfw B$ $=$ $ \xcV \{x \xfw b:$ $b \xbe B\}$ by
Definition \ref{Definition Sets} (page \pageref{Definition Sets}), (2).

If $a \xbe A,$ then $a \xbe B,$ moreover $a \xcg x,$ so $a \xbe x \xfw a
\xcc x \xfw B.$

Let $a \xbe x \xfw B,$ then there is $b \xbe B,$ $a \xbe x \xfw b.$
As $b \xbe B,$ $b \xcg y,$ $b \xcg z,$ so $a \xcg x,$ $a \xcg b \xcg y,$
$a \xcg b \xcg z,$ so $a \xbe A$
by transitivity.

 \xDH $ \xfw' $

(See above comment.)

Set $A':=min(A),$ $B':=min(B).$ $x \xfw' B' $ $=$ $min(\xcV \{x \xfw'
b:$ $b \xbe B' \}).$

Let $a \xbe A' \xcc A \xcc B,$ so there is $b' \xck a,$ $b' \xbe B',$ and
by $a \xbe A,$ $a \xcg x,$ so $a \xbe x \xfw b'.$

Suppose there is $a' \xbe \xcV \{x \xfw' b:$ $b \xbe B' \}$ $ \xcc $ $x
\xfw B,$ $a' <a.$
Then by (2.5.1) $a' \xbe A,$ contradicting minimality of a.

Conversely,
let $a \xbe min(\xcV \{x \xfw' b:$ $b \xbe B' \})$ $ \xcc $ $x \xfw B,$
then $a \xbe A$ by (2.5.1). Suppose there is
$a' <a,$ $a' \xbe x \xfw' y \xfw' z,$ so we may assume $a' \xbe A',$
then
$a' \xbe min(\xcV \{x \xfw' b:$ $b \xbe B' \}),$ as we just saw,
contradiction.

Thus, it works for $ \xfw',$ too.

 \xEj

 \xEj

 \xDH Distributivity for $ \xew,$ $ \xfw,$ $ \xew',$ $ \xfw' $

Let $ \xdx:=\{ \xcT,x,y,z, \xct \}.$

 \xEh
 \xDH $x \xew (y \xfw z)$ $=$ $(x \xew y) \xfw (x \xew z)$?

Then $y \xfw z=\{ \xct \},$ so $x \xew (y \xfw z)=\{x, \xcT \}.$

$y \xfw' z=\{ \xct \},$ so $x \xew' (y \xfw' z)=max(\{x, \xcT
\})=\{x\}.$

$x \xew y$ $=$ $x \xew z$ $=$ $\{ \xcT \},$ so $(x \xew y) \xfw (x \xew
z)$ $=$ $ \xdx.$

$x \xew' y$ $=$ $x \xew' z$ $=$ $\{ \xcT \},$ so $(x \xew' y) \xfw' (x
\xew' z)$ $=$ $min(\xdx)$ $=$ $\{ \xcT \}.$

So distributivity fails for both versions.

 \xDH $x \xfw (y \xew z)$ $=$ $(x \xfw y) \xew (x \xfw z)$?

$y \xew z$ $=$ $\{ \xcT \}$ $=$ $y \xew' z.$

$x \xfw \xcT $ $=$ $\{x, \xct \},$ $x \xfw' \xcT $ $=$ $\{x\}.$

$x \xfw y$ $=$ $\{ \xct \}$ $=$ $x \xfw' y,$ $x \xfw z$ $=$ $\{ \xct \}$
$=$ $x \xfw' z.$

$ \xct \xew \xct $ $=$ $ \xdx,$ $ \xct \xew' \xct $ $=$ $\{ \xct \}.$

So it fails again for both versions.
 \xEj

 \xDH $ \xDN $ and $ \xDN' $
 \xEh
 \xDH $ \xDN \xct $ $=$ $ \xcT $?

$ \xDN \xct $ $=$ $\{ \xcT \}$

$ \xDN' \xct $ $=$ $\{ \xcT \}$
 \xDH $ \xDN \xcT $ $=$ $ \xct $?

$ \xDN \xcT $ $=$ $ \xdx $

$ \xDN' \xcT $ $=$ $\{ \xct \}$

 \xDH $ \xDN \xDN x$ $=$ $x$?

Consider $ \xdx:=\{ \xcT,x',x,y, \xct \}$ with $x<x'.$

Then $ \xDN x=\{ \xcT,y\},$ $ \xDN' x=\{y\},$
$ \xDN (\xDN x)=\{ \xcT,x',x\},$
$ \xDN (\xDN' x)=\{ \xcT,x',x\},$
$ \xDN' (\xDN' x)=\{x' \},$
so it fails for both versions.

 \xDH $x \xew (\xDN y)$ $=$ $x \xDN y$?

$x \xDN y$ $=$ $\{a \xbe \xdx:$ $a \xck x$ and $a \xfB y\}.$

$ \xDN y$ $=$ $\{a \xbe \xdx:$ $a \xfB y\}.$

$x \xew (\xDN y)$ $=$ $ \xcV \{x \xew a:$ $a \xbe \xdx,$ $a \xfB y\}$
$=$ $\{b \xbe \xdx:$ $b \xck x$ and $b \xck a$ for some $a \xbe \xdx,$
$a \xfB y\}$ $=$
$\{b \xbe \xdx:$ $b \xck x$ and $b \xfB y\}$ by
Fact \ref{Fact Up} (page \pageref{Fact Up}).
 \xDH $x \xew (\xDN x)$ $=$ $ \xcT $?

$x \xew (\xDN x)$ $:=$ $ \xcV \{x \xew y:$ $y \xbe (\xDN x)\}$ $=$ $
\xcV \{x \xew y:$ $y \xfB x\}.$
Let $a \xbe x \xew y$ for $y \xfB x,$ then $a \xck x$ and $a \xck y,$ so
$a= \xcT.$

 \xDH $X \xew (\xDN X)$ $=$ $ \xcT $?

$X \xew (\xDN X)$ $:=$ $ \xcV \{x \xew y:$ $x \xbe X,$ $y \xbe (\xDN
X)\}$ $=$ $ \xcV \{x \xew y:$ $x \xbe X,$ $y \xfB x' $ for all $x' \xbe
X\}.$
Conclude as for (4.5).

 \xDH $x \xDN x$ $=$ $ \xcT $?

$x \xDN x$ $=$ $\{a \xbe \xdx:$ $a \xck x$ and $a \xfB x\}$ $=\{ \xcT \}$
$=$ $x \xDN' x.$
 \xDH $x \xfw (\xDN x)$ $=$ $ \xct $?

Consider $ \xdx:=(\xcT,a,b,c,ab, \xct \},$ with $a<ab,$ $b<ab.$

Then $ \xDN a=\{b,c, \xcT \},$ and $a \xfw (\xDN a)$ $=$ $ \xcV \{a \xfw
b,$ $a \xfw c,$ $a \xfw \xcT \}$ $=$
$\{ab, \xct,a\}$ $ \xEd $ $\{ \xct \}.$

$ \xDN' a$ $=$ $\{b,c\},$ $a \xfw' (\xDN' a)$ $=$ $min(\{ab, \xct \}
\xcv \{ \xct \})$ $=$ $\{ab\}$ $ \xEd $ $\{ \xct \},$
so it fails for both versions.

 \xDH $ \xDN $ is antitone: $X \xcc X' $ $ \xch $ $ \xDN X' \xcc \xDN X$

$a \xbe \xDN X' $ $ \xch $ $a \xfB x$ for all $x \xbe X',$ so $a \xfB x$
for all $x \xbe X$ $ \xch $ $a \xbe \xDN X.$

 \xDH $X \xcc \xDN \xDN X$

$ \xDN X$ $:=$ $\{a:$ $a \xfB x$ for all $x \xbe X\}.$

Let $x \xbe X,$ $a \xbe \xDN X.$ By $a \xbe \xDN X,$ $x \xfB a,$ so $x
\xbe \xDN \xDN X.$

 \xDH $ \xDN \xDN X \xcc X$ fails in general.

Consider $ \xdx $ $:=$ $\{ \xcT,$ a, $b,$ $c,$ $d,$ $ \xct \}$ with
$d<b,$ $d<c.$

Then $ \xDN \{b\}=\{ \xcT,a\},$ $ \xDN \{ \xcT,a\}=\{ \xcT,b,c,d\},$ so
$ \xDN \xDN \{b\} \xcC \{b\}.$

 \xEj

 \xEj

\efa

\bd

$\hspace{0.01em}$


\label{Definition Alternative}

We define alternative set operators, and argue in
Fact \ref{Fact Alternative} (page \pageref{Fact Alternative})  below
that they do not seem the right definitions.

 \xEh
 \xDH $ \xew_{1},$ $ \xew_{2}$
 \xEh
 \xDH
$X \xew_{1}Y$ $:=$ $ \xcS \{x \xew y:$ $x \xbe X,$ $y \xbe Y\},$
 \xDH
$X \xew_{2}Y$ $:=$ $ \xew (X \xcv Y)$
 \xEj
 \xDH $ \xfw_{1},$ $ \xfw_{2}$
 \xEh
 \xDH
$X \xfw_{1}Y$ $:=$ $ \xcS \{x \xfw y:$ $x \xbe X,$ $y \xbe Y\},$
 \xDH
$X \xfw_{2}Y$ $:=$ $ \xfw (X \xcv Y)$
 \xEj
 \xDH $ \xDN_{1}$

$ \xDN_{1}X$ $:=$ $\{a \xbe \xdx $: $a \xfB x$ for some $x \xbe X\}$
 \xEj

\ed

\bfa

$\hspace{0.01em}$


\label{Fact Alternative}

Consider again $ \xdx:=\{ \xcT,a,b, \xct \}$ with $a \xfB b,$ and
$X=\{a,b\} \xcc \xdx,$ and compare to
Fact \ref{Fact Versions-1} (page \pageref{Fact Versions-1}):

 \xEh
 \xDH $ \xew_{1},$ $ \xew_{2}$ applied to $ \xct,$ $X$

 \xEh
 \xDH $ \xew_{1}:$

Then by Definition 
\ref{Definition Alternative} (page 
\pageref{Definition Alternative}), (1.1),
$ \xct \xew_{1}X=\{a, \xcT \} \xcs \{b, \xcT \}=\{ \xcT \}.$
 \xDH $ \xew_{2}:$

Then by Definition 
\ref{Definition Alternative} (page 
\pageref{Definition Alternative}), (1.2),
$ \xct \xew_{2}X= \xew \{a,b, \xct \}=\{ \xcT \}.$
 \xEj

 \xDH $ \xfw_{1},$ $ \xfw_{2}$ applied to $ \xct,$ $X$

 \xEh
 \xDH $ \xfw_{1}:$

Then by Definition 
\ref{Definition Alternative} (page 
\pageref{Definition Alternative}), (2.1),
$ \xcT \xfw_{1}X=\{a, \xct \} \xcs \{b, \xct \}=\{ \xct \}.$
 \xDH $ \xfw_{2}:$

Then by Definition 
\ref{Definition Alternative} (page 
\pageref{Definition Alternative}), (2.2),
$ \xcT \xfw_{2}X= \xfw \{a,b, \xcT \}=\{ \xct \}.$
 \xEj

 \xDH $ \xDN_{1}$

 \xEh
 \xDH $ \xcT \xbe X$ $ \xch $ $ \xDN_{1}X= \xdx.$ (Trivial)

 \xDH In particular, $ \xDN_{1} \xdx = \xdx,$ which seems doubtful.

 \xDH $X \xcc \xDN_{1} \xDN_{1}X:$

Let $X \xEd \xCQ.$

By $ \xcT \xbe \xDN_{1}X$ and the above, $ \xDN_{1} \xDN_{1}X= \xdx.$

 \xDH $ \xDN_{1} \xDN_{1}X \xcc X$ fails in general.

Consider $ \xdx $ $:=$ $\{ \xcT,$ a, $b,$ $c,$ $d,$ $ \xct \}$ with
$d<b,$ $d<c.$

Then $ \xDN_{1}\{b\}=\{ \xcT,a\},$ $ \xDN_{1}\{ \xcT,a\}= \xdx,$ so $
\xDN_{1} \xDN_{1}\{b\} \xcC \{b\}.$

 \xDH $ \xDN_{1}$ so defined is not antitone

Consider $ \xdx $ $:=$ $\{ \xcT, \xct \},$ $X:=\{ \xct \},$ $X':= \xdx
,$ then $ \xDN_{1}X=\{ \xcT \},$ $ \xDN_{1}X' =\{ \xcT, \xct \}.$
 \xEj
 \xEj

Thus, the variants in
Definition \ref{Definition Alternative} (page \pageref{Definition Alternative})
do not seem adequate.
\section{
Elements and Sets with a Sign
}

\label{Section Sign}
\subsection{
Basic Idea
}

\efa

We will outline here a - to our knowledge, new - approach, and code the
last operation into the result, so the ``same'' result of two different
operations may look differently, and the difference will be felt in
further processing the result.

Basically, we give not only the result, as well as we can, but also an
indication, what the intended result is, ``what is really meant'', the ideal
- even if we are unable to
formulate it, for lack of an suitable element.

More precisely, if the result is a set $X,$ but what we really want is
$sup(X),$
which does not exist in the structure $ \xdx,$ we will have the result
with the
``sign'' sup, i.e., $sup(X),$ likewise inf and $inf(X),$
and further processing may take this into consideration.

In a way, it is a compromise. The full information gives all arguments and
operators, the basic information gives just the result, we give the result
with an indication how to read it.

The problem is not due to sets (instead of singletons) as rsults, as the
following example shows.

\be

$\hspace{0.01em}$


\label{Example New-Sign}

Consider $ \xdx:\{a,b,c,d,e,f,g,x\}$ with $g<c<x<a,$ $g<d<x<b,$ $g<f<x,$
$f<e.$

$a \xew b=x=c \xfw d,$ $a \xew b \xew e=f,$ $(c \xfw d) \xew e=f,$ $(c
\xew e) \xfw (d \xew e)=g \xfw g=g.$

If we note that we always went downward (as in the case $a \xew b$ etc.),
then
the result seems robust, whereas in the second case $(c \xfw d),$ we go
first upward,
then downward, and distributivity fails. Thus, adding a sign to $x$ could
be used as a warning for suitable further processing.

\ee

Consider now Example 
\ref{Example SupInf} (page 
\pageref{Example SupInf}), illustrated by
Diagram 
\ref{Diagram Inf/Sup} (page 
\pageref{Diagram Inf/Sup}), for $ \xew' $ and $ \xfw'.$
Note that $SUP\{x,x' \}$ and $INF\{x,x' \}$ do not exist, but they are
``meant''.

$a \xew' b$ is not a single element, $\{x,x' \}$ is the best we have, but
what we really
mean is something like $SUP\{x,x' \}.$ Of course, we could memorize the
arguments,
$ \xCf a$ and $b,$ and the operation, $ \xew',$ but then we have no
result, and things
become complicated when processing. So, we memorize $\{x,x' \},$ but add
the
``sign'' that $SUP\{x,x' \}$ was meant.

Likewise, $c \xfw' d$ is not a single element, but again $\{x,x' \}$ is
the best we have,
but this time, we mean rather $INF\{x,x' \}.$

\be

$\hspace{0.01em}$


\label{Example SupInf}

Let $ \xdx:=\{a,b,c,d,x,x',y,e,e',f,f' \}$ with

$e<c<x<a<f,$

$e<d<x' <b<f,$

$c<x' <a,$

$d<x<b,$

$e<y<f,$

$e' <x' <f',$

$e' <y<f',$

see Diagram \ref{Diagram Inf/Sup} (page \pageref{Diagram Inf/Sup}).
(The relations involving $ \xcT $ and $ \xct $ are not shown in the
diagram,
$ \xcT \xEd e,$ $ \xct \xEd f.)$

\ee

Consider now the slightly modified
Example 
\ref{Example SupInf-2} (page 
\pageref{Example SupInf-2}), illustrated by
Diagram 
\ref{Diagram Inf/Sup-2} (page 
\pageref{Diagram Inf/Sup-2}), for $ \xDN'.$

Here, $ \xDN' y$ is not a single element, but once again $\{x,x' \},$ and
what is
``really meant'' is $SUP\{x,x\}$ - which is absent again.

\be

$\hspace{0.01em}$


\label{Example SupInf-2}

Let $ \xdx:=\{a,b,c,d,x,x',y,e,f\}$ with

$e<c<x<a<f,$

$e<d<x' <b<f,$

$c<x' <a,$

$d<x<b,$

$y<a,$

$y<b,$

see Diagram \ref{Diagram Inf/Sup-2} (page \pageref{Diagram Inf/Sup-2}).
(The relations involving $ \xcT $ and $ \xct $ are not shown in the
diagram,
$ \xcT \xEd e,$ $ \xct \xEd f.)$

\ee

Thus:
 \xEh
 \xDH
In Example \ref{Example SupInf} (page \pageref{Example SupInf}), we have

$a \xew' b$ $=$ $\{x,x' \},$ more precisely $a \xew' b$ $=$ $sup\{y:$ $y
\xck a,$ $y \xck b\}$ $=$ $SUP\{x,x' \}$
- which does not exist, but we do as if, i.e., we give a ``label'' to
$\{x,x' \}.$

Reason:

We have $x<a,b,$ $x' <a,b,$ $SUP\{x,x' \}$ is the smallest $z$ such that
$z>x,$ $z>x',$ thus
$z<a,$ $z<b,$ but this $z$ does not exist.
 \xDH
Again in Example \ref{Example SupInf} (page \pageref{Example SupInf}), we have

$c \xfw' d$ $=$ $\{x,x' \},$ more precisely $c \xfw' d$ $=$ $inf\{y:$ $y
\xcg c,$ $y \xcg d\}$ $=$ $INF\{x,x' \},$
- which does not exist, but we do as if, i.e., we give a ``label'' to
$\{x,x' \}.$

Reason:

We have $x>c,d,$ $x' >c,d,$ $INF\{x,x' \}$ is the biggest $z$ such that
$z<x,$ $z<x',$ thus
$z>c,$ $z>d,$ but this $z$ does not exist.

 \xDH
In Example \ref{Example SupInf-2} (page \pageref{Example SupInf-2}), we have

$ \xDN' y=sup\{x:$ $x \xfB y\}=SUP\{x,x' \}.$

 \xEj

To summarize, we have $x,x' \xck sup\{x,x' \} \xck a,b$ and $c,d \xck
inf\{x,x' \} \xck x,x',$
and $x \xfB y,$ $x' \xfB y,$ $sup\{x,x' \} \xfB y$ - but
$inf\{x,x' \}$ and $sup\{x,x' \}$ need not exist.

More precisely, we have

\bfa

$\hspace{0.01em}$


\label{Fact InfSup}

 \xEh
 \xDH
$z \xck inf\{x,x' \}$ $ \xcj $ $z \xck x$ and $z \xck x' $

(Trivial.)
 \xDH
$z \xck sup\{x,x' \}$ $ \xcj $ $z \xck x$ or $z \xck x' $

(Trivial.)
 \xDH
$z \xcg inf\{x,x' \}$ $ \xcj $ $z \xcg x$ or $z \xcg x' $

(Trivial.)
 \xDH
$z \xcg sup\{x,x' \}$ $ \xcj $ $z \xcg x$ and $z \xcg x' $

(Trivial.)
 \xDH
$z \xfB x$ or $z \xfB $ $x' $ $ \xch $ $z \xfB inf\{x,x' \},$ but not
conversely:

$ \xch $: $z \xfK inf\{x,x' \}$ $ \xch $ $z \xfK x$ and $z \xfK x' $ by
(1).

Counterexample for the converse:
Consider $x,x',$ $inf\{x,x' \},z,u,u' $ with
$inf\{x,x' \} \xck x,$ $inf\{x,x' \} \xck x',$ $u<x,$ $u<z,$ $u' <x',$
$u' <z.$ (And $ \xct, \xcT,$ of course.)
Here, $x \xfK z,$ $x' \xfK z,$ but $inf\{u,u' \}$ does not exist, so $z
\xfB inf\{x,x' \}.$
 \xDH
$z \xfB sup\{x,x' \}$ $ \xch $ $z \xfB x$ and $z \xfB x',$ but not
conversely:

$ \xch $: Trivial.

Counterexample for the converse:
Consider $x,x',sup\{x,x' \},u,z,$ with
$x \xck sup\{x,x' \},$ $x' \xck sup\{x,x' \},$ $u<z,$ $u<sup\{x,x' \}.$
So $z \xfB x,$ $z \xfB x',$ but $z \xfK sup\{x,x' \}.$

 \xEj

Thus, there is not necessarily an equivalence for $ \xfB $ (nor for $ \xDN
'),$ though it
may hold in some cases, of course.

\efa

Consider now again in Example 
\ref{Example SupInf} (page 
\pageref{Example SupInf}),
see Diagram \ref{Diagram Inf/Sup} (page \pageref{Diagram Inf/Sup}),
$y \xew' \{x,x' \},$ $y \xfw' \{x,x' \},$ $ \xfB \{x,x' \},$ and $ \xDN
' \{x,x' \}$ to see the different
consequences for further operations. The differences are in $ \xcu $ vs. $
\xco $ (or
$ \xcA $ vs. $ \xcE).$

 \xEh
 \xDH $ \xew' $
 \xEh
 \xDH
$y \xew' sup\{x,x' \}$ $=$ $\{z:$ $z<y$ $ \xcu $ $(z<x$ $ \xco $ $z<x'
)\}$ $=$ $\{e',e\}$
 \xDH
$y \xew' inf\{x,x' \}$ $=$ $\{z:$ $z<y$ $ \xcu $ $(z<x$ $ \xcu $ $z<x'
)\}$ $=$ $\{e\}$
 \xEj
 \xDH $ \xfw' $
 \xEh
 \xDH
$y \xfw' sup\{x,x' \}$ $=$ $\{z:$ $z>y$ $ \xcu $ $(z>x$ $ \xcu $ $z>x'
)\}$ $=$ $\{f\}$
 \xDH
$y \xfw' inf\{x,x' \}$ $=$ $\{z:$ $z>y$ $ \xcu $ $(z>x$ $ \xco $ $z>x'
)\}$ $=$ $\{f',f\}$
 \xEj
 \xDH $ \xDN' $
 \xEh
 \xDH
$ \xDN' sup\{x,x' \}$ $=$ $\{a:$ $a \xfB x$ $ \xcu $ $a \xfB x' \}$ $=$
$\{ \xcT \}$
 \xDH
$ \xDN' inf\{x,x' \}$ $=$ $\{a:$ $a \xfB x$ $ \xco $ $a \xfB x' \}$ $=$
$\{ \xcT,$ $e' \}$
 \xEj
 \xEj

Basically, we remember the last operation resulting in an intermediate
result,
but even this is not always sufficient as the example in
Fact 
\ref{Fact Rules} (page 
\pageref{Fact Rules}), (3.1), failure of distributivity, shows:
The intermediate results $y \xfw' z,$ $x \xew' y,$ $x \xew' z$ are
singletons, so our idea
has no influence.

One could, as said, try to write everything down without intermediate
results,
but one has to find a compromise between correctness and simplicity.

\clearpage

\begin{diagram}

\label{Diagram Inf/Sup}
\index{Diagram Inf/Sup}


\unitlength0.8mm
\begin{picture}(150,180)(0,0)

\put(0,175){{\rm\bf Diagram Inf/Sup $ \xew', \xfw $ }}
\put(0,168){Recall that $INF\{x,x'\}$ and $SUP\{x,x'\}$ do not exist}

\put(40,2){e}
\put(40,5){\circle*{1}}
\put(40,5){\line(1,1){25}}
\put(40,5){\line(-1,1){25}}
\put(40,5){\line(2,1){50}}
\put(10,30){c}
\put(15,30){\circle*{1}}
\put(15,30){\line(0,1){100}}
\put(15,30){\line(1,1){75}}
\put(67,30){d}
\put(65,30){\circle*{1}}
\put(65,30){\line(0,1){100}}
\put(65,30){\line(-1,1){50}}

\put(40,52){$ INF\{x,x'\}=c \xfw d $}
\put(40,55){\circle*{1}}
\put(12,78){x}
\put(15,80){\circle*{1}}
\put(15,80){\line(1,1){50}}
\put(65,78){x'}
\put(65,80){\circle*{1}}

\put(90,52){e'}
\put(90,55){\circle*{1}}
\put(90,55){\line(-1,1){75}}
\put(90,55){\line(1,1){25}}

\put(40,102){$ SUP\{x,x'\}=a \xew' b $}
\put(40,105){\circle*{1}}
\put(12,130){a}
\put(15,130){\circle*{1}}
\put(15,130){\line(1,1){25}}
\put(67,130){b}
\put(65,130){\circle*{1}}
\put(65,130){\line(-1,1){25}}

\put(117,80){y}
\put(115,80){\circle*{1}}
\put(115,80){\line(-1,1){25}}
\put(115,80){\line(-1,-2){25}}
\put(115,80){\line(-1,2){25}}
\put(90,107){f'}
\put(90,105){\circle*{1}}

\put(40,147){f}
\put(40,155){\circle*{1}}
\put(40,155){\line(2,-1){50}}

\put(40,-10){{\bf \Large $\xcT$}}
\put(40,160){{\bf \Large $\xct$}}

\end{picture}

\end{diagram}

\vspace{4mm}

\clearpage

\clearpage

\begin{diagram}

\label{Diagram Inf/Sup-2}
\index{Diagram Inf/Sup-2}

\unitlength0.8mm
\begin{picture}(150,180)(0,0)

\put(0,175){{\rm\bf Diagram Inf/Sup $ \xDN' $ }}
\put(0,168){Recall that $INF\{x,x'\}$ and $SUP\{x,x'\}$ do not exist}

\put(40,2){e}
\put(40,5){\circle*{1}}
\put(40,5){\line(1,1){25}}
\put(40,5){\line(-1,1){25}}
\put(10,30){c}
\put(15,30){\circle*{1}}
\put(15,30){\line(0,1){100}}
\put(15,30){\line(1,1){50}}
\put(67,30){d}
\put(65,30){\circle*{1}}
\put(65,30){\line(0,1){100}}
\put(65,30){\line(-1,1){50}}

\put(40,52){$ INF\{x,x'\} $}
\put(40,55){\circle*{1}}
\put(12,78){x}
\put(15,80){\circle*{1}}
\put(15,80){\line(1,1){50}}
\put(65,78){x'}
\put(65,80){\circle*{1}}
\put(65,80){\line(-1,1){50}}


\put(40,102){$ SUP\{x,x'\}= \xDN' y $}
\put(40,105){\circle*{1}}
\put(12,130){a}
\put(15,130){\circle*{1}}
\put(15,130){\line(1,1){25}}
\put(67,130){b}
\put(65,130){\circle*{1}}
\put(65,130){\line(-1,1){25}}

\put(117,80){y}
\put(115,80){\circle*{1}}
\put(115,80){\line(-1,1){50}}
\put(115,80){\line(-2,1){100}}

\put(40,147){f}
\put(40,155){\circle*{1}}

\put(40,-10){{\bf \Large $\xcT$}}
\put(40,160){{\bf \Large $\xct$}}

\end{picture}

\end{diagram}

\vspace{4mm}

\clearpage

\br

$\hspace{0.01em}$


\label{Remark Sign-2}

The definitions using sup and inf are intuitively better than the
old definitions, i.e. those without sign.

However, even if they sometimes give better results than the old
definitions, they still do not always conform to the usual result
in complete partial orders - and this is probably irredeemably so,
the correct results simply are not there.

As an example, we re-consider the example in (4.8) of
Fact \ref{Fact Rules} (page \pageref{Fact Rules}):

$x \xfw (\xDN x)$ $=$ $ \xct $?

 \xEh

 \xDH According to the old definition

Consider $ \xdx:=(\xcT,a,b,c,ab, \xct \},$ with $a<ab,$ $b<ab.$

Then $ \xDN a=\{b,c, \xcT \},$ and $a \xfw (\xDN a)$ $=$ $ \xcV \{a \xfw
b,$ $a \xfw c,$ $a \xfw \xcT \}$ $=$
$\{ab, \xct,a\}$ $ \xEd $ $\{ \xct \}.$

$ \xDN' a$ $=$ $\{b,c\},$ $a \xfw' (\xDN' a)$ $=$ $min(\{ab, \xct \}
\xcv \xct \})$ $=$ $\{ab\}$ $ \xEd $ $\{ \xct \},$
so it fails for both versions.

 \xDH According to the new definition

Consider $ \xdx:=(\xcT,a,b,c,ab, \xct \},$ with $a<ab,$ $b<ab.$

In the new definition, $ \xDN a=sup\{b,c, \xcT \},$ and
$a \xfw \xDN a$ $=$ $inf\{y:$ $y \xcg a$ $ \xcu $ $y \xcg b$ $ \xcu $ $y
\xcg $ $c$ $ \xcu $ $y \xcg \xcT \}$ $=$ $\{ \xct \}.$

The new definition, however, fails for the following example:
$ \xdx:=\{ \xcT,y,x,b, \xct \}$ with $y<b,$ $x<b.$
Then $ \xDN x$ $=$ $\{y, \xcT \},$ $x \xfw \xDN x$ $=$ $inf\{z:$ $z \xcg
x$ $ \xcu $ $z \xcg y$ $ \xcu $ $z \xcg \xcT \}$ $=$ $inf\{b, \xct \}.$

 \xEj
\clearpage
\section{
Height and Size in Finite Partial Orders
}

\label{Section Height}

\er

We assume here a finite, strict, transitive partial order $ \xdo,$ with
relation $<.$ By abuse of language, $ \xdo $ will also be used for the set
of elements
of $ \xdo.$

Bottom $(\xcT)$ and top $(\xct)$ need not exist, neither $ \xew $ or $
\xfw,$ etc. When we
write those symbols, we assume that the elements do exist.

As in Definition \ref{Definition MinMax} (page \pageref{Definition MinMax})
$ \xfB $ will be used to say that two elements are incomparable:
$x \xfB y$ iff there is no $z,$ $z<x$ and $z<y,$ or only $ \xcT <x$ and $
\xcT <y.$
\subsection{
Basic Definitions
}

\label{Section Height-Basic}

\bd

$\hspace{0.01em}$


\label{Definition Height}

 \xEh
 \xDH
Let $x \xbe \xdx.$

Set $ht(x):=$ the length of the longest chain from $ \xcT $ to $x$ - where
we count
the number of $<$ in the chain. (If $ \xcT $ does not exist, take a
descending
chain, beginning in $x,$ of maximal length.)
 \xDH
This definition might seem arbitrary, why counting from the bottom, and
not
from the top? And, why counting from bottom or top, not both from bottom
and top?

Thus, we introduce an alternative definition for $ht(x):$

Let $b(x):=$ the length of the longest chain from $ \xcT $ to $x,$
$t(x):=$ the length of the longest chain from $x$ to $ \xct,$ and

$ht(x):= \frac{b(x)}{b(x)+t(x)}$.

Thus, $0 \xck ht(x) \xck 1,$ so, when adequate, we may interpret this
directly as
a probability.

Of course, this might be imprecise, so we may introduce a measure of
precision, e.g. for
$c(x)$ $:=$ the number of elements in $ \xdx $ comparable to $x,$

$pr(x):= \frac{c(x)}{card(\xdx)}.$

(See Remark \ref{Remark Uncertain} (page \pageref{Remark Uncertain}), (3)
for an alternative, and probably better, idea.)

 \xDH
Let $X \xcc \xdx.$

Define

$maxht(X)$ $:=$ $\{x \xbe X:$ $ \xcA x' \xbe X.ht(x) \xcg ht(x')\}$ and

$minht(X)$ $:=$ $\{x \xbe X:$ $ \xcA x' \xbe X.ht(x) \xck ht(x')\}.$

 \xEj

\ed

\br

$\hspace{0.01em}$


\label{Remark Height}

 \xEh
 \xDH
There is probably not a best choice of definition for all situations,
our aim is to indicate ways to proceed with incomplete information.

 \xDH
We pursue only the first definition of ht, as it seems the least
complicated
one, and we mainly want to illustrate the concept here.
 \xDH
Obviously, the height of an element is related to its ``size''.
Consider the powerset $ \xdp (X)$ over $X.$ A subset $X' $ will be bigger,
if it
sits higher in the $ \xcb $ relation. But we may also consider the number
(or set) of elements in $ \xdp (X)$ below $X'.$ The bigger this set, the
bigger $X' $ is.

This gives a different notion of size of an element in a
partial order: $size(x):=\{x':x' <x\}$ or $size(x):=card(\{x':x' <x\}).$

We may also consider a mixture of both approaches.

 \xEj

\er

In Remark 
\ref{Remark MoreStandard} (page 
\pageref{Remark MoreStandard}), we give an alternative definition
of a probability
using height.

\bfa

$\hspace{0.01em}$


\label{Fact Height}

 \xEh
 \xDH
$ht(\xcT)=0,$ $ht(\xct)>0.$
 \xDH
$ht(x) \xck ht(\xct)$ for all $x \xbe \xdx.$
 \xDH
We have $x<y$ $ \xcp $ $ht(x)<ht(y)$ for all $x,y \xbe \xdx.$
 \xDH
If $x$ and $y$ are $<$-incomparable, it does not necessarily follow that
$ht(x)=ht(y).$

(This is trivial, as seen e.g. in the example $ \xdx:=\{ \xcT,a,a',b,
\xct \}$ with
$a<a',$ so $ht(a')=2,$ $ht(b)=1,$ and $a',b$ are incomparable.)
 \xDH
$maxht(X)$ $=$ $maxht(max(X)),$ $minht(X)$ $=$ $minht(min(X))$

 \xEj

\efa

\bd

$\hspace{0.01em}$


\label{Definition ht-Versions}

 \xEh
 \xDH
$x \xew'' y:=maxht(x \xew y),$

$X \xew'' Y:=maxht(X \xew Y).$
 \xDH
$x \xfw'' y:=minht(x \xfw y),$

$X \xfw'' Y:=minht(X \xfw Y).$
 \xDH
$ \xDN'' x:=maxht(\xDN x),$

$ \xDN'' X:=maxht(\xDN X),$

$x \xDN'' y:=maxht(x \xDN y).$

 \xEj
We might also have chosen $x \xew' y$ instead of $x \xew y,$
etc., by Fact \ref{Fact Height} (page \pageref{Fact Height}), (5).

\ed

\be

$\hspace{0.01em}$


\label{Example ht-Versions}

Consider $ \xdx:=\{ \xcT,a,b,b',c, \xct \},$ with $b<b'.$

Then $ \xDN c=\{ \xcT,a,b,b' \},$ $ \xDN' c=\{a,b' \},$ and $ \xDN''
c=\{b' \},$ so we lose important
information, in particular, if we want to continue with Boolean
operations.

For this reason, the versions $ \xew'',$ $ \xfw'',$ $ \xDN'' $ should
be used with caution.

\ee

\br

$\hspace{0.01em}$


\label{Remark Uncertain}

 \xEh
 \xDH We may, for instance, define similarity between two points $x$ and
$y$
in a partial order, by the length of the longest common part of paths from
bottom to $x$ and $y.$
 \xDH Uncertainty of $x$ may be defined by the number of $y$ incomparable
with $x.$
 \xDH A probably better idea is as follows:

Let $X$ be the set of elements $y$ incomparable with $x.$ Consider a chain
$C \xcc X,$
which has maximal length, say $L.$
This gives an idea how much refinement is possible
in $X$ (and thus for the position of $x$ itself).
The bigger $L,$ the less precise (or certain) the value for $x$ is.

(If we were to learn more about the partial order, we might be
able to compare $x$ with all elements in $C,$ so $x$ might be below, or
above,
or in the middle of the chain $C.)$

 \xEj
\subsection{
Sequences
}

\label{Section Sequences}

\er

\be

$\hspace{0.01em}$


\label{Example Seq}

In the second example, we compensate a loss in the second coordinate
by a bigger gain in the first. Thus, the situation in the product might
be more complex that the combined situations of the elements of the
product.
 \xEh
 \xDH
Consider
$ \xdx $ $:=$ $\{0,1\}$ and $ \xdx' $ $:=$ $\{0',1' \}$ with the natural
orders. In $ \xdx,$ $ht(1)=1,$
in $ \xdx',$ $ht(1')=1.$

Order the sequences in $ \xdx \xDK \xdx' $ by the value of the sequences,
defined as
their sum.

So in $ \xdx \xDK \xdx',$ $ \xcT =(0,0')<(0,1')<(1,1')= \xct,$
$(0,0')<(1,0')<(1,1'),$
and $ht((1,1'))=2.$
 \xDH
Consider now
$ \xdx $ $:=$ $\{0,2\}$ and $ \xdx' $ $:=$ $\{0',1' \}$ with the natural
orders. In $ \xdx,$ $ht(2)=1,$
in $ \xdx',$ $ht(1')=1$ again.

Order the sequences in $ \xdx \xDK \xdx' $ again by the value of the
sequences, defined as
their sum.

So in $ \xdx \xDK \xdx',$ $ \xcT $ $=$ $(0,0')$ $<$ $(0,1')$ $<$
$(2,0')$ $<$ $(2,1')$ $=$ $ \xct,$ and $ht((2,1'))=3.$
 \xEj

\ee

Of course, we use here additional structure of the components, sum and
difference.

In general, we may consider rules like:

$ \xbs \xbs' < \xbt \xbt' $ iff $ \xbs' \xbs < \xbt' \xbt $ and

$ \xbs \xbs' < \xbt \xbt' $ iff $ \xbs \xbs' < \xbt' \xbt $ etc.

We might extend the comparison to sequences of different lengths by
suitable padding, e.g. $ \xBc 0,1,0 \xBe $ to $ \xBc 1,1 \xBe $ by appending
(e.g.) 0 to
$ \xBc 1,1 \xBe,$ resulting in $ \xBc 1,1,0 \xBe.$
\subsection{
Probability Theory on Partial Orders Using Height
}

\label{Section Proba}

We define two notions of size of a set here:
 \xEh
 \xDH the size of a set is the maximal height of its elements, in
Section \ref{Section P} (page \pageref{Section P}), and
 \xDH the size of a set is the sum of the heights of its elements, in
Remark \ref{Remark MoreStandard} (page \pageref{Remark MoreStandard}).
 \xEj

The first can be seen as a ``quick and dirty'' approach, the second
as a more standard one.

There probably is no unique best solution, it will depend on the situation
at hand.
\subsubsection{
Size of a Set as Maximal Height of its Elements
}

\label{Section P}

\bd

$\hspace{0.01em}$


\label{Definition P}

 \xEh
 \xDH
For $X \xcc \xdx,$ we set $ht(X):=max\{ht(x):x \xbe X\}.$

If we are interested in $sup(X),$ we might define $ht(sup(X)):=ht(X)+1,$
and

if we are interested in $inf(X),$ we might define $ht(inf(X))$ $:=$
$min\{ht(x):x \xbe X\}-1.$
 \xDH
We may define a relative height by $rht(x):= \frac{ht(x)}{ht(\xct)},$
and we have $0 \xck rht(x) \xck 1,$ which may be interpreted as the
probability of $x.$

Thus, we define $P(x):=rht(x),$ and $P(X)$ similarly for $X \xcc \xdx.$
 \xEj

\ed

\bfa

$\hspace{0.01em}$


\label{Fact Schnitt}

We have the following facts for the height for $ \xew $ and $ \xfw:$
 \xEh
 \xDH
$ht(X \xew X')$ $<$ $ht(X),$ $ht(X'),$
 \xDH
$ht(X),$ $ht(X')$ $<$ $ht(X \xfw X')$
 \xEj

\efa

\subparagraph{
Proof
}

$\hspace{0.01em}$


This is trivial, as any chain to $(X \xew X')$ may be continued to a
chain to $X$ and
$X'.$ The second property is shown analogously.
Alternatively, we may use Fact 
\ref{Fact Height} (page 
\pageref{Fact Height}), (3).

$ \xcz $
\\[3ex]

\br

$\hspace{0.01em}$


\label{Remark Powerset}

When we work with subsets of some powerset, we use, unless defined
otherwise,
$ \xcB $ for $<,$ $ \xcs $ for $ \xew',$ $ \xcv $ for $ \xfw',$ and $
\xDN' $ is set complement.

\er

\be

$\hspace{0.01em}$


\label{Example Schnitt}

Consider $ht(X)$ as defined in
Definition \ref{Definition Height} (page \pageref{Definition Height}), (1).

These examples show that $ht(X),$ $ht(X')$ may be arbitrarily bigger than
$ht(X \xew X'),$
and $ht(X \xfw X')$ may be arbitrarily bigger than $ht(X)$ and $ht(X').$

 \xEh
 \xDH
Let $ \xdx $ $:=$ $\{ \xCQ,$ $X_{1},$ $X_{2},$ $X'_{1},$ $X'_{2},$ $X,$
$X',$ $X \xcs X',$ $X \xcv X',$ $X'' \},$ with

$X:=\{a,a',b\},$ $X':=\{b,c,c' \},$ $X_{1}:=\{a\},$ $X_{2}:=\{a,a' \},$
$X'_{1}:=\{c\},$ $X'_{2}:=\{c,c' \}.$
Thus, $X \xcs X' =\{b\}=X'',$ and $ht(X \xcs X')=1,$ but $ht(X)=ht(X'
)=3.$

 \xDH
Let $ \xdx $ $:=$ $\{ \xCQ,$ $X_{1},$ $X_{2},$ $X,$ $X',$ $X \xcv X'
\},$ with

$X:=\{a,a' \},$ $X':=\{b,b' \},$ $X_{1}:=\{a,b\},$ $X_{2}:=\{a,a',b\}.$
Thus, $ht(X \xcv X')=3,$ but $ht(X)=ht(X')=1.$
 \xEj

$ \xcz $
\\[3ex]

\ee

\be

$\hspace{0.01em}$


\label{Example =1}

Some further examples:

Consider $ht(X)$ again as defined in
Definition \ref{Definition Height} (page \pageref{Definition Height}), (1),
and $P(X)$ $:=$ $ \frac{ht(X)}{ht(\xct)}$.

$P(x)+P(\xDN x):$

$P(x)+P(\xDN x)$ may be 1, but also $ \xCc 1$ or $ \xCe 1:$
 \xEh
 \xDH
$=1:$

Consider $ \xdx $ $:=$ $\{ \xcT,a,b, \xct \}.$

Then $ \xDN a=\{b, \xcT \},$ $ \xDN' a=\{b\},$ and $P(a)=P(b)=1/2.$
 \xDH
$<1:$

Consider $ \xdx $ $:=$ $\{ \xCQ,$ $\{a\},$ $\{b,d\},$ $\{a,b,c\},$
$\{a,b,c,d\}\}.$

Then $ \xDN \{a\}$ $=$ $\{ \xCQ,\{b,d\}\},$ $ \xDN' \{a\}=\{\{b,d\}\}.$
Thus $ht(\{a\})=ht(\xDN' \{a\})=1,$ $ht\{a,b,c,d\}=3,$ and $P(\{a\})=P(
\xDN' \{a\})=1/3,$ and
$1/3+1/3<1.$

 \xDH
$>1:$

Consider $ \xdx $ $:=$ $\{ \xCQ,$ $\{a\},$ $\{a,a' \},$ $\{b\},$ $\{b.b'
\},$ $\{a,a',b,b' \}\}.$

Then $ \xDN' \{\{a,a' \}\}$ $=$ $\{\{b,b' \}\},$ and
$ht\{a,a' \}$ $=$ $ht\{b,b' \}$ $=$ 2, $ht\{a,a',b,b' \}=3,$
so $P(\{a,a' \})=P(\xDN' \{a,a' \})=2/3,$ $2/3+2/3>1.$
 \xEj

\ee

We now consider independence, which we may define as usual:

\bd

$\hspace{0.01em}$


\label{Definition Ind1}

A and $B$ are independent iff $P(A \xew B)$ $=$ $P(A)*P(B).$

\ed

This, however, might be too restrictive, alternatives come to mind, e.g.

\bd

$\hspace{0.01em}$


\label{Definition Ind2}

A and $B$ are independent iff

$P(B):= \xba $ and

$P(A \xew B)/P(A)= \xba,$

or

$P(A \xew B)/P(A)< \xba,$
and $P((\xDN A) \xew B)/P(\xDN A) \xcg \xba,$

or

$P(A \xew B)/P(A)> \xba $ and
$P((\xDN A) \xew B)/P(\xDN A) \xck \xba,$

\ed

The best definition might also be domain dependent.

\br

$\hspace{0.01em}$


\label{Remark MoreStandard}

We turn to a more standard definition of size, based on point measure,
where the size of a point $x$ is again $ht(x),$ but the size of a set is
now the sum of the sizes of its points.

Alternatively, we may define for $X \xcc \xdx:$
 \xEh
 \xDH
$ \xbm (X)$ $:=$ $ \xbS \{ht(x):x \xbe X\},$
 \xDH
$P(X)$ $:=$ $ \frac{ \xbm (X)}{ \xbm (\xdx)}$
 \xEj

But we have similar problems as above with this definition, e.g. with
$P(x)$ and
$P(\xDN x),$ etc.:

If $x= \xcT $ or $x= \xct,$ then $P(x)+P(\xDN x)=1,$ but if
$x \xEd \xcT,$ and $x \xEd \xct,$ then $P(x)+P(\xDN x)<1,$ as $ \xct $
is missing.

Similarly, as $P(x \xew \xDN x)=0,$ we will often have $P(x \xfw \xDN x)$
$ \xEd $
$P(x)+P(\xDN x)-P(x \xew \xDN x).$

This, however is not due to incompleteness, as we can easily see by
considering complete partial orders. Consider e.g.
$ \xdx:=\{ \xcT,a,a',b,b', \xct \}$ with $a<a',$ $b<b'.$
Then $ \xDN a' =\{b,b' \},$ and $ht(a')=ht(b')=2,$ $ht(a)=ht(b)=1,$ $ht(
\xct)=3.$
$P(a')=2/9,$ $P(\xDN a')=3/9,$ $P(a' \xew \xDN a')=0,$ $P(a' \xfw \xDN
a')=3/9.$

Of course, for this definition of $P,$ considering a disjoint cover of
$ \xdx $ will have the desired property.

\er

The reader may also consider the ideas in
 \cite{DR15} on qualitative probability.
\section{
The Mean Value of Sets
}

\label{Section Mean}

We will use the maen value in
Chapter \ref{Chapter TRU} (page \pageref{Chapter TRU}),
see also Section \ref{Section General-6} (page \pageref{Section General-6}).

 \xEh

 \xDH Bare sets

Suppose we have just sets, without any additional structure, so we may
count the elements, and use the set constants like $ \xCQ,$ and
operations
like $ \xcs,$ $ \xcv,$ -, and thus also the symmetrical set difference
as a
measure of distance $A \xbD B:=(A-B) \xcv (B$-A), or its cardinality.

We now consider some variants.

 \xEh
 \xDH Use of $ \xbD,$ a candidate for mean value has $ \xbD $-minimal
distance
from A and $B.$

Let $A,B$ be disjoint, and consider $Z$ with small $ \xbD $-distance from
A and $B.$
Let $Z:= \xCQ.$ $ \xCQ \xbD A=A,$ $ \xCQ \xbD B=B.$ Let $Z:=A \xcv B.$ $Z
\xcD A=B,$ $Z \xcD B=A,$ so $ \xCQ $ and $A \xcv B$
are equivalent, which indicates that $ \xbD $ is not the right idea.

 \xDH Some more examples.

If $A \xcs B \xEd \xCQ,$ then $(A \xcs B) \xbD A=A$-B, $(A \xcs B) \xbD
B=B$-A, $(A \xcv B) \xbD A=B$-A, $(A \xcv B) \xbD B=A$-B,
$ \xCQ \xbD A=A,$ $ \xCQ \xbD B=B,$ so $A \xcs B$ and $A \xcv B$ are
equivalent, $ \xCQ $ is worse.

Take $A,B,C$ pairwise disjoint. then $ \xCQ \xbD A=A$ etc., $(A \xcv B
\xcv C) \xbD A=B \xcv C$ etc.,
so considering the distances individually (to A, then to $B,$ etc.),
$ \xCQ $ is better, but taking the union, they are equivalent.

 \xDH Interior and exterior average.

It seems more interesting to consider the interior and exterior distance,
where we minimize $Z-A_{i}$ for the interior average, and $A_{i}-Z$ for
the
exterior average.

In this definition, any $Z \xcd \xcV A_{i}$ is optimal for the exterior
average,
and any $Z \xcc \xcS A_{i}$ is optimal for the interior average. This
seems wrong.

 \xDH Combination of interior and exterior average

We suggest a combination of interior and exterior average, e.g. by a
lexicographic order (first interior, then exterior, or vice versa), or
by counting the interior twice, the exterior once, etc. The choice will
probably depend on the intention.

Some examples (we work here with counting the elements of the set
differences,
not directly with the sets) for $A:=\{a,b\}$ and $B:=\{b,c\}:$

 \xEh
 \xDH $Z:= \xCQ.$

$card(Z-A)=card(Z-B)=0,$

$card(A-Z)=card(B-Z)=2.$

 \xDH $Z:=\{b\}.$

$card(Z-A)=card(Z-B)=0,$

$card(A-Z)=card(B-Z)=1.$

 \xDH $Z:=\{a\}.$

$card(Z-A)=0,$ $card(Z-B)=1,$

$card(A-Z)=1,$ $card(B-Z)=2.$

 \xDH $Z:=\{a,b\}.$

$card(Z-A)=0,$ $card(Z-B)=1,$

$card(A-Z)=0,$ $card(B-Z)=1.$

 \xDH $Z:=\{a,c\}.$

$card(Z-A)=card(Z-B)=1,$

$card(A-Z)=card(B-Z)=1.$

 \xDH $Z:=\{a,b,c\}.$

$card(Z-A)=card(Z-B)=1,$

$card(A-Z)=card(B-Z)=0.$

 \xDH $Z:=\{a,b,c,d\}.$

$card(Z-A)=card(Z-B)=2,$

$card(A-Z)=card(B-Z)=0.$

 \xEj

Thus, by emphasis on the interior, (1.4.2) is the best (better than
(1.4.1) by
the exterior), by emphasis on the exterior, (1.4.6) is the best, better
than (1.4.7) by the interior.

 \xDH Further refinements are possible, e.g.

 \xEh
 \xDH
We may prefer those $Z$ with more equal distances to the $A_{i}.$
 \xDH We may consider the square of the distances, thus penalising big
differences.

 \xDH We may give some sets bigger weight, counting elements (and thus
difference) twice.

 \xEj

 \xDH Counting elements

An alternative approach is to divide the numbers of elements by the
number of sets, and chosing accordingly some elements from each set.
This does not seem very promising. (Problems with division, e.g. 2/3,
are probably not very serious.)

 \xEj

 \xDH Additional structure

The reader may have noticed that the situation for two sets is similar
to the semantic version of symmetric theory revision - except that we
have no distance.

Suppose we have a distance, so we can generalize symmetic theory
revision to more than two sets as follows:

Consider $A_{i},$ $i \xbe I.$ Define the average by $ \xcV \{(A_{i} \xfA (
\xcV \{A_{j}:j \xbe I,i \xEd j\})):i \xbe I\}$
where $A \xfA B:=\{b \xbe B:$ $ \xcE a_{b} \xbe A \xcA b' \xbe B \xcA a'
\xbe A(d(a_{b},b) \xck d(a',b'))\}.$

We might use multisets to give more weight to some sets.

 \xEj
\section{
Generalizing the Operations Used in Chapter \ref{Chapter TRU}
}

\label{Section General-6}
\subsection{
Operations Used in Chapter \ref{Chapter TRU}
}

In Chapter 
\ref{Chapter TRU} (page 
\pageref{Chapter TRU}), we use the definitions

 \xEh
 \xDH
$r_{i}$: value of $A_{i}$

 \xDH
$ \xbr_{i}$: reliability of $A_{i}$

 \xDH
$ \xbr c_{i}$: reliability of communication channel $i$

 \xDH
$m$ mean value of the $r_{i}$

 \xDH
$ \xbd_{i}$: distance between $m$ and $r_{i}$

 \xDH
$ \xbd $: mean value of the $ \xbd_{i}$

 \xDH
$t$: number of $r_{i}$ used to calculate $m$

 \xEj

We used the following operations in
Chapter \ref{Chapter TRU} (page \pageref{Chapter TRU})

 \xEh
 \xDH
Operations Variant 1:
 \xEh
 \xDH $m$ (mean value) of the $r_{i}$
 \xDH $ \xbd_{i}$ $=$ distance between $m$ and $r_{i}$
 \xDH $ \xbd $ $=$ mean value of all $ \xbd_{i}$
 \xDH adjusting $ \xbr_{i}$ using $ \xbd,$ $ \xbd_{i},$ and old $
\xbr_{i}$
 \xEj

 \xDH
Operations Variant 2:
 \xEh
 \xDH adjust $r_{i}$ using old $ \xbr_{i}$
 \xEj

 \xDH
Operations Variant 3:
 \xEh
 \xDH multiply old $m$ by $t$
 \xEj

 \xDH
Operations Variant 4:
 \xEh
 \xDH put $ \xbr_{i}$ in relation to $ \xbd_{i}$ (is $ \xbd_{i}$ small in
comparison to $ \xbr_{i}?)$

 \xEj

 \xDH
Operations for communication and own reliability:

 \xEh
 \xDH (serial) combination of two reliabilities, here $ \xbr_{i}$ and $
\xbr c_{i}$
 \xDH conversely, break down a modification of a combination of two
reliabilities to a modification of the individual reliabilities
(this should be an inverse operation to the first operation here)

 \xDH for $ \xbe_{i},$ see
Chapter \ref{Chapter TRU} (page \pageref{Chapter TRU}), nothing new

 \xEj

 \xEj

When we generalize from $ \xdR $ to more general structures, we will
probably first treat linear orders, then full power sets, and then
general partial orders. We give some examples.
\subsection{
Generalization From Values in $\xdR$ to Sets
}

Ideas (at least in first approach, let $ \xbr_{i},$ $ \xbr c_{i}$ will be
reals in $[0,1],$
otherwise some calculations seem difficult - if they are not real, we
transform them into a real using height etc.):

The following, however, seems an easy generalization for $ \xbr_{i}:$

Reliabilities (of agents or messages) will be multisets of the form
$\{r_{i}a_{i}:i \xbe I\}.$ $r_{i}$ will be a real value between -1 and
$+1,$ $a_{i}$ should be
seen as a ``dimension''.

This allows for easy adjustment, e.g. ageing over time, shifting
importance,
etc., as we will shortly detail now:
 \xEI
 \xDH the real values allow arbitrarily fine adjustments, it is not just
$\{-1,0,1\},$
 \xDH the dimensions allow to treat various aspects in different ways,
 \xDH for instance, we can introduce new agents with a totally ``clean
slate'',
0 in every dimension, or preset some dimensions, but not others,
 \xDH the uniform treatment of all dimensions in
is not
necessary, we can treat different dimensions differently, e.g., conflicts
between two agents in dimension $a_{i}$ need not touch dimension $a_{j},$
etc.
 \xEJ
Thus, we have arbitrarily many dimensions, with possibly different meaning
and
treatment, and within each dimension arbitrarily many values. This is not
a total order, but within each dimension, it is.

 \xEh
 \xDH Operations for Variant 1:
 \xEh
 \xDH mean value $m:$ this seems the main problem.

Consider the ``mean value of sets''. The best idea might be to
count the occurrence of elements in the sets considered,
and chose the set of those elements with the best count (or above a
certain
threshold).

Alternatively, given a notion of distance between sets, we might
chose those elements, which have the smallest distance from the
sets considered.

 \xDH $ \xbd_{i}$ we consider (a variant of) symmetrical difference $ \xbD
$
 \xEh
 \xDH $ \xbd_{i}$ is a number $(\xbD $ as number)
 \xDH $ \xbd_{i}$ is a set $(\xbD $ as set)
 \xEj
 \xDH $ \xbd $
 \xEh
 \xDH Case (1.2.1): classical
 \xDH Case (1.2.2): as in (1.1)
 \xEj
 \xDH Adjusting $ \xbr_{i}:$
 \xEh
 \xDH Case (1.2.1): evident.
 \xDH Case (1.2.2): put $ \xbD $ in relation to $U$ (with $ \xdp (U)$
considered).
 \xEj

 \xEj

 \xDH Operations for Variant 2:
 \xEh
 \xDH We should give weight to elements and perhaps non-elements, too.

 \xEj

 \xDH Operations for Variant 3:
 \xEh
 \xDH count multiple times
 \xEj

 \xDH Operations for Variant 4:
 \xEh
 \xDH transform both to numbers
 \xEj

 \xDH Operations for communication etc.:
 \xEh
 \xDH multiplication and breaking down.
 \xEj

 \xEj
\clearpage
\section{
Appendix
}

\label{Section PAR-Appendix}
\subsection{
The Core of a Set
}

The following remarks are only abstractly related to the main part
of this chapter. The concept of a core is a derivative concept to the
notion of a distance, and the formal approach is based on theory revision,
see e.g.  \cite{LMS01}, or  \cite{Sch18b},
section 4.3.

We define the core of a set as the subset of those elements which are
``sufficiently'' far away from elements which are NOT in the set.
Thus, even if we move a bit, we still remain in the set.

This has interesting applications. E.g., in legal reasoning, a witness may
not be very sure about colour and make of a car, but if he errs in one
aspect, this may not be so important, as long as the other aspect is
correct. We may also use the idea for a differentiation of truth
values, where a theory may be ``more true'' in the core of its
models than in the periphery, etc.

In the following, we have a set $U,$ and a distance $d$ between elements
of $U.$
All sets $X,Y,$ etc. will be subsets of $U.$ $U$ will be finite, the
intuition
is that $U$ is the set of models of a propositional language.

\bd

$\hspace{0.01em}$


\label{Definition Depth}

Let $x \xbe X \xcc U.$

(1) $depth(x)$ $:=$ $min\{d(x,y):$ $y \xbe U-X\}$

(2) $depth(X)$ $:=$ $max\{depth(x):$ $x \xbe X\}$

\ed

\bd

$\hspace{0.01em}$


\label{Definition Core}

Fix some $m \xbe \xdN,$ the core will be relative to $m.$
One might write $Core_{m},$ but this is not important here,
where the discussion is conceptual.

Define

$core(X)$ $:=$ $\{x \xbe X:$ $depth(x) \xcg depth(X)/m\}$

(We might add some constant like 1/2 for $m=2,$ so singletons have a
non-empty core - but this is not important for the conceptual
discussion.)

\ed

It does not seem to be easy to describe the core operator
with rules e.g. about set union, intersection, etc.
It might be easier to work with pairs (X, $depth(X)),$ but
we did not pursue this.

We may, however, base the notion of core on repeated application of
the theory revision operator $*$ (for formulas) or $ \xfA $ (for sets) as
follows:

Given $X \xcc U$ (defined by some formula $ \xbf),$ and $Y:=U-X$ (defined
by $ \xCN \xbf),$
the outer elements of $X$ (those of depth 1) are $Y \xfA X$ $(M(\xCN \xbf
* \xbf)).$
The elements of depth 2 are $(Y \xcv (Y \xfA X)) \xfA (X-(Y \xfA X)),$
M($((\xCN \xbf) \xco (\xCN \xbf * \xbf))$ $*$ $(\xbf \xcu \xCN (
\xCN \xbf * \xbf))$) respectively, etc.

We make this formal.

\bfa

$\hspace{0.01em}$


\label{Fact Core}

 \xEh
 \xDH The set version

Consider $X_{0},$ we want to find its core.

Let $Y_{0}$ $:=$ $U-X_{0}$

Let $Z_{0}$ $:=$ $Y_{0} \xfA X_{0}$

Let $X_{1}$ $:=$ $X_{0}-Z_{0}$

Let $Y_{1}$ $:=$ $Y_{0} \xcv Z_{0}$

Continue $Z_{1}:=$ $Y_{1} \xfA X_{1}$
etc. until it becomes constant, say $Z_{n}=X_{n}$

Now we go back: $Core(X_{0})$ $:=$ $X_{n} \xcv  \Xl  \xcv X_{n/2}$

 \xDH The formula version

Consider $ \xbf_{0},$ we want to find its core.

Let $ \xbq_{0}$ $:=$ $ \xCN \xbf_{0}$

Let $ \xbt_{0}$ $:=$ $ \xbq_{0}* \xbf_{0}$

Let $ \xbf_{1}$ $:=$ $ \xbf_{0} \xcu \xCN \xbt_{0}$

Let $ \xbq_{1}$ $:=$ $ \xbq_{0} \xco \xbt_{0}$

Continue $ \xbt_{1}:=$ $ \xbq_{1}* \xbf_{1}$
etc. until it becomes constant, say $ \xbt_{n}= \xbf_{n}$

Now we go back: $Core(\xbf_{0})$ $:=$ $ \xbf_{n} \xco  \Xl  \xco
\xbf_{n/2}$

 \xEj


\section{
Motivation: Ethics and philosophy of Law
}

\label{Section ETH}
\subsection{
Introduction
}

\efa

We mention here some aspects of philosophy of law, in particular, where
they are related to other subjects of this text, or other work by the
author.
\subsection{
Haack's Criticism of Probability In Law
}

These are comments on Susan Haack's criticism of probabilistic approaches
to legal reasoning.
Page numbers refer to  \cite{Haa14}.
See  \cite{Sta18c}, too.

See  \cite{Haa14}, $p.$ 62, in shorthand:

\xEn
 \xDH (a)

evidential quality is not necessarily linearly ordered.

 \xDH (b)

$p(\xbf)+p(\xCN \xbf)=1,$ but in the case of no or very weak evidence
neither might be warranted.

 \xDH (c)

By the product rule of probability, the probability of combined evidence
is never stronger than the individual probabilities are -
but combined evidence might be stronger.
\xEp

We think that (a) is the deepest objection, (b) and (c) are about
details of application, and not about applicability in principle of
probability
theory to evidence. The criticism of (b) and (c) may also be directed
against evidence in natural science, and this is obviously wrong.

 \xEI
 \xDH
On (a).

Assigning numbers or places in a total order seems indeed rather arbitrary
in many cases.

There are approaches to generalized probability theory which address
such questions,
see  \cite{Leh96},  \cite{DR15}.

In addition, nonmonotonic logic can be seen as qualitative reasoning about
size, another way of doing generalized probability theory.
See here e.g.  \cite{GS16},
in particular sections 5.3, 6.4, 6.5, and chapter 11 there,
and also  \cite{Sch95-3}.

This objection was also the initial reason to develop
Chapter \ref{Chapter PAR} (page \pageref{Chapter PAR}).

We should also mention here that some comparisons might seem unethical,
e.g. comparing the ``value'' of human life,
without comitting to equal value. We do not know how to treat this
formally.

 \xDH
On (b).

This seems the same problem as in Intuitionistic Logic (which can best be
seen as constructive logic, one has a proof, a counterexample, or
neither).

Fermat's conjecture was right or wrong. But for a long time, no
one had a proof either way.

It is the difference between what holds, and what we know to hold. This
should not be confused.

 \xDH
On (c).

This might be the most difficult point, as it has so many interpretations,
e.g.
 \xEh
 \xDH
do we speak about a hypothesis, how a scenario might have developped,
e.g. how the perpetrator might have entered the house,
 \xDH
do we speak about a chain of observations, the more detailed they are,
the more they allow to differentiate between hypotheses
(compare to an experimentum crucis in science), and the less likely it
happened
by chance
(compare to the $5 \xbs $ rule in physics),
 \xDH
do we speak about uncertain observations, which may support each other
to a global probability, though taken individually, the exactness of all
elements might be relatively weak

etc. etc.
 \xEj

Moreover, one should separate the quality of an explanation from its
likelihood. A detailed explanation is better than a less detailed one,
as an explanation, even though it is less likely. ``It happened'' is very
likely, but worthless as an explanation.

Perhaps one should compare only explanations of similar quality by their
likelihood.

 \xEJ

Probability theory might in these cases be more complicated to apply, or
not be the right level of abstraction,
without being wrong in principle, $ \xfI $ think.

Abstract approaches are important for two reasons:
First, they allow to isolate reasoning from arbitrary influences, second,
if they work with intermediate results (as, e.g., probabilities),
they simplify reasoning.
\subsection{
Remarks on Various Aspects of Philosophy of Law
}

 \xEh
 \xDH
A basic principle is equality, justice as fairness.

This, however, is a necessary, but not sufficient condition. Consider a
society where each year the first child born this year is sacrificed to
the gods. We will hardly consider this system as a decent legal system.

 \xDH
First, following Kant, we have to
separate aims and things as they are (Sein und Sollen).

This has led some people to reject a possible world semantics (here,
the set of ``good'' worlds) for
obligation (deontic logic). We do not share this, and think that
a set of possible worlds (or systems of such sets) has no fixed
interpretation.
It may describe
what we think possible, what we think good, etc. We have to be clear
about the meaning, but this is a different thing. We may even have
several such systems simultaneously, e.g. for different moral or legal
systems, they may be indexed, etc. In addition, we may look at coherence
properties between different such systems.

 \xDH
The next major distinction is between ``natural'' law and positive law.
The first has to do what we ``feel'' to be right (``Rechtsempfinden'' in
German
law), it certainly depends on the cultural context, and probably has its
roots in animal behaviour and feeling - e.g., animals seems to be able
to have a bad conscience. The cultural context might also be a history
of past aberrations, which we try to avoid in future.
It might be difficult to describe,
and we refer the reader to
Section \ref{Section NEU} (page \pageref{Section NEU}),
where we discuss the limits of
language in the context of neuroscience. Positive law is what is written
in legal texts, or established in the tradition of legal reasoning.

Conflicts between natural and positive law are a major subject of the
philosophy of law, see e.g. the
Radbruch formula:

``The conflict between justice and the reliability of the law should be
solved in
favour of the positive law,
law enacted by proper authority and power, even in cases where it is
injust in
terms of content and purpose,
except for cases where the discrepancy
between the positive law and justice
reaches a level so
unbearable that the statute has to make way for justice because it has to
be
considered''erroneous law".
It is impossible to draw a sharper line of demarcation between cases of
legal
injustice and statutes that
are applicable despite their erroneous content; however, another line of
demarcation can be drawn with
rigidity: Where justice is not even strived for, where equality, which is
the
core of justice, is
renounced in the process of legislation, there a statute is not just
'erroneous
law', in fact is
not of legal nature at all. That is because law, also positive law, cannot
be
defined otherwise as a rule, that is precisely intended to serve justice."

We see here that philosophy of law works (perhaps has to work) with
somewhat imprecise notions, in this case with a ``distance'' between
natural and positive law. Distance and size are fundamental notions
of non-monotonic reasoning and logics, so a connection between
philosophy of law and non-monotonic reasoning seems evident.
Recall here also the origins of formal theory revision in legal
thinking,
 \cite{AGM85}.

 \xEh

 \xDH Natural law, ethics, and morality

One should perhaps invest more thought to clarify above imprecise notions
of natural laws, ``Rechtsempfinden'', etc.

Our background will be one of restraint - laws should not try to
regulate too much.

Consider the law that anyone who insults the prophet (Allah) should be
killed. Not everyone believes in Allah, so this law cannot claim universal
validity. We have to find a general property which excludes such laws.
Perhaps a look at Rawls helps. To exclude cases like ``I feel extremely
extremely bad if not everyone (except myself) feels very bad'' to valuate
a political system, he excludes such artificial constructions. In our
example, Allah is for someone who is not Muslim as real as Snow White.
So, it is an artificial construction, and laws should abstain from
working with artificial constructions ``out of thin air''.

(The strength of convictions is a bad criterion whether to make
a principle a law, as religious convictions show.)

Conversely, any law based on ideas and emotions which all people share,
perhaps even some animals, might be considered well-founded.

Moreover, ethics is perhaps too burdened with absolute notions like
(absolute) good, evil, God, to be a good guide for the more pragmatic
law and its philosophy.

A more general comment: we should not work with incremental distances
(if $x$ is within the field of law, and $y$ is close to $x,$ then so is
$y),$
but always measure from the point of departure, to avoid excesses and
paradoxa.

 \xDH
The distinction between law and morality.
This does not seem to be the same as the distinction between
positive and natural law.

We may think some behaviour to be immoral, often in a sexual or religious
context, without feeling it is against
natural law - if we are tolerant enough to do so.

It may be good measure of tolerance and liberalism of a society, if
people can live with this distinction.

A few remarks on moral systems:
It might be necessary to differentiate the areas of moral judgements
and their relations to laws, consider e.g.:
 \xEh
 \xDH judgements about possessions, my house, my garden, my car  \Xl.
 \xDH cohesion of the family, like marital fidelity
 \xDH sexuality, like exhibitionism
 \xDH personal insult, other attacks on social status
 \xDH racism
 \xDH doubts or attacks of religious beliefs

etc.
 \xEj

 \xEj

 \xDH
The next distinction is between consequentialism and deontologism.

The first describes good or bad results of actions, the second good or
bad actions - as in the Ten Commandements.

The first may lead in excesses to ``ends justify means'', where a surgeon
is authorized to slaughter one patient to use his organs to save
several other patients.

The second sees some actions as intrinsically bad, condemning
also tyrannicide.

Again, as in the Radbruch formula, a compromise seems necessary, involving
again some (abstract) notions of size and distance.

Analogical reasoning in comparing cases has to be done carefully, so we
do not justify excesses.

 \xDH
Case based reasoning, i.e. using prototypes, will not escape above
distinction between the quality of actions and results. Its reasoning
is a special case of analogical reasoning, see
Chapter \ref{Chapter ANA} (page \pageref{Chapter ANA}).

 \xDH
Critical rationalism sees (positive) laws as experimental. Laws are made
to have a certain effect on society, but we cannot be totally sure about
this effect (society changes, the behaviour of judges is not totally
predictable), they might have to be revised, improved, etc.

 \xDH ``Dignity of men'' (Menschenwuerde)

This is a fundamental notion of the German constitution.
It has no clear definition, its meaning has changed over time.
In the sense of German constitution every human being has Menschenwuerde,
one cannot lose it, not even, Hitler, Stalin, etc. lost it through their
acts.

It seems absurd to base a constitution on an unclear notion, but this
also leaves space for development through interpretation - see also
Point \ref{Point Sexing} (page \pageref{Point Sexing})  in
Section \ref{Section NEU} (page \pageref{Section NEU})  for
nonverbal, implicit knowledge.

 \xDH
Equilibrium

Some legal systems seem to try to preserve a certain equilibrium,
if person A has done damage to person $B,$ then person A has to compensate
person $B.$ This idea is present in the Ur-Nammu code, the oldest known
law code,
ca. 2100 - 2050 BC, Sumer.

(Note that a positive disturbance of the equilibrium is not punished:
if $I$ give you 100.- (without any motive), then no one will punish me.)

 \xDH
Further remarks on consequentialism

In general, acting with bad consequences is considered worse than
not acting to prevent (the same kind of) bad consequences: pushing a
person
in a wheelchair over a cliff is worse than not preventing him to roll
over the cliff.

In the context of autonomous cars, there are many questions concerning
consequences. Sometimes, they seem far-fetched, not only because they
will seldom arise, but also because human beings might act in
unpredictable
ways, without incurring any legal punishment. Examples: shall the
driver rather kill an old man than a young child, when killing seems
inevitable. (Strangely, questions about safety of the computer and
communications systems against hacking, which are probably a bigger
problem, seem neglected. Likewise, the ``horizon'' of actions (consequences
within the next second are certainly important, those after one year
probably not) seems little considered.

Note also that comparing the ``value'' of the life of a human being with
other values is regularly done, also in civil life. Chosing the patient
to be given a life saving transplant, or an extremely costly or limited
other treatment, is based on criteria like age, chances of survival, etc.
It seems difficult to proceed otherwise (chosing randomly does NOT seem
better!). Building a motorway may always be done better and safer, but
we have to limit the costs somewhere.

 \xEj
\vspace{3mm}


\vspace{3mm}

\clearpage

$ \xCO $

$ \xCO $
\markboth{\centerline{\scriptsize Counterfactuals}}
{\centerline{\scriptsize Counterfactuals}}

$ \xCO $
\chapter{
Composing Logics and Counterfactual Conditionals
}

\label{Section CFC}

\label{Chapter CFC}

$ \xCO $
\section{
Introduction
}

 \xEh

 \xDH
We re-consider the Stalnaker/Lewis semantics for counterfactual
conditionals.
E.g., ``in situation $ \xbs,$ if $ \xbf $ were the case, then $ \xbq $
would be the case, too''.
The idea is to change $ \xbs $ minimally so that $ \xbf $ holds, and see
if then $ \xbq $
holds. E.g., the sun is shining, but if it were to rain, we would use
an umbrella. A non-minimal change might be to consider a place with very
strong winds, or we carry objects in both hands, and cannot hold an
umbrella, etc.

This somehow suggests that we may consider all possible situations and
choose
those diverging minimally from $ \xbs $ such that $ \xbf $ holds. But we
have no
table of all situations in our head, instead we have to construct suitable
situations (in the episodic memory), combining $ \xbs $ with past
experiences,
etc. Such constructions will be influenced by frequencies of experiences,
etc., so the result is not an objective look-up, but rather a subjective
construction and evaluation. The actual reasoning process is more
complicated
than the Stalnaker/Lewis semantics suggests.

 \xDH Basic entities and operations

The basic entities are scenarios or pictures. Scenarios
have a complicated structure, and no atoms,
see Section \ref{Section NEU} (page \pageref{Section NEU}).
Usually, we cannot use them as
they are, but have to cut them up, and combine them with other
basic scenarios (or their parts). The combination itself may be
a complicated process, resulting in a multi-neuron path between
two scenarios.
 \xDH Controle

The search for suitable scenarios is (usually) a complex active process,
the combination might lead to an error, and backtracking. Attention,
experience, desires etc. may be important guiding forces.

In a dream, we may combine the picture of a elefant with that of wings.
Controle rules this out.

There will usually not be a unique possible combination,
but not all are useful.

 \xEj
\section{
Composition of Logics
}

\label{Section Compos}

Consider e.g. a preferential logic which describes a situation, and now
we want the situation to evolve over time (as in  \cite{GR17}).

We may construct one, complicated, logic to cover all aspects, or,
decompose the overall picture into two or more logics, each covering some
aspect or aspects.

This is not a deep philosophical problem, but a practical one. We can
compare
the situation to a programming task: make one big main program, or
decompose the main program into a small main program and several functions
(or subroutines). The advantage of decomposition is that smaller parts
are easier to understand, check, and re-use in other contexts.

On the downside, there are several problems to solve:
 \xEh
 \xDH we have to determine how to cut the problem into smaller parts in a
useful way,
 \xDH we have to determine how the parts communicate with each other,
 \xDH is there a main logic, and a hierarchy of auxiliary logics, or are
the tasks (and thus the logics) more equally organised?
 \xDH will one logic wait for the answer of another logic before it can
continue its own task, as it needs the answer, or can it just start the
other logic (synchrone vs asynchrone cooperation)?

Suppose we have common variables, and the logics work in parallel.
Anything may happen, the behaviour is basically unpredictable. It is like
a program for seat reservation or bank transfers without momentary
locking the status. The actual behaviour depends on run-time properties
(speed
etc.).

 \xDH if one logic is not satisfied with the result of another logic, what
will it do? Can it re-start the second logic with different arguments,
will it start a third logic? Will one logic present several answers,
from which another logic may choose?
 \xDH if logics $L$ and $L' $ have the same syntactic operator, say $*,$
is the
meaning precisely the same in both logics?
 \xEj

These problems are usually not trivial, as anyone who ever did more
important
programming tasks can testify. We might call them problems of
``logic engineering''.

The first problem is influenced by the type of logics already existing, or
which seem relatively easy to construct.

The second problem depends on the necessary communication. An efficient,
but often not very clear solution is to use (in programming) common
variables, but one tends to forget which parts influence their values.
A usually better solution is to mention the communicating variable
explicitly, stating if communication is only in one direction, or in
both directions. In short, we have to find an efficient, but also safe,
interface between different logics. (The interface may also be dynamic:
in one situation, we have to use ``common variables'', in another
situation, modifying just a few ``variables'' might be sufficient.)

It does not seem that there will be many general regularities for such
logics, except trivial ones, e.g., if we have specificity, then, if
logic $L$ provides a more precise argument to logic $L',$ then $L' $ will
probably give a better answer.
\subsection{
Examples: Temporal Logic and Anankastic Conditionals
}

D. Gabbay, G. Rozenberg, and co-authors discuss in  \cite{GR17}
combinations of temporal logic with extension-based logics (here
argumentation theory) and counting of certain events.
It seems that using temporal logic as a ``master logic'',
and the extension based logic, or counting certain events, as
secondary logics are useful.

Further examples are (usual) counterfactual conditionals (see below)
and anankastic
conditionals, see e.g.  \cite{Sab14} for the latter. In the
Stalnaker/Lewis
semantics for counterfactual conditionals, the present and the
hypothetical
situation are described in the same logic (and language). But this need
not
be the case. If not, we need a ``higher'' logic which puts models from
different logics together, and chooses the closest ones according to some
criterion. In anankastic conditionals, example: ``If you want to go London,
you have to take the Eurostar train.'', we have an initial situation $S,$
a desired final situation $S',$ and a means to go from $S$ to $S',$ $M,$
describing an action.
Here, the initial situation is, implicitly, some place in western Europe.
This will be described in some logic $L.$ The destination will be
described
in some, perhaps different, logic $L'.$ The part describing the means is
probably the most complicated one. Of course, one might first fly to
Alaska,
and then to London, the suggested choice is supposed to be the simplest,
cheapest, etc. But even if we are in Paris at the start, the Eurostar
might
not always be the best solution. If we live close to Charles de Gaulle
airport,
and need to go close to Heathrow airport, flying might well be more
convenient.
So, the choice of the means or action is probably the most complicated
part of
the reasoning.
\section{
Human Reasoning and Counterfactual Conditionals
}

We started our investigation by looking into the semantics of
counterfactual conditionals (CFC), and contrasting this with human
reasoning:
we certainly have no list of all possible worlds in our brain, from which
we choose the closest. Our reasoning is more flexible and
constructive.

Here, composing situations might be the master logic, and remembering
situations an auxiliary logic. The probably fundamental difference between
logic
and working of the brain will mostly be handled in the auxiliary logic.
Before we take a more general look at the human brain and its functioning,
we discuss very shortly the human memory, as it presents already many
problems.
(Note that, first, we are no neuroscientists, and, second, even
neuroscientists
have many conjectures, but much less established facts. This is, of
course,
due to the complexity of the subject.)

Human reasoning with counterfactual conditionals is much less regular
than the formal approach in philosophical logic. It is not a procedure
of simple choice by distance, but an active construction, a dialogue
between different requirements. We might see it as a puzzle, where
in addition, the tiles have to be cut to fit together.
\subsection{
Introduction
}

We discuss now our ideas how human beings think: not only in propositions
and
logical operators, or in
models, but also in pictures, scenarios, prototypes, etc. On the neural
level,
such pictures correspond to groups of neurons.

We will be (necessarily) vague, on the meaning level, as well as the
neural
level. This vagueness results in flexibility,
the price to pay are conceptual difficulties.

We have three types of objects:
 \xEh
 \xDH pictures or groups (of neurons),
 \xDH connections,
 \xDH attention.
 \xEj

Groups will be connected areas of the brain, corresponding to some
picture,
groups may connect to other groups with different types of connections,
which
may be positive or negative. Finally, attention
focusses on groups or parts of groups, and their connections, or part
thereof.

Groups do not necessarily correspond to nodes of graphs, as they are not
atomic, and they can combine to new groups. Attention may hide
contradictions,
so the whole picture may contain contradictions.

We use the word ``group'' to designate
 \xEI
 \xDH
on the physiological level a (perhaps only momentarily) somehow connected
area of the brain, they may be formed and dissolved dynamically,
 \xDH
on the meaning level a picture, scene (in the sense of conscious scene),
a prototype (without all the connotations the word ``prototype'' might
have),
any fragment of information. It need not be complete with all important
properties, birds which fly, etc., it might be a robin sitting on a branch
in sunshine,
just any bit of information, abstract, concrete, mixture of both,
whatever.
 \xEJ
\subsection{
Human Reasoning and Brain Structure
}

\label{Section HumanReason}
\subsubsection{
Pre-semantics and Semantics
}

In logic, a sentence like ``it rains'', or ``if it were to rain, $ \xfI $
would
take an umbrella'' has a semantics, which describes a corresponding state
in
the world.

We describe here what happens in our brain - according to our hypothesis -
and what corresponds to this ``brain state'' in the real world.
Thus, what we do is to describe an intermediate step between
an expression in the language, and the semantics in the world.

For this reason, we call this intermediate step a pre-semantics.

In other words,
real semantics interpret language and logic in (an abstraction of) the
world. Pre-semantics is an abstraction of (the functioning of) the brain.
Of course, the brain is ``somehow'' connected to the world, but this
would then be a semantics of (the functioning of) the brain.
Thus, this pre-semantics is an intermediate step between language and the
world.
\subsubsection{
The Elements of Human Reasoning
}

In the following, we concentrate on episodes, pictures, etc., as they seem
to be
the natural structure to consider counterfactual conditionals. They also
illustrate well the
deep differences with the way formal logics work.
For more details, see Section 
\ref{Section NEU} (page 
\pageref{Section NEU})  below.

It seems that humans (and probably other animals) often think in
episodes, scenarios, prototypes, pictures, etc.,
which are connected by association, reasoning, developments, etc.
We do not seem to think only in propositions, models, properties, with the
help of logical operators, etc.

For simplicity, we call all above episodes, scenarios or pictures
``pictures''. Pictures can be complex, represent developments
over time, can be combined, analysed, etc., they need not be precise,
may be inconsistent, etc. We try to explain this, looking simultanously
at the thoughts, ``meanings'' of the pictures,
and the underlying neural structures and processes.

We imagine these scenarios etc. to be realised on the neural level by
neurons or groups of neurons, and the connections between pictures also by
neurons, or bundles of neurons.

To summarize, we have

 \xEh
 \xDH Pictures, on the
 \xEI
 \xDH
meaning level, they correspond to thoughts, scenarios, pictures, etc.
 \xDH
neural level, they correspond to neurons, clusters of connected neurons,
etc.
 \xEJ
 \xDH Connections or paths, on the
 \xEI
 \xDH
meaning level, they correspond to associations, deductions, developments,
coherences within an episode, etc.
 \xDH
neural level, they correspond to neurons, bundles of more or less
parallel neurons, connecting groups of neurons, etc.
 \xEJ
 \xEj

All that follows is relative to this assumption about human reasoning.
\paragraph{
Pictures: the Language and the Right Level of Abstraction
}

 \xEh
 \xDH
The Level of Meaning

A picture can be complex, we can see it as composed of sub-pictures.
A raven eats a piece of cheese, so there are a raven, a piece of cheese,
perhaps some other objects in this picture. The raven has a beak, etc.

It is not clear where the ``atomic'' components are. On the neural level, we
have
single neurons, but they might not have meaning any more.
In addition, they will usually not be accessible to conscious reasoning.

To solve this problem with atoms, we postulate that there are no atomic
pictures, we can
always decompose and analyse. For our purposes, this seems the best
way out of the dilemma.
In one context, the raven eating the cheese is the right level of
abstraction,
we might be interested in the behaviour of the raven. In a different
context, it might be the feathers, the beak of the animal, or the taste
of the cheese. Thus, there is no uniform adequate level of abstraction, it
depends on the context.

(The ``meaning'' of a neuron, i.e. the conditions under which it
fires, might be quite complicated. This is true even for relatively
low-level
neurons, i.e. close to sensory input: Visual information originating in
the
retina travels through the lateral geniculate nucleus in the thalamus to
the
visual cortex, first to V1. V1 itself is decomposed into 6 layers. Even
cells in V1 receive feedback from higher-level areas like V4, which
cover bigger and more complex receptive fields than those covered
(directly)
by V1 cells. This feedback can modify and shape responses of V1 cells.
Thus,
we may imagine such V1 cells to ``say'' something like: ``I see an edge in my
part of the retina, but context (sent by higher-level cells) thinks it is
unlikely, still  \Xl., etc.'' So, such cells may express rather complex
situations.
See  \cite{Wik17e},  \cite{Geg11}, etc.

The organisation of V1 in ``hypercolumns''
going through the layers of V1
is very interesting, see the work by D. H. Hubel and
T. Wiesel for which they were awarded a Nobel prize. (Roughly) edge
detecting
cells for one ``spot'' of the retina are grouped together, and moreoever,
the direction of the edge which is detected changes continually. Thus,
neighbouring ``spots'' have neighbouring groups of edge detectors, and, say,
0 degree detector sits close to 30 degree detector, 60 degree detector is
farther away from the 0 degree detector, etc., in a cyclic way
(``pinwheel'').
This reminds of Hamming distances, where ``spots'' and degrees are the
dimensions of the distance, and raises the question if other parts of
the brain are organised similarly, with conceptually close information
coded in neighbouring cells. The importance of neighbourhoods is seen,
e.g.,
in the common-neighbor-rule (CNR), according to which a pair of neurons
is more likely to be connected the more common neighbors it shares, see
e.g.
 \cite{AIZ16}.)

 \xDH
Relation to Models

Pictures will usually not be any models in the logical sense.
There need not be any language defined, some parts may be complex and
elaborate,
some parts may be vague or uncertain, or only rough sketches, different
qualities like visual, tactile, may be combined. Pictures may also be
inconsistent.

 \xDH
The Neural Level

On the neural level, pictures will usually be realised by groups of
neurons,
consisting perhaps of several thousands neurons. Those groups will have an
internal
coherence, e.g. by strong internal positive links among their neurons.
But they will not necessarily have a ``surface'' like a cell wall
to which other cells or viruses may attach. The links between groups
go (basically) from
all neurons of group 1 to all neurons of group 2. There is no exterior vs.
interior, things are more flexible.

If group 1 ``sees'' (i.e. is positively connected to) all neurons of group
2,
then group 2 is the right level of abstraction relative to group 1 (and
its
meaning). (In our example, group 1 looks at the behaviour of the raven.)
If
group 1a ``sees'' only a subgroup of group 2 (e.g. the feathers of the
raven),
either by having particularly strong connections to this subgroup, or
by having negative connections to the rest of group 2, then
this subgroup is the right level of abstraction relative to group 1a.
Thus, the ``right'' level of abstraction is nothing mysterious, and does
not depend on our speaking about the pictures, but is given by the
activities of the neuron groups themselves.

(We neglect that changing groups of neurons may represent
the same pictures.)

 \xDH
The Conceptual Difficulties of this Idea

This description has certain conceptual difficulties.

 \xEh
 \xDH In logic, we have atoms (like propositional variables), from which
we
construct complex propositions with the use of operators like $ \xcu,$ $
\xco,$ etc.
Here, our description is ``bottomless'', we have no atoms, and can
always look inside.
 \xDH There is no unique adequate level of abstraction to think about
pictures. The right granularity depends on the context, it is
dynamic.
 \xDH The right level of abstraction on the neural level is not given by
our
thoughts about
pictures, but by the neural system itself. Other groups determine the
right granularity.
 \xDH Groups of neurons have no surface like a cell or a virus do, there
is
no surface from which connections arise.
Connections go from everywhere.
 \xEj

 \xDH
Summary

 \xEh
 \xDH
``Pictures'' on the meaning level correspond to (coherent) groups of
neurons.
 \xDH
There are no minimal or atomic pictures and groups, they can always
be decomposed (for our purposes). Single neurons might not have
any meaning any more.
 \xDH
Conversely, they can be composed to more complex pictures and groups.
 \xDH
Groups of neurons have no surface, connections to other groups are
from the interior.

 \xEj

 \xEj
\subsubsection{
Connections or Paths Between Neuron Groups
}

We sometimes call connections paths.

Connections may correspond to many different things on the meaning level.
They may be:
 \xEh
 \xDH arbitrary associations, e.g. of things which happened at the same
moment,
 \xDH inferences, classical or others,
 \xDH connections between related objects or properties, like between
people
and their ancestors, animals of the same kind, etc.,
 \xDH developments over time, etc.
 \xEj

Again, there are some conceptual problems involved.

 \xEh
 \xDH As for pictures, it seems often (but perhaps less dramatically)
difficult to
find atomic connections. If group $N_{1}$ is connected via path $P$ to
group
$N_{2},$ but $N'_{1}$ is a sub-group of $N_{1},$ connected via a subset
$P' $ of $P$ to sub-group
$N'_{2}$ of $N_{2},$ then it may be reasonable to consider $P' $ as a
proper path itself.
 \xDH If, e.g., the picture describes a development over time, with
single pictures at time $t,$ $t',$ etc. linked via paths expressing
developments,
then we have paths
inside the picture, and the picture itself may be considered a path
from beginning to end.

Thus, paths may be between pictures, or internal to pictures, and there is
no
fundamental distinction between paths and pictures.
It depends on the context.
More abstractly, the
whole path is a picture, in more detail, we have paths between single
pictures, ``frames'', as in a movie.

 \xEj

Remember: Everything is just suitably connected neurons!
\subsubsection{
Operations on Pictures and Connections
}

To simplify, we will pretend that operations are composed of
cutting and composing. We are aware that this is probably
artificial, and, more generally, an operation takes one or more
pictures (on the meaning level) or groups (on the neural level),
and constructs one or more new pictures or groups.

Before we describe our ideas, we discuss attention.
\paragraph{
Attention
}

$ \xDB $

An additional ingredient is ``attention''.
We picture attention as a light which shines on some areas of the brain,
groups of neurons, perhaps only
on parts of those areas, and their connections, or only parts of the
connections.
Attention allows, among other things, to construct a seemingly coherent
picture
by focussing only on parts of the picture, which are coherent.
In particular, we might focus our attention on coherences, e.g., when we
want to consolidate a theory, or incoherences, when we want to attack a
theory. Focussing on coherences might hide serious flaws in a theory,
or our thinking in general. In context $ \xCf A,$ we might focus on $ \xba
,$ in context
$B,$ on $ \xbb,$ etc.
As we leave attention deliberately unregulated, changes in attention may
have very ``wild'' consequences.

Activation means that the paths leading to the picture become more active,
as well as the internal paths of the picture.
Thus, whereas memory (recent use) automatically increases activity,
attention
is an active process.
Conversely, pictures which are easily accessible (active paths
going there), are more in the focus of our attention.
E.g., we are hungry, think of a steak (associative memory), and
focus our attention on the fridge where the steak is.

Attention originates in the ``$ \xfI $'' and its aims and desires.
Likewise, ``accessibility'' is relative to the ``$ \xfI $'' - whatever
that means.
(This is probably a very simplistic picture, but suffices here.
We conjecture that the ``$ \xfI $'' is an artifact, a dynamic
construction,
with no clear definition and boundary. The ``$ \xfI $'' might be just as
elusive
as atomic pictures.)
Attention is related to our aims (find food, avoid dangers, etc.)
and allows to focus on certain pictures (or parts of pictures) and
paths.
\paragraph{
Operations
}

$ \xDB $

Consider again the picture of a raven eating a piece of cheese.

We might focus our attention on the raven, and neglect the cheese. It is
just
a raven, eating something, or not. So the connection to the raven part
will
be stronger (positive), to the cheese part weaker positive and/or stronger
negative.
Conversely, we might never have seen a raven eat a piece of cheese.
But we can imagine a raven, also a raven eating something, and a piece
of cheese, and can put these pictures together. This may be more or less
refined, adjusting the way the cheese lies on the ground, the raven
pecks at it, etc. It is not guaranteed that the picture is consistent,
and we might also adjust the picture ``on the fly'' to make it consistent
or plausible.
(When composing ``raven'' with ``cheese'' and ``pecking'', the order might be
important: Composing ``raven'' first with ``pecking'' and the result
``raven $+$ pecking'' with ``cheese'', or ``raven'' with ``pecking $+$ cheese''
might a priori give different results.)

It is easy to compose a picture of an elephant with the picture of wings,
and to imagine an elephant with small wings which hovers above the ground.
Of course, we know that this is impossible under normal circumstances.
There is no reality check in dreams, and a flying elephant is quite
plausible.

This is all quite simple (in abstract terms), and everyone has done it.
Details need to be filled in by experimental psychologists and
neuroscientists.
Obviously, these problems are related to planning.
\subparagraph{
Remarks on Composition
}

$ \xDB $

This might be a good place to elaborate our remarks of
Section 
\ref{Section Compos} (page 
\pageref{Section Compos})  in the context of the functioning of the
brain.

The working of the brain has often been described as ``experts talking
to each other'', where the experts are different areas of the brain.
So, we have a modular structure, and communication between the
different modules. Why this structure? Is this only through evolution,
which has added more structures to the brain, or can we see a different
reason
for this modularity? We think so.

Consider human experts discussing a situation, e.g. medical experts,
a cardiologist, and a specialist for infectious disease, discussing
a patient suffering from an unknown disease. They will discuss with
each other every idea they have, but first try to come to some possible
diagnoses, and then discuss the result of their thoughts. If one talks
too early, the other might interrupt him: ``let me think  \Xl.''.
Too much communication may disturb reflection.
This is due to the fact that attention (here triggered by communication)
might disturb a reflection process. More complicated reflection might
involve
considering rare situations, whose ``signal'' is weak, and this weak signal
may easily be drowned by ``loud'' signals, e.g. from communication.
Converse processes are concentration, attention, etc.

We have here an auxiliary process: active search for arguments, situations
to consider. Both the reasoning itself, and the auxiliary process may be
disturbed by ``ouside noise''. Thus, temporary isolation of reasoning
modules may be helpful. The price to pay for the flexibility of the brain,
shifting attention, various influences, is a necessary control over our
reasoning processes.
It seems complicated to cover all these aspects in a single logic,
especially
as many aspects will be dynamic, and there will be different processes
running in parallel, with differing attentions, etc.
Little seems to be known about control processes in human reasoning.
A very interesting aspect is discussed in  \cite{WSFR02}. It is
argued there that processing fluency is hedonically marked. Fluid
processing elicits positive affective responses, visible in increased
activity of the ``smiling muscle''. This would be a very high level control
mechanism.
It seems reasonable to ask if this
is related to the fact that e.g. physicists emphasize the beauty of
``right''
theories, and consider the esthetic quality of a theory as an indication
of its correctness. (The theory ``flows''.)
If we want to model human reasoning, we will have to model such auxiliary
processes, attention, etc., too.
\subparagraph{
Comparison to Operations in Logic
}

$ \xDB $

Consider an implication $A \xcp B$ in logic. To apply it, we have to look
for
$ \xCf A.$ Consider a picture $ \xba \xcp \xbb,$ which allows to be
completed with $ \xba.$
But often, this will not be that precise. We can perhaps complete
with some $ \xba',$ as there is no unique place to ``dock'' something.
Perhaps
we can complete with $ \xba,$ $ \xba',$ $ \xba'',$ etc., recall also
that there is no
surface to present all the docking possibilities, and no atoms in a useful
sense. Suppose we have
a picture $ \xba \xcp \xbb \xcp \xbg $ describing some 3 stage
development. Perhaps
a picture $ \xbd $ may dock to $ \xba,$ but later, we see that it does
not fit
with $ \xbb.$ So, we have local coherence, but not global coherence.
A simplified analogon is perhaps putting together a puzzle.

We conjecture that we have active search for fitting pictures,
composition itself, and then evaluation, i.e. checking for global fit.
This seems quite complicated and it seems difficult to find
general principles, analogue to logical properties, which govern
such processes. (Language allows a different kind of flexibilty, based
on categories of words, irrespective of their meaning.)
\subsection{
Counterfactual Conditionals
}
\subsubsection{
The Stalnaker/Lewis Semantics For Counterfactual Conditionals
}

Stalnaker and Lewis, see e.g.
 \cite{Sta68},  \cite{Lew73},
gave a very elegant semantics to counterfactual conditionals, based on
minimal change.
To give meaning to the sentence ``if it were to rain, $ \xfI $ would take
an
umbrella'', we look at all situations (models) where it rains, and
which are minimally different from the present situation. If $ \xfI $ take
an
umbrella in all those situations, then the sentence is true.
E.g., situations where there is hurricane - and $ \xfI $ will therefore
not take an
umbrella - will, usually, be very different from the present situation.

This idea is very nice, but we do not think this way. First, we have no
catalogue of all possible worlds in out head. Thus, we will have to
compose the situations to consider from various fragments. Second,
classical reasoning, taken for granted in usual semantics, has an
``inference cost''. E.g., when reasoning about birds, we might know that
penguins are birds, but they might be too ``far fetched'', and forgotten.

It seems that human beings reason in pictures, scenes, perhaps prototypes,
but in relatively vague terms. We try to use a more plausible
model of this reasoning, based on neural systems, to explain
counterfactual conditionals. But the basic Stalnaker/Lewis idea is upheld.

Our ideas, developped
in Section \ref{Section HumanReason} (page \pageref{Section HumanReason})
are very rudimentary, all details are left open. Still, we think
that it is a reasonable start.
We might be overly flexible in our concepts, but it is probably easier
to become more rigid later, than inversely.

We did not discuss how the various choices are made between different
possibilities. On the one side, an overly rigid attention or memory
might prevent flexibility, on the other side, too much flexibility might
result in chaos and not enough focus. The brain ``needs to roam'',
but with a purpose.

The Stalnaker/Lewis semantics is a passive procedure, there is a list of
models, and we chose the ``best'' or ``closest'' with a certain property.
Distance is supposed to be given.
Our pre-semantics is much more active, we construct, disassemble,
chose using several criteria. Thus, it is not surprising that control
of the procedure is complex (and not discussed here).
\subsubsection{
The Umbrella Scenario
}

We apply our ideas to counterfactuals.

Note that the Stalnaker/Lewis semantics hides all problems
in the adequate notion of distance, so we should not expect miracles
from our approach.

``If it were to rain, $ \xfI $ would take an umbrella.''

We have the following present situations, where the sentence is
uttered.

 \xEh
 \xDH
Case 1: The ``normal'' case. No strong wind, $ \xfI $ have at least one hand
free to
hold an umbrella, $ \xfI $ do not want to get wet, etc.
 \xDH
Case 2: As case 1, but strong wind.
 \xDH
Case 3: As case 1, but $ \xfI $ carry things, and cannot hold an umbrella.
 \xEj

We have the following pictures in our memory:
 \xEh
 \xDH
Picture 1: Normal weather, it rains, we have our hands free, but forgot
the umbrella, and get soaked.
 \xDH
Picture 2: As picture 1, but have umbrella, stay dry.
 \xDH
Picture 3: Rain, strong wind, use umbrella, umbrella is torn.
 \xDH
Picture 4: As picture 1, but carry things, cannot hold umbrella, get
soaked.
 \xDH
Picture 5: The raven eating a piece of cheese.
 \xEj

Much background knowledge goes into our treatment of counterfactuals.
For instance, that a strong wind might destroy an umbrella (and that
the destruction of an umbrella in picture 3 is not due to some
irrelevant aspect), that we need
at least one hand free to hold an umbrella, that we want to stay dry,
that we cannot change the weather, etc.

First, we actively (using attention) look for pictures which have
something to
do with
umbrellas. Thus, in all cases, picture 5 is excluded.

Next, we look at pictures which support using an umbrella, and those which
argue against this. This seems an enormous amount of work, but our
experience tells us that a small number of scenarios usually give
the answer. There are already strong links to those scenarios.

Case 1: Pictures 3 and 4 do not apply - they are too distant in the
Stalnaker/Lewis terminology.
So we are left with pictures 1 and 2. As we want to stay dry, we choose
picture 2.
Now, we combine case 1 with picture 2 by suitable connections, and
``see'' the imagined picture where we use an umbrella and stay dry.

Case 2: Pictures 1 and 4 do not apply. $ \xfI $ would prefer to stay dry,
but
a torn umbrella does not help. In addition, $ \xfI $ do not want my
umbrella
to be torn. Combining case 2 with picture 3 shows that the umbrella
is useless, so $ \xfI $ do not take the umbrella.

Case 3: Only picture 4 fits, $ \xfI $ combine and see that $ \xfI $ will
get wet,
but there is nothing $ \xfI $ can do.
\subsubsection{
A Tree Felling Scenario
}

Consider the sentence:

``If $ \xfI $ want to fell that tree, $ \xfI $ would hammer a pole into the
ground,
and tie a rope between tree and pole, so the tree cannot fall on the
house.''

(This is an anankastic conditional, see e.g.
 \cite{Sab14}, but the lingistic problems need not bother us.
By the way, the following statement: ``If you want to jump to the moon, you
should wait for a clear night with full moon, so you do not miss it.''
might be
fun to look at.)

We have the present situation where the tree stands close to the
house, there are neither rope nor pole, nor another solid tree where we
could
anchor the rope, and we do not want to fell the tree.

We have

 \xEI
 \xDH Picture 1 of a pole being hammered into the ground - for instance,
we remember this from camping holidays.

 \xDH Picture 2 of a rope tied to a tree and its effect - for instance,
we once fastened a hammock between two young trees and saw the effect,
bending the trees over.

 \xDH Picture 3 of someone pulling with a rope on a big tree - it did not
move.
 \xEJ

We understand that we need a sufficiently strong force to prevent the
tree from falling on the house.

Pictures 2 and 3 tell us that a person pulling on the tree, or a rope
tied to a small tree will not be sufficient.

As there is no other sturdy tree around, we have to build a complex
picture. We have to cut up the hammock Picture 2 and the tent Picture 1,
using the rope part from Picture 2, the pole part from Picture 1.
It is important that the pictures are not atomic.
Note that we can first compose the situation with the pole part, and the
result with the rope part, or first the situation with rope part, and the
result
with the pole part,
or first combine the pole part with the
rope part, and then with the situation.
It is NOT guaranteed that the outcome of the different ways will be the
same.
When there are more pictures to consider, even the choice of the
pictures might depend on the sequence.
\subsection{
Comments
}

There are many aspects we did not treat.
We established a framework only.

 \xEh
 \xDH Usually, there are many pictures to choose. How do we make the
choice?
 \xDH How do we cut pictures?
 \xDH How do we determine if a combined picture is useful?
 \xDH Attention can hide inconsistencies, or focus on inconsistencies,
how do we decide?
 \xDH Are all these processes on one level, or is it an interplay
between different levels (execution and control)?
 \xDH These processes seem arbitrary, but we are quite successful, so
there
must be a robust procedure to find answers.
 \xEj

Some of the answers will lie in the interplay between (active) attention
and more passive memory (more recent and more frequently used pictures and
processes are easier accessible).
Recall here Edelman's insight, see e.g.
 \cite{Ede89},  \cite{Ede04}, that there are
parallels between the brain and the immune
system, both working with selection from many possibilities.
We assume that we
have many candidates of the same type, so we have a population from
which to chose. We chose the best, and consider this set
for the properties of those combined areas.

It is natural to combine the ideas of the hierarchy in
 \cite{GS16}, chapter 11 there,
with our present ideas. Exceptional classes, like penguins,
are only loosely bound to regular classes, like birds; surprise
cases even more loosely.
\section{
More Formal Remarks
}

We do not have elementary propositions, nor operators like $ \xcu $ etc.
Instead, we have (groups of) neurons, and connections between them.
The possibility of ``pruning'', see below, captures the fact that
pictures/situations are not elementary. Conversely, it is possible
to combine pictures.

Attention (on the level of active search, pruning, combining)
considers choice and operations on situations. On the level
of evaluation, attention considers the result of the operations.

We give now a simple abstraction of networks of neurons.
The basics are common knowledge, see e.g.
 \cite{GLP17}.

\bd

$\hspace{0.01em}$


\label{Definition Neuron-Body}

The body of a neuron is a counter, which counts positive (excitatory)
inputs
into the neuron, subtracts negative (inhibitory) inputs, if the result
is above a certain (individual for this neuron, but static value)
threshold,
the neuron become active (fires), otherwise, it stays dormant.

Consequently, we have a simple property:
If a neuron is active, increasing the positive input and/or decreasing
the negative input will keep it active. Conversely, if a neuron
is dormant, decreasing the positive input and/or increasing
the negative input will keep it dormant.

\ed

\bd

$\hspace{0.01em}$


\label{Definition Arrows-Between-Neurons}

Neuron bodies can be connected via arrows. Each arrow has an integer
value, which may change over time. For simplicity, we assume that there
is only one arrow between two neurons (per direction). (If there are
several,
we code this by modifying the value.)

\ed

Arrows are dynamic over time:

 \xEh
 \xDH
New arrows may be created, and arrows may disappear.
 \xDH
The values of arrows may change over time.
 \xEj

In particular:
 \xEh

 \xDH Recent use:
Use of a positive connection between neuron $ \xCf a$ and neuron $b$ may
increase the value of the arrow $a \xcp b,$ lack of use may decrease the
value. This results in stronger bonds for frequent situations, but
also a certain rigidity.

 \xDH Hebb's rule, see Definition 
\ref{Definition Hebb} (page 
\pageref{Definition Hebb}):
If neurons $ \xCf a$ and $b$ are activated at the same time, e.g. by
positive
arrows $c \xcp a$ and $c \xcp b,$ then the connection between $ \xCf a$
and $b$ is
strengthened (i.e. arrows $a \xcp b$ and $b \xcp a).$

 \xDH Attention:
Attention is a more active way of modifying arrows, but like past
use, it may modify the values of arrows in both directions (and also
create new ones).
See e.g.  \cite{Auf17}.
(Despite the word ``attention'', we do not think that consciousness is
a necessary condition for attention. Even simple animals need to focus,
to escape from predators over search for food, etc.)

 \xEj

More aspects:
 \xEh
 \xDH Active search:
Attention can focus on some aspects, e.g. rain, umbrella in our example,
of a picture, and search for other pictures with the rain/umbrella
element.
A positive connection is made to such situations.

 \xDH Pruning:
Attention may neglect some aspects of a situation, and make e.g. negative
connections to those aspects.

 \xDH Combining:
Attention may combine two or more situations by making positive
connections between them, making them components of a more
complex situation.

 \xDH Evaluation:
The result of these operations may be evaluated again.
The criteria will be (in an incomplete list)
 \xEI
 \xDH are important parts of the situation we found neglected (e.g.,
there was a strong wind, which destroyed the umbrella)?
 \xDH is the constructed situation sufficiently coherent?
 \xDH does it seem necessary to search for competing situations?
 \xEJ
Evaluation may lead to backtracking, new attempts, etc.

 \xEj
\section{
Appendix - Some Remarks on Neurophilosophy
}

\label{Section NEU}
\subsection{
Introduction
}

We give here a very short summary of aspects of human reasoning which are
important in our context. Human reasoning and the functioning of the brain
are extremely complex, and largely still unknown, we only indicate
roughly some aspects.
(There is a vast literature, and we just mention some we looked into:
 \cite{Sta17d},  \cite{Rot96},  \cite{Chu89},
 \cite{Chu86},  \cite{Chu07},  \cite{Wik17a},
 \cite{Wik17b},  \cite{Wik17c},  \cite{Wik17d},
 \cite{CCOM08},  \cite{HM17},  \cite{KPP07},
 \cite{ZMM15},  \cite{Wik17e},  \cite{Geg11},
 \cite{AIZ16},  \cite{Pul13}.)

We think some - even rudimentary understanding - is important in our
context
for the following reason: The more a logic might seem close
to human reasoning, the more differences to human reasoning might be
important. For instance, the Stalnaker/Lewis semantics for counterfactual
conditionals (CFC's),
see  \cite{Sta68},  \cite{Lew73},
is intuitively very attractive, so we might be tempted to
see it as describing actual human reasoning. This may have serious
consequences. In court, a defendant might say ``if $ \xfI $ had done  \Xl.
then  \Xl.''
in good faith, following his own reasoning,
but a judge familiar with the theory of formal
CFC's might come to a different conclusion and accuse
the defendant of lying.
See e.g.  \cite{Wik16a} and  \cite{IEP16} for
different legal systems.
We will come back to CFC's below.

Just as classical logic is not a description of actual human
reasoning, concepts of philosophical logic need not correspond directly
to the way we think.

Thus, our remarks are also a warning against hasty conclusions.
But, of course, we may speculate on the (neural) naturalness of concepts
like ``distance'', which have an analogon in the brain, the strength of
the connection
between areas of the brain, or between groups of neurons.

The author is absolutely no expert on neuroscience. In addition, it seems
that
recent research has concentrated on the structure and function of single
or
small numbers of nerve cells, and somewhat neglected the overall picture
of how our brain works.
Thus, there does not seem to exist an abstract summary of present
knowledge
about human reasoning on the neural level.
Perhaps, one should more dare to be wrong, but
incite criticism, and thus advance our knowledge?
\subsection{
Details
}
\subsubsection{
Basics
}

The basic unity of a nervous system are neurons, consisting
of dendrites, core, and axon. Signals travel from the axon of neuron $
\xCf A$ via a
synapse, the connection, to a dendrite of neuron $B,$ etc.
Usually, a neuron has one axon and several dendrites. The axon of cell
$ \xCf A$ may be connected via synapses to the dendrites of several
neurons $B,$ $B',$
etc.

A synapse may be positive or negative, excitatory or inhibitory. Suppose
neuron $B$ is in exited level $b,$ and it receives a signal from neuron $
\xCf A$ via
an excitatory synapse, then $B$ goes to level $b' >b;$ if the synapse is
inhibitory, it goes to $b' <b.$ As $B$ might receive several signals,
(very
roughly) the sum of incoming signals, positive or negative, determines
$b' $-b.

More precisely, if, within a
certain time interval, the sum of positive signals $ \xbS^{+},$ i.e. from
positive
synapses, arriving at the dendrites of a given neuron
is sufficiently bigger than the sum of negative signals $ \xbS^{-},$ i.e.
from negative synapses,
arriving at the dendrites of the same neuron, the neuron is activated and
will
fire, i.e. send
a signal via its axon to other neurons. This is a 0/1 reaction, it will
fire or
not, and always with the same strength. (If $ \xbS^{+}$ is much bigger
than $ \xbS^{-},$ the
neuron may fire with a higher frequency. We neglect this here.)

Note that negative synapses are ``related'' to negation, but are not
negation
in the usual sense (nor negative arrows in defeasible inheritance
systems).
They rather express (roughly) ``contradict each other'' like ``black'' and
``yellow'' do.

An extremely important fact is Hebb's rule
(see  \cite{Heb49}):

\bd

$\hspace{0.01em}$


\label{Definition Hebb}

When neurons $ \xCf A$ and $B$ are
simultaneously activated, and are connected via some synapse, say from $
\xCf A$ to $B,$
then
the connection is strenthened, i.e. the weight of the synapse is
increased, and thus the future influence of $ \xCf A$ on $B$ is increased.
This property is also expressed by: ``fire together, wire together''.
The dynamic history is thus remembered as an association between neurons
(or groups of neurons).
\subsubsection{
The ``Meaning'' and Dynamics of the Activity of a Neuron
}

\ed

If a photoreceptor cell in the eye is excited (by light or pressure,
etc.),
it will always send the message ``light''. Hair cells in the ear detect
sounds,
they send the message ``sound''. These determine the different qualia, light
and
sound.

In general, things are not so simple. First, activities of the brain
usually
involve many, perhaps thounds of neurons, which work together as a
(strongly interconnected) group
of neurons. Second, this activity may involve in moment $ \xCf a$ a neuron
group $ \xCf A,$
in moment $b$ neuron group $B,$ etc. (A good example is a cloud which
``sits''
seemingly stationary on top of a mountain in a strong wind. As a matter of
fact, single water vapor molecules are pushed upwards over the mountain,
they condense as they cool down, they reflect light, and become visible.
When
they descend again, they warm, evaporate, and ``disappear'' on the other
side
of the mountain. New water molecules follow, so the overall picture is
static, the components which create the picture change all the time.)
This also results in the extreme flexibility of the brain. Usually, the
death of one neuron may be compensated by another neuron.

Thus, in general, the ``meaning'' of an active (i.e. firing or close to
firing)
neuron or a group of active neurons
is defined by the context within the present active network of the brain.
This is called the functional role semantics of neurons or groups
of neurons. The state of the brain (activities, strengths of synapses,
etc.)
may be seen as an extremely complex vector, and its transformation
from one state to the other as a vector transformation. Thus, the
mathematics of dynamical systems seems a promising approach to
brain activity.

Consider a picture or scene coded (at present, and static for simplicity)
by the activity of some group of neurons. E.g., we look through the window
an the garden. Then:

 \xEh

 \xDH
There are no atoms, we can always analyze parts of the picture even
further. Of course, we have neurons as ``atoms'' on the neural level, but
they have lost their meaning, which exists only in the context.
Thus, there are no atoms of thought.

 \xDH
The group of neurons has no ``surface''. One group of neurons is not
``seen'' by another group of neurons with presenting some opaque surface,
but, usually, connections between the two groups go as well between the
``surfaces'' - how ever they may be defined - as between the interior
neurons.

There is no ``right'' level of abstraction or granularity to consider this
picture - it depends on the connection to other groups of
neurons. Fix a neuron group $G.$
One (other) group of neurons may be connected to, ``sees'' all neurons of
$G,$
another group of neurons may be connected to, ``sees'' only a
subset of $G.$

Suppose we see a raven eating a bit of cheese in the garden. If we are
interested in ravens, we will focus our attention on different aspects
of the scenario, than if we want to buy cheese, and this reminds us not to
forget.

 \xDH In addition,
the connections to other groups of neurons may themselves be complex, and
may consist again of many neurons, they are not simple operators
like $ \xcu,$ $ \xCN $ etc. as between words in a language.
So, groups of neurons are connected to other groups of neurons
via groups of neurons, and the same considerations as above apply
to the connecting groups.

 \xEh
 \xDH The use of a group of neurons makes this group easier accessible,
and strengthens its internal coherence.
Thus, the normal case becomes stronger.
 \xDH The use of a neural connection strengthens this connection.
Again, the normal connections become stronger.
Both properties favour learning, but may also lead to overly
rigid thinking and prejudice.
Note that this is the opposite of basic linear logic, where the use of
an argument may consume it.
 \xDH When two groups of neurons, $N_{1}$ and $N_{2}$ are activated
together,
this strengthens the connection between $N_{1}$ and $N_{2}$ by Hebb's
rule,
see Definition \ref{Definition Hebb} (page \pageref{Definition Hebb}),
and e.g.  \cite{Pul13}.
This property establishes associations. When $ \xfI $ hear a roar in the
jungle,
and see an attacking tiger, next time, $ \xfI $ will think ``tiger'' when $
\xfI $ hear
a roar, even without seeing the tiger.

 \xDH A longer connection may be weaker. For instance, penguins are an
abnormal subclass of birds. Going from birds to penguins will not be
via a strong connection (though $penguin \xcp bird$ is a classical
inference).
If Tweety is a penguin, we might access
Tweety only by detour through penguin. Thus, Tweety is ``less''
bird than the raven which $ \xfI $ saw in my garden.
Consequently, the subset relation involved in the properties of many
nonclassical logics
has a certain ``cost'', and the resulting properties
(e.g. $X \xcc Y \xcp \xbm (Y) \xcs X \xcc \xbm (X)$ for basic preferential
logic, similar
properties for theory revision, update, counterfactual conditionals)
cannot always be expected.
 \xEj

We summarize:

 \xEI
 \xDH
Connections are made of single axon-synapse-dendrite tripels, connecting
one neuron to another, or many such tripels, bundels, or composed
bundels. Again, it seems useful to say that connections can usually be
decomposed into sub-connections.
 \xDH
Connections can be via excitatory or inhibitory synapses, the former
activate the downstream neuron, the latter de-activate the downstream
neuron.
 \xDH
Connections can have very different meanings.
 \xDH
Connections can be interior to groups, or between groups.
 \xDH
There usually is no clear distinction between groups and connections.

 \xEJ

 \xDH
It is important to note fundamental differences to (formal) languages:

There is no ``right'' level of abstraction or granularity to consider this
picture - it depends on the connection to other groups of
neurons.

The connections to other groups of neurons are themselves complex, and
may consist again of many neurons, they are not simple operators
like $ \xcu,$ $ \xCN $ etc.

 \xEj
\subsubsection{
Organisation of the Brain
}

 \xEh
 \xDH Recursiveness

The processing of information via connections is not linear, but
recursive, even in relatively basic (visual) circuits.

Consequently, there must be some mechanism preventing wild oscillations
and
uncontrolled reinforcement,
see also
Chapter \ref{Chapter TRU} (page \pageref{Chapter TRU}).

Thus, a hypothesis might enhance lower level attention to certain aspects,
and
help decide about truth of the hypothesis, which, by its organisation,
the lower level center might be unable to do on its own.

 \xDH Different areas of the brain

There are different areas for different tasks of the brain. E.g., there
are areas for language processing, and there is a semantic memory
for facts and concepts, an episodic memory for events, experiences,
scenes, etc. The latter allows to construct new scenes from old ones, etc.
Scenes will be memorised by (intraconnected) groups of active neurons,
consisting of perhaps thousands of neurons.

The brain has grown in the course of evolution, but it is not an
``organic'' growth, it is rather like a shanty town, where old parts are
still
being used, or cooperating with newer parts.

Most important, the brain has to be sufficiently connected to the
(natural and social) world to be useful.

 \xDH
There are different types of memory in the human brain.

A look at the literature on human memory, e.g.,
(\cite{Wik17a},  \cite{Wik17b},  \cite{Wik17c},  \cite{Wik17d},
 \cite{CCOM08},  \cite{HM17},  \cite{KPP07},  \cite{ZMM15})
shows that:
 \xEh
 \xDH There are several memories, and
some tasks and problems due to lesions allow to differentiate
between different memories.

There are e.g. short-term and long-term memory,
within long-term memory implicit memory (unconcious, concerning skills
etc.) vs
explicit memory (conscious, declarative), within declarative memory
episodic memory (events, experiences, scenarios, pictures) vs. semantic
memory
(facts, concepts),
etc.

We are interested here in episodic and semantic memory.
Episodic memory concerns episodes within a context, ``stories'', semantic
memory
concerns principles and facts independent of context.

As the word ``semantics'' is heavily used in logics, we will call semantic
memory ``conceptual memory'' here.

Episodic memory consists of (intraconnected) groups of
active neurons, of perhaps thousands of neurons.

 \xDH
If episodic memory were based on conceptual memory (a number of connected
conceptual entities over time), or vice versa (the common feature of a
number
of episodes), then failure of one system would also cause failure of the
other system. This is not the case.
On the other hand, it seems unlikely that both systems coexist without
connections.
The exact connections and independencies seem unclear.
Of course, this is a fundamental problem, and extremely important to the
relation between the world, our brain, and our language.

 \xDH
There seem to be memory structures in the strict sense (where information
is
stored), and auxiliary structures, e.g. for storing (writing) and
accessing
(reading) information. But this separation might not be strict.
Acessing memory may be top-down (e.g. ``where did $ \xfI $ park
the car?'') or bottom-up (e.g. image of the car near a tree).

 \xDH
Memory activities might involve several centers of the brain, e.g.
visual memory might involve the centers for processing visual information,
likewise for auditory memory, etc. These questions are not settled.

 \xDH
There are several models of conceptual memory, e.g. network models, and
feature
models. The first are types of neuronal networks, the latter close to
defeasible inheritance networks. -
The location of the conceptual memory in the brain is not clear. It might
be a collection of functionally and anatomically distinct systems.

 \xDH
Episodic memory allows to construct new scenes from past scenes,
insertions,
blending, cutting, etc. It is not clear where this happens, in the
episodic
memory itself, or in a ``higher'' structure, or both? Is there a control
mechanism which supervises the result (and is perhaps asleep when we
dream)? -
The parietal cortex and the hippocampus seem to be involved in episodic
memory.

 \xDH

Learning is believed to be a Hebbian process, see
Definition \ref{Definition Hebb} (page \pageref{Definition Hebb}).

 \xEj

 \xDH The brain does not contain 1-1 images of the world

Neither the present image of the world nor the memory is a simple
1-1 image. In both cases, past experiences, desires, attention, etc.
form and deform the image.

We cannot exclude that there are aspects of the world which are, in
principle, inaccessible to our thoughts. Our existence proves that our
brain is sufficiently adapted to the world to deal with it. It seems
difficult
to go beyond this scepticism.

 \xEj
\subsubsection{
The (limited) Role of Language
}

 \xEh
 \xDH

\label{Point Sexing}

Obviously, not all knowledge is coded by language.
Sexing (determining the sex of) chicken is a famous example. Experts (here
in
sexing) may be very efficient, but unable to express their knowledge by
words.

 \xDH
Even scientific theories seem to have non-verbal aspects.
 \xDH
In legal reasoning, even fundamental ideas may be expressed by words with
unclear meaning. A famous example is the German constitution which has as
basic concept ``dignity of man''. This concept has a long history, but no
clear meaning. At first sight, it seems absurd to base a constitution on
an unclear concept, leaving a wide margin for interpretation, but the
authors of the constitution may have been confident that judges and law
makers will know how to apply it (as experts know how to sexe chicken).
Thus, ``dignity of man'' is perhaps best seen as a pointer (or label) to
ways to interpret it, as a collection of prototypes.
 \xDH
The (philosophical) notion of an ideal might be seen in the same way
(and not as a list of properties which fails, as is well known).
 \xDH
A more general problem is to describe brain states which are principally
unreachable by language - if they exist.
(And, even more generally, are there brainstates A and $B,$ which have
only
limited communication possibilities?)

 \xDH
Language may help to structure, stabilise, and refine knowledge: wine
connaisseurs use seemingly
bizarre expressions to describe their experiences.

 \xDH
Expressions of our languages have no direct semantics in the world,
but first in brain states, and those brain states somehow correspond
to the world. Thus the first meaning of expressions is a kind of
``pre-semantics'' (not between formal language and formal semantics, but
a brain activity between natural language and the ``real world'').

 \xEj
\subsubsection{
Comparison to Defeasible Inheritance Networks
}

For an overview of defeasible inheritance, see e.g.
 \cite{Sch97-2}.

Note that several neurons may act together as an amplifier for a neuron:
Say $N_{1}$ connects positively to $N_{2}$ and $N_{3},$ and $N_{2}$ and
$N_{3}$ each connect
positively to $N_{4},$ and $N_{1}$ is the only one to connect to $N_{2}$
and $N_{3},$ then
any signal from $N_{1}$ will be doubled in strength when arriving at
$N_{4},$
in comparison to a direct signal from $N_{1}$ to $N_{4}.$
 \xEh
 \xDH Consequently, direct links do not necessarily
win over indirect paths. If, in
addition, $N_{1}$ is connected negatively to $N_{4},$ then the signal from
$N_{1}$ to $N_{4}$
is 2 for positive value (indirect via $N_{2}$ and $N_{3})$ and -1 for
negative
value (direct to $N_{4}),$ so the indirect paths win.
 \xDH By the same argument, longer paths may be better.

On the other hand, a longer path has more possibilities of interference
by other signals of opposite polarity.

So length of path is no general criterion,
contrary to inheritance systems, where connections correspond to ``soft''
inclusions.
 \xDH The number of paths of the same polarity is important, in
inheritance
systems, it is only the existence.
 \xDH There is no specificity criterion, and no preclusion.
 \xDH In inheritance systems, a negative arrow may only be at the end of a
path, it cannot continue through a negative arrow. Systems of neurons
are similar: If a negative signal has any effect, it prevents the
receiving neuron from firing, so this signal path is interrupted.
 \xDH Neurons act directly sceptically - there is no branching into
different extensions.
 \xEj
\clearpage
\section{
Acknowledgements
}

The author is much indebted to C.v.d. Malsburg, FIAS, Frankfurt,
for patient advice, and to D. Gabbay for discussions.

$ \xCO $

$ \xCO $
\markboth{\centerline{\scriptsize Analogical Reasoning}}
{\centerline{\scriptsize Analogical Reasoning}}

$ \xCO $
\chapter{
A Comment on Analogical Reasoning
}

\label{Section ANA}

\label{Chapter ANA}

$ \xCO $
\section{
Introduction
}
\subsection{
Overview
}

Our idea originated from the remarks of Section 2.4 of
 \cite{SEP13} and the scepticism expressed there to find a logic for
analogical reasoning, see
Section \ref{Section Scept} (page \pageref{Section Scept})  below.

In a way, our (positive) reply is a form of cheating: we avoid the
problem, and
push it and the solution into a suitable choice function - as is done for
Counterfactual Conditionals, Preferential Structures, etc. We then
have a characterisation machinery we can just take off the shelf, and we
have results for ANY choice function.

We present in the rest of this introduction (largely verbatim, only
punctually
slightly modified) excerpts
from  \cite{SEP13}, see also  \cite{SEP19c}, to set the stage
for
the answer in
Section \ref{Section Idea} (page \pageref{Section Idea}).
\subsection{
Section 2.2 of \cite{SEP13}
}

(These remarks concern $p.$ 5-7 of Section 2.2
of  \cite{SEP13}.)

\bd

$\hspace{0.01em}$


\label{Definition}

An analogical argument has the following
form:

1. $S$ is similar to $T$ in certain (known) respects.

2. $S$ has some further feature $Q.$

3. Therefore, $T$ also has the feature $Q,$ or some feature $Q*$
similar to $Q.$

(1) and (2) are premises. (3) is the conclusion of the argument.
The argument form is
inductive in the sense that the conclusion is not guaranteed to follow
from the
premises.

\ed

$S$ and $T$ are referred to as the source domain and target domain,
respectively. A domain is
a set of objects, properties, relations and functions, together
with a set of accepted
statements about those objects, properties, relations and
functions. More formally, a
domain consists of a set of objects and an interpreted set of
statements about them. The
statements need not belong to a first-order language, but to keep
things simple, any
formalizations employed here will be first-order. We use
unstarred symbols (a, $P,$ $R,$ $f)$ to
refer to items in the source domain and starred symbols $(a*,$ $P*,$
$R*,$ $f*)$ to refer to
corresponding items in the target domain.

\bd

$\hspace{0.01em}$


\label{Definition Mapping}

Formally, an analogy between $S$ and $T$ is a one-to-one mapping
between objects,
properties, relations and functions in $S$ and those in $T.$

\ed

J. M. Keynes,
in  \cite{Key21}, introduced
some helpful terminology:
 \xEh
 \xDH
Positive analogy.

Let $P$ stand for a list of accepted propositions $P_{1}, \Xl,$ $P_{n}$
about
the source domain $S.$
Suppose that the corresponding propositions $P_{1}*$ $, \Xl,P_{n}*$,
abbreviated as $P*,$ are
all accepted as holding for the target domain $T,$ so that $P$ and $P*$
represent accepted
(or known) similarities. Then we refer to $P$ as the positive
analogy.

 \xDH
Negative analogy.

Let $ \xCf A$ stand for a list of propositions $A_{1}, \Xl,$ $A_{r}$
accepted as
holding in $S,$ and $B*$ for
a list $B_{1}*, \Xl,$ $B_{s}*$ of propositions holding in $T.$ Suppose
that the
analogous
propositions $A*$ $=$ $A_{1}*, \Xl,$ $A_{r}*$ fail to hold in $T,$ and
similarly
the propositions $B$ $=$
$B_{1}, \Xl,$ $B_{s}$ fail to hold in $S,$ so that $ \xCf A,$ $ \xCN A*$
and $ \xCN B,$ $B*$ represent
accepted (or
known) differences. Then we refer to $ \xCf A$ and $B$ as the negative
analogy.

 \xDH
Neutral analogy.

The neutral analogy consists of accepted propositions about $S$ for
which it is not
known whether an analogue holds in $T.$

 \xDH
Hypothetical analogy.

The hypothetical analogy is simply the proposition $Q$ in the
neutral analogy that is
the focus of our attention.

 \xEj

These concepts allow us to provide a characterization for an
individual analogical
argument that is somewhat richer than the original one.

\bd

$\hspace{0.01em}$


\label{Definition Aug}

(Augmented representation)

Correspondence between SOURCE (S) and TARGET (T)

 \xEh

 \xDH
Positive analogy:

$P$ $ \xcj $ $P*$
 \xDH
Negative analogy:

$ \xCf A$ $ \xcj $ $ \xCN A*$

and

$ \xCN B$ $ \xcj $ $B*$
 \xDH
Plausible inference:

$Q$ $ \xcj $ $Q*$
 \xEj

An analogical argument may thus be summarized:
It is plausible that $Q*$ holds in the target because of certain
known (or
accepted) similarities with the source domain, despite certain
known (or
accepted) differences.
\subsection{
Section 2.4 of \cite{SEP13}
}

\label{Section Scept}

\ed

Scepticism:

Of course, it is difficult to show that no successful analogical
inference rule will ever be
proposed. But consider the following candidate, formulated using
the concepts of the schema
in Definition \ref{Definition Aug} (page \pageref{Definition Aug})
and taking us only a short step beyond that basic
characterization.

\bd

$\hspace{0.01em}$


\label{Definition Scept}

Suppose $S$ and $T$ are the source and target domains. Suppose $P_{1}, \Xl
,$ $P_{n}$
represents the positive analogy, $A_{1}, \Xl,$ $A_{r}$ and $ \xCN B_{1},
\Xl,$ $ \xCN B_{s}$
represent the (possibly
vacuous) negative analogy, and $Q$ represents the hypothetical
analogy. In the
absence of reasons for thinking otherwise, infer that $Q*$ holds in
the target domain
with degree of support $p$ $>$ 0, where $p$ is an increasing function
of $n$ and a
decreasing function of $r$ and $s.$

(Definition \ref{Definition Scept} (page \pageref{Definition Scept})
is modeled on the straight rule for enumerative
induction and inspired by Mill's
view of analogical inference, as described
in  \cite{SEP13} above. We use the
generic phrase ``degree of
support'' in place of probability, since other factors besides the
analogical argument may
influence our probability assignment for $Q*.)$

\ed

The schema in
Definition \ref{Definition Scept} (page \pageref{Definition Scept})
justifies too
much.

So, how do we chose the ``right one''?

\br

$\hspace{0.01em}$


\label{Remark Ana}

The author was surprised to find a precursor to his concept of
homogenousness,
see Chapter \ref{Chapter HOM} (page \pageref{Chapter HOM}), and
 \cite{Sch97-2},  \cite{GS16},
in the work of J. M. Keynes,
 \cite{Key21}, quoted in Section 4.3 of  \cite{SEP13}.
\section{
The Idea
}

\label{Section Idea}

\er

We now describe the idea, and compare it to other ideas in philosophical
and AI related logics.

But first, we formalize above ideas into a definition.

\bd

$\hspace{0.01em}$


\label{Definition Ana}

Let $ \xdl $ be an alphabet.
 \xEh
 \xDH
Let $ \xdl_{ \xba } \xcc \xdl,$ and $ \xba: \xdl_{ \xba } \xcp \xdl $ an
injective function,
preserving the type of symbol, e.g.,
 \xEI
 \xDH
if $x \xbe \xdl_{ \xba }$ stands for an object of the universe, then so
will $ \xba (x)$
 \xDH
if $X \xbe \xdl_{ \xba }$ stands for a subset of the universe, then so
will $ \xba (X)$
 \xDH
if $P(.) \xbe \xdl_{ \xba }$ stands for an unary predicate of the
universe,
then so will $ \xba (P)(.)$
 \xDH
etc., also for higher symbols, like $f: \xdp (U) \xcp \xdp (U),$ $U$ the
universe.
 \xEJ
 \xDH
Let $ \xdf_{ \xba }$ a subset of the formulas formed with symbols from $
\xdl_{ \xba }.$

For $ \xbf \xbe \xdf_{ \xba },$ let $ \xba (\xbf)$ be the obvious
formula constructed from $ \xbf $
with the function $ \xba.$
 \xDH
We now look at the truth values of $ \xbf $ and $ \xba (\xbf),$ $v(\xbf
)$ and $v(\xba (\xbf)).$
In particular, there may be $ \xbf $ s.t. $v(\xbf)$ is known, $v(\xba (
\xbf))$ not, and
we extrapolate that $v(\xbf)=v(\xba (\xbf)),$ this is then the
analogical
reasoning based on $ \xba.$

More precisely:
 \xEh
 \xDH
There may be $ \xbf $ s.t. $v(\xbf)$ is not known, $v(\xba (\xbf))$
is known or not, such
$ \xbf $ do not interest us here.

Assume in the following that $v(\xbf)$ is known.
 \xDH
$v(\xbf)$ and $v(\xba (\xbf))$ are known, and $v(\xbf)=v(\xba (
\xbf)).$ The set of such $ \xbf $
is the positive support of $ \xba,$ denoted $ \xba^{+}.$
 \xDH
$v(\xbf)$ and $v(\xba (\xbf))$ are known, and $v(\xbf) \xEd v(\xba
(\xbf)).$ The set of such $ \xbf $
is the negative support of $ \xba,$ denoted $ \xba^{-}.$
 \xDH
$v(\xbf)$ is known, $v(\xba (\xbf))$ is not known. The set of such $
\xbf $
is denoted $ \xba^{?}.$

The ``effect'' of $ \xba $ is to conjecture, by analogy, that $v(\xbf)=v(
\xba (\xbf))$ for
such $ \xbf.$
 \xEj
 \xEj
Intuitively, $ \xba^{+}$ strengthens the case of $ \xba,$ $ \xba^{-}$
weakens it - but these
need not be the only criteria, see also
 \cite{SEP13} and  \cite{SEP19c}.

Such $ \xba $ - or a modification thereof - are the basic concepts of
analogy. Source and destination of $ \xba $ often describe aspects of the
``world'',
$ \xba $ itself need not correspond to anything in the world (like a
common cause),
but may be merely descriptional. Moreover, there may also be
``meta-analogies'' between analogies.

Let $ \xda $ be a set of functions $ \xba $ as defined in
Definition \ref{Definition Ana} (page \pageref{Definition Ana}).

We may close $ \xda $ under combinations, as illustrated in the following
Example \ref{Example Combi} (page \pageref{Example Combi}), or not.

\ed

\be

$\hspace{0.01em}$


\label{Example Combi}

Consider $ \xba,$ $ \xba'.$

Let $x,x',P,Q$ $ \xbe $ $ \xdl_{ \xba }= \xdl_{ \xba' },$ $Q(x),Q(x')
\xbe \xba^{?}, \xba'^{?}.$
 \xEh
 \xDH
$ \xba $ works well for $x,$ but not for $x':$
$P(x)= \xba (P)(x),$ $P(x') \xEd \xba (P)(x'),$
so $P(x) \xbe \xba^{+},$ $P(x') \xbe \xba^{-},$
 \xDH
$ \xba' $ works well for $x',$ but not for $x:$
$P(x')= \xba' (P)(x'),$ $P(x) \xEd \xba' (P)(x),$
so $P(x) \xbe \xba'^{-},$ $P(x') \xbe \xba'^{+}.$
 \xEj

Let further $ \xba (Q)(x) \xEd \xba' (Q)(x)$ and $ \xba (Q)(x') \xEd
\xba' (Q)(x').$

What shall we do, should we chose one, $ \xba $ or $ \xba',$ for
guessing, or
combine $ \xba $ and $ \xba' $ to $ \xba'',$ chosing $ \xba'' = \xba $
for expressions with $x,$
and $ \xba'' = \xba' $ for expressions with $x',$ more precisely
$ \xba'' (Q)(x):= \xba (Q)(x),$ and $ \xba'' (Q)(x'):= \xba' (Q)(x')$
?

\ee

The idea is now to push the choice of suitable $ \xba \xbe \xda $ into a
relation $ \xeb,$
expressing quality of the analogy.
E.g., in Example 
\ref{Example Combi} (page 
\pageref{Example Combi}), $ \xba'' \xeb \xba $ and $ \xba
'' \xeb \xba' $ - for historical
reasons, smaller elements will be ``better''.

Usually, this ``best'' relation will be partial only, and
there will be many ``best'' $f.$
Thus, it seems natural to conclude the properties which hold
in ALL best $f.$

\bd

$\hspace{0.01em}$


\label{Definition Valid}

Let $ \xda $ be a set of functions as described in
Definition 
\ref{Definition Ana} (page 
\pageref{Definition Ana}), and $ \xeb $ a relation on $ \xda $
(expressing ``better''
analogy wrt. the problem at hand).

We then write $ \xda \xcm_{ \xeb } \xbf $ iff $ \xbf $ holds in all $ \xeb
-$best $ \xba \xbe \xda.$

(This is a sketch only, details have to be filled in according to the
situation considered.)
\subsection{
Discussion
}

\ed

This sounds like cheating: we changed the level of abstraction, and
packed the question of ``good'' analogies into the $ \xeb $-relation.

But when we look at the Stalnaker-Lewis semantics of counterfactual
conditionals, see  \cite{Sta68},  \cite{Lew73},
the preferential semantics for non-monotonic reasoning and
deontic logic, see e.g.  \cite{Han69},  \cite{KLM90},  \cite{Sch04}, 
\cite{Sch18},
the distance semantics for theory revision, see e.g.  \cite{LMS01},
 \cite{Sch04},
this is a
well used ``trick'' we need not be ashamed of.

In above examples, the comparison was between possible worlds, here it is
between usually more complicated structures (functions), yet this is no
fundamental difference.

But even if we think that there is a element of cheating in our idea, we
win something: properties which hold in ALL preferential structures,
and which may be stronger for stronger relations $ \xeb,$ see
Fact \ref{Fact Representation} (page \pageref{Fact Representation})  below.
\subsection{
Problems and Solutions
}

 \xEh
 \xDH

In the case of infinitely many $ \xba $'s we might have a definability
problem, as the resulting best guess might not be definable any more -
as in the case of preferential structures,
see e.g.  \cite{Sch04}.

 \xDH

Abstract treatment of representation problems for abovementioned logics
works with arbitrary sets, so we have a well studied machinery for
representation results for various types of relations of ``better''
analogies - see e.g.
 \cite{LMS01},  \cite{Sch04},  \cite{Sch18}.

To give the reader an idea of such representation results, we mention
some,
slightly simplified (we neglect the multitude of copies for simpler
presentation).

\bd

$\hspace{0.01em}$


\label{Definition Representation}

(1) Let again $ \xeb $ be the relation, and $ \xbm (X):=\{x \xbe X: \xCN
\xcE x' \xbe X.x' \xeb x\},$

(2) $ \xeb $ is called smooth iff for all $x \xbe X,$ either $x \xbe \xbm
(X)$ or there is
$x' \xbe \xbm (X),$ $x' \xeb x,$

(3) $ \xeb $ is called ranked iff for all $x,y,z,$ if neither $x \xeb y$
nor $y \xeb x,$ then
if $z \xeb x,$ then $z \xeb y,$ too, and, analogously, if $x \xeb z,$ then
$y \xeb z,$ too.

\ed

We then have e.g.

\bfa

$\hspace{0.01em}$


\label{Fact Representation}

 \xEh
 \xDH
General and transitive relations are characterised by

$(\xbm \xcc)$ $ \xbm (X) \xcc X$

and

$(\xbm PR)$ $X \xcc Y$ $ \xcp $ $ \xbm (Y) \xcs X \xcc \xbm (X)$
 \xDH
Smooth and transitive smooth relations are characterised
by $(\xbm \xcc),$ $(\xbm PR),$ and the additional property

$(\xbm CUM)$ $ \xbm (X) \xcc Y \xcc X$ $ \xcp $ $ \xbm (X)= \xbm (Y)$
 \xDH
Ranked relations are characterised by $(\xbm \xcc),$ $(\xbm PR),$
and the additional property

$(\xbm =)$ $X \xcc Y,$ $ \xbm (Y) \xcs X \xEd \xCQ $ $ \xcp $ $ \xbm (X)=
\xbm (Y) \xcs X.$
 \xEj

For more explanation and details, see e.g.
 \cite{Sch18}, in particular Table 1.6 there.

\efa

 \xEj

$ \xCO $

$ \xCO $
\markboth{\centerline{\scriptsize Homogenousness}}
{\centerline{\scriptsize Homogenousness}}

$ \xCO $
\chapter{
Relevance and Homogenousness
}

\label{Chapter HOM}

\label{Chapter REL}
\section{
Introduction
}

Erledigt bis 5.2.2 einschliesslich am 21.4.22

We keep this introduction short and refer the reader for more details to
 \cite{GS16}, Chapter 11, and
 \cite{Sch18b}, Chapter 5.
\subsection{
The Principle of Homogenousness
}

\bd

$\hspace{0.01em}$


\label{Definition Principle}

The principle of homogenousness can be stated, informally, as:

``Given an set $X$ of cases, and $X' \xcc X,$ the elements of $X' $ usually
behave as the
elements of $X$ do.''

\ed

The strict version (without the ``usually'') is obviously wrong, but the
soft
version (as above) is an
extremely useful hypothesis, and
it seems impossible for living beings even to survive without it.

The principle of homogenousness is implicit in many systems of
NML.

It seems impossible to understand the intuitive justification of the
downward
version in the default (in the sense of Reiter) and the defeasible
inheritance variants of NML, without accepting a default (in the intuitive
sense) version of homogenousness: If, in $X,$ normally $ \xbf $ holds, why
should
$ \xbf $ normally hold in a subset $X' $ of $X$ (unless $X' $ is ``close
to'' $X)$ - if we
do not accept some form of homogenousness? Of course, by the very
principle of
non-monotonicity, this need not be the case for all $X' \xcc X,$ as $X' $
might
just be the set of exceptions.
(See the discussion in  \cite{Sch97-2}.)

We are mainly interested in non-monotonic logic, whose consequence
relation
is often written $ \xcn,$ and, by their very nature, $X \xcn \xbf $ does
not impl;y
$X' \xcn \xbf $ - though it will ``often'' hold.
Thus, for some properties, the ``border'' between $X$ and $X' $ is
relevant, for many it is not.

Depending on the strength of the underlying non-monotonic logic $ \xcn,$
we
have ``hard'' second order properties, like: if $X$ differs little in size
from $Y,$
and $X \xcn \xbf,$ then $Y \xcn \xbf.$ We want to go beyond these hard
second order
properties, and look at reasonable other second order properties and
their relation to each other. We will argue semantically, and abstract
size
will be central to our work. Many properties will be default coherence
properties between $X,$ $Y,$ etc.

We have now the following levels of reasoning:

 \xEh
 \xDH Classical logic:

monotony, no exceptions, clear semantics

 \xDH Preferential logic:

small sets of exceptions possible, clear semantics, strict rules about
exceptions, like $(\xbm CUM),$ no other restrictions

 \xDH Meta-Default rules (Homogenousness):

They have the form: $ \xba \xcn \xbb,$ and even if $ \xba \xcu \xba'
\xcN \xbb $ in the nonmonotonic
sense of (2), we prefer those models where $ \xba \xcu \xba' \xcn \xbb,$
but exceptions are
possible by nonmonotonicity itself, as, e.g., $ \xba \xcu \xba' \xcn \xCN
\xbb $ in (2).

We minimize those exceptions, and resolve conflicts whenever possible,
as in Fact \ref{Fact TweetySize} (page \pageref{Fact TweetySize}),
by using the same principle as in level (2): we keep
exception sets small. This is summarized in the specificity criterion.
 \xEj

(We might add a modified length of path criterion as follows:
Let $x_{0} \xcp x_{1} \xcp  \Xl  \xcp x_{n},$ $x_{i} \xcp y_{i},$ $x_{i+1}
\xcP y_{i}.$ We know by
Fact 
\ref{Fact TweetySize} (page 
\pageref{Fact TweetySize})  that $x_{0}< \Xl <x_{n},$ then any shorter
chain $a \xcp  \Xl  \xcp b$ has a
shorter possible size reduction (if there are no other chains, of
course!),
and we can work with this. This is the same concept as in
 \cite{Sch18e}, section 4.)

This has again a clear (preferential) semantics, as our characterisations
are abstract, see e.g.  \cite{Sch18a}.

Inductive reasoning is the upward version, the problem is to find the
cases where it holds.

Analogical reasoning is the upward version going from $X$ to $X \xcv X'.$

The author recently discovered
(reading  \cite{SEP13}, section 4.3)
that J. M. Keynes's
Principle of the Limitation of Independent Variety,
see  \cite{Key21} expresses essentially the same idea
as homogenousness.
(It seems, however, that the epistemological aspect, the naturalness
of our concepts, is missing in his work.)
By the way,  \cite{SEP13} also mentions
``inference pressure'' (in section 3.5.1)
discussed in
 \cite{Sch97-2}, section 1.3.4, page 10.
Thus, the ideas are quite interwoven.

``Uniformity of nature'' is a concept similar to our homogenousness, see
 \cite{Wik18c}, and  \cite{Sim63}. The former is, however,
usually applied
to changes in time and place, and not to subsets, as we do.
\subsection{
Contributions of the Present Text
}

 \xEh
 \xDH
Our main formal contribution here is to analyse various size relations
between sets,
see Section \ref{Section FormalProp} (page \pageref{Section FormalProp}),
in particular when this relation is
generated itself by a relation $ \xeb $ between
elements, see Section 
\ref{Section Filt-Pref} (page 
\pageref{Section Filt-Pref}),
- similarly to Definition 2.6 and Fact 2.7 in
 \cite{Sch97-2}.

These relations can then be used as semi-quantitative distances
between sets, to calculate the ``seriousness'' of homogeneity
violations.

Ideas and proofs are elementary.

 \xDH
Conceptually, we consider
 \xEh
 \xDH Background logic, often non-monotonic, in some fixed language

 \xDH Reference classes, usually a subset of the formulas in the fixed
language.
They may be closed under operations like $ \xcs,$ but not necessarily

 \xDH Notion of consistency of the background logic

 \xDH Specificity to solve conflicts in the background logic between
properties of
different reference classes, often based on a notion of distance

 \xDH We do not treat things like: related classes, properties, this is
left
to ``knowledge engineering''.
 \xEj

It is important to carefully select the predicates treated.

For instance, if $X$ is the set of all vertebrate,
$Y$ the set of all mammals, $Y' $ the set of all cats, then it is
plausible
that elements of $Y' $ have many more additional properties to those valid
in $X$
than elements of $Y$ do. Taxonomies are made exactly for this purpose,
and good natural categories behave this way.

Compare the author's favorite example of bad categories:

\be

$\hspace{0.01em}$


\label{Example 33}

Enumerate
the objects of the universe, and consider the class
of those objects, whose number ends by 3.
Now consider the subclasses of all objects whose number ends by 33, by
333,
etc. We will not expect these concepts to have reasonable properties -
apart from trivial ones.

\ee

 \xEj

The separation of reference classes and properties existed e.g.
in KL-ONE.

\bd

$\hspace{0.01em}$


\label{Definition Sim-Th}

 \xEh
 \xDH
A similarity structure consists of a set of models or set of sets of
models (for
some fixed language),
called possible reference classes - we do not assume any closure
conditions -,
and a model or set of models (not in the set of possible
reference classes) $T,$ called the target class, a set of formulas in the
fixed
language, called the possible properties, and a partial order $ \xck $ on
the
reference classes relative to the target class. (Intuitively, this partial
order chooses the ``best'' reference classes wrt. the target class.) For
simplicity, assume that there are no infinite descending chains in the
order.

 \xDH
A similarity theory consists of a similarity structure, together with
a procedure to solve contradictions. More precisely, if two or more
reference classes, all $ \xck $-optimal, contradict each other about $
\xbf,$ we need
to know what to do.

Several possibilities come to mind (there might be still others):

 \xEh
 \xDH branch into different ``extensions'' of $T,$ with, e.g. $T \xcm \xbf $
and $T \xcm \xCN \xbf,$
 \xDH give no information about $T$ and $ \xbf $ (direct scepticism)
 \xDH if $ \xbf $ permits more than two truth values, like numerical
values,
find a compromise (which may depend on the number of votes for the
different values).

This is also possible for non-monotonic logics, where (instead of
above models, we have now sets of models for reference and target classes)
$R \xcm \xeA \xbf $ (meaning: almost everywhere in $R,$ $ \xbf $ holds)
and $R' \xcm \xeA \xCN \xbf,$
we conclude for the target class $T \xcm \xea \xbf,$ meaning: the subset
of $T$
where $ \xbf $ holds, has medium size.
 \xEj

 \xDH
A homogenousness structure is a similarity structure (the set version),
with
$ \xcc $ as basic relation $(\xcc $ may be strict, or ``soft'', with
exceptions, with
some underlying theory of exceptions). Possible reference classes have
to be $ \xcc $-above the target class, i.e. $T \xcc R$ for each $R.$ The
partial order
is again $ \xcc,$ perhaps with some embellishments to solve more
conflicts, see
Chapter 11, Formal Construction,
in  \cite{GS16}.
(E.g., if we cannot compare by $ \xcc,$ we may resort to some
variant of path length.)

The more specific superclass will win, e.g., if $T \xcc R \xcc R',$ then
$R$ will
win over $R'.$

 \xEj
\section{
Formal Properties
}

\label{Section FormalProp}
\subsection{
Basic Definitions and Facts
}

\ed

\br

$\hspace{0.01em}$


\label{Remark Trivial}

Note that we use in the proof of
Fact \ref{Fact NoChange-2} (page \pageref{Fact NoChange-2})
a - to the author - new way of writing down proofs which
makes them almost mechanical, comparable to elementary
school maths, eliminating terms on both sides of an
expression. The idea is to write down all sets involved,
this is tedious, but the rest is trivial.

\er

\bd

$\hspace{0.01em}$


\label{Definition Filter}

Let $X \xEd \xCQ.$
 \xEh
 \xDH
$ \xdf (X) \xcc \xdp (X)$ is called a filter on $X$ iff

(F1) $X \xbe \xdf (X),$ $ \xCQ \xce \xdf (X)$

(F2) $A \xcc B \xcc X,$ $A \xbe \xdf (X)$ $ \xch $ $B \xbe \xdf (X)$

(F3) $A,B \xbe \xdf (X)$ $ \xch $ $A \xcs B \xbe \xdf (X)$ (finite
intersection suffices here)

 \xDH
If there is $A \xcc X$ such that $ \xdf (X)=\{A' \xcc X:$ $A \xcc A' \},$
we say that $ \xdf (X)$
is the (principal) filter generated by $ \xCf A.$
For historical reasons, we will often note this A $ \xbm (X).$
 \xDH
$ \xdi (X) \xcc \xdp (X)$ is called an ideal on $X$ iff

(I1) $X \xce \xdi (X),$ $ \xCQ \xbe \xdi (X)$

(I2) $A \xcc B \xcc X,$ $B \xbe \xdi (X)$ $ \xch $ $A \xbe \xdi (X)$

(I3) $A,B \xbe \xdi (X)$ $ \xch $ $A \xcv B \xbe \xdi (X)$ (finite union
suffices here)

 \xEj

Intuitively, filters over $X$ contain big subsets of $X,$ ideals small
subsets.

Thus, (F1), (F2), (I1), (I2) are natural properties, (F3) and (I3)
give algebraic strength. We sometimes emphasize the number of times the
latter (and other properties to be discussed below) were used in a proof.

\ed

\bd

$\hspace{0.01em}$


\label{Definition Correspondence}

Let $X \xEd \xCQ.$

If $ \xdf (X)$ is a filter over $X,$ then

$\{A \xcc X:$ $X-A \xbe \xdf (X)\}$ is the corresponding ideal $ \xdi
(X),$

and, conversely

if $ \xdi (X)$ is an ideal over $X,$ then

$\{A \xcc X:$ $X-A \xbe \xdi (X)\}$ is the corresponding filter $ \xdf
(X).$

When we go from filter to ideal to filter over fixed $X,$ we always mean
the
corresponding structure.

Given $ \xdf (X)$ and the corresponding $ \xdi (X),$ we set

$ \xdm (X):=\{A \xcc X:$ $A \xce \xdf (X) \xcv \xdi (X)\}$ - the set of
medium size subsets.

\ed

We now define several coherence properties - properties relating
filters and ideals over different base sets $X,$ $Y,$ etc.

\bd

$\hspace{0.01em}$


\label{Definition Coherence}

(Coh1) $X \xcc Y$ $ \xch $ $ \xdi (X) \xcc \xdi (Y).$

(Coh2) $X \xbe \xdf (Y),$ $A \xbe \xdi (Y)$ $ \xch $ $A \xcs X \xbe \xdi
(X)$

(Coh2a) $Z,Z' \xbe \xdi (B)$ $ \xch $ $Z-Z' \xbe \xdi (B-Z')$

(Coh-RK) $X \xcc Y,$ $X \xce \xdi (Y),$ $A \xbe \xdi (Y)$ $ \xch $ $A \xcs
X \xbe \xdi (X)$

(Coh1) is, by the intuition of an ideal, a very natural property, and
will not be mentioned in proofs.

For principal filters, we define:

$(\xbm PR)$ $X \xcc Y$ $ \xch $ $ \xbm (Y) \xcs X \xcc \xbm (X)$

$(\xbm CUM)$ $ \xbm (Y) \xcc X \xcc Y$ $ \xch $ $ \xbm (X)= \xbm (Y)$

$(\xbm RK)$ $X \xcc Y,$ $ \xbm (Y) \xcs X \xEd \xCQ $ $ \xch $ $ \xbm (Y)
\xcs X= \xbm (X)$

Working with principal filters gives us easy examples, drawing simple
diagrams is often sufficient as an (idea for) a proof.

\ed

\bfa

$\hspace{0.01em}$


\label{Fact 22a}

(Coh2) and (Coh2a) are equivalent.

\efa

\subparagraph{
Proof
}

$\hspace{0.01em}$


``$(Coh2a) \xch (Coh2)$'':

Let $X \xbe \xdf (Y),$ $A \xbe \xdi (Y),$ then $Y-X \xbe \xdi (Y),$ so $A
\xcs X=A-(Y-X) \xbe \xdi (Y-(Y-X))= \xdi (X)$

``$(Coh2) \xch (Coh2a)$'':

$Z,Z' \xbe \xdi (B)$ $ \xch $ $B-Z' \xbe \xdf (B)$ $ \xch $ $Z-Z' =Z \xcs
(B-Z') \xbe \xdi (B-Z')$

$ \xcz $
\\[3ex]

We now show further coherence properties in
Fact \ref{Fact Delta} (page \pageref{Fact Delta})  through
Fact \ref{Fact TweetySize} (page \pageref{Fact TweetySize}).

\bfa

$\hspace{0.01em}$


\label{Fact Delta}

(1) $A,A' \xcc B,$ $A \xbD A' \xbe \xdi (B),$ $A \xbe \xdf (B)$ $ \xcp $
$A' \xbe \xdf (B).$

(2) $A,A' \xcc B,$ $A \xbD A' \xbe \xdi (B),$ $A \xbe \xdi (B)$ $ \xcp $
$A' \xbe \xdi (B).$

\efa

\subparagraph{
Proof
}

$\hspace{0.01em}$


(1) $A \xbD A' \xbe \xdi (B)$ $ \xcp $ $(B-(A \xcv A')) \xcv (A \xcs A'
)$ $=$ $B-(A \xbD A') \xbe \xdf (B).$
By $A \xbe \xdf (B),$ $A \xcs ((B-(A \xcv A')) \xcv (A \xcs A'))$ $=$ $A
\xcs A' \xbe \xdf (B),$ so $A' \xbe \xdf (B).$

(2) Note that $A \xbD A' $ $=$ $(B-A) \xbD (B-A').$ Thus consider $B$-A,
$B-A',$ and apply (1).

\bfa

$\hspace{0.01em}$


\label{Fact NoChange}

Let (Coh1) and (Coh2a) hold.

Let $X \xcc Y,$ $X \xbD X' \xbe \xdi (X \xcv X'),$ then:

(1) $X \xbe \xdi (Y)$ iff $X' \xbe \xdi (Y \xcv X')$

(2) $X \xbe \xdf (Y)$ iff $X' \xbe \xdf (Y \xcv X')$

(3) $X \xbe \xdm (Y)$ iff $X' \xbe \xdm (Y \xcv X')$

\efa

\subparagraph{
Proof
}

$\hspace{0.01em}$


By $X \xcc Y$ and (Coh1) $X \xbD X' \xbe \xdi (Y \xcv X').$

(1)

``$ \xch $'': $X \xbe \xdi (Y) \xcc \xdi (Y \xcv X'),$ $(X \xbD X')
\xbe \xdi (X \xcv X') \xcc \xdi (Y \xcv X'),$ so $X' \xbe \xdi (Y \xcv
X')$
by Fact \ref{Fact Delta} (page \pageref{Fact Delta}).

``$ \xci $'': $X' -Y \xcc X' \xbe \xdi (Y \xcv X'),$ so by (Coh2a)
$X=X-(X' -Y) \xbe \xdi ((Y \xcv X')-(X' -Y))= \xdi (Y).$

(2)

``$ \xci $'': $X' \xbe \xdf (Y \xcv X'),$ $X \xbD X' \xbe \xdi (X \xcv
X') \xcc \xdi (Y \xcv X'),$ so by
Fact \ref{Fact Delta} (page \pageref{Fact Delta})
$X \xbe \xdf (Y \xcv X'),$ so $X \xbe \xdf (Y).$

``$ \xch $'': $X \xbe \xdf (Y)$ $ \xch $ $Y-X \xbe \xdi (Y) \xcc \xdi (Y
\xcv X').$
$(Y-X) \xbD (Y-X')$ $=$ $X \xbD X' $ $ \xbe $ $ \xdi (X \xcv X')$ $ \xcc
$ $ \xdi (Y \xcv X').$
Thus, $Y-X' \xbe \xdi (Y \xcv X')$
by Fact \ref{Fact Delta} (page \pageref{Fact Delta}), so
$X' =(Y \xcv X')-(Y-X') \xbe \xdf (Y \xcv X').$

(3)

Suppose $X \xbe \xdm (Y),$ but $X' \xbe \xdi (Y \xcv X'),$ then $X \xbe
\xdi (Y),$ contradiction. The
other cases are analogous.

$ \xcz $
\\[3ex]

The strategy is essentially the same for the following Facts. We go
up to $Y \xcv X \xcv X',$ and again down to $Y \xcv X$ of $Y \xcv X',$
using $X \xbD X' \xbe \xdi (X \xcv X' \xcv Y),$ and thus $X \xbe \xdi (X
\xcv X' \xcv Y)$ iff $x' \xbe \xdi (X \xcv X' \xcv Y).$

Roughly, the argument is that changing things a little has no
influence, see Fact \ref{Fact Delta} (page \pageref{Fact Delta}).

We give now alternative proofs, which are rather mechnical, but
follow the same strategy.

\bfa

$\hspace{0.01em}$


\label{Fact NoChange-2}

Let $X-Y \xbe \xdi (X),$ $X \xbD X' \xbe \xdi (X \xcv X'),$ then:
 \xEh
 \xDH $X \xbe \xdi (Y \xcv X)$ iff $X' \xbe \xdi (Y \xcv X')$

 \xDH $X \xbe \xdf (Y \xcv X)$ iff $X' \xbe \xdf (Y \xcv X')$

 \xDH $X \xbe \xdm (Y \xcv X)$ iff $X' \xbe \xdm (Y \xcv X')$
 \xEj

\efa

\subparagraph{
Proof
}

$\hspace{0.01em}$


We simplify notation, using $+$ and - for set union and difference,
e.g. $-X+X' -Y$ will mean $(X' -X)$-Y.

Define

$A:=+X+X' +Y$

$B:=+X+X' -Y$

$C:=+X-X' +Y$

$D:=+X-X' -Y$

$E:=-X+X' +Y$

$F:=-X+X' -Y$

$G:=-X-X' +Y$

$H:=-X-X' -Y$

Let

$X:=ABCD$ (for $A+B+C+D$ etc.)

$X':=ABEF$

$Y:=ACEG.$

Then

$X-Y=BD$

$X \xcv X' =ABCDEF$

$X \xbD X' =CDEF$

$Y \xcv X=ABCDEG$

$Y \xcv X' =ABCEFG$

$(Y \xcv X)-X=EG$

$(Y \xcv X')-X' =CG$

The prerequisites are:

(1) $BD \xbe \xdi (ABCD)$

(2) $CDEF \xbe \xdi (ABCDEF)$

 \xEh

 \xDH We have to show
$X \xbe \xdi (Y \xcv X)$ iff $X' \xbe \xdi (Y \xcv X'),$ i.e.

(3) $ABCD \xbe \xdi (ABCDEG)$

iff

(4) $ABEF \xbe \xdi (ABCEFG)$

(3) $ \xch $ (4):

By (3), ABCD $ \xbe $ $ \xdi (ABCDEG)$ $ \xcc $ $ \xdi (ABCDEFG),$
by (2), CDEF $ \xbe $ $ \xdi (ABCDEF)$ $ \xcc $ $ \xdi (ABCDEFG),$
so ABCDEF $ \xbe $ $ \xdi (ABCDEFG),$
so ABEF $ \xcc $ ABCEF $ \xbe $ $ \xdi (ABCEFG).$

(4) $ \xch $ (3):

Analogously:

By (4), ABEF $ \xbe $ $ \xdi (ABCEFG)$ $ \xcc $ $ \xdi (ABCDEFG),$
by (2), CDEF $ \xbe $ $ \xdi (ABCDEF)$ $ \xcc $ $ \xdi (ABCDEFG),$
so ABCDEF $ \xbe $ $ \xdi (ABCDEFG),$
so ABCD $ \xcc $ ABCDE $ \xbe $ $ \xdi (ABCDEG).$

 \xDH (Similarly)
$X \xbe \xdf (X \xcv Y)$ iff $X' \xbe \xdf (Y \xcv X'),$ or
$(X \xcv Y)-X$ $ \xbe $ $ \xdi (X \xcv Y)$ iff $(Y \xcv X')-X' $ $ \xbe $
$ \xdi (Y \xcv X'),$ i.e.

(3) EG $ \xbe $ $ \xdi (ABCDEG)$ iff (4) CG $ \xbe $ $ \xdi (ABCEFG).$

$(3) \xch (4):$

By (3), EG $ \xbe $ $ \xdi (ABCDEFG),$ by (2) CDEF $ \xbe $ $ \xdi
(ABCDEFG),$
so CDEFG $ \xbe $ $ \xdi (ABCDEFG),$ and CEFG $ \xbe $ $ \xdi (ABCEFG),$
and CG $ \xbe $ $ \xdi (ABCEFG).$

$(4) \xch (3):$

By (4), CG $ \xbe $ $ \xdi (ABCDEFG),$ by (2) CDEF $ \xbe $ $ \xdi
(ABCDEFG),$
so CDEFG $ \xbe $ $ \xdi (ABCDEFG),$ and CDEG $ \xbe $ $ \xdi (ABCDEG),$
and EG $ \xbe $ $ \xdi (ABCDEG).$

 \xDH As above:
$X \xbe \xdm (Y \xcv X)$ iff $X' \xbe \xdm (Y \xcv X').$

Suppose e.g. $X \xbe \xdm (Y \xcv X)$ and $X' \xbe \xdi (Y \xcv X'),$
then $X \xbe \xdi (Y \xcv X),$ contradiction.

 \xEj

$ \xcz $
\\[3ex]

\br

$\hspace{0.01em}$


\label{Remark NoChange-3}

Consider the proof of (1) above.
The important properties are:
the right hand side of properties (3) and (4) differ in $D$ (in (3)) and
$F$
$(in(4)).$
Property (2) has $D$ and $F$ on both sides, CDEF $ \xbe $ $ \xdi
(ABCDEF),$ which
allows (after taking unions) to eliminate both $D$ and $F.$

Similar properties hold below in
Fact \ref{Fact NoChange-3} (page \pageref{Fact NoChange-3}).

\er

\bfa

$\hspace{0.01em}$


\label{Fact NoChange-3}

Let (1) $X-Y \xbe \xdi (X),$ (2) $Y \xbD Y' \xbe \xdi (Y \xcv Y'),$ then:

 \xEh

 \xDH $X \xbe \xdi (Y \xcv X)$ iff $X \xbe \xdi (Y' \xcv X)$

 \xDH $X \xbe \xdf (Y \xcv X)$ iff $X \xbe \xdf (Y' \xcv X)$

 \xDH $X \xbe \xdm (Y \xcv X)$ iff $X \xbe \xdm (Y' \xcv X)$

 \xEj

\efa

\subparagraph{
Proof
}

$\hspace{0.01em}$


$A:=+X+Y+Y' $

$B:=+X+Y-Y' $

$C:=+X-Y+Y' $

$D:=+X-Y-Y' $

$E:=-X+Y+Y' $

$F:=-X+Y-Y' $

$G:=-X-Y+Y' $

$H:=-X-Y-Y' $

Let

$X:=ABCD$

$Y:=ABEF$

$Y':=ACEG.$

Then

$X-Y=CD$

$X-Y' =BD$

$Y \xbD Y' =BCFG$

$Y \xcv Y' =ABCEFG$

$Y \xcv X=ABCDEF$

$Y' \xcv X=ABCDEG$

$(Y \xcv X)-X$ $=$ EF

$(Y' \xcv X)-X$ $=$ EG

The prerequisites are:

(1) $CD \xbe \xdi (ABCD)$

(2) $BCFG \xbe \xdi (ABCEFG)$

 \xEh
 \xDH We have to show $X \xbe \xdi (Y \xcv X)$ iff $X \xbe \xdi (Y' \xcv
X)$
i.e., that (3) and (4) are equivalent.

(3) ABCD $ \xbe $ $ \xdi (ABCDEF)$

(4) $ABCD \xbe \xdi (ABCDEG)$

(3) $ \xch $ (4):

BCFG $ \xbe $ $ \xdi (ABCEFG)$ $ \xcc $ $ \xdi (ABCDEFG)$

ABCD $ \xbe $ $ \xdi (ABCDEF)$ $ \xcc $ $ \xdi (ABCDEFG)$

so

ABCDFG $ \xbe $ $ \xdi (ABCDEFG),$ so ABCD $ \xbe $ $ \xdi (ABCDE)$ $ \xcc
$
$ \xdi (ABCDEG)$

(4) $ \xch $ (3):

BCFG $ \xbe $ $ \xdi (ABCEFG)$ $ \xcc $ $ \xdi (ABCDEFG)$

ABCD $ \xbe $ $ \xdi (ABCDEG)$ $ \xcc $ $ \xdi (ABCDEFG)$, so

ABCDFG $ \xbe $ $ \xdi (ABCDEFG),$ so
ABCD $ \xbe $ $ \xdi (ABCDE)$ $ \xcc $ $ \xdi (ABCDEF)$

 \xDH We have to show $X \xbe \xdf (Y \xcv X)$ iff $X \xbe \xdf (Y' \xcv
X),$ i.e. that
$(Y \xcv X)-X$ $ \xbe $ $ \xdi (Y \xcv X)$ iff $(Y' \xcv X)-X$ $ \xbe $ $
\xdi (Y' \xcv X),$
i.e.,

(3) EF $ \xbe $ $ \xdi (ABCDEF)$ iff (4) EG $ \xbe $ $ \xdi (ABCDEG).$

$(3) \xch (4):$

By (3) EF $ \xbe $ $ \xdi (ABCDEFG),$ by (2) BCFG $ \xbe $ $ \xdi
(ABCDEFG),$ so
BCEFG $ \xbe $ $ \xdi (ABCDEFG),$ so EG $ \xcc $ BCEG $ \xbe $ $ \xdi
(ABCDEG).$

$(4) \xch (3):$
By (4) EG $ \xbe $ $ \xdi (ABCDEFG),$ by (2) BCFG $ \xbe $ $ \xdi
(ABCDEFG),$ so
BCEFG $ \xbe $ $ \xdi (ABCDEFG),$ so EF $ \xcc $ BCEF $ \xbe $ $ \xdi
(ABCDEF).$

 \xDH $X \xbe \xdm (Y \xcv X)$ iff $X \xbe \xdm (Y' \xcv X):$

As the corrsponding result for $ \xdm $ in
Fact \ref{Fact NoChange-2} (page \pageref{Fact NoChange-2}).

 \xEj

$ \xcz $
\\[3ex]

\bfa

$\hspace{0.01em}$


\label{Fact Subset}

If $X \xbe \xdf (X'),$ then $(X \xcs A \xbe \xdf (X)$ $ \xcj $ $X' \xcs A
\xbe \xdf (X'))$

\efa

\subparagraph{
Proof
}

$\hspace{0.01em}$


``$ \xch $'': $X' -X \xbe \xdi (X'),$ $X-A \xbe \xdi (X) \xcc \xdi (X'
),$ so
$(X' -X) \xcv (X-A) \xbe \xdi (X')$ $ \xch $ $X' -((X' -X) \xcv (X-A))=A
\xbe \xdf (X').$

``$ \xci $'': $X' \xcs A \xbe \xdf (X'),$ $X' -X \xbe \xdi (X')$ $ \xch
$ $X \xcs A=(X' \xcs A) \xcs X \xbe \xdf (X).$

Second proof:

Set
$B:=(X' -X)$-A,
$C:=(X' -X) \xcs A,$
$D:=X$-A,
$E:=X \xcs A.$

We want to show:
if $D \xcv E \xbe \xdf (B \xcv C \xcv D \xcv E),$ then $(E \xbe \xdf (E
\xcv D)$ $ \xcj $ $E \xcv C \xbe \xdf (B \xcv C \xcv D \xcv E)),$
equivalently
if $B \xcv C \xbe \xdi (B \xcv C \xcv D \xcv E),$ then $D \xbe \xdi (E
\xcv D)$ $ \xcj $ $B \xcv D \xbe \xdi (B \xcv C \xcv D \xcv E))$

Let $B \xcv D \xbe \xdi (B \xcv C \xcv D \xcv E),$ by $C \xbe \xdi (B \xcv
C \xcv D \xcv E),$ $B \xcv C \xcv D \xbe \xdi (B \xcv C \xcv D \xcv E),$
so $D \xbe \xdi (D \xcv E).$

Conversely, let $D \xbe \xdi (D \xcv E) \xcc \xdi (B \xcv C \xcv D \xcv
E),$ by $B \xcv C \xbe \xdi (B \xcv C \xcv D \xcv E),$
$B \xcv D \xbe \xdi (B \xcv C \xcv D \xcv E).$

$ \xcz $
\\[3ex]

\bfa

$\hspace{0.01em}$


\label{Fact TweetySize}

Let $(X-Y) \xbe \xdi (X),$ $(X \xcs Z) \xbe \xdi (X),$ $(Y-Z) \xbe \xdi
(Y),$ then $X \xbe \xdi (X \xcv Y).$

Thus, if $ \xCN \xbf $ holds mostly in $X,$ $ \xbf $ holds mostly in $Y,$
most of
$X$ is in $Y,$ then $X<Y$
(see Definition \ref{Definition <} (page \pageref{Definition <})),
and in inheritance notation, if $X \xcp Y,$ $X \xcP Z,$ $Y \xcp Z,$ then
$X<Y,$ and $Y \xcP X.$

\efa

\subparagraph{
Proof
}

$\hspace{0.01em}$


$A:=+X+Y+Z$

$B:=+X+Y-Z$

$C:=+X-Y+Z$

$D:=+X-Y-Z$

$E:=-X+Y+Z$

$F:=-X+Y-Z$

$G:=-X-Y+Z$

$H:=-X-Y-Z$

Let

$X:=ABCD$

$Y:=ABEF$

$Z:=ACEG.$

Then

$X-Y=CD$

$X \xcs Z=AC$

$Y-Z=BF$

$X \xcv Y=ABCDEF.$

The prerequisites are:

CD $ \xbe $ $ \xdi (ABCD)$ $ \xcc $ $ \xdi (ABCDEF)$

AC $ \xbe $ $ \xdi (ABCD)$ $ \xcc $ $ \xdi (ABCDEF)$

BF $ \xbe $ $ \xdi (ABEF)$ $ \xcc $ $ \xdi (ABCDEF),$ so

ABCD $ \xcc $ ABCDF $ \xbe $ $ \xdi (ABCDEF).$

$ \xcz $
\\[3ex]
\subsubsection{
Remarks on Principal and Relation Generated Filters
}

\bfa

$\hspace{0.01em}$


\label{Fact Mu}

Let $ \xdf (A):=\{A' \xcc A:$ $ \xbm (A) \xcc A' \}$ the principal filter
over $ \xCf A$ generated
by $ \xbm (A),$ then the corresponding $ \xdi (A)=\{A' \xcc A:A' \xcs \xbm
(A)= \xCQ \},$
and $ \xdm (A)=\{A' \xcc A:$ $A' \xcs \xbm (A) \xEd \xCQ,$ and $ \xbm (A)
\xcC A' \}.$
$ \xcz $
\\[3ex]

\efa

\bfa

$\hspace{0.01em}$


\label{Fact Coher}

Let the filters be principal filters.

(1) (Coh1) is equivalent to $(\xbm PR).$

(2) $(\xbm Cum)$ implies (Coh2), and $(Coh1)+(Coh2)$ imply $(\xbm Cum).$

(3) $(\xbm RK)$ implies (Coh-RK), and $(Coh1)+(Coh$-RK) imply $(\xbm
RK).$

\efa

\subparagraph{
Proof
}

$\hspace{0.01em}$


(1) $(\xbm PR)$ $ \xch $ (Coh1):
$A \xbe \xdi (X)$ $ \xcp $ $A \xcs \xbm (X)= \xCQ $ $ \xcp $ $A \xcs \xbm
(Y)= \xCQ $ $ \xcp $ $A \xbe \xdi (Y).$

(Coh1) $ \xch $ $(\xbm PR):$
Suppose there is $X \xcc Y$ such that $(\xbm PR)$ fails, so $ \xbm (Y)
\xcs X \xcC \xbm (X),$ then
$X- \xbm (X) \xbe \xdi (X),$ but $(X- \xbm (X)) \xcs \xbm (Y) \xEd \xCQ,$
so $X- \xbm (X) \xce \xdi (Y).$

(2) $(\xbm CUM)$ $ \xch $ (Coh2):
Let $A,B \xbe \xdi (X),$ $A \xcs B= \xCQ,$ so
$ \xbm (X-B)= \xbm (X)$ $ \xch $ $A \xbe \xdi (X- \xCf B).$

$(Coh1)+(Coh2)$ $ \xch $ $(\xbm CUM):$
Let $ \xbm (X) \xcc Y \xcc X.$
$X- \xCf Y,$ $Y- \xbm (X) \xbe \xdi (X),$ and $(X-Y) \xcs (Y- \xbm (X))=
\xCQ,$ so
$Y- \xbm (X) \xbe \xdi (X-(X-Y))= \xdi (Y),$ so $ \xbm (Y) \xcc \xbm (X).$
$ \xbm (X) \xcc \xbm (Y)$ follows from (Coh1)

(3) $(\xbm RK)$ $ \xch $ (Coh-RK):
Let $A \xbe \xdi (Y)$ $ \xcp $ $A \xcs \xbm (Y)= \xCQ $ $ \xcp $ $A \xcs
\xbm (X)= \xCQ $ $ \xcp $ $A \xbe \xdi (X).$

$(Coh1)+(Coh$-RK) $ \xch $ $(\xbm RK):$
Let $X \xcc Y$ and $X \xce \xdi (Y),$ $(Y- \xbm (Y)) \xbe \xdi (Y),$ so by
(Coh-RK) $(Y- \xbm (Y)) \xcs X \xbe \xdi (X),$
so $(X-(Y- \xbm (Y))) \xbe \xdf (X)$ and $ \xbm (X)$ $ \xcc $ $(X-(Y- \xbm
(Y)))$ $=$ $ \xbm (Y) \xcs X,$ but by
(Coh1) $ \xbm (Y) \xcs X \xcc \xbm (X).$

$ \xcz $
\\[3ex]

We now consider filters generated by preferential structures.

\bd

$\hspace{0.01em}$


\label{Definition Preference}

Let $X \xEd \xCQ,$ $ \xeb $ a binary relation on $X,$
we define for $ \xCQ \xEd A \xcc X$

$ \xbm (A)$ $:=$ $\{x \xbe A:$ $ \xCN \xcE x' \xbe A.x' \xeb x\}$

(This is simplified definition, without ``copies'', see e.g.
 \cite{Sch18} for the full picture.)

We assume in the sequel that for any such $X$ and $ \xCf A,$
$ \xbm (A) \xEd \xCQ.$

We define the following standard properties for the relation $ \xeb:$

(1) Transitivity (trivial)

(2) Smoothness

If $x \xbe X- \xbm (X),$ there there is $x' \xbe \xbm (X).x' \xeb x$

(3) Rankedness

If neither $x \xeb x' $ nor $x' \xeb x,$ and $x \xeb y$ $(y \xeb x),$ then
also $x' \xeb y$ $(y \xeb x').$

(Rankedness implies transitivity.)

See, e.g. Chapter 1 in  \cite{Sch18a}.

\ed

\br

$\hspace{0.01em}$


\label{Remark Pref}

(Simplified)

$(\xbm PR)$ characterizes general preferential structures,

$(\xbm CUM)$ characterizes smooth preferential structures,

$(\xbm RK)$ characterizes ranked preferential structures.

See  \cite{Sch18} for details.
\subsection{
More Detailed Size Comparisons
}

\label{Section Size-N-Relative}

\er

We consider here more detailed size comparisons, and comparisons relative
to
some given set $X.$

\bd

$\hspace{0.01em}$


\label{Definition <}

Given $X,$ $ \xdf (X)$ (and corresponding $ \xdi (X),$ $ \xdm (X)),$ and
$A,B \xcc X,$ we define:

 \xEh
 \xDH
$X \xer A$ iff $X$ is a small subset of A, i.e. $X \xbe \xdi (A),$

 \xDH
$X \xer_{n}A$ iff there are $A_{i},1 \xck i \xck n-1$ such that $X \xer
A_{1} \xer A_{2} \xer  \Xl  \xer A_{n-1} \xer A,$

The index $n$ says how much smaller $X$ is compared to $ \xCf A.$

 \xDH
$A<_{X}B$ $: \xcj $ $A \xbe \xdi (X),$ $B \xbe \xdf (X)$
 \xDH
$A<_{X}' B$ $: \xcj $

(a) $B \xbe \xdf (X)$ and $A \xbe \xdi (X) \xcv \xdm (X)$

or

(b) $B \xbe \xdm (X)$ and $A \xbe \xdi (X)$

 \xDH
If $X=A \xcv B,$ we write $A<B$ and $A<' B,$ instead of $A<_{X}B$ and
$A<_{X}' B.$

(Note: $X \xbe \xdi (X \xcv Y)$ $ \xcj $ $X \xbe \xdi (X \xcv Y)$ $ \xcu $
$Y \xbe \xdf (X \xcv Y)$
 \xDH
$X<_{n}A$ iff there are $A_{i},1 \xck i \xck n-1$ such that
$X<A_{1}<A_{2}< \Xl <A_{n-1}<A.$
 \xEj

\ed

\br

$\hspace{0.01em}$


\label{Remark <}

$X \xbe \xdi (X \xcv Y)$ $ \xch $ $Y \xbe \xdf (X \xcv Y),$ but not
necessarily the converse.

\er

\subparagraph{
Proof
}

$\hspace{0.01em}$


$X \xbe \xdi (X \xcv Y)$ $ \xch $ $(X \xcv Y)-X \xbe \xdf (X \xcv Y),$ and
$(X \xcv Y)-X \xcc Y,$ so $Y \xbe \xdf (X \xcv Y).$

For the converse: Consider $X=Y,$ then $Y \xbe \xdf (X \xcv Y),$ but $X
\xce \xdi (X \xcv Y).$

$ \xcz $
\\[3ex]

Note that case $(4)(b)$ of the definition is impossible if $X=A \xcv B:$
By Remark 
\ref{Remark <} (page 
\pageref{Remark <}), if $A \xbe \xdi (A \xcv B),$ then
$B \xbe \xdf (A \xcv B).$

We will sometimes count applications of (I3) and (Coh2) - equivalently
(Coh2a) -,
see Definition \ref{Definition Filter} (page \pageref{Definition Filter})  and
Definition \ref{Definition Coherence} (page \pageref{Definition Coherence}),
using the
notation $(\xCf ns),$ for $n$ applications of smaller (``s'' for ``size'' or
``smaller'').

\bfa

$\hspace{0.01em}$


\label{Fact <}

$<$ is transitive

\efa

\subparagraph{
Proof
}

$\hspace{0.01em}$


Let $X<Y<Z,$ so $X \xbe \xdi (X \xcv Y)$ and $Y \xbe \xdi (Y \xcv Z).$ We
have to show $X \xbe \xdi (X \xcv Z).$

Consider $X \xcv Y \xcv Z,$ then by $X \xbe \xdi (X \xcv Y),$ $X \xbe \xdi
(X \xcv Y \xcv Z).$ By the same argument,
$Y \xbe \xdi (X \xcv Y \xcv Z),$ thus $Y-(X \xcv Z) \xbe \xdi (X \xcv Y
\xcv Z).$ As $(X \xcv Y \xcv Z)-(Y-(X \xcv Z))=X \xcv Z,$
and $X \xcs (Y-(X \xcv Z))= \xCQ,$ $X \xbe \xdi (X \xcv Z)$ by
Definition 
\ref{Definition Coherence} (page 
\pageref{Definition Coherence}), (Coh2a).
$ \xcz $
\\[3ex]

\bfa

$\hspace{0.01em}$


\label{Fact Size-1}

(See  \cite{Sch18}.)

 \xEh

 \xDH
$A \xer B \xer C$ $ \xch $ $A \xer C$ (without any (ns))

 \xDH
$A \xer B \xer A$ is impossible

 \xDH
$A \xcc B<C$ $ \xch $ $A<C$ (1s)

 \xDH
$A<B \xcc C$ $ \xch $ $A<C$ (without any (ns))

 \xDH
$X<Y<Z$ $ \xch $ $X<Z$ (1s)

 \xDH
$X<Y<X$ is impossible (1s)

 \xDH
$X<Y<Z<U$ $ \xch $ $X<U$ (2s)

 \xDH
$X_{0}< \Xl <X_{n}$ $ \xch $ $X_{0}<X_{n}$ by $((n-1)*s)$

 \xDH
$X<Y$ $ \xch $ $X \xcs Y \xer Y$ (1s)

 \xDH
$X \xcs Y \xer X \xcv Y,$ $X-Y \xer X$ $ \xch $ (1s) $X \xcs Y \xer Y$
 \xEj

\efa

\subparagraph{
Proof
}

$\hspace{0.01em}$


 \xEh

 \xDH
By (I2) or (Coh1).

 \xDH
By (1).

 \xDH
By $B \xer B \xcv C$ and $B-A \xcc B,$ we have $B-A \xer B \xcv C$ by $(I
\xcc).$
So by (Coh2a) $A=B-(B-A) \xer B \xcv C-(B-A) \xcc A \xcv C,$ the latter by
(Coh1).

 \xDH
$A \xer A \xcv B \xcc A \xcv C$ $ \xch $ $A \xer A \xcv C$ by (Coh1).

 \xDH
We have $X \xer X \xcv Y,$ $Y \xer Y \xcv Z,$ so $X \xer X \xcv Y \xcv Z$
and $Y \xer X \xcv Y \xcv Z,$ so
$Y-(X \xcv Z) \xer X \xcv Y \xcv Z.$
So (1s) $X=X-(Y-(X \xcv Z)) \xer X \xcv Y \xcv Z-(Y-(X \xcv Z))=X \xcv Z.$

 \xDH
By (6) and (7).

 \xDH
By $X \xer X \xcv Y,$ $Y \xer Y \xcv Z,$ $Z \xer X \xcv U,$ so $Y \xcv Z
\xer X \xcv Y \xcv Z \xcv U$ (1s).
So $Y \xcv Z-X \xcv U \xcc Y \xcv Z \xer X \xcv Y \xcv Z \xcv U.$
$X \xcv U=X \xcv Y \xcv Z \xcv U-(Y \xcv Z-X \xcv U),$ $X=X-(Y \xcv Z-X
\xcv U).$
So $X \xer X \xcv U$ (1s).

 \xDH
Analogous to (7).

 \xDH
$X<Y$ $ \xch $ $X \xer X \xcv Y,$ so $X-Y \xer X \xcv Y=Y \xcv (X- \xCf
Y),$ and $X=(X \xcs Y) \xcv (X- \xCf Y).$
Thus (1s) $X \xcs Y \xer Y.$

 \xDH
$X-Y \xer X \xcc X \xcv Y$ $ \xch $ (1s) $X \xcs Y=X \xcs Y-(X-Y) \xer X
\xcv Y-(X-Y)=Y.$
 \xEj

$ \xcz $
\\[3ex]

We mention the following without proofs, it is very close to
Fact \ref{Fact NoChange-2} (page \pageref{Fact NoChange-2}).

\bfa

$\hspace{0.01em}$


\label{Fact Subset-2}

$(Coh1)+(Coh2)$ imply:

(1)
Let $X \xbe \xdf (X'),$ then $(X \xcs A \xbe \xdf (X)$ $ \xcj $ $X' \xcs
A \xbe \xdf (X'))$

(2)
Let $X' \xbe \xdf (X),$ $Y' \xbe \xdf (Y),$ then the following four
conditions
are equivalent:

$X<Y,$ $X' <Y,$ $X<Y',$ $X' <Y' $

\efa

\bfa

$\hspace{0.01em}$


\label{Fact Trans-Rank}

If (Coh-RK) holds, then $<' $ is transitive.

\efa

\subparagraph{
Proof
}

$\hspace{0.01em}$


First, if $A \xbe \xdi (A \xcv B),$ then $B \xbe \xdf (A \xcv B),$ so we
need not consider
case $(4)(b)$ in Definition \ref{Definition <} (page \pageref{Definition <}).

Second, we will use repeatedly:

$(*)$ If $RSTU \xbe \xdm (RSTUVW),$ and $RSVW \xbe \xdf (RSTUVW),$ so $TU
\xbe \xdi (RSTUVW),$ then
by (Coh-RK) $TU \xbe \xdi (RSTU),$ and thus also $T \xbe \xdi (RST)$ and
$U \xbe \xdi (RSU)$ by
(Coh2).

Consider

$A:=+X+Y+Z$

$B:=+X+Y-Z$

$C:=+X-Y+Z$

$D:=+X-Y-Z$

$E:=-X+Y+Z$

$F:=-X+Y-Z$

$G:=-X-Y+Z$

$H:=-X-Y-Z$

and

$X:=ABCD$

$Y:=ABEF$

$Z:=ACEG.$

Then

$X \xcv Y=ABCDEF$

$X \xcv Z=ABCDEG$

$Y \xcv Z=ABCEFG.$

There are 4 cases to consider:

\xEn

 \xDH (a)

$X \xbe \xdi (X \xcv Y),$ $Y \xbe \xdf (X \xcv Y),$
$Y \xbe \xdi (Y \xcv Z),$ $Z \xbe \xdf (Y \xcv Z)$

 \xDH (b)

$X \xbe \xdm (X \xcv Y),$ $Y \xbe \xdf (X \xcv Y),$
$Y \xbe \xdm (Y \xcv Z),$ $Z \xbe \xdf (Y \xcv Z)$

 \xDH (c)

$X \xbe \xdm (X \xcv Y),$ $Y \xbe \xdf (X \xcv Y),$
$Y \xbe \xdi (Y \xcv Z),$ $Z \xbe \xdf (Y \xcv Z)$

 \xDH (d)

$X \xbe \xdi (X \xcv Y),$ $Y \xbe \xdf (X \xcv Y),$
$Y \xbe \xdm (Y \xcv Z),$ $Z \xbe \xdf (Y \xcv Z)$

\xEp

In each case we show $Z \xbe \xdf (X \xcv Z),$ $X \xce \xdf (X \xcv Z),$
i.e.

$ACEG \xbe \xdf (ABCDEG)$ and $ABCD \xce \xdf (ABCDEG).$

The proofs are elementary and tedious.

\xEn

 \xDH Case (a)

This was done already in
Fact \ref{Fact Size-1} (page \pageref{Fact Size-1}), (5).

 \xDH Case (b)

$X \xbe \xdm (X \xcv Y),$ $Y \xbe \xdf (X \xcv Y),$
$Y \xbe \xdm (Y \xcv Z),$ $Z \xbe \xdf (Y \xcv Z),$ thus:

$ABCD \xbe \xdm (ABCDEF),$ $ABEF \xbe \xdf (ABCDEF),$
$ABEF \xbe \xdm (ABCEFG),$ $ACEG \xbe \xdf (ABCEFG).$

We show

$ACEG \xbe \xdf (ABCDEG),$ $ABCD \xce \xdf (ABCDEG).$

By applying $(*)$ twice, we have $CD \xbe \xdi (ABCD)$ and $BF \xbe \xdi
(ABEF),$ thus
$B \xbe \xdi (ABE)$ too.

Thus $BCD \xbe \xdi (ABCDE) \xcc \xdi (ABCDEG),$ so $AEG \xbe \xdf
(ABCDEG),$ so $ACEG \xbe \xdf (ABCDEG).$

Suppose $ABCD \xbe \xdf (ABCDEG),$ so $EG \xbe \xdi (ABCDEG),$ so by $BCD
\xbe \xdi (ABCDEG),$
$BCDEG \xbe \xdi (ABCDEG),$ so $BCEG \xbe \xdi (ABCEG) \xcc \xdi
(ABCEFG),$ so $CG \xbe \xdi (ABCEFG),$ and
$ABEF \xbe \xdf (ABCEFG),$ contradiction.

 \xDH Case (c)

$X \xbe \xdm (X \xcv Y),$ $Y \xbe \xdf (X \xcv Y),$
$Y \xbe \xdi (Y \xcv Z),$ $Z \xbe \xdf (Y \xcv Z),$ thus:

$ABCD \xbe \xdm (ABCDEF),$ $ABEF \xbe \xdf (ABCDEF),$
$ABEF \xbe \xdi (ABCEFG),$ $ACEG \xbe \xdf (ABCEFG).$

By applying $(*)$, we have $CD \xbe \xdi (ABCD),$ thus $C \xbe \xdi
(ABC)$ and $D \xbe \xdi (ABD).$
Thus, by $ABEF \xbe \xdi (ABCEFG),$ $ABCEF \xbe \xdi (ABCEFG),$ and $ABCE
\xbe \xdi (ABCEG).$
Thus, $B \xbe \xdi (BG),$ $D \xbe \xdi (ABD),$ and $BD \xbe \xdi
(ABCDEG),$ so $ACEG \xbe \xdf (ABCDEG).$

On the other hand, $ABCE \xbe \xdi (ABCEG) \xcc \xdi (ABCDEG)$ and by $D
\xbe \xdi (ABD),$
$ABCD \xbe \xdi (ABCDEG).$

 \xDH Case (d)

$X \xbe \xdi (X \xcv Y),$ $Y \xbe \xdf (X \xcv Y),$
$Y \xbe \xdm (Y \xcv Z),$ $Z \xbe \xdf (Y \xcv Z),$ thus:

$ABCD \xbe \xdi (ABCDEF),$ $ABEF \xbe \xdf (ABCDEF),$
$ABEF \xbe \xdm (ABCEFG),$ $ACEG \xbe \xdf (ABCEFG).$

By applying $(*)$, we have $BF \xbe \xdi (ABEF),$ thus $B \xbe \xdi
(ABE)$ and $F \xbe \xdi (AEF).$

By $ABCD \xbe \xdi (ABCDEF)$ and $F \xbe \xdi (AEF),$ we have $ABCDF \xbe
\xdi (ABCDEF),$ so
$(**)$ $ABCD \xbe \xdi (ABCDE),$ so $BD \xbe \xdi (ABCDEG)$ and $ACEG \xbe
\xdf (ABCDEG).$

Suppose $ABCD \xbe \xdf (ABCDEG),$ then $EG \xbe \xdi (ABCDEG),$ so $E
\xbe \xdi (ABCDE),$
contradicting $(**).$

\xEp

$ \xcz $
\\[3ex]
\subsection{
Filters Generated by Preferential Relations
}

\label{Section Filt-Pref}

When we discuss $ \xeb $ on $U,$ and $<_{X},$ $<,$ $<_{X}',$ $<' $ for
subsets of $U,$ we
implicitly mean the filters, ideals, etc. generated by $ \xbm $ on subsets
of $U,$
as discussed in Fact \ref{Fact Mu} (page \pageref{Fact Mu}),
or a relation $ \xeb $ on $U,$ as defined in
Definition \ref{Definition Preference} (page \pageref{Definition Preference}).

We give some examples.

\be

$\hspace{0.01em}$


\label{Example <-1}

 \xEh
 \xDH Let $ \xeb $ not be transitive.

Let $z \xeb y \xeb x,$ then $\{x\}<_{\{x,y\}}\{y\},$
$\{y\}<_{\{y,z\}}\{z\},$ but $\{x\} \xEc_{\{x,z\}}\{z\},$ as
$ \xbm (\{x,z\})=\{x,z\}.$ However, $\{x\}<_{\{x,y,z\}}\{z\}.$
 \xDH Let $ \xeb $ be transitive.

In Case (1), add $z \xeb x,$ then $\{x\} \xeb_{\{x,z\}}\{z\}.$

 \xDH Let $ \xeb $ again be transitive.

Consider $A=\{a\},$ $C=\{c\},$ $B=\{b_{i}:i< \xbo \} \xcv \{b\},$ and
$b_{i} \xeb a,$ $c \xeb b,$ $b_{i+1} \xeb b_{i}.$
$ \xeb $ is transitive. Then $ \xbm (A \xcv B)=\{b\},$ $ \xbm (B \xcv
C)=\{c\},$ $ \xbm (A \xcv C)=\{a,c\}.$
So $A<_{A \xcv B}B,$ $B<_{B \xcv C}C,$ but not $A<_{A \xcv C}C.$ However,
$ \xbm (A \xcv B \xcv C)=C,$ so
$A<_{A \xcv B \xcv C}C.$

 \xDH Let $ \xeb $ again be transitive.

Let $X=\{x\} \xcv \{x_{i}:i< \xbo \},$ $Y=\{y\},$ $Z=\{z\} \xcv
\{z_{i}\},$ and $y \xeb x,$ $z_{0} \xeb y,$ $x_{0} \xeb z,$ so $z_{0} \xeb
x$
by transitivity (and $x_{j} \xeb x_{i}$ for $j>i$ etc.).

We have $ \xbm (X \xcv Y)=\{y\},$ $ \xbm (Y \xcv Z)=\{z\},$ $ \xbm (X \xcv
Z)= \xCQ,$ so
$X<_{X \xcv Y}Y,$ $Y<_{Y \xcv Z}Z,$ but $X \xEc_{X \xcv Z}Z$ and $X
\xEc_{X \xcv Y \xcv Z}Z.$

 \xDH
Again, $ \xeb $ is transitive, in addition, $ \xeb $ is smooth.

Consider
$X:=\{x_{2},x_{3},x_{4}\},$ $Y:=\{x_{1},x_{2},y\},$
$x_{4} \xeb x_{2},$ $y \xeb x_{3},$ $y \xeb x_{1}$ (the transitivity
condition is empty).

Then
$ \xbm (X)=\{x_{3},x_{4}\},$ $\{x_{3}\} \xbe \xdm (X),$ $\{x_{2}\} \xbe
\xdi (X),$ $\{x_{2}\}<'_{X}\{x_{3}\}.$

$ \xbm (Y)=\{x_{2},y\},$ $\{x_{2}\} \xbe \xdm (Y),$ $\{x_{1}\} \xbe \xdi
(Y),$ $\{x_{1}\}<'_{Y}\{x_{2}\}.$

Let $x_{1},x_{3} \xbe Z,$ is $\{x_{1}\}<'_{Z}\{x_{3}\}?$

If $y \xbe Z,$ $\{x_{1}\},\{x_{3}\} \xbe \xdi (Z).$

If $y \xce Z,$ $\{x_{1}\},\{x_{3}\} \xbe \xdm (Z).$

So, in both cases, $\{x_{1}\} \xEc'_{Z}\{x_{3}\},$ as they have the same
size.

 \xEj

\ee

\be

$\hspace{0.01em}$


\label{Example Absolute}

$<$ is neither upward nor downward absolute.
Intuitively, in a bigger set, formerly big sets might become small,
conversely, in a smaller set, formerly small sets might become big.

Let $A,B \xcc X \xcc Y.$ Then

(1) $A<_{X}B$ does not imply $A<_{Y}B$

(2) $A<_{Y}B$ does not imply $A<_{X}B$

(1): Let $Y:=\{a,b,c\},$ $X:=\{a,b\},$ $c \xeb b \xeb a.$ Then
$\{a\}<_{X}\{b\},$ but
both $\{a\},\{b\} \xbe \xdi (Y).$

(2): Let $Y:=\{a,b,c\},$ $X:=\{a,c\},$ $c \xeb b \xeb a,$ but NOT $c \xeb
a.$
Then $\{a\}<_{Y}\{c\},$ but both $\{a\},\{c\} \xbe \xdm (X).$

\ee

For homogenousness, we chose violation in a comparatively smaller subset.
As said above, this corresponds to the
non-monotonicity idea, and, intuitively, going from a big set to a very
small
set, more things can happen. The smaller a subset, the less likely
homogenousness is expected.

We then have a construction similar to defeasible inheritance as
metatheory, so overall a coherent approach on object and meta level.
\clearpage
\section{
Application to Argumentation
}


\subsection{
Introduction
}
\paragraph{
Abstract Description
}

Argumentation is about putting certain objects together.
The interested reader might compare this to the constructions in
Section \ref{Section NEU} (page \pageref{Section NEU})  and in
Chapter \ref{Chapter CFC} (page \pageref{Chapter CFC}).

There are three things to consider:
 \xEh
 \xDH the objects themselves, and their inner structure (if they have any)
-
this inner structure may be revealed successively, or be immediately
present,
 \xDH rules about how to put them together,
 \xDH avoid certain results (contradictions) in the resulting pattern.
 \xEj

To help intuition, we picture as result of an argumentation, an
inheritance
network the agents can agree on.

This network may consist of strict
and defeasible rules only, with no elements or sets it is applied to.
Think of the argumentation going on when writing a book about
medical diagnosis. This will not be about particular cases, but about
strict and default rules. ``Sympton $x$ is usually a sign of illness $y,$
but there are the following exceptions:  \Xl''
In addition, the network might contain cycles. There is nothing wrong with
cycles.
Mathematics is full of cycles, equivalences and their proofs. But consider
also the following: We work in the set of adult land mammals. ``Most
elefants
weigh more than 1 ton.'' ``Most elements (i.e. adult land mammals) which
weigh more than 1 ton are elefants.'' There is nothing in principle wrong
with
this either - except, in reality, we forgot perhaps about hippopotamus
etc.

Arguments need not be contradictions to what exists already. They can be
confirmations, elaborations, etc. For instance, we might have the default
rule that birds fly, and clarify that penguins don't fly. This is not a
contradiction, but an elaboration.
\paragraph{
The structure of the objects
}

Facts are either so simple that a dispute seems unreasonable. Or, they are
a combination of basic facts and (default) rules, like, what $ \xfI $ see
through
my microscope is really there, and not an artifact of some speck of
dust on the lenses. For simplicity, facts will be basic, undisputable
facts.

Expert opinion may be considered a default rule, where details stay
unexplained, perhaps even unexplainable by the expert himself.

Rules (classical or defaults) have three aspects:
 \xEh
 \xDH the rule itself,
 \xDH the application of the rule,
 \xDH the result of the application of the rule.
 \xEj

Classical rules cannot be contested. We can contest their application,
i.e. one of their prerequisites, or their result, and, consequently,
their application. We can confirm their result by different means,
likewise, their application.

Default rules are much more complicated, but not fundamentally different.
Again, we can attack their
application, by showing that one of the prerequisites does not hold, or,
that we are in an (known) exceptional case. We can attack the conclusion,
and, consequently, the rule, or its application. In particular, we may
attack the conclusion, without attacking the application or the rule
itself,
by arguing that we are in a surprising exceptional case - and perhaps try
to find a new set of exceptions. We can attack the default rule itself,
as in the case of ``normally, tigers are vegans''. We can confirm a rule
by confirming its conclusion, or adding a new rule, which gives the same
result. We can elaborate a default rule, by adding an exception set,
stating
that all exceptions are known, and give the list of exceptions, etc.
We can stop homogenousness (downward inheritance) e.g. for Quakers which
are Republicans, we stop inheriting pacifism (or its opposite).
This is not a contradiction to the default itself, but to the
downward inheritance of the default (or to homogenousness) by
meta-default,
to be precise.

Obviously, the more we add (possible) properties to the objects (here
default
rules), the more we
can attack, elaborate, confirm.

In the following section, we describe our general picture:
 \xEh
 \xDH there is an
arbiter which checks for consistency, and directs the discussion,
 \xDH how to
handle classical arguments and resulting contradictions,
 \xDH how to handle default arguments.
 \xEj
\subsection{
The Classical Part
}

We suppose there is an arbiter, whose role is to check consistency, and to
authorise participants to speak.

If the arbiter detects an inconsistency, then he points out the
``culprits'',
i.e. minimal inconsistent sets. As he detects inconsistencies immediately,
the last argument will be in all those sets. The last argument need not
be the problem, it might be one of the earlier arguments.

He asks all participants if they wish to retract one of the arguments
involved in at least one minimal inconsistent set.
(They have to agree unanimously on such retraction.)
If there is no minimal
inconsistent set left, the argumentation proceeds with the ``cleaned''
set of arguments, as if the inconsistency did not arise.
Of course, arguments which were based on some of the retracted arguments
are now left ``hanging in the air'', and may be open to new attacks.

If not, i.e. at least one minimal inconsistent set is left, the
participants
can defend (and attack) the arguments involved in those sets.
The arbiter will chose the argument to be attacked/defended.
See Example 
\ref{Example Symmetrical} (page 
\pageref{Example Symmetrical})  below.
Suppose $ \xba $ is the argument chosen, then a defense will try to prove
or
argue for $ \xba,$ an attack will try to prove or argue for $ \xcx \xCN
\xba,$ i.e.
it is possible or consistent that $ \xCN \xba.$
In particular, an attacker might try to prove $ \xcT,$ or some other
unlikely
consequence of $ \xba $ (and some incontested $ \xbb $'s), and he need
not begin
with some $ \xba \xcp \xbg,$ it might be a more roundabout attack.

If at least one minimally inconsistent set is left with all elements
defended, then there is a deadlock, and the arbiter declares failure.

Consider

\be

$\hspace{0.01em}$


\label{Example Symmetrical}

We argue semantically. Let $A:=\{x,a\},$ $B:=\{x,b\},$ $C:=\{x,c\},$
$Y:=\{a,b,c\}.$
Let $Y$ be the last set added. For $A,B,C,$ the situation is symmetrical.
Let $Z \xEd Z' $ be $A,B,$ or $C,$ then $Y \xcs Z \xEd \xCQ,$ but $Y \xcs
Z \xcs Z' = \xCQ,$ $Z \xcs Z' =\{x\},$ etc.
Moreover, $A \xcs B \xcc C,$ etc. Thus, $ \xCf A$ and $B$ together are an
argument for $C,$ etc.,
so they argue for each other, and there is no natural way to chose any
of $A,B,C$ to be attacked. Thus, it is at the discretion of the parties
involved (or the arbiter) to chose the aim of any attack - apart from $Y,$
which is not supported by any of $A,B,C.$ Still, $Y$ might in the end be
the strongest argument.

We may add $D,$ $E,$ with $D:=\{x,d\},$ $Y:=\{a,b,c,d\}$ etc., the example
may be extended
to arbitrarily many sets.

\ee

At any moment, any argument can be attacked, not only if an inconsistency
arises. We may continue an argumentation, even if not all minimally
inconsistent subsets are treated as yet, but the arbiter has to keep track
of
them, and of the use of their elements. They and their consequences may
still be questioned.
\subsection{
Defaults
}
\paragraph{
The classical part of defaults
}

We see defaults primarily not as rules, but as relatively complicated
classical constructions, which we may see as objects for the moment.
The default character is in applying those objects, not in the
objects themselves.

We follow here the theory described in Chapter 11 of
 \cite{GS16}.

In our view, a (semantical) default $(\xCf X:Y)$ says:

 \xEh
 \xDH ``most'' elements of $X$ are in $Y,$
 \xDH there may be exception sets $X_{1},$ $X_{2},$ etc. of $X,$ where the
elements
are ``mostly'' not in $Y$ (but $X_{1} \xcv X_{2} \xcv  \Xl.$ has to be a
``small'' subset of $X),$
 \xDH in addition, there may be a ``very small'' subset $X' \xcc X,$ which
contains
``surprise elements'' (i.e. not previously known exceptions), which are not
in $Y,$
 \xDH in addition, we may require that subsets of $X$ ``normally''
behave in a homogenous way.
 \xEj

The notions of ``most'', ``small'' etc. are left open, a numerical
interpretation
suffices for the intuition. These notions are discussed in depth e.g. in
 \cite{GS08f} and  \cite{GS10}.

Introducing a default has to result in a (classically) consistent
theory. E.g., it must not be the case that $ \xcA x \xbe X.x \xce Y,$ this
contradicts
the first requirement about defaults (and any reasonable interpretation
of ``most'').
\paragraph{
The default part of defaults
}

This leads to a hierarchy as defined in
Section 11.4.1 of  \cite{GS16}.
We use the hierarchy to define the $ \xCf use$
of the defaults.

To use the standard example with birds, penguins, fly, we proceed as
follows. Suppose we introduce a bird $x$ into the discussion.
We try to put $x$ as low as possible in the hierarchy, i.e. into the
set of birds, but not into any known exception set, and much less into
any ``surprise'' set. Only (classical) inconsistency, as checked by the
arbiter, may force us to climb higher.
Thus, unless there is a contradiction, we let $x$ fly.
\paragraph{
Attacks against defaults and their conclusions
}

Classical rules are supposed to be always true. Thus, classical
rules themselves cannot be attacked, and an attack against a
classical conclusion has to be an attack against one of its
prerequisites.

Attacks against defaults can be attacks against
 \xEh
 \xDH the rule itself,
 \xDH one of the prerequisites,
 \xDH membership in or not in one of the exception sets,
 \xDH membership in or not in the surprise set,
 \xDH perhaps even the notions of size involved,
 \xDH etc.
 \xEj

Each component of a default rule may be attacked.
\subsection{
Comments
}

We assume that there
is no fundamental difference between facts and conclusions:
Usually, we were told facts, remember facts, have read facts, observed
facts (perhaps with the help of a telescope etc.). These things
can go wrong. Situations where things are obvious, and no error
seems humanly possible, will not be contradicted.
\paragraph{
Auxiliary elements
}

We now introduce some auxiliary elements which may help in the
argumentation.

 \xEh
 \xDH ``$ \xfI $ agree.''

This makes an error in this aspect less likely, as both parties agree -
but
still possible!

 \xDH ``$ \xfI $ confirm.''

$ \xfI $ am very certain about this aspect.

 \xDH
Expert knowledge:

Expert knowledge and its conclusions act as ``black box defaults'', which
the expert himself may be unable to analyse. Other experts (in the same
field)
will share the conclusion. (This is simplified, of course.)

(One way to contest an expert's
conclusion is to point out that he neglected an aspect of the situation,
which is outside his expertise. His ``language of reasoning'' is too poor
for the situation.)

 \xDH
The arbiter may ask questions.

 \xEj
\paragraph{
Examples of attacks
}

 \xEh
 \xDH
Defaults:

Normally, there is a bus line number 1 running every 10 minutes between 10
and
11 in the morning.

Attack: No, the conclusion is wrong.

Question: Why?

Elaboration:

 \xEh
 \xDH
No, the default is wrong (e.g.: it is line number 2 running every 10
minutes).
 \xDH
Yes, but this is not homogenous, i.e. does not break down to subsets,
and we know more. (For instance, we know that today is Tuesday or
Wednesday,
and it runs that often only Monday, Thursday, Friday, Saturday, Sunday -
but
we do not know this, only that is does not apply to all days of the week.)

 \xDH
Yes, but today is an exception, and we know this. (e.g., we know that
today is Tuesday, and we know that Tuesday is an exception.)
(In addition, there might be exceptional Tuesdays, Christmas market day,
etc.  \Xl)

 \xDH
Yes, but $ \xfI $ do not know why this is an exception. (This is a
surprise case,
$ \xfI $ know about different days, but today should not be an exception,
still
$ \xfI $ was just informed that it does not hold today.) We do not attack
the
default, nor the applicability - but agree that it fails here.

 \xEj

 \xDH
Classical conclusions:

From $ \xCf A$ and $B,$ $C$ follows classically.

Attack: $C$ does not hold.

Question: Why?

Elaboration:

 \xEh
 \xDH
$ \xCf A$ does not hold or $B$ does not hold, but $ \xfI $ do not know
which.

 \xDH
$ \xCf A$ does not hold.

 \xDH
$B$ does not hold.

 \xDH
$ \xCf A$ does not hold, and $B$ does not hold.

 \xEj

 \xDH
Fact: $ \xCf A$ holds.

Attacks: No, $ \xCf A$ does not hold.

Question: Why?

Elaboration:

 \xEh
 \xDH
You remember incorrectly.

 \xDH
You did not observe well.

 \xDH
Your observation tools do not work.
 \xDH
You were told something wrong.
 \xDH
etc.
 \xEj

 \xDH
Expert knowledge, expert concludes that $ \xCf A.$

Attack: $ \xCf A$ does not hold.

Question: Why?

Elaboration:

The situation involves aspects where you are not an expert. It is beyond
your language. (Of course, the expert can ask for elaboration  \Xl.)

 \xEj

$ \xCO $

We did not treat here:
 \xEh
 \xDH when it is necessary to remember not only the result of an argument,
but also the way it was reached,
(compare this to
Chapter \ref{Chapter PAR} (page \pageref{Chapter PAR}), where we
remembered whether we approximated from above or from below.
 \xDH the usual ``dirty tricks'' of political argumentation like:

- changing focus,

- attack unimportant details, etc.
 \xEj
\subsection{
Various other Applications
}

 \xEh
 \xDH Inheritance:

We refer the reader to
Section 5.8 of  \cite{Sch18b},
which contains a detailed discussion,
and just add some remarks.

Apart from any formal reasons, there might be philosophical or even
pragmatic arguments to choese one way or the other for

 \xEh

 \xDH reference classes: a user might think boolean combinations
of reference classes natural, or surprising, this may influence
our decision,

 \xDH we may consider, beyond specificity, length of path, in particular
if we have proof that the arrows in the path describe size relations
(see Fact \ref{Fact TweetySize} (page \pageref{Fact TweetySize})),

 \xDH the decision for extensions or direct scepticism might depend on
whether
the problem reflects a lack of information, or rather too much
contradictory
information.

 \xEj

 \xDH

We developed similar ideas in Chapter 11 of
 \cite{GS16},
and refer the reader there.

 \xDH
Analogical reasoning:
See Chapter \ref{Chapter ANA} (page \pageref{Chapter ANA}).

 \xEj
\vspace{3mm}


\vspace{3mm}

\vspace{3mm}


\vspace{3mm}

\vspace{3mm}


\vspace{3mm}

\clearpage

$ \xCO $
\markboth{\centerline{\scriptsize Truth and Reliability}}
{\centerline{\scriptsize Truth and Reliability}}

$ \xCO $
\chapter{
A Reliability Theory of Truth
}

\label{Section TRU}

\label{Chapter TRU}

$ \xCO $
\section{
Introduction: Motivation, Example and Basic Idea
}
\subsection{
Motivation
}

Our motivation is not to detect inconsistencies in present theories of
truth, and how to remedy them, but to separate truth from falsity in a
flood of information.

The problem is acerbated by a strategy to destroy truth as an important
criterion in political and other discussions. Jonathan Rauch's
important book  \cite{Rau21},
discusses these efforts in detail, culminating
perhaps in Steve Bannon's ``\Xl. flood the zone with shit.''
(Of course, other countries' behaviour is not better, see China, Russia,
etc.)

Similar problems appear in the myths surrounding the Covid pandemic,
where rumours without the slightest factual foundation abound.

Thus, we think, it is very important to have a theory that tries to help
distinguish facts from myths (or worse), leading perhaps even to
algorithms which help to sort today's flood of (dis-)information.
\subsection{
Example and Basic Idea
}

We continue with a simple example.

\be

$\hspace{0.01em}$


\label{Example Tru-1}

Suppose we want to know the temperature in a room. We have four
thermometers,
and no other way to know the temperature.

$T_{1}$ one says 20 $C,$ $T_{2}$ says 19 $C,$ $T_{3}$ 21 $C,$ and $T_{4}$
says 30 $C.$ Thus, $T_{4}$
reports an exceptional value, and we doubt its reliability.

How do we model this? A simple idea is as follows: Each $T_{i}$ is given a
reliability $ \xbr (T_{i})$ between 0 (totally unreliable) and 1 (totally
reliable). At
the
beginning, each $ \xbr (T_{i})$ is a neutral value, say 0.5.
We now calculate the mean value, $90/4=$ 22.5. As we start with equal
reliability,
each $T_{i}$ is given the same weight 0.5. We see now that $T_{4}$ is
exceptional, and
adjust reliabilities, e.g. $ \xbr (T_{4})=1/3,$ and $ \xbr (T_{i})=2/3$
for the other $i.$
If, in the next moment, all $T_{i}$ give again the same data, we will
adjust
the mean value, by counting the values for $T_{1}$ to $T_{3}$ twice, the
value for
$T_{4}$ once, and divide by 7, resulting in $(120+30)/7$ $=$ 21.43. Etc.

\ee

This is our basic idea. It seems a reasonable way to treat contradictory
numerical information, and some variant is probably used in many
``real life'' situations where we need some information, cannot trust
absolutely any single source, but need the information, e.g. to act.

We do not doubt that there is some ``real'' temperature of the room, but
this is
irrelevant, as we cannot know it. We have to do with what we know, but
are aware that additional information might lead us to revise our
estimate.

There are a number of ways to elaborate, modify, and apply to different
situations.

 \xEh
 \xDH

First, we work here with numerical values, both the data and reliabilities
are
real numbers, so is the mean value. We want to do more. We want to work
with
totally ordered sets instead of reals, at least for the data, then with
partial orders complete under sup and inf, perhaps complement, and,
finally,
with arbitrary partial orders. So, we have to try to adapt our data
and operations in some way or the other to the limited possibilities of
the
structure at hand. We do not claim that our suggestions are the only or
best
ways to proceed, the ``right'' way may also depend on the situation.
We note here which operations on the reals we use, and refer the reader to
Chapter \ref{Chapter PAR} (page \pageref{Chapter PAR})
for adaptations to weaker structures.

 \xDH

Then, even the numerical case need not have a unique solution. For
instance,
when calculating the mean value, we might give less (or more!) weight
to exceptional values, without considering reliabilities. This could be
done,
e.g., by calculating first the usual mean, and then ``pull'' the
exceptional values closer to the mean, and calculate the mean value again.
In above example, once we calculated the mean value, 22.5, we note the
exceptional difference between 22.5 and 30, modify 30 to 27.5, and
calculate
the mean of $\{19,20,21,27.5\},$ etc.

We will not follow all such possibilies, but use abstract functions,
here ``mean'', in Example \ref{Example Tru-1} (page \pageref{Example Tru-1})
we have $mean(\{19,20,21,30\})=22.5.$ We will, however,
indicate the translation of all variants discussed in detail to less
rich domains.

 \xDH

The communication channels may have a reliability, too. So the message
which
arrives has a combined reliability of the agent's reliability and the
messages reliability. How do we calculate this combination, and
conversely,
when adjusting the overall reliability, how do we adjust the individual
ones of agent and channel?

 \xDH

In our example, we have one agent which listens and calculates, the other
agents
measure and send values to the ``central'' agent. Moreover, the sending is
synchronised. We might also have situations where the measuring and
calculating
agents are the same, and the messages are broadcast.

 \xDH

The measuring agents might have an estimate about their own reliability,
and communicate this with their data. E.g., in above Example,
a thermometer may have different precisions for different temperature
ranges, e.g., very good from 10 to 30 degrees $C,$ from 0 to 10, and 30 to
40
not so good, etc.

 \xDH

The agents may have opinions about the reliabilities of other agents,
think of politicians who consider each others crooks, so the data may not
only
be ``facts'', but also reliabilities of other agents.

 \xDH

The history should perhaps be preserved beyond the individual
reliabilities.
Suppose we measured above temperature repeatedly (and the temperature is
supposed to be constant), then, in order to calculate the mean over time,
we need to memorize past results or mean values in some way.

 \xEj

Agents may be people, devices like thermometers, theories, etc.
Sometimes, it is more adequate to see reliability as degree of competence,
for instance for moral questions.
Messages may be numbers, but also statements, like the earth is flat.
The formal treatment of such cases is discussed in
Chapter \ref{Chapter PAR} (page \pageref{Chapter PAR}).

A human agent may be a good chemist, but a poor mathematician, so his
reliability varies with the subject. For simplicity, we treat this
agent as two diffent agents, $A$-Chemist, $A$-Mathematician, etc.

Philosophical theories of truth are often mainly about contradictions,
in the tradition of the liar paradox
(see Chapter \ref{Chapter YAB} (page \pageref{Chapter YAB})).
Our approach is very different. We create on the fly new truth values,
they do not stand for ``true'' and ``false'', but for more or less reliable,
and whenever we need a new value of reliability, we create it.

We will say more about the comparison of out idea to other theories of
truth
below.

Note that a theory and corresponding algorithms to help decide between
reliable
und unreliable information
are particularly important in the present flood of misinformation.
Similarly, stock markets need good algorithms to differentiate between
changes based on underlying facts and mere contagion of behaviour
between agents.
\section{
In more Detail
}

We now address above points.
\subsection{
The Basic Scenario With some Features Added
}
\subsubsection{
The Basic Scenario with History Added
}

Agents $A_{i}$ send numerical values $r_{i}$ to a central evaluation agent
$E.$
These are the only messages sent. Each agent $A_{i}$ has a real value
reliability
$ \xbr_{i} \xbe [0,1]$ which is determined by $E.$ 0 stands for total
unreliability,
1 for total reliability. At the outset, each $ \xbr_{i}$ will have the
neutral
value 0.5. (The agents will not know their reliabilities, nor those of
other agents.)

We will indicate the (additional for Variant 2 upward and the following
sections) operations needed,
and which will have to be adapted in non-numerical settings.

At a given time, agents $A_{i}$ send their $r_{i}$ to $E.$ This is done
synchonously.
Once all $r_{i}$ are received by $E:$

 \xEh

 \xDH Variant 1:

$E$ calculates the mean (average) $m$ of all $r_{i}.$ The closer the
individual $r_{i}$
is to $m,$ the more reliable $r_{i}$ is considered, the better $
\xbr_{i}.$
More precisely:
Let $ \xbd_{i}$ be the distance from $r_{i}$ to $m,$ and let $ \xbd $ be
the mean of all $ \xbd_{i}.$
The better $ \xbd_{i}$ is in comparison to $ \xbd,$ the more reliable
$A_{i}$ seems to be.

We calculate the new $ \xbr_{i}$ by a suitable function: $ \xbr'_{i}:=f_{
\xbr }(\xbr_{i}, \xbd, \xbd_{i}).$
If $ \xbd_{i}$ is better than $ \xbd,$ we increase $ \xbr_{i},$ if not,
we decrease $ \xbr_{i}.$
The precise details will not matter, and depend also on the context.
(We might, e.g., have a minimal threshold of discrepancy, below which we
do nothing.)

Operations:
 \xEh
 \xDH $m$ (mean value) of the $r_{i}$
 \xDH $ \xbd_{i}$ $=$ distance between $m$ and $r_{i}$
 \xDH $ \xbd $ $=$ mean value of all $ \xbd_{i}$
 \xDH adjusting $ \xbr_{i}$ using $ \xbd,$ $ \xbd_{i},$ and old $
\xbr_{i}$
 \xEj

 \xDH Variant 2:

$E$ uses the old $ \xbr_{i}$ already to give different weight to the
$r_{i}.$
E.g., if $ \xbr_{i}$ is twice as good as $ \xbr_{j},$ we may count $r_{i}$
twice (and $r_{j}$ once),
to give it more weight, as $A_{j}$ has a ``bad reputation'', and $A_{i}$ a
good one.
The rest is the same as in Variant 1.

Operations:

adjust $r_{i}$ using old $ \xbr_{i}$

 \xDH Variant 3:

As in Variant 1 or 2, but we assume we have already earlier measurements
of the same entity (assumed constant), so we have already an ``old'' $m,$
which
summarizes the old data, the history. The ``inertia'' $t$ of the old $m$
should
express the number of $r_{i}$ which went into the calculation of the old
$m.$

Thus, e.g., we enter the old $m$ $t$ times, just as we would enter $t$ new
$r_{i}.$
We may modify, e.g. give the old $m$ more or less weight, etc.

In the same way, we may give the old $ \xbr_{i}' s$ more or less weight.

This way, we may also treat asynchronous arrival of messages from
different
agents. Some precaution against receiving repeated messages from the same
agent might be necessary.

Operations:
 \xEh
 \xDH purely administrative: count numer of $r_{i}$
 \xDH multiply old $m$ by $t$
 \xEj

 \xDH Variant 4:

For this variant, we need to put the $ \xbr_{i}$ in relation to the values
$r_{i}$ (and $m).$
Suppose, e.g., that an agent $A_{i}$'s $r_{i}$ should be within $10 \xET
$ of $m$ if $ \xbr_{i}$ is 0.9.
The better $ \xbr_{i},$ the more $r_{i}$ should be close to $m.$
We might then decide to decrease $ \xbr_{i},$ if $ \xbd_{i}$ is too big
for $ \xbr_{i},$ etc.
Details, again, are not important, the relation of $ \xbr_{i}$ to $
\xbd_{i},$ and thus
between reliability and data, is important.

Operations:

put $ \xbr_{i}$ in relation to a difference between $m$ and $r_{i}$

 \xEj
\subsubsection{
Hypotheses About own Reliability and Reliability of Communication
Channels
}

An agent may have a hypothesis about the reliability of his own message.
E.g., a human being might caution that his expertise is not very good in
a certain field, or that he feels very confident. A thermometer may have
a temperature range where it is very precise, and outside this range it
may
be less so - and it may ``know'' about it.
Thus, the message has two parts, data, and presumed reliability, say $
\xbe_{i}.$
Consider above Variant 4. If $ \xbd_{i}$ is not too big in relation to
$ \xbe_{i},$ $E$ may renounce on decreasing $ \xbr_{i},$ as the agent was
aware of the
problem. In addition, $E$ may give less weight to $r_{i},$ see Variant 2.

Communication channels may have a reliability, too, say $ \xbr c_{i}.$
The value $r_{i}$ has now combined reliability of $ \xbr_{i}$ and $ \xbr
c_{i}.$
The simplest way to combine them might be multiplication, and it
should probably not be bigger than $min\{ \xbr_{i}, \xbr c_{i}\}.$
Conversely, if we want to modify the combined reliability, we have to
decide how to modify both parts. Multiplication by a common
factor seems a simple way to proceed.

Operations:
 \xEh
 \xDH (serial) combination of two reliabilities, here $ \xbr_{i}$ and $
\xbr c_{i}$
 \xDH conversely, break down a modification of a combination of two
reliabilities to a modification of the individual reliabilities
(this should be an inverse operation to the first operation here)

 \xEj
\subsection{
Broadcasting and Messages About Reliability of other Agents
}
\subsubsection{
Broadcasting
}

One problem with broadcasting (anyone may send messages to anyone) is that
it may lead to contradictory or self-supporting cycles.
E.g. agent A sends a message to $B$ whereupon $B$ sends a message to A
amplifying A's message and so forth. A need not see that it is just
A's own message coming back stronger. Adding history to the messages
solves the problem. Suppose A sends the message $ \xBc A,r \xBe,$ expressing
that
it is
a message from A, and $B$ sends the message expressing that it is a reply
$ \xBc A,r,B,r'  \xBe,$ then A sees that the message originated from A, and
will
not send
it again. Likewise, $B$ might not send it to A, as A ``saw'' it already.

If $E$ listens in to all messages, $E$ can detect such cycles, and react
accordingly, e.g. contract the whole group to a single agent, or
neglecting
the whole group, if it is infighting.

The problem of oscillations is a common one, and it might be interesting
to see
how the brain avoids them.

Operations:

Note that we do not need any new operations as the new elements are
about control.
\subsubsection{
Messages about Reliabilities of other Agents
}

Messages about reliabilities of other agents may easily be destructive (as
can
be seen in politics!). Let $ \xbe_{i,j}$ be a message from agent $i$ about
the
reliability of agent $j.$

Suppose $A_{i}$ sends $ \xbe_{i,j}$ to $E.$

First, how much weight should $E$ give to $ \xbe_{i,j}?$ It should
probably not be
totally neglected, but history (which $E$ should store) should matter.
E.g., if $A_{i}$ and $A_{j}$ support each other positively, this might
(but need not)
be cronyism, if they do so negatively (e.g. $A_{i}$ and $A_{j}$ say that
the other
is unreliable), it might be a case of infighting.

The problem is easy to see, but there is probably no general solution,
only answers to particular cases.

If we allow such messages to be broadcast, they might end in positive or
negative cycles, which an ``umpire'' should detect and prevent.

Operations:

Again, we do not need any new operations as the new elements are
about control.
\subsubsection{
Operations On the Contents of the Messages
}

We do not discuss operations on the contents of the messages beyond this
short remark. E.g., the reliability of a value in the interval $[10,20]$
should be at least as good as individual reliabilities for $x \xbe
[10,20],$
the same applies for the reliability of $ \xbf \xco \xbq $ in relation to
the
reliabilities for $ \xbf $ and $ \xbq.$
\section{
Discussion
}

\br

$\hspace{0.01em}$


\label{Remark Extensions}

The following extensions seem possible:
 \xEI
 \xDH Actions and animals:

We can apply similar reasoning to actions. The action
of a monkey (the agent) which sees a lion and climbs a tree to safety is
``true'',
or, better, adequate.

 \xDH Values:

Values, obligations, ``natural laws'' (in the sense of philosophy of law)
are
subjective. Still, some influences are
known, and we can try to peel them off. Religion, politics, personal
history, influence our ideas about values. One can try to find the
``common'' and ``reasonable'' core of them. For instance, religious extremism
tends to produce ruthless value systems, so we might consider religious
extremists as less reliable about values.
 \xEJ

\er

Our approach is very pragmatic, a method,
and takes its intuition from e.g. physics, where a theory is
considered true - but revisably so! - when there is ``sufficient''
confirmation, by experiments, support from other theories, etc.

Automatic trading in financial markets has to consider some aspects of
our ideas: one should caution against excessive feedback, as it might
generate unfounded fluctuations.

Many human efforts are about establishing reliability of humans or
devices. An egineer or physician has to undergo exams to assure that
he is competent, a bridge has to meet construction standards, etc.
All this is not infallible, experts make mistakes, new, unknown
possibilities of failure may appear - we just try to do our best.

Our ideas are examples how it can be done,
but no definite solutions. The exact choice is perhaps not so
important, as long as there is a process of permanent adjustment.
This process has proven extremely fruitful in science, and deserves
to be seen as a powerful method, if not to find truth, at least to find
``sufficient'' information.

From an epistemological point of view, our position is that of
``naturalistic epistemology'', and we need not decide between
``foundationalism'' and ``coherentism'', the interval $[0,1]$ has
enough space to maneuvre between more and less foundational
information. See e.g.  \cite{Sta17c}.

Our approach has some similarities with the utility approach,
see the discussion in  \cite{BB11}, the chapter on utility.
An assumption, though false, can be useful: if you think a lion is
outside,
and keep the door closed, this is useful, even if, in fact, it is a tiger
which is outside. ``A lion is outside'' is false, but sufficiently true.
We think that this shows again that truth should not be seen as something
absolute, but as something we can at best approximate; and, conversely,
that it is not necessary to know ``absolute truth''.
We go beyond utility, as improvement is implicit in our approach.
Of course, approximation may only be an illusion generated by the fact
that we develop theories which seem to fit better and better, but whether
we approach reality and truth, or, on the contrary, move away from reality
and
truth, we cannot know.

There are many things we did not consider, e.g. if
more complicated, strongly connected, structures have stronger inertia
against
adjustment.

We use meta-information (reliability - which, importantly, is not binary,
not just true/false) to avoid mistakes, in our example of measurements.
We further use control information to detect cycles and group
behaviour, to avoid further mistakes. Of course, this is all very
primitive, and further elaborations are possible and necessary,
sometimes depending on the type and environment of the data.
The question is whether the method is adequate, there are no
completeness and correctness properties to be discovered -
it is about methods, not logics, there are no axioms etc.

We have an
example structure which handles these problems very well: our brain.
Attacks, negative values of reliability, correspond to inhibitory
synapses,
positive values, support, to excitatory synapses. Complex, connected
structures
with loops are created all the time without uncontrolled feedback.
It is perhaps not sufficiently clear how this works, but it must work!
(The ``matching inhibition'' mechanism seems to be a candidate. See also
 \cite{OL09} for a discussion of the ``cooperation'' of excitatory and
inhibitory inputs of a neuron. E.g., excitation may be followed closely by
inhibition, thus explaining the suppression of such feedback.
The author is indebted to Ch. von der Malsburg, FIAS, for these hints.)
Our theories about the
world survive some attack (inertia), until ``enough is enough'', and we
switch
emphasis. The brain's mechanisms for attention can handle this.
\section{
Philosophical Background
}
\subsection{
The Coherence and Correspondence Theories of Truth
}

See  \cite{Sta17a} for an overview for the coherence theory,
and  \cite{Sta17b} for an overview for the correspondence theory.
The latter contains an extensive bibliography, and we refer the
reader there for more details on the correspondence theory.

We think that the criticisms of the coherence theory of truth are
peripheral,
but the criticism of the correspondence theory of truth is fundamental.

The criticism of the correspondence theory, that we have no direct access
to reality, and have to do with our limitations in observing and thinking,
seems fundamental to the author. The discussion whether there are some
``correct'' theories our brains are unable to formulate, is taken seriously
by physicists, likewise the discussion, whether e.g. Quarks are real, or
only helpful ``images'' to understand reality, was taken very seriously.
E.g. Gell Mann was longtime undecided about it, and people perhaps just
got used to them. We don't know what reality is, and it seems we will
never know.
See also discussions in neurophilosophy,
 \cite{Sta17d} for a general introduction.

On the other side, two main criticisms of the coherence theory can be
easily countered, in our opinion.
See e.g.  \cite{Rus07} and  \cite{Tha07} for objections to
coherence theory.
Russell's objection, that $ \xbf $ and $ \xCN \xbf $
may both be consistent with a given theory, shows just that ``consistency''
is the wrong interpretation of ``coherence'', and it also leaves open the
question which logic we work in. The objection that the background
theory against which we check coherence is undefined, can be countered
with a
simple
argument: Everything. In ``reality'', of course, this is not the case.
If we have a difficult physical problem, we will not ask our baker, and
even if he has an opinion, we will not give it much consideration.
Sources of information are assessed, and only ``good'' sources (for the
problem at hand!) will be considered.
(Thus, we also avoid the postmodernist trap: there are standards of
``normal reasoning'' whose values have been shown in unbiased everyday
life, and against which standards of every society have to be compared.
No hope for the political crackpots here!)

Our approach will be a variant of the coherence theory,
related ideas were also expressed by  \cite{Hem35} and  \cite{Neu83}.

We can see our approach in the tradition of relinquishing absoluteness:
 \xEI
 \xDH
The introduction of axiom systems made truth relative to axioms.
 \xDH
Nonmonotonic reasoning allowed for exceptions.
 \xDH
Our approach treats uncertainty of information, and our potential
inability to know reality.
 \xEJ
\subsection{
A Short Comparison of Our Approach to Other Theories
}

 \xEh

 \xDH
Our approach is not about discovery, only about evaluating information.

 \xDH
In contrast to many philosophical theories of truth,
we do not treat paradoxa, as done e.g. in
 \cite{Kri75} or  \cite{BS17},
we assume statements to be ``naive'' and free from semantic problems.

We do treat cycles too, but they are simpler, and we take care not to go
through them repeatedly. In addition, our structures are assumed to be
finite.

 \xDH
On the philosophical side, we are probably closest to the discourse
theory of the Frankfurt School,
in particular to the work by J. Habermas and K. O. Apel
(as we discovered by chance!),
see e.g.
 \cite{Wik18b},  \cite{Sta18b},  \cite{Hab73},
 \cite{Hab90},  \cite{Hab96},  \cite{Hab01},  \cite{Hab03}.

Importantly, they treat with the basically same methods problems of
truth and ethics, see our
Remark \ref{Remark Extensions} (page \pageref{Remark Extensions})  below.

We see three differences with their approach.
 \xEh
 \xDH
A minor difference: We also consider objects like thermometers as agents,
not only human beings, thus eliminating some of the subjectivity.
 \xDH
A major difference: We use feedback to modify reliability of agents and
messages. Thus, the $ \xcA $-quantifier over participating agents in the
Frankfurt
School is attenuated to those considered reliable.
 \xDH
Conversely, their discourse theory is, of course, much more developed than
our approach.
 \xEj

Thus, an integration of both approaches seems promising.

 \xDH
Articles on trust, like
 \cite{BBHLL10} or  \cite{BP12}, treat different, more subtle,
and perhaps less
fundamental, problems.
A detailed overview over trust systems is given in
 \cite{SS05}.

We concentrate on logics, cycles, and composition of values by
concatenation.
Still, our approach is in methods, but not in motivation, perhaps closer
to the
basic ideas of trust systems,
than to those of theories of truth, which often concentrate on paradoxa.

Articles on trust will often describe interesting ideas about details of
coding,
e.g.  \cite{BP12} describes how
to code a set of numerical values by an interval (or, equivalently, two
values).

 \xDH
Basic argumentation systems, see e.g.  \cite{Dun95},
will not distinguish between arguments of different quality.
Argumentation systems with preferences,
see e.g.  \cite{MP13}, may do so, but they do not seem to
propagate conflict and confirmation backwards to the source of arguments,
which
is an essential part of our approach. This backward
propagation also seems a core part of any truth theory
in our spirit. Such theories have to be able to learn
from past errors and successes.

 \xDH
Is this a Theory of Truth?

The author thinks that, yes, though
we hardly mentioned truth in the text.

Modern physics are perhaps the best attempt to find out what
``reality'' is, what ``truly holds''. We had the development of physics in
mind,
reliability of experiments, measurements, coherence of theories (forward
and backward influence of reliabilities), reputation of certain
physicists,
predictions, etc. Of course, the present text is only a very rough sketch,
we see it as a first attempt, providing some highly flexible ingredients
for a
more complete theory in this spirit.
 \xEj

$ \xCO $

$ \xCO $
\markboth{\centerline{\scriptsize Yablo's Paradox}}
{\centerline{\scriptsize Yablo's Paradox}}

$ \xCO $
\chapter{
Remarks on Yablo's Paradox
}

\label{Section YAB}

\label{Chapter YAB}

$ \xCO $
%
%
\clearpage
\section{
Introduction
}

Unless stated otherwise, we work in propositional logic, with disjunctive
normal
forms, i.e. formulas of the type $ \xcO \xcU.$ Formulas may, however, be
infinite.
\subsection{
Overview
}

After some definitions, we show in
Section \ref{Section Rab} (page \pageref{Section Rab})  that a conjecture in
 \cite{RRM13} is wrong.

In Section \ref{Section Remarks} (page \pageref{Section Remarks}),
we discuss basic contradictions, cells,
and give an example of basic reasoning about
contradictory sequences,
see Section \ref{Section Complicated} (page \pageref{Section Complicated}).

Section 
\ref{Section Yablo-Analysis} (page 
\pageref{Section Yablo-Analysis})  sees are detailed analysis of
Yablo's construction, some aspects of his construction
are hidden behind its elegance. This leads to the
concepts of ``head'', ``knee'', and ``foot'', and then
to ``saw blades'' in
Section \ref{Section Saw-Blades-All} (page \pageref{Section Saw-Blades-All}).
Section \ref{Section Or-And} (page \pageref{Section Or-And})
generalizes Yablo's construction to
arbitrary formulas of the type $ \xcO \xcU $ (disjunctive normal
forms), and offers a number of easy variations of such
structures by modifying the order of the graph.

Section 
\ref{Section Saw-Blades-All} (page 
\pageref{Section Saw-Blades-All})  uses our idea of finer analysis
of contradictory cells to build somewhat different
structures - though the distinction is blurred by the
necessarily recursive construction of contradictions.

We do not go in a straight line for the representation problem,
but rather collect some ideas.
We hope they are useful building blocks for a solution of the
representation problem.

The author of the present text did not study the literature
systematically. So,
if some examples are already discussed elsewhere, the author would ask to
be
excused for not quoting previous work.
\subsection{
Basic Definitions and Results
}

\label{Section Definitions}

We start with some notation and a trivial fact:

\bd

$\hspace{0.01em}$


\label{Definition Cont}

 \xEh
 \xDH $Inc(\xbf)$ will stand for (classical propositional) inconsistency
of $ \xbf $
 \xDH $Cont(\xbf)$ for ``contradictory'', i.e. $Inc(\xbf)$ $ \xcu $
$Inc(\xCN \xbf)$

(by abuse of language, $ \xcu $ etc. will be used in object and meta
language).
 \xEj

\ed

We then have the trivial result

\bfa

$\hspace{0.01em}$


\label{Fact Cont}

 \xEh
 \xDH
$Cont(\xbf)$ iff $Cont(\xCN \xbf)$

 \xDH
$Cont(\xbf \xcu \xbq)$ iff $Inc(\xbf \xcu \xbq)$ $ \xcu $ $Inc(\xCN (
\xbf \xcu \xbq)).$

$Inc(\xCN (\xbf \xcu \xbq))$ iff $Inc(\xCN \xbf \xco \xCN \xbq)$ iff
$Inc(\xCN \xbf)$ $ \xcu $ $Inc(\xCN \xbq)$

$Inc(\xbf)$ $ \xch $ $Inc(\xbf \xcu \xbq),$ but (obviously) $Inc(\xbf
\xcu \xbq)$ $ \xcu $ $ \xCN Inc(\xbf)$ $ \xcu $ $ \xCN Inc(\xbq)$
is possible.

Thus, $Cont(\xbf)$ $ \xcu $ $Inc(\xCN \xbq)$ $ \xch $ $Cont(\xbf \xcu
\xbq),$
but neither
$Cont(\xbf)$ nor $Cont(\xbq)$ follow from $Cont(\xbf \xcu \xbq).$

In particular, $Cont(\xbf)$ $ \xcu $ TRUE $ \xcj $ $Cont(\xbf \xcu
TRUE).$

 \xDH
$Cont(\xbf \xco \xbq)$ iff $Inc(\xbf \xco \xbq)$ $ \xcu $ $Inc(\xCN (
\xbf \xco \xbq)).$

$Inc(\xbf \xco \xbq)$ iff $Inc(\xbf)$ $ \xcu $ $Inc(\xbq),$

$Inc(\xCN (\xbf \xco \xbq))$ iff $Inc(\xCN \xbf \xcu \xCN \xbq),$ so
$Inc(\xCN \xbf)$ $ \xch $ $Inc(\xCN (\xbf \xco \xbq)).$

Thus, $Cont(\xbf)$ $ \xcu $ $Inc(\xbq)$ $ \xch $ $Cont(\xbf \xco \xbq
).$

In particular, $Cont(\xbf)$ $ \xcu $ FALSE $ \xcj $ $Cont(\xbf \xco
FALSE)$

 \xDH
Consequently, adding or eliminating a branch evaluating to TRUE with $
\xcu $
will not change the ``contradictory'' status, neither will adding a branch
evaluating to FALSE with $ \xco.$

This is important for simplifications of a diagram.

 \xEj

\efa

Definition \ref{Definition Basics} (page \pageref{Definition Basics}),
Definition \ref{Definition Downward} (page \pageref{Definition Downward}), and
Definition 
\ref{Definition Yablo-Structure} (page 
\pageref{Definition Yablo-Structure}),
are taken mostly from  \cite{RRM13}.

\bd

$\hspace{0.01em}$


\label{Definition Basics}

 \xEh
 \xDH
Given a (directed or not) graph $G,$ $V(G)$ will denote its set of
vertices, $E(G)$ its set of edges. In a directed graph, $xy \xbe E(G)$
will denote
an arrow from $x$ to $y,$ which we also write $x \xcp y,$
if $G$ is not directed, just a line from $x$ to $y.$

We often use $x,y,$ or $X,Y,$ etc. for vertices.
 \xDH
A graph $G$ is called transitive iff $xy,yz \xbe E(G)$ implies $xz \xbe
E(G).$
 \xDH
Given two directed graphs $G$ and $H,$ a homomorphism from $G$ to $H$ is a
function
$f:V(G) \xcp V(H)$ such that, if $xy \xbe E(G),$ then $f(x)f(y) \xbe
E(H).$
 \xDH
Given a directed graph $G,$ the underlying undirected graph is defined as
follows: $V(U(G)):=V(G),$ $xy \xbe E(U(G))$ iff $xy \xbe E(G)$ or $yx \xbe
E(G)),$ i.e., we
forget the orientation of the edges. Conversely, $G$ is called an
orientation of $U(G).$
 \xDH
$S,$ etc. will denote the set of propositional variables of some
propositional language $ \xdl,$ $S^{+},$ etc. the set of its formulas.
$ \xct $ and $ \xcT $ will be part of the formulas.
 \xDH
Given $ \xdl,$ $v$ will be a valuation, defined on $S,$ and extended to
$S^{+}$
as usual - the values will be $\{0,1\},$ $\{ \xct, \xcT \},$ or so.
$[s]_{v},$ $[ \xba ]_{v}$ will denote the valuation of $s \xbe S,$ $ \xba
\xbe S^{+},$ etc.
When the context is clear, we might omit the index $v.$
 \xDH
$d$ etc. will be a denotation assignment, or simply denotation, a function
from $S$ to $S^{+}.$

Note that $d$ need not have any logical meaning, it is an
arbitrary function.

We sometimes abbreviate, e.g. $d(x)=y \xcu \xCN z$ will be written $x=y
\xcu \xCN z,$ etc.

The arrows in the graph $G_{S,d}$ below will point to the variables in
$d(s),$
not to $d(s)$ or so.
 \xDH
A valuation $v$ is acceptable on $S$ relative to $d,$ iff for all $s \xbe
S$
$[s]_{v}=[d(s)]_{v},$ i.e. iff $[s \xcr d(s)]_{v}= \xct.$
(When $S$ and $d$ are fixed, we just say that $v$ is acceptable.)
 \xDH
A system $(S,d)$ is called paradoxical iff there is no $v$ acceptable for
$S,$ $d.$
 \xDH
Given $S,$ $d,$ we define $G_{S,d}$ as follows:
$V(G_{S,d}):=S,$ $ss' \xbe E(G_{S,d})$ iff $s' \xbe S$ occurs in $d(s).$

If there are no arrows originating in $s,$ then $d(s)$ is equivalent
to $ \xcT $ or $ \xct.$

For clarity, one might write $d(s)$ next to $s$ in $G_{S,d},$ but this
would
further complicate the graphs. But, of course, $d(s)$ is essential
for the comprehension.

 \xDH
A directed graph $G$ is dangerous iff there is a paradoxical system
$(S,d),$
such that $G$ is isomorphic to $G_{S,d}.$
 \xEj

\ed

\bcom

$\hspace{0.01em}$


\label{Comment Basics}

Note that (9) and (11) give very different representation problems,
(11) offers much more freedom, as justified by
Fact \ref{Fact Cont} (page \pageref{Fact Cont})  (4).

In (11), we are given only the variables ocurring in $d(x),$ and
may build up any formula with them.
Fact 
\ref{Fact Cont} (page 
\pageref{Fact Cont})  (4) may thus offer ways to simplify a problem,
e.g. by
interpreting $x \xcp a$ as $a \xco \xCN a,$ $a \xcu \xCN a,$ etc.,
whenever it is possible
to give classical truth values.

This suggests a strategy of pre-processing: if we want to examine whether
a structure has a contradictory interpretation,
chose suitable classical
truth values whenever possible (e.g. if the structure below $x$ is finite,
has finite depth, is a tree, etc.) to simplify the problem.
Of course, finding possible classical truth values is dual to finding
contradictory truth value, $Cont(x)$ above.

\ecom

\bd

$\hspace{0.01em}$


\label{Definition Downward}

Let $G$ be a directed graph, $x,x' \xbe V(G).$
 \xEh
 \xDH
$x' $ is a successor of $x$ iff $xx' \xbe E(G).$

$succ(x):=\{x':$ $x' $ is a successor of $x\},$
 \xDH
Call $x' $ downward from $x$ iff there is a path from $x$ to $x',$
i.e. $x' $ is in the transitive closure of the succ operator.
 \xDH
Let $[x \xcp ]$ be the subgraph of $G$ generated by $\{x\} \xcv \{x':x' $
is downward from $x\},$
i.e. $V([x \xcp ]):=\{x\} \xcv \{x':x' $ is downward from $x\},$ and $x'
\xcp x'' \xbe E([x \xcp ])$ iff
$x',x'' \xbe V([x \xcp ]),$ and $x' x'' \xbe E(G).$
 \xEj

\ed

\bd

$\hspace{0.01em}$


\label{Definition Yablo-Structure}

For easier reference, we define the Yablo structure,
see e.g.  \cite{RRM13}.

Let $V(G):=(Y_{i}:i< \xbo \},$ $E(G):=\{Y_{i}Y_{j}:i,j< \xbo,$ $i<j\},$
and
$d(Y_{i}):= \xcU \{ \xCN Y_{j}:i<j\}.$

$(Y_{i} \xbe S$ for a suitable language.)

\ed

We see immediately the following simple, but very important fact:

\bfa

$\hspace{0.01em}$


\label{Fact NotClass}

 \xEh

 \xDH
The logic as used in Yablo's construction is not compact.

 \xDH
It is impossible to construct a Yablo-like structure with classical logic.
 \xEj

\efa

\subparagraph{
Proof
}

$\hspace{0.01em}$


 \xEh
 \xDH
Trivial.

(Take $\{ \xcO \{ \xbf_{i}:i \xbe \xbo \}\} \xcv \{ \xCN \xbf_{i}:i \xbe
\xbo \}.$ This is obviously inconsistent,
but no finite subset is.).

 \xDH
Take an acyclic graph, and interpret it as in Yablo's construction.
Wlog., we may assume the graph is connected. Suppose it shows that
$x_{0}$ cannot be given a truth value. Then the set of formulas showing
this does not have a model, so it is inconsistent. If the formulas were
classical, it would have a finite, inconsisten subset, $ \xbF.$ Define
the depth
of a formula as the shortest path from $x_{0}$ to this formula.
There is a (finite) $n$ such that all formulas in $ \xbF $ have depth $
\xck n.$
Give all formulas of depth $n$ (arbitrary) truth values, and work
upwards using truth functions. As the graph is acyclic, this is possible.
Finally, $x_{0}$ has a truth value.

Thus, we need the infinite $ \xcU / \xcO.$
 \xEj

$ \xcz $
\\[3ex]

\br

$\hspace{0.01em}$


\label{Remark Depth}

By the same argument as in the second half of (2) above, we see that we
need
infinite descending
chains to obtain Yablo's Paradox.

\er

\bn

$\hspace{0.01em}$


\label{Notation PlusMinus}

 \xEh
 \xDH
As shorthand, we will sometimes use:

$x+$ will mean that $x$ is true, likewise $ \xCf x-$ that $x$ is false, $x
\xCL $ that
$x$ is contradictory, i.e. it cannot have a truth value in the structure
considered.
 \xDH
$x \xcp y$ will mean that $y$ occurs positively in $d(x),$ e.g. $d(x)=y
\xcu \xCN z,$
in the same example we would write $x \xcP z.$
 \xDH
$x \xch_{ \xCL }y$ (or $x \xcp_{ \xCL }y)$ stands for $d(x)=y \xcu \xCN
y.$
 \xEj

\en

We begin with some trivialities, just to remind the reader.

\br

$\hspace{0.01em}$


\label{Remark Trivial-1}

 \xEh
 \xDH When we construct a structure, we may have e.g. the choice of
constructing two or three branches. If we construct an example
(and not all cases), we can chose as we like - from the outside so to say.
Once
we did chose a structure, we are not free any more, we have to follow all
branches in a $ \xcA $ situation - inside, we are not free any more.
 \xDH
When we want to show that $ \xbf = \xbf' \xco \xbf'' $ is contradictory,
we have to show
that both $ \xbf' $ and $ \xbf'' $ are contradictory.

When we want to show that $ \xbf = \xbf' \xcu \xbf'' $ is contradictory,
it suffices to show
that one of $ \xbf' $ or $ \xbf'' $ is contradictory (or both together).
 \xDH
If the structure has no infinite descending chains, then there is a
consistent valuation:

We may give arbitrary values to the bottom elements, and calculate
upwards, using the truth functions.
 \xDH
We have to show that for some node $x$ in the structure, assigning
TRUE to $x$ leads to a contradiction, and assigning FALSE to $x$ will
also lead to a contradiction.
 \xDH
We do not need constants $ \xct,$ $ \xcT.$

Instead of assigning $ \xct $ to some node $y,$ we may introduce a new
node $y' $
and define $y:=y' \xco \xCN y',$ similarly for $ \xcT.$

 \xEj
\section{
Comments on Rabern et al., \cite{RRM13}
}

\label{Section Rab}
\subsection{
Introduction
}

\er

This section is a footnote to  \cite{RRM13}.
 \cite{RRM13} is perhaps best
described as a graph theoretical analysis of
Yablo's construction, see  \cite{Yab82}.
We continue this work.

To make the present paper self-contained, we repeat the
definitions of  \cite{RRM13}. To keep it short, we
do not repeat ideas and motivations of  \cite{RRM13}.
Thus, the reader should probably be familiar with or have a copy of
 \cite{RRM13} ready.

All graphs etc. considered will be assumed to be cycle-free, unless said
otherwise.
\subsubsection{
Overview
}

 \xEh
 \xDH
Section \ref{Section Definitions} (page \pageref{Section Definitions})  contains
most of the definitions we use,
many are taken from  \cite{RRM13}.
 \xDH
In Section 
\ref{Section Conjecture} (page 
\pageref{Section Conjecture}), we show that
conjecture 15 in  \cite{RRM13} is wrong.
This conjecture says that a directed graph $G$ is dangerous iff
every homomorphic image of $G$ is dangerous.
(The definitions are given in
Definition 
\ref{Definition Basics} (page 
\pageref{Definition Basics}), (3) and (11).)

To show that the conjecture is wrong, we
modify the Yablo construction, see
Definition 
\ref{Definition Yablo-Structure} (page 
\pageref{Definition Yablo-Structure}), slightly in
Example \ref{Example YG'} (page \pageref{Example YG'}), illustrated in
Diagram \ref{Diagram YG'} (page \pageref{Diagram YG'}), show that it is
still dangerous in Fact \ref{Fact YG'} (page \pageref{Fact YG'}),
and collaps it
to a homomorphic image in
Example \ref{Example Homomorphismus} (page \pageref{Example Homomorphismus}).
This homomorphic image is not dangerous, as
shown in Fact \ref{Fact Integer} (page \pageref{Fact Integer}).
 \xDH
In Section 
\ref{Section 24} (page 
\pageref{Section 24}), we discuss implications of Theorem 24
in  \cite{RRM13} - see the
paragraph immediately after the proof of the theorem
in  \cite{RRM13}.
This theorem states
that an undirected graph $G$ has a dangerous orientation iff it
contains a cycle.
(See Definition 
\ref{Definition Basics} (page 
\pageref{Definition Basics})  (4) for orientation.)

We show that for any simply connected directed graph $G$ - i.e., in the
underlying
undirected graph $U(G),$ from any two vertices $X,Y,$ there is at most one
path
from $X$ to $Y,$
see Definition 
\ref{Definition Simply} (page 
\pageref{Definition Simply})  - and for
any denotation $d$ for $G,$ we find an acceptable valuation for $G$ and
$d.$

The proof consists of a mixed induction, successively assigning values
for the $X,$ and splitting up the graph into ever smaller independent
subgraphs. The independence of the subgraphs relies essentially on the
fact that $G$ (and thus also all subgraphs of $G)$ is simply
connected.
 \xDH
In Section 
\ref{Section Various} (page 
\pageref{Section Various}), we discuss some modifications and
generalizations of the Yablo structure.
Example \ref{Example NotN} (page \pageref{Example NotN})
considers trivial modifications of the Yablo structure.
In Fact \ref{Fact Finite-Branching} (page \pageref{Fact Finite-Branching})
(see also Fact \ref{Fact NotClass} (page \pageref{Fact NotClass}))
we show that
infinite branching is necessary for a graph
being dangerous, and
Example 
\ref{Example Procrastination} (page 
\pageref{Example Procrastination})  shows why
infinitely many finitely branching points cannot
replace infinite branching - there is an infinite
``procrastination branch''.

Our main result here is in
Fact \ref{Fact TransAndNot} (page \pageref{Fact TransAndNot}), where we
show that in Yablo-like structures,
the existence of an acceptable valuation is strongly
related to existence of successor nodes,
where $X' $ is a successor of $X$ in a directed graph $G,$ iff $X \xcp X'
$ in $G,$
or, written differently, $XX' \xbe E(G),$ the set of edges in $G.$

 \xEj

\bd

$\hspace{0.01em}$


\label{Definition AndNot}

 \xEh
 \xDH
Call a denotation $d$ $ \xcU \xCN $ or $ \xcU -$ iff all $d(X)$ have the
form $d(X)= \xcU \{ \xCN X_{i}:i \xbe I\}$ -
as in the Yablo structure.
 \xDH
The dual notation $ \xcU +$ expresses the analogous case with $+$ instead
of $ \xCN,$
i.e. $d(X)= \xcU \{X_{i}:i \xbe I\}.$
 \xDH
We will use $ \xCN $ and - for negation, and $+$ when we want to emphasize
that
a formula is not negated.
 \xEj

\ed

\br

$\hspace{0.01em}$


\label{Remark Empty}

Note that we interpret $ \xcU $ in the strict sense of $ \xcA,$ i.e., $
\xCN \xcU \{ \xCN X_{i}:i \xbe I\}$
means that there is at least one $X_{i}$ which is true. In particular,
if $d(X)= \xcU \{ \xCN X_{i}:i \xbe I\},$ and $[X]=[d(X)]= \xcT,$ then
$d(X)$ must contain a propositional variable, i.e. cannot be composed only
of $ \xcT $ and $ \xct,$ so there is some arrow $X \xcp X' $ in the
graph.

Thus, if in the corresponding graph $succ(X)= \xCQ,$ $d$ is
of the form $ \xcU -,$ $v$ is an acceptable valuation for $d,$ then
$[X]_{v}= \xct.$

Dually, for $ \xcU +,$ $ \xCN \xcU \{X_{i}:i \xbe I\}$
means that there is at least one $X_{i}$ which is false.

Thus, if in the corresponding graph $succ(X)= \xCQ,$ $d$ is
of the form $ \xcU +,$ $v$ is an acceptable valuation for $d,$ then
$[X]_{v}= \xcT.$
\subsection{
A Comment on Conjecture 15 in \cite{RRM13}
}

\label{Section Conjecture}

\er

We show in this section that conjecture 15 in
 \cite{RRM13} is wrong.

\bd

$\hspace{0.01em}$


\label{Definition Contiguous}

Call $ \xdx \xcc \xdZ $ (the integers) contiguous iff for all $x,y,z \xbe
\xdZ,$ if $x<y<z$ and
$x,z \xbe \xdx,$ then $y \xbe \xdx,$ too.

\ed

\bfa

$\hspace{0.01em}$


\label{Fact Integer}

Let $G$ be a directed graph, $V(G)= \xdx $ for some contiguous $ \xdx,$
and
$x_{i}x_{j} \xbe E(G)$ iff $x_{j}$ is the direct successor of $x_{i}.$

Then for any denotation $d$:

 \xEh
 \xDH
$d(x)$ may be (equivalent to) $x+1,$ $ \xCN (x+1),$ $ \xcT,$ or $ \xct.$

If $d(x)= \xcT $ or $ \xct,$ we abbreviate $d(x)=c,$ $c',$ etc. (c for
constant).
 \xEj
If $v$ is acceptable for $d,$ then:
 \xEh
 \xDH
If $d(x)=c,$ then $d(x-1)=c' $ (if $x-1$ exists in $ \xdx).$
 \xDH
if $d(x)=(x+1),$ then $[x]_{v}=[x+1]_{v}$

if $d(x)= \xCN (x+1),$ then $[x]_{v}= \xCN [x+1]_{v}$
 \xDH
Thus:
 \xEh
 \xDH
If $d(x)=c$ for some $x,$ then for all $x' <x$ $d(x')=c' $ for some $c'
.$
 \xDH
We have three possible cases:
 \xEh
 \xDH
$d(x)=c$ for all $x \xbe \xdx,$
 \xDH
$d(x)=c$ for no $x \xbe \xdx,$
 \xDH
there is some maximal $x' $ s.t. $d(x')=c,$ so $d(x'') \xEd c' $ for all
$x'' >x'.$
 \xEj
 \xEI
 \xDH
In the first case, for all $x,$ if $d(x)$ is $ \xcT $ or $ \xct,$ then
the valuation
for $x$ starts anew, i.e. independent of $x+1,$ and continues to $x-1$
etc.
according to (2).
 \xDH
in the second case, there is just one acceptable valuation:
we chose some $x \xbe \xdx,$ and $[x]_{v}$ and propagate the value up and
down
according to (2)
 \xDH
in the third case, we work as in the first case up to $x',$ and treat the
$x'' >x' $ as in the second case.
 \xDH
Basically, we work downwards from constants, and up and down beyond the
maximal constant. Constants interrupt the upward movement.
 \xEJ
 \xEj
 \xDH
Consequently, any $d$ on $ \xdx $ has an acceptable valuation $v_{d},$ and
the graph
is not dangerous.

(The present fact is a special case of
Fact 
\ref{Fact SimplyConnected} (page 
\pageref{Fact SimplyConnected}), but it seems useful to discuss a
simple
case first.)
 \xEj

\efa

\be

$\hspace{0.01em}$


\label{Example YG'}

We define now a modified Yablo graph $ \xCf YG',$ and a corresponding
denotation $d,$
which is paradoxical.

We refer to Fig.3 in  \cite{RRM13}, and
Diagram \ref{Diagram YG'} (page \pageref{Diagram YG'}).

 \xEh
 \xDH
The vertices (and the set $S$ of language symbols):

We keep all $Y_{i}$ of Fig.3 in  \cite{RRM13},
and introduce new vertices $(Y_{i},Y_{j},Y_{k})$ for $i<k<j.$
(When we write $(Y_{i},Y_{j},Y_{k}),$ we tacitly assume that $i<k<j.)$
 \xDH
The arrows:

All $Y_{i} \xcp Y_{i+1}$ as before.
We ``factorize'' longer arrows through new vertices:
 \xEh
 \xDH
$Y_{i} \xcp (Y_{i},Y_{j},Y_{i+1})$
 \xDH
$(Y_{i},Y_{j},Y_{k}) \xcp (Y_{i},Y_{j},Y_{k+1})$
 \xDH
$(Y_{i},Y_{j},Y_{j-1}) \xcp Y_{j}$
 \xEj
See Diagram \ref{Diagram YG'} (page \pageref{Diagram YG'}).
 \xEj

We define $d$ (instead of writing $d((x,y,z))$ we write $d(x,y,z)$ -
likewise $[x,y,z]_{v}$ for $[(x,y,z)]_{v}$ below):
 \xEh
 \xDH
$d(Y_{i}):= \xCN Y_{i+1} \xcu \xcU \{ \xCN (Y_{i},Y_{j},Y_{i+1}):$ $i+2
\xck j\}$

(This is the main idea of the Yablo construction.)
 \xDH
$d(Y_{i},Y_{j},Y_{k})$ $:=$ $(Y_{i},Y_{j},Y_{k+1})$ for $i<k<j-1$
 \xDH
$d(Y_{i},Y_{j},Y_{j-1})$ $:=$ $Y_{j}$
 \xEj

Obviously, $ \xCf YG' $ corresponds to $S$ and $d,$ i.e. $YG' =G_{S,d}.$

\ee

\bfa

$\hspace{0.01em}$


\label{Fact YG'}

$ \xCf YG' $ and $d$ code the Yablo Paradox:

\efa

\subparagraph{
Proof
}

$\hspace{0.01em}$


Let $v$ be an acceptable valuation relative to $d.$

Suppose $[Y_{1}]_{v}= \xct,$ then $[Y_{2}]_{v}= \xcT,$ and
$[Y_{1},Y_{k},Y_{2}]_{v}= \xcT $ for $2<k,$ so
$[Y_{k}]_{v}= \xcT $ for $2<k,$ as in Fact 
\ref{Fact Integer} (page 
\pageref{Fact Integer}), (2).
By $[Y_{2}]_{v}= \xcT,$ there must be $j$ such that
$j=3$ and $[Y_{3}]_{v}= \xct,$ or $j>3$ and $[Y_{2},Y_{j},Y_{3}]_{v}=
\xct,$ and as in
Fact \ref{Fact Integer} (page \pageref{Fact Integer}), (2)
again, $[Y_{j}]_{v}= \xct,$
a contradiction.

If $[Y_{1}]_{v}= \xcT,$ then as above for $[Y_{2}]_{v},$ we find $j \xcg
2$ and $[Y_{j}]_{v}= \xct,$ and
argue with $Y_{j}$ as above for $Y_{1}.$

Thus, $ \xCf YG' $ with $d$ as above is paradoxical, and $ \xCf YG' $ is
dangerous.

$ \xcz $
\\[3ex]

\be

$\hspace{0.01em}$


\label{Example Homomorphismus}

We first define $ \xCf YG'' $:
$V(YG''):=\{ \xBc Y_{i} \xBe:i< \xbo \},$ $E(YG''):=\{ \xBc Y_{i} \xBe  \xcp 
\xBc Y_{i+1} \xBe:i<
\xbo \}.$

We now define the homomorphism from $ \xCf YG' $ to $ \xCf YG''.$
We collaps for fixed $k$ $Y_{k}$ and all $(Y_{i},Y_{j},Y_{k})$ to
$ \xBc Y_{k} \xBe,$ more precisely,
define $f$
by $f(Y_{k}):=f(Y_{i},Y_{j},Y_{k}):= \xBc Y_{k} \xBe $ for all suitable $i,j.$

Note that $ \xCf YG' $ only had arrows between ``successor levels'',
and we have now only arrows from $ \xBc Y_{k} \xBe $ to $ \xBc Y_{k+1} \xBe,$
so $f$ is a
homomorphism,
moreover, our structure $ \xCf YG'' $ has the
form described in Fact \ref{Fact Integer} (page \pageref{Fact Integer}),
and is not dangerous, contradicting
conjecture 15 in  \cite{RRM13}.

\clearpage

\begin{diagram}

\label{Diagram YG'}
\index{Diagram YG'}

\unitlength0.65mm
\begin{picture}(150,180)(0,0)

\put(0,175){{\rm\bf Diagram YG' }}

\put(0,160){$Y_1$}
\put(140,160){$ \xBc Y_1 \xBe $}
\put(2,158){\line(0,-1){32}}
\put(143,158){\line(0,-1){33}}
\put(3,158){\line(3,-4){25}}
\put(4,158){\line(4,-3){45}}
\put(5,158){\line(5,-2){80}}

\put(0,120){$Y_2$}
\put(2,118){\line(0,-1){32}}
\put(143,118){\line(0,-1){33}}
\put(20,120){{\footnotesize $(Y_1,Y_3,Y_2)$}}
\put(50,120){{\footnotesize $(Y_1,Y_4,Y_2)$}}
\put(80,120){{\footnotesize $(Y_1,Y_5,Y_2)$}}
\put(140,120){$ \xBc Y_2 \xBe $}
\put(3,118){\line(5,-2){80}}
\put(4,118){\line(3,-1){105}}
\put(27,118){\line(-3,-4){25}}
\put(55,118){\line(-3,-4){25}}
\put(85,118){\line(-3,-4){25}}

\put(0,80){$Y_3$}
\put(2,78){\line(0,-1){32}}
\put(143,78){\line(0,-1){33}}
\put(20,80){{\footnotesize $(Y_1,Y_4,Y_3)$}}
\put(50,80){{\footnotesize $(Y_1,Y_5,Y_3)$}}
\put(80,80){{\footnotesize $(Y_2,Y_4,Y_3)$}}
\put(110,80){{\footnotesize $(Y_2,Y_5,Y_3)$}}
\put(140,80){$ \xBc Y_3 \xBe $}
\put(3,78){\line(5,-2){80}}
\put(27,78){\line(-3,-4){25}}
\put(55,78){\line(-3,-4){25}}
\put(80,79){\line(-2,-1){73}}
\put(115,78){\line(-3,-2){50}}

\put(0,40){$Y_4$}
\put(2,38){\line(0,-1){32}}
\put(143,38){\line(0,-1){33}}
\put(20,40){{\footnotesize $(Y_1,Y_5,Y_4)$}}
\put(50,40){{\footnotesize $(Y_2,Y_5,Y_4)$}}
\put(80,40){{\footnotesize $(Y_3,Y_5,Y_4)$}}
\put(140,40){$ \xBc Y_4 \xBe $}
\put(27,38){\line(-3,-4){25}}
\put(56,38){\line(-3,-2){51}}
\put(80,39){\line(-2,-1){73}}

\put(0,0){$Y_5$}
\put(140,0){$ \xBc Y_5 \xBe $}



\end{picture}

\end{diagram}

\vspace{4mm}

\ee

This is just the start of the graph, it continues downward through
$ \xbo $ many levels.

The lines stand for downward pointing arrows.
The lines originating from the $Y_{i}$ correspond to the negative
lines in the original Yablo graph, all others are simple positive
lines, of the type $d(X)=X'.$

The left part of the drawing represents the graph YG', the
right hand part the collapsed graph, the homomorphic image YG".

Compare to Fig.3 in  \cite{RRM13}.
\clearpage
\subsection{
A Comment on Theorem 24 of \cite{RRM13}
}

\label{Section 24}

We comment in this section on the meaning of theorem 24
in  \cite{RRM13}.

\bd

$\hspace{0.01em}$


\label{Definition Symbols}

Fix a denotation $d.$

Let $s(X):=s(d(X))$ be the set of $s \xbe S$ which occur in $d(X).$

Let $r(X) \xcc s(X)$ be the set of relevant $s,$ i.e. which influence
$[d(X)]_{v}$ for
some $v.$
E.g., in $(\xba \xco \xCN \xba) \xcu \xba',$ $ \xba' $ is relevant, $
\xba $ is not.

\ed

\bd

$\hspace{0.01em}$


\label{Definition Simply}

 \xEh
 \xDH
Let $G$ be a directed graph.
For $X \xbe G,$ let the subgraph $C(X)$ of $G$ be the connected component
of $G$ which
contains $X:$
$X \xbe V(C(X)),$ and
$X' \xbe V(C(X))$ iff there is a path in $U(G)$ from $X$ to $X',$
together with the induced edges of $G,$ i.e., if $Y,Y' \xbe V(C(X)),$ and
$YY' \xbe E(G),$
then $YY' \xbe E(C(X)).$
 \xDH
$G$ is called a simply connected graph iff for all $X,Y$ in $G,$ there is
at most
one path in $U(G)$ from $X$ to $Y.$

(One may debate if a loop $X \xcp X$ violates simple connectedness, as we
have the paths $X \xcp X$ and $X \xcp X \xcp X$ - we think so. Otherwise,
we exclude loops.)
 \xDH
Two subgraphs $G',$ $G'' $ of $G$ are disconnected iff there is no path
from any
$X' \xbe G' $ to any $X'' \xbe G'' $ in $U(G).$
 \xEj

\ed

\bfa

$\hspace{0.01em}$


\label{Fact Unconnected}

Let $G,$ $d$ be given, $G=G_{S,d}.$

If $G',$ $G'' $ are two disconnected subgraphs of $G,$ then they can be
given
truth values independently.

\efa

\subparagraph{
Proof
}

$\hspace{0.01em}$


Trivial, as the subgraphs share no propositional variables.
$ \xcz $
\\[3ex]

\bfa

$\hspace{0.01em}$


\label{Fact SimplyConnected}

Let $G$ be simply connected, and $d$ any denotation, $G=G_{S,d}.$
Then $G,d$ has an acceptable valuation.

\efa

\subparagraph{
Proof
}

$\hspace{0.01em}$


This procedure assigns an acceptable valuation to $G$ and $d$ in several
steps.

More precisely, it is an inductive procedure, defining $v$ for more and
more elements, and cutting up the graph into diconnected subgraphs.
If necessary, we will use unions for the definition of $v,$ and the common
refinement for the subgraphs in the limit step.

The first step is a local step, it tries to simplify $d(X)$ by looking
locally
at it, propagating [X] to $X' $ with $X' \xcp X$ if possible, and erasing
arrows
from and to $X,$ if possible. Erasing arrows decomposes the graph into
disconnected subgraphs, as the graph is simply connected.

The second step initializes an arbitrary value $X$ (or, in step (4), uses
a value determined in step (2)), propagates the value to $X' $ for $X'
\xcp X,$
erases the arrow $X' \xcp X.$ Initialising $X$ will have repercussions on
the
$X'' $ for $X \xcp X'',$ so we chose a correct possibility for the $X'' $
(e.g.,
if $d(X)=X'' \xcu X''',$ setting $[X]= \xct,$ requires to set $[X''
]=[X''' ]= \xct,$ too),
and erase the arrows $X \xcp X''.$ As $G$ is simply connected, the only
connection between the different $C(X'')$ is via $X,$ but this was
respected
and erased, and they are now independent.

 \xEh
 \xDH
Local step
 \xEh
 \xDH
For all $X' \xbe s(X)-r(X):$

 \xEh
 \xDH
replace $X' $ in $d(X)$ by $ \xct $ (or, equivalently, $ \xcT),$
resulting in
logically equivalent $d' (X)$ $(s(X')$ might now be empty),

 \xDH
erase the arrow $X \xcp X'.$

Note that $C(X')$ will then be disconnected from $C(X),$ as $G$ is simply
connected.
 \xEj
 \xDH
Do recursively:

If $s(d(X))= \xCQ,$ then $d(X)$ is equivalent to $ \xct $ (or $ \xcT)$
(it might also be
$ \xct \xcu \xcT $ etc.),
so $[X]_{v}=[d(X)]_{v}= \xct $ (or $ \xcT)$ in any acceptable valuation,
and $[d(X)]_{v}$
is independent of $v.$

 \xEh
 \xDH
For $X' \xcp X,$ replace $X$ in $d(X')$ by $ \xct $ (or $ \xcT)$ $(s(X'
)$ might now be empty),
 \xDH
erase $X' \xcp X$ in $G.$

$X$ is then an isolated point in $G,$ so its truth value is independent
of the other truth values (and determined already).
 \xEj
 \xEj

 \xDH
Let $G'' $ be a non-trivial (i.e. not an isolated point) connected
component
of the original graph $G,$ chose $X$ in $G''.$
If $X$ were already fixed as $ \xct $ or $ \xcT,$ then $X$ would have
been isolated
by step (1). So $[X]_{v}$ is undetermined so far.
Moreover, if $X \xcp X' $ in $G'',$ then $d(X')$ cannot
be equivalent to a constant value either, otherwise, the arrow $X \xcp X'
$ would
have been eliminated already in step (1).

Chose arbitrarily a truth value for $d(X),$ say $ \xct.$

 \xEh
 \xDH
Consider any $X' $ s.t. $X' \xcp X$ (if this exists)
 \xEh
 \xDH
Replace $X$ in $d(X')$ with that truth value, here $ \xct.$

 \xDH
Erase $X' \xcp X$

As $G'' $ is simply connected, all such $C(X')$ and $C(X)$ are now
mutually
disconnected.
 \xEj
 \xDH
Consider simultanously all $X'' $ s.t. $X \xcp X''.$
(They are not constants, as any $X'' \xbe V(G'')$ must be a propositional
variable.)

 \xEh
 \xDH
Chose values for all such $X'',$ corresponding to $[X]_{v}=[d(X)]_{v}$
$(= \xct $ here).

E.g., if $d(X)=X'' \xcu X''',$ and the value for $X$ was $ \xct,$ then
we have to chose
$ \xct $ also for $X'' $ and $X'''.$

This is possible independently by
Fact \ref{Fact Unconnected} (page \pageref{Fact Unconnected}),
as the graph $G'' $ is simply connected, and
$X$ is the only connection between the different $X'' $
 \xDH
Erase all such $X \xcp X''.$

$X$ is now an isolated point, and
as $G'' $ is simply connected, all $C(X'')$ are mutually disconnected,
and disconnected
from all $C(X')$ with $X' \xcp X,$ considered in (2.1).
 \xEj
 \xEj
The main argument here is that we may define $[X'' ]_{v}$ and $[X'''
]_{v}$ for all
$X \xcp X'' $ and $X \xcp X''' $ independently, if we respect the
dependencies resulting
through $X.$
 \xDH
Repeat step (1) recursively on all mutually disconnected fragments
resulting
from step (2).
 \xDH
Repeat step (2) for all $X'' $ in (2.2), but instead of the free choice
for $[X]_{v}$ in (2), the choice for the $X'' $
has already been made in step (2.2.1),
and work with this choice.
 \xEj
$ \xcz $
\\[3ex]
\clearpage
\section{
Remarks on Contradictory Structures
}

\label{Section Remarks}
\subsection{
Introductory Comments
}

\label{Section Various}

\be

$\hspace{0.01em}$


\label{Example NotN}

We discuss here some very simple examples, all modifications of the
Yablo structure.

Up to now, we considered graphs isomorphic to (parts of) the natural
numbers
with arrows pointing to bigger numbers. We consider now other cases,
and describe the underlying graphs. Arrows are understood as negative.
 \xEh
 \xDH
Consider the negative numbers (with 0), arrows pointing again to
bigger numbers.
Putting $+$ at 0, and - to all other elements is an acceptable valuation.
 \xDH
Consider a tree with arrows pointing to the root. The tree may be
infinite.
Again $+$ at the root, - at all other elements is an acceptable valuation.
 \xDH
Consider an infinite tree, the root with $ \xbo $ successors $x_{i},$ $i<
\xbo,$ and
from each $x_{i}$ originating a chain of length $i$ as in Fig. 10
of  \cite{RRM13}, putting $+$ at the end of the branches, and -
everywhere else
is an acceptable valuation.
 \xDH
This trivial example shows that an initial segment of a Yablo construction
can again be a Yablo construction.

Instead of considering all $Y_{i},$ $i< \xbo,$ we consider $Y_{i},$ $i<
\xbo + \xbo,$
extending the original construction in the obvious way.
 \xEj

\ee

\bfa

$\hspace{0.01em}$


\label{Fact Finite-Branching}

Let $G$ be loop free and finitely forward branching, i.e. for any $s,$
there are
only finitely many $s' $ such that $s \xcp s' $ in $G.$ Then $G$ is not
dangerous.

(d may be arbitrary, not necessarily of the $ \xcU -$ form, i.e. $ \xcU
\xCN x_{i}.)$

See also Fact \ref{Fact NotClass} (page \pageref{Fact NotClass}).

\efa

\subparagraph{
Proof
}

$\hspace{0.01em}$


Let $d$ be any assignment corresponding to $G.$ Then $d(s)$ is a finite,
classical
formula. Replace $[s]_{v}=[d(s)]_{v}$ by the classical formula $
\xbf_{s}:=s \xcr d(s).$
Then any finite number of $ \xbf_{s}$ is consistent.

Proof: Let $ \xbF $ be a finite set of such $ \xbf_{s},$ and $S_{ \xbF }$
the set of $s$ occurring
in $ \xbF.$ As $G$ is loop free, and $S_{ \xbF }$ finite, we may
initialise the minimal $s \xbe S_{ \xbF }$ (i.e. there is no $s' $
such that $s \xcp s' $ in the part of $G$ corresponding to $ \xbF)$ with
any truth values, and propagate the truth
values upward according to usual valuation rules.
This shows that $ \xbF $ is consistent, i.e. we have constructed a
(partial) acceptable valuation for $d.$

Extend $ \xbF $ by classical compactness, resulting in a total acceptable
valuation for $d.$

(In general, in the logics considered here, compactness obviously does not
hold:
Consider $\{ \xCN \xcU \{Y_{i}:i< \xbo \} \xcv \{Y_{i}:i< \xbo \}.$
Clearly, every finite subset is
consistent, but the entire set is not.)

$ \xcz $
\\[3ex]

\bfa

$\hspace{0.01em}$


\label{Fact TransAndNot}

Let $G$ be transitive, and $d$ be of the type $ \xcU -.$
 \xEh
 \xDH
If $ \xcE X.$ $(succ(X) \xEd \xCQ $ and $ \xcA X' \xbe succ(X).succ(X')
\xEd \xCQ),$ then
$d$ has no acceptable valuation.

Let acceptable $v$ be given, $[.]$ is for this $v.$

Case 1: $[X]= \xct.$ So for all $X' \xbe succ(X)$ $[X' ]= \xcT,$ and
there is such $X',$ so
(either by the prerequisite $succ(X') \xEd \xCQ,$ or by
Remark \ref{Remark Empty} (page \pageref{Remark Empty}))
$ \xcE $ $X'' \xbe succ(X').[X'' ]= \xct,$ but $succ(X') \xcc succ(X),$
a contradiction.

In abbreviation:
$X^{+}$ $ \xcp_{ \xcU -}$ $X'^{-}$ $ \xcp_{ \xcU -}$ $X''^{+}$

Case 2: $[X]= \xcT.$ So $ \xcE X' \xbe succ(X).[X' ]= \xct,$ so $ \xcA
X'' \xbe succ(X').[X'' ]= \xcT,$
and by prerequisite $succ(X') \xEd \xCQ,$ so there is such $X'',$
so by Remark 
\ref{Remark Empty} (page 
\pageref{Remark Empty})  $succ(X'') \xEd \xCQ,$ so
$ \xcE $ $X''' \xbe succ(X'').[X''' ]= \xct,$ but $succ(X'') \xcc
succ(X'),$ a contradiction.

$X^{-}$ $ \xcp_{ \xcU -}$ $X'^{+}$ $ \xcp_{ \xcU -}$ $X''^{-}$ $ \xcp_{
\xcU -}$ $X'''^{+}$

(Here we need Remark \ref{Remark Empty} (page \pageref{Remark Empty})
for the additional step from $X'' $ to $X''' $.)

 \xDH
Conversely:

Let
$ \xcA X$ $(succ(X)= \xCQ $ or $ \xcE X' \xbe succ(X).succ(X')= \xCQ):$

By Remark \ref{Remark Empty} (page \pageref{Remark Empty}),
if $succ(Y)= \xCQ,$ then for any
acceptable valuation, $[Y]= \xct.$ Thus, if there is $X' \xbe succ(X),$
$succ(X')= \xCQ,$
$[X' ]= \xct,$ and $[X]= \xcT.$

Thus, the valuation defined by $[X]= \xct $ iff $succ(X)= \xCQ,$ and $
\xcT $ otherwise is
an acceptable valuation. (Obviously, this definition is free from
contradictions.)
 \xEj

$ \xcz $
\\[3ex]
\subsection{
Basics For a more Systematic Investigation
}
\subsubsection{
Some Trivialities
}

\efa

We will work here with disjunctive normal forms, i.e. with formulas of the
type
$a:= \xcO \{ \xcU a_{i}:i \xbe I\},$ where $a_{i}:=\{a_{i,j}:j \xbe
J_{i}\},$ and the $a_{i,j}$ are propositional
variables or negations thereof.

\bfa

$\hspace{0.01em}$


\label{Fact DNF}

Let $a:= \xcO \{ \xcU a_{i}:i \xbe I\},$ where $a_{i}:=\{a_{i,j}:j \xbe
J_{i}\},$ and the $a_{i,j}$ are propositional
variables or negations thereof.

 \xEh

 \xDH

Let $F:= \xbP \{a_{i}:i \xbe I\}.$

Then $ \xCN a$ $=$ $ \xcO \{ \xcU \{ \xCN a_{i,j}:$ $a_{i,j} \xbe
ran(f)\}:f \xbe F\}.$

(By the laws of distributivity.)

 \xDH

Contradictions will be between two formulas only, one a propositional
variable, the other the negation of the former.

 \xEj

$ \xcz $
\\[3ex]

\efa

For illustration, we develop
Example 
\ref{Example Distributivity} (page 
\pageref{Example Distributivity}), to
see how this works.

This particularly useful as an ilustration for
Section \ref{Section Complicated} (page \pageref{Section Complicated}).

\be

$\hspace{0.01em}$


\label{Example Distributivity}

Consider the following situation:

$x=(a \xcu b) \xco (c \xcu d)$

$a=(aa \xcu ab) \xco (ac \xcu ad)$

$b=(ba \xcu bb) \xco (bc \xcu bd)$

$c=(ca \xcu cb) \xco (cc \xcu cd)$

$d=(da \xcu db) \xco (dc \xcu dd)$

 \xEh

 \xDH Conjunction

 \xEh

 \xDH $a \xcu b$ $=$ $[(aa \xcu ab) \xco (ac \xcu ad)]$ $ \xcu $ $[(ba
\xcu bb) \xco (bc \xcu bd)]$ $=$

$[(aa \xcu ab) \xcu (ba \xcu bb)]$ $ \xco $ $[(aa \xcu ab) \xcu (bc \xcu
bd)]$ $ \xco $
$[(ac \xcu ad) \xcu (ba \xcu bb)]$ $ \xco $ $[(ac \xcu ad) \xcu (bc \xcu
bd)]$
\vspace{2mm}

a has 2 components, $a_{1}=(aa \xcu ab)$ and $a_{2}=(ac \xcu ad),$
analogously
$b_{1}=(ba \xcu bb)$ and $b_{2}=(bc \xcu bd).$

Distributivity results in choice functions in the components:
$(a_{1} \xcu b_{1})$ $ \xco $ $(a_{1} \xcu b_{2})$ $ \xco $ $(a_{2} \xcu
b_{1})$ $ \xco $ $(a_{2} \xcu b_{2}).$

 \xDH $c \xcu d$ $=$ $[(ca \xcu cb) \xco (cc \xcu cd)]$ $ \xcu $ $[(da
\xcu db) \xco (dc \xcu dd)]$ $=$

$[(ca \xcu cb) \xcu (da \xcu db)]$ $ \xco $ $[(ca \xcu cb) \xcu (dc \xcu
dd)]$ $ \xco $
$[(cc \xcu cd) \xcu (da \xcu db)]$ $ \xco $ $[(cc \xcu cd) \xcu (dc \xcu
dd)]$

 \xEj

 \xDH Negation

 \xEh
 \xDH
$ \xCN x$ $=$ $ \xCN ((a \xcu b) \xco (c \xcu d))$ $=$ $ \xCN (a \xcu b)
\xcu \xCN (c \xcu d)$ $=$ $(\xCN a \xco \xCN b)$ $ \xcu $ $(\xCN c \xco
\xCN d)$ $=$

$(\xCN a \xcu \xCN c)$ $ \xco $ $(\xCN a \xcu \xCN d)$ $ \xco $ $(\xCN
b \xcu \xCN c)$ $ \xco $ $(\xCN b \xcu \xCN d)$
\vspace{2mm}

Negation works with distributivity, thus
with choice functions, here in the 2 components $x_{1}=a \xcu b,$ $x_{2}=c
\xcu d.$

The elements of the components stay the same, only the sign changes.

 \xDH
$ \xCN a$ $=$ $ \xCN ((aa \xcu ab) \xco (ac \xcu ad))$ $=$ $ \xCN (aa \xcu
ab) \xcu \xCN (ac \xcu ad)$ $=$
$(\xCN aa \xco \xCN ab)$ $ \xcu $ $(\xCN ac \xco \xCN ad)$ $=$

$(\xCN aa \xcu \xCN ac)$ $ \xco $ $(\xCN aa \xcu \xCN ad)$ $ \xco $ $(
\xCN ab \xcu \xCN ac)$ $ \xco $ $(\xCN ab \xcu \xCN ad)$
 \xDH
$ \xCN b$ $=$ $ \xCN ((ba \xcu bb) \xco (bc \xcu bd))$ $=$ $ \xCN (ba \xcu
bb) \xcu \xCN (bc \xcu bd)$ $=$
$(\xCN ba \xco \xCN bb)$ $ \xcu $ $(\xCN bc \xco \xCN bd)$ $=$

$(\xCN ba \xcu \xCN bc)$ $ \xco $ $(\xCN ba \xcu \xCN bd)$ $ \xco $ $(
\xCN bb \xcu \xCN bc)$ $ \xco $ $(\xCN bb \xcu \xCN bd)$
 \xDH
$ \xCN c$ $=$ $ \xCN ((ca \xcu cb) \xco (cc \xcu cd))$ $=$ $ \xCN (ca \xcu
cb) \xcu \xCN (cc \xcu cd)$ $=$
$(\xCN ca \xco \xCN cb)$ $ \xcu $ $(\xCN cc \xco \xCN cd)$ $=$

$(\xCN ca \xcu \xCN cc)$ $ \xco $ $(\xCN ca \xcu \xCN cd)$ $ \xco $ $(
\xCN cb \xcu \xCN cc)$ $ \xco $ $(\xCN cb \xcu \xCN cd)$
 \xDH
$ \xCN d$ $=$ $ \xCN ((da \xcu db) \xco (dc \xcu dd))$ $=$ $ \xCN (da \xcu
db) \xcu \xCN (dc \xcu dd)$ $=$
$(\xCN da \xco \xCN db)$ $ \xcu $ $(\xCN dc \xco \xCN dd)$ $=$

$(\xCN da \xcu \xCN dc)$ $ \xco $ $(\xCN da \xcu \xCN dd)$ $ \xco $ $(
\xCN db \xcu \xCN dc)$ $ \xco $ $(\xCN db \xcu \xCN dd)$

 \xEj

 \xDH Conjunction of negations

$ \xCN a \xcu \xCN c$ $=$

[ $(\xCN aa \xcu \xCN ac)$ $ \xco $ $(\xCN aa \xcu \xCN ad)$ $ \xco $ $(
\xCN ab \xcu \xCN ac)$ $ \xco $ $(\xCN ab \xcu \xCN ad)$ ] $ \xcu $

[ $(\xCN ca \xcu \xCN cc)$ $ \xco $ $(\xCN ca \xcu \xCN cd)$ $ \xco $ $(
\xCN cb \xcu \xCN cc)$ $ \xco $ $(\xCN cb \xcu \xCN cd)$ ] $=$
\vspace{2mm}

[ $(\xCN aa \xcu \xCN ac)$ $ \xcu $ $(\xCN ca \xcu \xCN cc)$ ] $ \xco $
[ $(\xCN aa \xcu \xCN ac)$ $ \xcu $ $(\xCN ca \xcu \xCN cd)$ ] $ \xco $

[ $(\xCN aa \xcu \xCN ac)$ $ \xcu $ $(\xCN cb \xcu \xCN cc)$ ] $ \xco $
[ $(\xCN aa \xcu \xCN ac)$ $ \xcu $ $(\xCN cb \xcu \xCN cd)$ ] $ \xco $

[ $(\xCN aa \xcu \xCN ad)$ $ \xcu $ $(\xCN ca \xcu \xCN cc)$ ] $ \xco $
[ $(\xCN aa \xcu \xCN ad)$ $ \xcu $ $(\xCN ca \xcu \xCN cd)$ ] $ \xco $

[ $(\xCN aa \xcu \xCN ad)$ $ \xcu $ $(\xCN cb \xcu \xCN cc)$ ] $ \xco $
[ $(\xCN aa \xcu \xCN ad)$ $ \xcu $ $(\xCN cb \xcu \xCN cd)$ ] $ \xco $

[ $(\xCN ab \xcu \xCN ac)$ $ \xcu $ $(\xCN ca \xcu \xCN cc)$ ] $ \xco $
[ $(\xCN ab \xcu \xCN ac)$ $ \xcu $ $(\xCN ca \xcu \xCN cd)$ ] $ \xco $

[ $(\xCN ab \xcu \xCN ac)$ $ \xcu $ $(\xCN cb \xcu \xCN cc)$ ] $ \xco $
[ $(\xCN ab \xcu \xCN ac)$ $ \xcu $ $(\xCN cb \xcu \xCN cd)$ ] $ \xco $

[ $(\xCN ab \xcu \xCN ad)$ $ \xcu $ $(\xCN ca \xcu \xCN cc)$ ] $ \xco $
[ $(\xCN ab \xcu \xCN ad)$ $ \xcu $ $(\xCN ca \xcu \xCN cd)$ ] $ \xco $

[ $(\xCN ab \xcu \xCN ad)$ $ \xcu $ $(\xCN cb \xcu \xCN cc)$ ] $ \xco $
[ $(\xCN ab \xcu \xCN ad)$ $ \xcu $ $(\xCN cb \xcu \xCN cd)$ ]
\vspace{2mm}

$ \xCN a$ and $ \xCN c$ have 4 components each, we chose in the 4
components.

 \xEj
\subsubsection{
Cells and Contradictions
}

\ee

\bd

$\hspace{0.01em}$


\label{Definition Cell}

 \xEh
 \xDH
A cell is a set of labelled paths with the following properties:
 \xEh
 \xDH they have a common origin (this is not essential, it simplifies
the definition slightly).
 \xDH for each pair of paths there is a point $x,$ where they diverge,
 \xDH for each pair of paths, once they diverged, they meet again,
but will not diverge again.
 \xEj

 \xDH
A contradictory cell is a cell such that there is at least one
valuation and one pair of paths whose valued versions contradict
each other.
 \xEj

\ed

\be

$\hspace{0.01em}$


\label{Example Cell}

 \xEh
 \xDH
The following are cells:
 \xEh
 \xDH $x \xcp y,$ $x \xcP y$
 \xDH $x \xcp y \xcp z,$ $x \xcp z$
 \xDH $x \xcP y \xcP z,$ $x \xcP z$
 \xDH $x \xcP y \xcP z \xcp w$, $x \xcP z \xcp w$
 \xDH $w \xcp x \xcP y \xcP z,$ $w \xcp x \xcP z$
 \xDH $u \xcp x \xcP x' \xcp x'' \xcp z,$
$u \xcp x \xcp x'' \xcp z,$
$u \xcp y \xcP y' \xcp y'' \xcp z,$
$u \xcp y \xcp y'' \xcp z$

Note that this cell is composed of two sub-cells, which are in parallel.
 \xEj
 \xDH
The following are not cells, as they diverge again:
 \xEh
 \xDH $x \xcp y \xcp z,$ $x \xcP y \xcP z$
 \xDH $x \xcP y \xcP z \xcP y' \xcP z',$ $z \xcP z \xcP z' $
 \xEj
These cells are serially connected.
 \xDH
In Example 
\ref{Example Value} (page 
\pageref{Example Value}), the following are cells:
 \xEh
 \xDH
$x_{0} \xcP x_{1} \xcp x_{2},$ $x_{0} \xcp x_{2}$,
 \xDH
$x_{0} \xcP x_{1} \xcp x_{2} \xcp x_{3},$ $x_{0} \xcp x_{2} \xcp x_{3}$,
 \xDH
$x_{0} \xcP x_{1} \xcp x_{2} \xcp x_{3} \xcp x_{4},$ $x_{0} \xcp x_{2}
\xcp x_{3} \xcp x_{4}$, $x_{0} \xcp x_{4}$
 \xDH
$x_{2} \xcp x_{3},$ $x_{2} \xcP x_{3}$
 \xDH
$x_{0} \xcp x_{2} \xcp x_{3},$ $x_{0} \xcp x_{2} \xcP x_{3}$
 \xDH
and the following is not a cell

$x_{0} \xcP x_{1} \xcp x_{2} \xcP x_{3},$ $x_{0} \xcp x_{2} \xcp x_{3}$

(after meeting at $x_{2},$ the paths continue differently).
 \xEj
 \xEj

\ee

\bfa

$\hspace{0.01em}$


\label{Fact Cell-1}

Let $ \xbs $ and $ \xbt $ be two paths with same origin, which meet again
at some $x$
and are contradictory.
If $v(\xbs_{0})=v(\xbt_{0}),$ and $v(\xbs_{x}) \xEd v(\xbt_{x}),$ then
the number of negative
arrows in $ \xbs $ between $ \xbs_{0}$ and $ \xbs_{x}$ modulo 2 is unequal
to the number of negative
arrows in $ \xbt $ between $ \xbt_{0}$ and $ \xbt_{x}$ modulo 2.

\efa

The following example is very important, and the basis for the original
Yablo construction, as well as our ``saw blade'' construction.

\be

$\hspace{0.01em}$


\label{Example Simple-Cells}

See also Example \ref{Example Paths} (page \pageref{Example Paths}).

We consider now some simple, contradictory cells. They should not only
be contradictory for the case $x+,$ but also be a potential start for
the case $x-,$ without using more complex cells.

For this, we order the complexity of the cases by $(1)<(2.1)<(2.3)$ below,
(2.2) is not contradictory, so it is excluded.

See Diagram \ref{Diagram ContCell} (page \pageref{Diagram ContCell}).

 \xEh
 \xDH The cell with 2 arrows.

It corresponds to the formula
$d(x)=y \xcu \xCN y,$ graphically, it has a positive and a negative arrow
from
$x$ to $y,$ so exactly one of $ \xba $ and $ \xbb $ is negative.

If $x$ is positive, we have a contradiction.

If $x$ is negative, however, we have a problem. Then, we have
$d(x)=y \xco \xCN y.$ Let $ \xba $ be the originally positive path, $ \xbb
$ the
originally negative path. Note that $ \xba $ is now negative, and $ \xbb $
is positive.
The $ \xba $ presents no problem, as $y$ is positive,
and we can append the same construction to $y,$ and have a contradiction.
However, $ \xbb $ has to lead to a contradiction, too, and, as we will not
use more complicated cells, we face the same problem again, $y$ is
negative. So we have an ``escape path'', assigning $ \xcT $ to all elements
in one
branch.

(Consider
$x_{0} \xch_{ \xCN }x_{1} \xch_{ \xCN }x_{2} \xch_{ \xCN }x_{3}$  \Xl,
setting all $x_{i}:=-$ is a consistent valuation.
So combining this cell with itself does not result in a contradictory
structure.)

Of course, appending at $y$ a Yablo Cell (see below, case (2.3))
may be the beginning of a contradictory structure, but this is
``cheating'', we use a more complex cell.

 \xDH Cells with 3 arrows.

Note that the following examples are not distinguished in the graph!

Again, we want a contradiction for $x$ positive, so we need an $ \xcu $ at
$x,$
$d(x)$ (or $x,$ we abbreviate) and have the possibilities
(up to equivalence) $x=x' \xcu y,$ $x=x' \xcu \xCN y,$ $x= \xCN x' \xcu
y,$ $x= \xCN x' \xcu \xCN y.$

Again, if $x$ is negative, all paths $ \xba:x \xcp y,$ $ \xbb:x \xcp x'
$ (or $ \xbb \xbg:x \xcp x' \xcp y)$
have to lead to a contradiction.

 \xEh
 \xDH one negative arrow, with $d(x)= \xCL x' \xcu \xCL y$
 \xEh
 \xDH
$x \xcp x' \xcP y,$ $x \xcp y,$ corresponding to $x=x' \xcu y,$ so if $x$
is negative, this
is $ \xCN x' \xco \xCN y.$

But, at $y,$ this possible paths ends, and we have the same situation
again,
with a negative start, as in Case (1).

 \xDH
$x \xcp x' \xcp y,$ $x \xcP y$

Here, we have again a positive path to $y,$ through $x',$ so both $x' $
and $y$
will be negative, and neither gives a start for a new contradiction.
 \xDH
$x \xcP x' \xcp y,$ $x \xcp y$

This case is analogous to case (2.1.1).
 \xEj

(Similar arguments apply to more complicated cells with an even number
of negative arrows until the first branching point - see
Remark \ref{Remark YC} (page \pageref{Remark YC})  below
and the discussion of the Yablo construction,
Section \ref{Section Yablo-Analysis} (page \pageref{Section Yablo-Analysis}).

 \xDH 2 negative arrows: not contradictory.

 \xDH The original type of contradiction in Yablo's construction

$x \xcP x' \xcP y,$ $x \xcP y.$

This will be discussed in detail below in
Section \ref{Section Yablo-Analysis} (page \pageref{Section Yablo-Analysis}).
But we see already that both paths, $x \xcP x' $ and $x \xcP y$ change
sign,
so $x' $ and $y$ will be positive, appending the same type of cell
at $x' $ and $y$ solves the problem (locally), and offers no escape.

 \xEj
 \xEj

\ee

\bd

$\hspace{0.01em}$


\label{Definition YC}

 \xEh
 \xDH
We call the contradiction of the type (2.3) of
Example \ref{Example Simple-Cells} (page \pageref{Example Simple-Cells}), i.e.
$x \xcP x' \xcP y,$ $x \xcP y,$
a Yablo Cell, YC,
and sometimes $x$ its head, $y$ its foot, and $x' $ its knee.

 \xDH
We sometimes abbreviate a Yablo Cells simply by $ \xeA,$ without going
into any further details.

 \xEj

If we combine Yablo Cells, the knee for one cell may become the head
for another Cell, etc.

See Diagram \ref{Diagram YC} (page \pageref{Diagram YC}).

\clearpage

\begin{diagram}

\label{Diagram YC}
\index{Diagram YC}

\unitlength0.7mm
\begin{picture}(150,100)(0,0)

\put(0,80){{\rm\bf Yablo Cells }}

\put(15,65){Head}
\put(39,65){Knee}
\put(15,35){Foot}


\put(20,60){$x$}
\put(20,40){$y$}
\put(26,61){\line(1,0){12}}
\put(32,60){\line(0,1){2}}
\put(22,43){\line(0,1){14}}
\put(21,50){\line(1,0){2}}
\put(24,43){\line(1,1){16}}
\put(31,51.5){\line(1,-1){1.5}}


\put(40,60){$x'$}


\put(75,65){Head}
\put(93,65){Knee=Head'}
\put(75,35){Foot}
\put(120,65){Knee'}
\put(95,35){Foot'}


\put(80,60){$x$}
\put(80,40){$y$}
\put(86,61){\line(1,0){12}}
\put(92,60){\line(0,1){2}}
\put(82,43){\line(0,1){14}}
\put(81,50){\line(1,0){2}}
\put(84,43){\line(1,1){16}}
\put(91,51.5){\line(1,-1){1.5}}


\put(100,60){$x'$}
\put(100,40){$y'$}
\put(106,61){\line(1,0){12}}
\put(112,60){\line(0,1){2}}
\put(102,43){\line(0,1){14}}
\put(101,50){\line(1,0){2}}
\put(104,43){\line(1,1){16}}
\put(111,51.5){\line(1,-1){1.5}}


\put(120,60){$x''$}

\end{picture}

\end{diagram}

\vspace{10mm}

\clearpage

\begin{diagram}

\label{Diagram ContCell}
\index{Diagram ContCell}

\unitlength1.0mm
\begin{picture}(150,100)(0,0)

\put(0,80){{\rm\bf Contradictory Cells }}


\put(20,60){$x$}
\put(20,40){$y$}
\put(19,43){\line(0,1){14}}
\put(22,43){\line(0,1){14}}

\put(15,50){$\xba$}
\put(24,50){$\xbb$}

\put(50,60){$x$}
\put(50,40){$y$}
\put(56,61){\line(1,0){12}}
\put(51,43){\line(0,1){14}}
\put(54,43){\line(1,1){16}}
\put(70,60){$x'$}

\put(47,50){$\xba$}
\put(62,64){$\xbb$}
\put(64,50){$\xbg$}

\end{picture}

\end{diagram}

\vspace{10mm}

\clearpage

\begin{diagram}

\label{Diagram Remark-YC}
\index{Diagram Remark-YC}

\unitlength1.0mm
\begin{picture}(150,100)(0,0)

\put(0,80){{\rm\bf See Remark $\ref{Remark YC}$ }}

\put(10,60){Head}
\put(42,57){Knee}
\put(15,35){Foot}


\put(20,60){$x$}
\put(20,40){$y$}
\put(26,61){\line(1,0){12}}
\put(32,60){\line(0,1){2}}
\put(22,43){\line(0,1){14}}
\put(21,50){\line(1,0){2}}
\put(24,43){\line(1,1){16}}
\put(31,51.5){\line(1,-1){1.5}}


\put(40,60){$x'$}
\put(60,60){$z$}

\put(46,61){\line(1,0){12}}
\put(52,60){\line(0,1){2}}

\put(22,63){\line(0,1){6}}
\put(62,63){\line(0,1){6}}
\put(22,69){\line(1,0){40}}
\put(42,68){\line(0,1){2}}

\end{picture}

\end{diagram}

\vspace{10mm}

\ed

\br

$\hspace{0.01em}$


\label{Remark YC}

 \xEh
 \xDH
The distinction between $x' $ and $y,$ i.e. between knee and foot, is
very important. In the case $x+,$ we have at $y$ a complete contradiction,
at $x',$ we have not yet constructed a contradiction. Thus, if we have
at $x' $ again an $ \xcu $ (as at $x),$ this becomes an $ \xco,$ and we
have to
construct a contradiction for all $x' \xcp z$ (or $x' \xcP z),$ not only
for
$x' \xcP y.$ Otherwise, we have an escape possibility for $x+.$ Obviously,
the contradiction need not be immediate at $z,$ the important property
is that ALL paths through all $z$ lead to a contradiction, and the
simplest
way is to have the contradiction immediately at $z$ - as in Yablo's
construction, and our saw blades.

See Diagram \ref{Diagram Remark-YC} (page \pageref{Diagram Remark-YC}).

See also the discussion in
Fact \ref{Fact Infin-Branching} (page \pageref{Fact Infin-Branching})  and
the construction of saw blades in
Section \ref{Section Saw-Blades} (page \pageref{Section Saw-Blades}).

 \xDH
As we work for a contradiction in the case x-, too, the simplest
way to achieve this is to have a negative arrow $x \xcP x',$ and at $x' $
again
an $ \xcu.$ This gives a chance to construct a contradiction at $x'.$
Of course, we have to construct a contradiction at $y,$ too, as in
the case $x-,$ we have an $ \xco $ at $x.$

Of course, we may have branches originating at $x',$ which all lead to
contradictions in the case $x-,$ so $x \xcp x' $ (resp. $ \xco $ at $x')$
is
possible, too.

But, for the simple construction, we need $ \xcP $ between $x$ and $x',$
and
$ \xcu $ at $x'.$ And this leads to the construction of contradictions
for all $x' \xcp z$ (or $x' \xcP z)$ as just mentioned above.

 \xEj
\clearpage
\subsubsection{
Further Comments On the Yablo Construction and Variations
}

\label{Section Yablo-Analysis}

\er

We start with a Yablo Cell (YC) $x_{0} \xcP x_{1} \xcP x_{2},$ $x_{0} \xcP
x_{2}.$

Note that in both cases, initialising $x_{0}$ with $+$ or with -,
the paths $x_{0} \xcP x_{1} \xcP x_{2},$ $x_{0} \xcP x_{2}$ are
contradictory, but in the case
$+,$ with $ \xcu $ at $x_{0},$ we ``feel'' the contradiction, in the case -
with $ \xco $ at $x_{0},$
the contradiction is irrelevant.

We are mainly interested in solutions which may be repeated without
any modification.

 \xEh

 \xDH

If $x_{0}+,$ then $x_{1}-$ and $ \xco.$ Thus, to have a contradiction at
$x_{0}+,$
any $y$ with $x_{1} \xcP y$ (or $x_{1} \xcp y)$ has to be contradictory,
either directly
at $x_{0}$ (as is the case for $x_{1} \xcP x_{2}),$ or in some other way,
for say $x_{0}'.$

If we repeat the latter construction with $x_{0}',$ then we may construct
an escape path $x_{0} \Xl x_{0}'  \Xl x_{0}''  \Xl $ for the $ \xco $'s,
i.e. an infinite choice
for the $ \xco $'s, never leading to a contradiction.

\be

$\hspace{0.01em}$


\label{Example OR}

 \xEh
 \xDH
$x_{0} \xcP x_{1} \xcP x_{1}' \xcP x_{3} \xcP x_{3}' $  \Xl., $x_{0}
\xcP x_{2},$ $x_{1} \xcP x_{2},$ $x_{1}' \xcP x_{4},$ $x_{3} \xcP x_{4}$
etc.
with $x_{0} \xcP x_{1} \xcP x_{1}' \xcP x_{3} \xcP x_{3}' $  \Xl. the
escape route.
(The $x_{1} \xcP x_{1}' $ serve to have at $x_{1}' $ again an $ \xcu).$
 \xDH
$x_{0} \xcP x_{1} \xcP x_{3} \xcP x_{5}$  \Xl., $x_{0} \xcP x_{2},$
$x_{1} \xcP x_{2},$ $x_{1} \xcP x_{4},$ $x_{3} \xcP x_{4},$ $x_{3} \xcP
x_{6},$ $x_{5} \xcP x_{6}$ etc.
with $x_{0} \xcP x_{1} \xcP x_{3} \xcP x_{5}$  \Xl. the escape route.

This is similar to (1), the escape is constructed at $x_{1},$ $x_{3}$ is
an $ \xcu,$
but we have an escape again at $x_{5},$ etc.
 \xEj
 \xEI
 \xDH
We repair the new possibilities $x_{1} \xcP y$ by direct contradictions
$x_{0} \xcP y,$
avoiding procrastination recursively leading to other contradictions.

(We indroduced a new arrow $x_{1} \xcP y,$ and have to repair the new
possibility for the OR-case in $x_{1}$ by the arrow $x_{0} \xcP y.)$
 \xEJ

\ee

 \xDH

If we simplify, we might use the arrow $x_{1} \xcP x_{2},$ instead of
starting anew e.g.
as above with $x_{1} \xcP x_{3} \xcP x_{4},$ $x_{1} \xcP x_{4}$ in case
(2) of
Example \ref{Example OR} (page \pageref{Example OR}).

We have two possibilities:

 \xEh
 \xDH

Use $x_{1} \xcP x_{2}$ for the ``long'' arrow, add $x_{1} \xcP x_{3} \xcP
x_{2}.$

This leads to the problem that $x_{2}$ becomes a smallest element, and we
may insert truth values bottom to top:

$x_{0} \xcP x_{1} \xcP x_{3} \xcP x_{4} \xcP x_{5}$  \Xl., $x_{1} \xcP
x_{2},$ $x_{3} \xcP x_{2},$ $x_{4} \xcP x_{2},$ $x_{5} \xcP x_{2},$ etc.

So, this solution is not ``sustainable''. (Of course, we may continue below
$x_{2}$ as for $x_{0}.$ But this is unnecessarily complicated.)

 \xDH

Use $x_{1} \xcP x_{2}$ for the ``short'' arrow, add $x_{1} \xcP x_{3},$
$x_{2} \xcP x_{3},$ etc.

This is the original Yablo construction.

 \xEh

 \xDH

As the truth value of $x_{3}$ must not be definable, $x_{3}$ has to be the
head
of a YC, so there must be $x_{3} \xcP x_{4} \xcP x_{5},$ $x_{3} \xcP
x_{5},$ each again heads of $YC$'s.

 \xEI
 \xDH
This forces infinite depth, so we cannot fill values from below.

 \xEJ

But as $x_{2}$ is the head of a YC, by
(1.2) above,
we have to add arrows $x_{2} \xcP x_{4}$ and $x_{2} \xcP x_{5}.$

Recursively, this forces us again by
(1.2) above
to add arrows $x_{1} \xcP x_{4}$ and $x_{1} \xcP x_{5},$ and then
to add arrows $x_{0} \xcP x_{4}$ and $x_{0} \xcP x_{5}.$

 \xEI
 \xDH
This forces infinite branching (width).
 \xEJ

See also Example \ref{Example YA} (page \pageref{Example YA}).

 \xDH

Note that it is not necessary to keep the whole construction once
created, it is enough to keep suitable fragments cofinally often.

 \xEj

 \xEj

 \xDH

Adding constants:

Suppose we have $x \xcP x',$ $x \xcP x'',$  \Xl  and add the arrow $x
\xcP TRUE.$
Then $x$ becomes FALSE (because of $ \xcP).$

Suppose we have $x \xcP x',$ $x \xcP x'',$  \Xl  and add the arrow $x
\xcP FALSE.$
Then $x$ is as if the new arrow does not exist, except when there are no
$x' $ etc.,
then $x$ becomes TRUE (because of $ \xcP).$

Of course, this propagates upward (and downward).

 \xDH
We emphasize again the importance of
 \xEh
 \xDH repairing added arrows

and
 \xDH avoiding procrastination, see also
Example \ref{Example Procrastination} (page \pageref{Example Procrastination})
 \xEj

 \xEj

\be

$\hspace{0.01em}$


\label{Example YA}

$x_{0} \xcP x_{1} \xcP x_{2} \xcP x_{3} \xcP x_{4},$ $x_{0} \xcP x_{2},$
$x_{0} \xcP x_{3},$

$x_{1} \xcP x_{3},$ $x_{1} \xcP x_{4},$

$x_{2} \xcP x_{4},$

$x_{3} \xcP x_{4},$

add $x_{0} \xcP x_{4}$ in the last step.

%
%
%
%
%
%
%
%
%
%
%
%
%
%
%
%
%
%
%
%
%



\ee

The following
Example 
\ref{Example Procrastination} (page 
\pageref{Example Procrastination})  shows that
infinitely many finitely branching points cannot always
replace infinite branching - there is an infinite
``procrastination branch'' or ``escape branch''.
This modification of the Yablo structure has one acceptable
valuation for $Y_{1}:$

\be

$\hspace{0.01em}$


\label{Example Procrastination}

Let $Y_{i},$ $i< \xbo $ as usual, and introduce new $X_{i},$ $3 \xck i<
\xbo.$

Let $Y_{i} \xcp Y_{i+1},$ $Y_{i} \xcp X_{i+2},$ $X_{i} \xcp Y_{i},$ $X_{i}
\xcp X_{i+1},$ with

$d(Y_{i}):=$ $ \xCN Y_{i+1} \xcu X_{i+2},$ $d(X_{i}):= \xCN Y_{i} \xcu
X_{i+1}.$

If $Y_{1}= \xct,$ then $ \xCN Y_{2} \xcu X_{3},$ by $X_{3},$ $ \xCN Y_{3}
\xcu X_{4},$ so, generally,

if $Y_{i}= \xct,$ then $\{ \xCN Y_{j}:$ $i<j\}$ and $\{X_{j}:$ $i+1<j\}.$

If $ \xCN Y_{1},$ then $Y_{2} \xco \xCN X_{3},$ so if $ \xCN X_{3},$
$Y_{3} \xco \xCN X_{4},$ etc., so, generally,

if $ \xCN Y_{i},$ then $ \xcE j(i<j,$ $Y_{j})$ or $ \xcA j\{ \xCN X_{j}:$
$i+1<j\}.$

Suppose now $Y_{1}= \xct,$ then $X_{j}$ for all $2<j,$ and $ \xCN Y_{j}$
for all $1<j.$
By $ \xCN Y_{2}$ there is $j,$ $2<j,$ and $Y_{j},$ a contradiction, or $
\xCN X_{j}$ for all $3<j,$
again a contradiction.

But $ \xCN Y_{1}$ is possible, by setting $ \xCN Y_{i}$ and $ \xCN X_{i}$
for all $i.$

Thus, replacing infinite branching by an infinite number of finite
branching does not work for the Yablo construction, as we can always chose
the
``procrastinating'' branch.

See Diagram 
\ref{Diagram Procrastination} (page 
\pageref{Diagram Procrastination}).

\ee

We can turn this into a trivial little fact:

\bfa

$\hspace{0.01em}$


\label{Fact Procrastination}

If, at each stage, we leave some work undone, we construct an
escape path, following which never finish work.
(Obvious, or look at the proof of Koenig's Infinity Lemma (in set
theory).)

We have to be careful, however. This applies to an ever branching
situation,
but not to a set of branches given at the outset, where we eliminate
one branch after the other. Consider the classic example, where we have
$ \xbo $ many branches of height $n$ for branch $n,$ branching directly
off the root,
which we eliminate one after
the other. There is no escape branch here.

$ \xcz $
\\[3ex]

\clearpage

\begin{diagram}

\label{Diagram Procrastination}
\index{Diagram Procrastination}

\unitlength1.0mm
\begin{picture}(150,180)(0,0)

\put(0,175){{\rm\bf Diagram for Example \ref{Example Procrastination}}}

\put(0,160){$Y_1$}
\put(2,158){\line(0,-1){32}}
\put(3,158){\line(3,-4){23}}
\put(0,132){\line(1,0){4}}

\put(0,120){$Y_2$}
\put(2,118){\line(0,-1){32}}
\put(25,120){$X_3$}
\put(27,118){\line(0,-1){32}}
\put(3,118){\line(3,-4){23}}
\put(26,118){\line(-3,-4){23}}
\put(0,92){\line(1,0){4}}
\put(6,94){\line(4,-3){3}}

\put(0,80){$Y_3$}
\put(2,78){\line(0,-1){32}}
\put(25,80){$X_4$}
\put(27,78){\line(0,-1){32}}
\put(3,78){\line(3,-4){23}}
\put(26,78){\line(-3,-4){23}}
\put(0,52){\line(1,0){4}}
\put(6,54){\line(4,-3){3}}

\put(0,40){$Y_4$}
\put(2,38){\line(0,-1){32}}
\put(25,40){$X_5$}
\put(27,38){\line(0,-1){32}}
\put(3,38){\line(3,-4){23}}
\put(26,38){\line(-3,-4){23}}
\put(0,12){\line(1,0){4}}
\put(6,14){\line(4,-3){3}}

\put(0,0){$Y_5$}
\put(25,0){$X_6$}

\end{picture}

\end{diagram}

\vspace{4mm}

\clearpage

\clearpage
\subsubsection{
More Complicated Systems of Contradictions
}

\label{Section More}

\efa

We consider now more complex cells, built up, as before,
from negative arrows and conjunctions.

\bn

$\hspace{0.01em}$


\label{Notation NotOr}

We may write $x-,$ $x \xco,$ $x- \xco,$ etc., to abbreviate that the
truth value of $x$
is negative, we have an $'' \xco'' $ at $x,$ both hold, etc.

This is useful to find one's way through the different cases and
valuations.

\en

We start again from $x_{0}.$ If $x_{0}-,$ then $x_{0} \xco,$ and each
branch starting
at $x_{0}$ has to lead directly or indirectly to a contradiction, i.e.
a contradiction cell.

We concentrate on $x_{0}+,$ so we need a contradiction at $x_{0}.$
 \xEh
 \xDH
In the simplest case, with no branching except at $x_{0},$
we have a contradiction: $x_{0} \xcP x_{1} \xcP x_{2},$ $x_{0} \xcP
x_{2},$ the Yablo Cell.
See Diagram \ref{Diagram C1} (page \pageref{Diagram C1}).

 \xDH
The following situation is hardly more complicated:
$x_{0} \xcP x_{1} \xcP x_{2} \xcP x_{3},$ $x_{0} \xcP y_{1} \xcP x_{3}.$
See Diagram \ref{Diagram C2} (page \pageref{Diagram C2}).

 \xDH

In the situation of the Yablo construction, we branch at
$x_{1},$ but as $x_{1}- \xco,$ we do not know if $x_{1}-$ holds
because of $x_{2}+,$ or because of $x_{2}' +,$ so we have two branches
originating in $x_{0},$ which both have to be contradictory.
See Diagram \ref{Diagram C3} (page \pageref{Diagram C3}).

 \xDH
Suppose we do not branch at $x_{1},$ but at $x_{2},$ and have no
contradiction at $x_{2}:$

$x_{0} \xcP x_{1} \xcP x_{2} \xcP x_{3},$ $x_{2} \xcP x_{3}',$ $x_{0}
\xcP y_{1} \xcP x_{3}.$
See Diagram \ref{Diagram C4} (page \pageref{Diagram C4}).

If $x_{0}+,$ then $x_{2}+$ by $x_{0} \xcP x_{1} \xcP x_{2},$ and $x_{3}+$
by
$x_{0} \xcP y_{1} \xcP x_{3}.$ But, as $x_{2}+= \xCN x_{3} \xcu \xCN
x_{3}',$ we have a
contradiction, because $x_{2}+ \xcu.$
So we do not need some $x_{0} \xcP y_{1}' \xcP x_{3}'.$
 \xDH
Our investigation is asymmetrical, as we concentrate on the
$x_{i}' s.$ For instance, if we were to branch at
$y_{1},$ then the branch leading not to $x_{3}$ would also have to
be contradictory, e.g. by:

$x_{0} \xcP x_{1} \xcP x_{2} \xcP x_{3},$ $x_{2} \xcP x_{3}',$ $x_{0}
\xcP y_{1} \xcP x_{3},$ $y_{1} \xcP x_{3}'.$
See Diagram \ref{Diagram C5} (page \pageref{Diagram C5}).

 \xDH
Take this one step further:

$x_{0} \xcP x_{1} \xcP x_{2} \xcP x_{3} \xcP x_{4},$ $x_{3} \xcP x_{4},$
$x_{3} \xcP x_{4}',$ $x_{2} \xcP x_{3}',$ $x_{0} \xcP x_{4},$ $x_{0}
\xcP x_{4}'.$
See Diagram \ref{Diagram C6} (page \pageref{Diagram C6}).

$x_{0} \xcP x_{4}$ alone would not be enough, we have to add $x_{0} \xcP
x_{4}',$
as $x_{3}- \xco,$ and we do not know which of $x_{4}+$ or $x_{4}' +$
holds.
 \xDH
So we check formulas with alternating quantifiers for contradictions.
The author does not know if there are known facts about this question.

 \xEj

\clearpage

\begin{diagram}

\label{Diagram C1}
\index{Diagram C1}

\unitlength1.0mm
\begin{picture}(150,100)(0,0)

\put(0,80){{\rm\bf C1 }}

\put(10,20){\circle*{1}}
\put(30,20){\circle*{1}}
\put(50,20){\circle*{1}}

\put(12,20){\line(1,0){16}}
\put(32,20){\line(1,0){16}}
\put(20,21){\line(0,-1){2}}
\put(40,21){\line(0,-1){2}}

\put(10,18){\line(0,-1){8}}
\put(50,18){\line(0,-1){8}}
\put(10,10){\line(1,0){40}}
\put(30,11){\line(0,-1){2}}

\put(8,23){$x_0$}
\put(28,23){$x_1$}
\put(48,23){$x_2$}

\end{picture}

\end{diagram}

\clearpage

\begin{diagram}

\label{Diagram C2}
\index{Diagram C2}

\unitlength1.0mm
\begin{picture}(150,100)(0,0)

\put(0,80){{\rm\bf C2 }}

\put(10,40){\circle*{1}}
\put(30,40){\circle*{1}}
\put(50,40){\circle*{1}}
\put(70,40){\circle*{1}}

\put(40,20){\circle*{1}}

\put(12,40){\line(1,0){16}}
\put(32,40){\line(1,0){16}}
\put(52,40){\line(1,0){16}}
\put(20,41){\line(0,-1){2}}
\put(40,41){\line(0,-1){2}}
\put(60,41){\line(0,-1){2}}

\put(24,29){\line(2,3){1}}
\put(54,30){\line(2,-3){1}}

\put(12,38){\line(3,-2){26}}
\put(42,21){\line(3,2){26}}

\put(8,43){$x_0$}
\put(28,43){$x_1$}
\put(48,43){$x_2$}
\put(68,43){$x_3$}

\put(38,17){$y_1$}

\end{picture}

\end{diagram}

\clearpage

\begin{diagram}

\label{Diagram C3}
\index{Diagram C3}

\unitlength1.0mm
\begin{picture}(150,100)(0,0)

\put(0,80){{\rm\bf C3 }}

\put(10,40){\circle*{1}}
\put(30,40){\circle*{1}}
\put(50,20){\circle*{1}}
\put(50,60){\circle*{1}}

\put(12,40){\line(1,0){16}}
\put(20,41){\line(0,-1){2}}

\put(12,42){\line(2,1){36}}
\put(29,52){\line(1,-2){1}}
\put(12,38){\line(2,-1){36}}
\put(29,28){\line(1,2){1}}

\put(32,42){\line(1,1){16}}
\put(39,51){\line(1,-1){2}}
\put(32,38){\line(1,-1){16}}
\put(40,28){\line(1,1){2}}

\put(5,39){$x_0$}
\put(33,39){$x_1$}
\put(52,19){$x_2$}
\put(52,59){$x_2'$}

\end{picture}

\end{diagram}

\clearpage

\begin{diagram}

\label{Diagram C4}
\index{Diagram C4}

\unitlength1.0mm
\begin{picture}(150,100)(0,0)

\put(0,80){{\rm\bf C4 }}

\put(10,40){\circle*{1}}
\put(30,40){\circle*{1}}
\put(50,40){\circle*{1}}

\put(30,60){\circle*{1}}

\put(70,60){\circle*{1}}
\put(70,20){\circle*{1}}

\put(12,40){\line(1,0){16}}
\put(32,40){\line(1,0){16}}
\put(20,41){\line(0,-1){2}}
\put(40,41){\line(0,-1){2}}

\put(12,42){\line(1,1){16}}
\put(19,51){\line(1,-1){2}}
\put(20,41){\line(0,-1){2}}
\put(40,41){\line(0,-1){2}}

\put(52,38){\line(1,-1){16}}
\put(52,42){\line(1,1){16}}
\put(59,51){\line(1,-1){2}}
\put(59,29){\line(1,1){2}}
\put(59,51){\line(1,-1){2}}
\put(59,29){\line(1,1){2}}
\put(20,41){\line(0,-1){2}}
\put(40,41){\line(0,-1){2}}

\put(32,60){\line(1,0){36}}
\put(50,61){\line(0,-1){2}}

\put(5,39){$x_0$}
\put(28,37){$x_1$}
\put(53,39){$x_2$}
\put(72,19){$x_3'$}
\put(72,59){$x_3$}
\put(28,63){$y_1$}

\end{picture}

\end{diagram}

\clearpage

\begin{diagram}

\label{Diagram C5}
\index{Diagram C5}

\unitlength1.0mm
\begin{picture}(150,100)(0,0)

\put(0,80){{\rm\bf C5 }}

\put(10,40){\circle*{1}}
\put(30,40){\circle*{1}}
\put(50,40){\circle*{1}}

\put(30,60){\circle*{1}}

\put(70,60){\circle*{1}}
\put(70,20){\circle*{1}}

\put(12,40){\line(1,0){16}}
\put(32,40){\line(1,0){16}}

\put(12,42){\line(1,1){16}}

\put(52,38){\line(1,-1){16}}
\put(52,42){\line(1,1){16}}

\put(32,60){\line(1,0){36}}

\put(30,62){\line(0,1){8}}
\put(30,70){\line(1,0){50}}
\put(80,70){\line(0,-1){50}}
\put(72,20){\line(1,0){8}}
\put(79,40){\line(1,0){2}}

\put(50,61){\line(0,-1){2}}
\put(59,51){\line(1,-1){2}}
\put(59,29){\line(1,1){2}}
\put(19,51){\line(1,-1){2}}
\put(20,41){\line(0,-1){2}}
\put(40,41){\line(0,-1){2}}

\put(5,39){$x_0$}
\put(28,37){$x_1$}
\put(53,39){$x_2$}
\put(68,16){$x_3'$}
\put(72,59){$x_3$}
\put(25,59){$y_1$}

\end{picture}

\end{diagram}

\clearpage

\begin{diagram}

\label{Diagram C6}
\index{Diagram C6}

\unitlength1.0mm
\begin{picture}(150,100)(0,0)

\put(0,80){{\rm\bf C6 }}

\put(10,40){\circle*{1}}
\put(30,40){\circle*{1}}
\put(50,40){\circle*{1}}
\put(70,40){\circle*{1}}

\put(70,20){\circle*{1}}

\put(90,60){\circle*{1}}
\put(90,20){\circle*{1}}

\put(12,40){\line(1,0){16}}
\put(32,40){\line(1,0){16}}
\put(52,40){\line(1,0){16}}
\put(20,41){\line(0,-1){2}}
\put(40,41){\line(0,-1){2}}
\put(60,41){\line(0,-1){2}}

\put(52,38){\line(1,-1){16}}
\put(72,38){\line(1,-1){16}}
\put(72,42){\line(1,1){16}}

\put(12,41){\line(4,1){76}}
\put(49,51){\line(1,-4){0.5}}

\put(10,38){\line(0,-1){28}}
\put(90,18){\line(0,-1){8}}
\put(10,10){\line(1,0){80}}
\put(50,11){\line(0,-1){2}}

\put(59,29){\line(1,1){2}}
\put(79,29){\line(1,1){2}}
\put(79,51){\line(1,-1){2}}

\put(5,39){$x_0$}
\put(28,37){$x_1$}
\put(48,43){$x_2$}
\put(73,39){$x_3$}

\put(68,16){$x_3'$}

\put(93,59){$x_4$}
\put(93,19){$x_4'$}

\end{picture}

\end{diagram}

\clearpage
\section{
Valuations and Cycles
}

Usually, it is not necessary to write down the details of valuations,
and we may avoid additional notation. Here, we treat somewhat complicated
cases, where this notation is useful.

Our main result here is
Proposition \ref{Proposition NoWay-2} (page \pageref{Proposition NoWay-2}),
which illustrates that we cannot achieve impossibility of
valuation with finite depth formulas.
We know this, but our result is more constructive.
We show that any attempt to construct such a set of formulas without
consistent valuation will construct a cycle of uneven length
of contradictory paths, which is impossible, basically due to
the fact that we have 2 truth values.
\subsection{
Paths
}

\bd

$\hspace{0.01em}$


\label{Definition Value}

Let $ \xdg $ be a graph as in the Yablo structure.

 \xEh

 \xDH Label of a node

Call $d(x)$ as in  \cite{RRM13} the label of node $x.$

 \xDH Labelled arrow

We work with $ \xcO \xcU.$

Consider the example $d(x)=(y \xcu z) \xco (\xCN y \xcu \xCN z) \xco (y
\xcu u).$ We have two
positive instances of $y,$ and one negative instance. The number of
instances
is not important, but the fact that we have a positive and a negative
instance is.

Thus, we will distinguish (as before) the positive arrow $x \xcp y,$
the negative arrow $x \xcP y,$ and (new) the situation where we have
a positive and a negative arrow from $x$ to $y,$ written as $x \xcp_{ \xCL
}y,$
see also Diagram \ref{Diagram ContCell} (page \pageref{Diagram ContCell}).

 \xDH Labelled path

Let $ \xbs $ be a path as usual in the graph $ \xdg.$

$ \xbs_{0},$ $ \xbs_{1},$ etc. will be the nodes on the path.

(We are not very consistent here, sometimes $ \xbs_{0},$ $ \xbs_{1}$ etc.
will also
be different paths - context will tell.)

As the graph is free from cycles, $ \xbs_{0},$ $ \xbs_{n},$ $ \xbs_{x},$ $
\xbs_{ \xCc x},$ $ \xbs_{ \xck x},$ $ \xbs_{ \xCe x},$ $ \xbs_{ \xcg x}$
will all be defined if $x$ is a node on the path. (The latter will
indicate the path (or its elements) up to $x$ without $x,$ etc.)

We define the labelled path from the path $ \xbs $ inductively.

$ \xbs_{0}$ will be the first element of the labelled path, too.

Let the labelled path be defined up to $ \xbs_{n}.$

If $ \xbs_{n} \xcp \xbs_{n+1},$ then
$ \xbs_{0}$  \Xl. $ \xbs_{n} \xcp \xbs_{n+1}$ is the labelled path up to
$ \xbs_{n+1}.$

If $ \xbs_{n} \xcP \xbs_{n+1},$ then
$ \xbs_{0}$  \Xl. $ \xbs_{n} \xcP \xbs_{n+1}$ is the labelled path up to
$ \xbs_{n+1}.$

If $ \xbs_{n} \xcp_{ \xCL } \xbs_{n+1},$ then we split into two different
labelled paths:
$ \xbs_{0}$  \Xl. $ \xbs_{n} \xcp \xbs_{n+1}$ is the first labelled path
up to $ \xbs_{n+1},$ and
$ \xbs_{0}$  \Xl. $ \xbs_{n} \xcP \xbs_{n+1}$ is the second labelled path
up to $ \xbs_{n+1}.$

Thus, $ \xcp_{ \xCL }$ does not occur any more.

 \xDH Valued paths

We define the valued paths from a labelled path $ \xbs $ by induction from
an
initial value $+$ or - for $ \xbs_{0},$ by propagation.

Let $v(\xbs_{0}):=+$ or -, corresponding to the initial value.

Let $v(\xbs_{0}) \Xl v(\xbs_{n})$ be defined.

 \xEh

 \xDH Case 1: $v(\xbs_{n})=+.$

If $ \xbs_{n} \xcp \xbs_{n+1},$ then $v(\xbs_{n+1})=+,$

If $ \xbs_{n} \xcP \xbs_{n+1},$ then $v(\xbs_{n+1})=-.$

 \xDH Case 2: $v(\xbs_{n})=-.$

If $ \xbs_{n} \xcp \xbs_{n+1},$ then $v(\xbs_{n+1})=-,$

If $ \xbs_{n} \xcP \xbs_{n+1},$ then $v(\xbs_{n+1})=+.$

 \xEj

Recall that $ \xbs_{n} \xcp \xbs_{n+1}$ or $ \xbs_{n} \xcP \xbs_{n+1},$
but never both in labelled paths.

 \xDH We denote $ \xbs^{+}$ the valued path $ \xbs $ beginning with $v(
\xbs_{0})=+,$ and
$ \xbs^{-}$ the valued path $ \xbs $ beginning with $v(\xbs_{0})=-.$

We call $ \xbs^{+}$ the opposite of $ \xbs^{-},$ and vice versa.

 \xEj

\ed

\bfa

$\hspace{0.01em}$


\label{Fact Value1}

 \xEh

 \xDH If $v_{1}(\xbs_{0})$ is the opposite of (contradicts) $v_{2}(
\xbs_{0}),$ then so do
$v_{1}(\xbs_{n})$ to $v_{2}(\xbs_{n})$ for all $n.$

(Proof by induction.)

 \xDH If $ \xbs_{ \xcg x}= \xbt_{ \xcg x}$ and $v(\xbs_{x})$ contradicts
$v' (\xbt_{x}),$ then it also holds
for all $y>x.$

(By (1).)

 \xDH If $v(\xbs_{x})$ contradicts $v' (\xbt_{x}),$ then $v^{-}(
\xbs_{x})$ contradicts $v'^{-}(\xbt_{x}),$ too.
 \xEj

\efa

\be

$\hspace{0.01em}$


\label{Example Value}

Consider a graph with 5 nodes, $x_{0},$ $x_{1},$ $x_{2},$ $x_{3},$
$x_{4}.$ $x_{4}$ will not have any
successors.

Let $d(x_{0})= \xCN x_{1} \xco x_{2} \xco x_{4},$ $d(x_{1})=x_{2},$
$d(x_{2})=x_{3} \xcu \xCN x_{3},$ $d(x_{3})=x_{4}.$
For brevity, we will
sometimes also write $x_{0}= \xCN x_{1} \xco x_{2}$ etc.
 \xEh
 \xDH
The graph as in the Yablo structure is $x_{0} \xcp x_{1} \xcp x_{2} \xcp
x_{3} \xcp x_{4},$ $x_{0} \xcp x_{2},$ $x_{0} \xcp x_{4}.$

 \xDH
The graph with labelled arrows is $x_{0} \xcP x_{1} \xcp x_{2} \xcp_{ \xCL
}x_{3} \xcp x_{4}$, $x_{0} \xcp x_{2}$, $x_{0} \xcp x_{4}$.
 \xDH
The labelled paths are

$ \xbs^{0}:$ $x_{0} \xcP x_{1} \xcp x_{2} \xcp x_{3} \xcp x_{4}$,

$ \xbs^{1}:$ $x_{0} \xcP x_{1} \xcp x_{2} \xcP x_{3} \xcp x_{4}$,

$ \xbs^{2}:$ $x_{0} \xcp x_{2} \xcp x_{3} \xcp x_{4}$,

$ \xbs^{3}:$ $x_{0} \xcp x_{2} \xcP x_{3} \xcp x_{4}$,

$ \xbs^{4}:$ $x_{0} \xcp x_{4}.$

 \xDH
The valued paths are
 \xEh
 \xDH
$ \xbs^{+}:$

$ \xbs^{0+}:$ $x^{+}_{0}$ $ \xcP x^{-}_{1}$ $ \xcp x^{-}_{2}$ $ \xcp
x^{-}_{3} \xcp x^{-}_{4}$,

$ \xbs^{1+}:$ $x^{+}_{0}$ $ \xcP x^{-}_{1}$ $ \xcp x^{-}_{2}$ $ \xcP
x^{+}_{3} \xcp x^{+}_{4}$,

$ \xbs^{2+}:$ $x^{+}_{0}$ $ \xcp x^{+}_{2}$ $ \xcp x^{+}_{3} \xcp
x^{+}_{4}$,

$ \xbs^{3+}:$ $x^{+}_{0}$ $ \xcp x^{+}_{2}$ $ \xcP x^{-}_{3} \xcp
x^{-}_{4}$,

$ \xbs^{4+}:$ $x^{+}_{0}$ $ \xcp x^{+}_{4}$.
 \xDH
$ \xbs^{-}:$

$ \xbs^{0-}:$ $x^{-}_{0}$ $ \xcP x^{+}_{1}$ $ \xcp x^{+}_{2}$ $ \xcp
x^{+}_{3} \xcp x^{+}_{4}$,

$ \xbs^{1-}:$ $x^{-}_{0}$ $ \xcP x^{+}_{1}$ $ \xcp x^{+}_{2}$ $ \xcP
x^{-}_{3} \xcp x^{-}_{4}$,

$ \xbs^{2-}:$ $x^{-}_{0}$ $ \xcp x^{-}_{2}$ $ \xcp x^{-}_{3} \xcp
x^{-}_{4}$,

$ \xbs^{3-}:$ $x^{-}_{0}$ $ \xcp x^{-}_{2}$ $ \xcP x^{+}_{3} \xcp
x^{+}_{4}$,

$ \xbs^{4-}:$ $x^{-}_{0}$ $ \xcp x^{-}_{4}$.

 \xEj

 \xEj
\subsection{
Discussion of a Non-recursive $\xcO \xcU$ Construction
}

\label{Section Complicated}

\ee

\bfa

$\hspace{0.01em}$


\label{Fact Value2}

 \xEh

 \xDH No triangles (this is a special case of (2))

Let $ \xbr, \xbs, \xbt $ be labelled paths such that once they meet,
they continue
the same way.
So, if $ \xbr,$ $ \xbs $ meet at $x_{ \xbr, \xbs },$ then $ \xbr_{ \xcg
x_{ \xbr, \xbs }}= \xbs_{ \xcg x_{ \xbr, \xbs }},$ etc.

Let $v_{ \xbr }(\xbr),$ $v_{ \xbs }(\xbs),$ $v_{ \xbt }(\xbt)$ be
valued versions of $ \xbr, \xbs, \xbt.$

Then it is impossible that $v_{ \xbr }(\xbr_{x_{ \xbr, \xbs }})$
contradicts $v_{ \xbs }(\xbs_{x_{ \xbr, \xbs }}),$
$v_{ \xbs }(\xbs_{x_{ \xbs, \xbt }})$ contradicts $v_{ \xbt }(\xbt_{x_{
\xbs, \xbt }}),$
$v_{ \xbt }(\xbt_{x_{ \xbt, \xbr }})$ contradicts $v_{ \xbr }(\xbr_{x_{
\xbt, \xbr }}).$

Case 1: $x_{ \xbr, \xbs }=x_{ \xbs, \xbt }$ $(=x_{ \xbr, \xbt }$ by
prerequisite). As we have only two
values, this is impossible.

Case 2: Assume wlog that $x_{ \xbr, \xbs }$ is ``above'' $x_{ \xbs, \xbt
}.$
$v_{ \xbr }(\xbr_{x_{ \xbr, \xbs }})$ contradicts $v_{ \xbs }(\xbs_{x_{
\xbr, \xbs }})$ by prerequisite, but by (2)
$v_{ \xbr }(\xbr_{x_{ \xbs, \xbt }})$ contradicts $v_{ \xbs }(\xbs_{x_{
\xbs, \xbt }}),$ so
$v_{ \xbt }(\xbt_{x_{ \xbs, \xbt }})$ cannot contradict both
$v_{ \xbr }(\xbr_{x_{ \xbs, \xbt }})$ and $v_{ \xbs }(\xbs_{x_{ \xbs,
\xbt }}),$ as we have only two values.

 \xDH No loops of odd length

We argue as in (3).

Let $ \xbs^{0},$  \Xl  $ \xbs^{2n}$ be paths with valuations $v^{0},$  \Xl
 $v^{2n}$ such that $v^{i}(\xbs^{i})$
contradicts $v^{i+1}(\xbs^{i+1})$ modulo $2n+1.$ This is impossible.
Assume the contrary.

Let e.g. $ \xbs^{0}$ meet $ \xbs^{1}$ at $x_{ \xbs^{0}, \xbs^{1}},$ and
$v^{0}(\xbs^{0}_{x_{ \xbs^{0}, \xbs^{1}}})$ contradict
$v^{1}(\xbs^{1}_{x_{ \xbs^{0}, \xbs^{1}}}),$ and from $x_{ \xbs^{0},
\xbs^{1}}$ on, $ \xbs^{0}$ and $ \xbs^{1}$ are identical.

Argue for $ \xbs^{1}$ and $ \xbs^{2}$ the same way. $x_{ \xbs^{1}
\xbs^{2}}$ may be below $x_{ \xbs^{0}, \xbs^{1}},$ above,
or identical. When we consider $x_{ \xbs^{0} \xbs^{1} \xbs^{2}}$ $:=$
$min(x_{ \xbs^{0}, \xbs^{1}},x_{ \xbs^{1}, \xbs^{2}}),$
then $ \xbs^{0}, \xbs^{1}, \xbs^{2}$ will be identical from $x_{ \xbs^{0}
\xbs^{1} \xbs^{2}}$ on, etc.

Finally, there is a point $z$ where all $ \xbs^{i}$ are identical, and by
(2),
$v^{i}(\xbs^{i}_{z})$ contradicts $v^{i+1}(\xbs^{i+1}_{z}),$ etc. Assume
e.g. wlog.
$v^{0}(\xbs^{0}_{z})=+,$ $v^{1}(\xbs^{1}_{z})=-,$ $v^{2}(
\xbs^{2}_{z})=+,$ etc., and $v^{2n}(\xbs^{2n}_{z})=+,$
contradiction.

 \xEj

\efa

We may have arbitrarily many paths pairwise contradictory, as the
following
example shows. This is impossible within one cell.

\be

$\hspace{0.01em}$


\label{Example Tower2}

Let $ \xbs_{0}:x_{0} \xcP x_{1} \xcp x_{2} \xcP x_{3} \xcp x_{4},$
$ \xbs_{1}:x_{0} \xcP x_{1} \xcp x_{2} \xcp x_{4},$
$ \xbs_{2}:x_{0} \xcp x_{2} \xcP x_{3} \xcp x_{4},$
$ \xbs_{3}:x_{0} \xcp x_{2} \xcp x_{4},$

then $ \xbs_{0}$ contradicts $ \xbs_{1}$ in the lower part, $ \xbs_{2}$
and $ \xbs_{3}$ in the upper part,
$ \xbs_{1}$ contradicts $ \xbs_{2}$ and $ \xbs_{3}$ in the upper part,
$ \xbs_{2}$ contradicts $ \xbs_{3}$ in the lower part.

Obviously, this may be generalized to $2^{ \xbo }$ paths.

\ee

The following example is for illustration, the general solution is
in Proposition \ref{Proposition NoWay-2} (page \pageref{Proposition NoWay-2}).

See also Example 
\ref{Example Distributivity} (page 
\pageref{Example Distributivity})
for illustration.

\be

$\hspace{0.01em}$


\label{Example 4-case}

Suppose we work with DNF, and have solved the distributions, so
$d(x)= \xda \xco \xdb \xco \xdc \xco \xdd,$ where $ \xda = \xcU \{a_{i}:i
\xbe I_{a}\},$ $ \xdb = \xcU \{b_{i}:i \xbe I_{b}\},$ $ \xdc = \xcU
\{c_{i}:i \xbe I_{c}\},$
$ \xdd = \xcU \{d_{i}:i \xbe I_{d}\}.$
As $d(x)$ is a disjunction, and, by prerequisite, $d(x)$ has to be
contradictory,
$ \xda $ etc. have to be contradictory for $x+,$ thus there
have to $A,a \xbe \{a_{i}:i \xbe I_{a}\}$ which are contradictory (recall:
contradictions
are always between two elements, see
Fact \ref{Fact DNF} (page \pageref{Fact DNF})).
Likewise, there must be such $B,b,$ $C,c,$ $D,d.$

As shown in Fact \ref{Fact Value1} (page \pageref{Fact Value1}),
$A,a$ are contradictory, iff their opposites
$A^{-},a^{-}$ are contradictory.

To make x- contradictory, every choice (of opposites) in $ \xda, \xdb,
\xdc, \xdd $
has to contradictory
(see Example 
\ref{Example Distributivity} (page 
\pageref{Example Distributivity})),
in particular, every choice
in $\{A^{-},a^{-}\},$ $(B^{-},b^{-}\},$ $\{C^{-},c^{-}\},$
$\{D^{-},d^{-}\}$ has to be contradictory.

Thus,

$\{A^{-},a^{-}\},$ $(B^{-},b^{-}\},$ $\{C^{-},c^{-}\},$ $\{D^{-},d^{-}\}$
each are contradictory by prerequisite
(for $x+)$

and all choice sets for x-

$\{A^{-},B^{-},C^{-},D^{-}\},$ $\{A^{-},B^{-},C^{-},d^{-}\},$  \Xl,
$\{a^{-},b^{-},c^{-},D^{-}\},$ $\{a^{-},b^{-},c^{-},d^{-}\}$

have to be made contradictory.

We show that this leads to loops of contradictions of odd length,
contradicting
Fact \ref{Fact Value2} (page \pageref{Fact Value2}), (2).
For notational simplicity, we denote $A^{-}$ by A (or, using again
Fact 
\ref{Fact Value1} (page 
\pageref{Fact Value1})), so we have to chose a contradiction in all
$\{A,B,C,D\},$ $\{A,B,C,d\},$  \Xl, $\{a,b,c,D\},$ $\{a,b,c,d\}$

See Diagram \ref{Diagram 4-case} (page \pageref{Diagram 4-case}).

The lines there connect contradictory couples.

The vertical lines are the contradictions originating from $x+.$

For reasons of symmetry, we may assume that $\{A,B\}$ are contradictory -
recall again Fact \ref{Fact DNF} (page \pageref{Fact DNF}).

We may continue by adding the contradictions $\{A,C\},$ or $\{B,C\}$ as in
the
diagram, or $\{a,C\},$ or $\{C,D\}.$ The other cases are symmetrical.

 \xEh
 \xDH
Fix now $\{B,C\}.$ The choices in $\{A,b,C,D\}$ and $\{A,b,C,d\}$ together
are
impossible, as we will show by examining the cases.

(Note that in the set $\{A,b,C\}$ all nodes $A,b,C$ have even distance
from A. This is the basic reason for the fact that adding xD and $x' d$
leads to a cycle of uneven length: $A \Xl xDdx'  \Xl A.$ This is
elaborated
in Proposition 
\ref{Proposition NoWay-2} (page 
\pageref{Proposition NoWay-2}).)

Chosing $\{A,b\},$ $\{A,C\},$ $\{b,C\}$ are all impossible, as
they lead to loops of length 3, e.g. $A-b-B- \xCf A.$

(We simplify notation.)

The possible choices for AbCD are thus AD, bD, CD, and for AbCd
Ad, bd, Cd.
 \xEh
 \xDH
Assume the choice AD for AbCD. Then the choice of Ad is impossible by the
loop
ADdA,
the choice of bd by the loop ADdbBA, the choice of Cd by ADdCBA.

 \xDH
Assume bD for AbCD. Then Ad is impossible by AdDbBA, bd by bdDb,
Cd by BCdDbB (or ABCdDbBA)

 \xDH
Assume CD for AbCD. Then Ad is impossible by AdDCBA, bd by bdDCBb,
Cd by CDdC.

 \xEj

So $\{B,C\}$ is impossible.

 \xDH
Consider the case of $\{C,D\}$ added as contradictory:
We show that AbCd and AbcD together are impossible.

Ab, Cd, cD are impossible.

For AbCd, we consider $AC,Ad,bC,bd$ as candidates for contradiction.

For AbcD, we consider $Ac,AD,bc,bD$ as candidates for contradiction.
 \xEh

 \xDH
Assume AC: Ac is impossible, AD because of ACDA, bc because of ABbcCA,
bD because of ACDbBA.

 \xDH
Assume Ad: AD is impossible, Ac because of AdDCcA, bc because of AdDCcbBA,
bD because of AdDbBA.

 \xDH
Assume bC: bc is impossible, AC because of AcCbBA, AD because of ADCbBA,
bc because of bCDb.

 \xDH
Assume bd: bD is impossible, Ac because of AcCDdbBA, AD because of ADdbBA,
bc because of bdDCcb.

 \xEj

So $\{C,D\}$ is impossible.

 \xDH
The other cases are similar.

 \xEj

We first checked this and other examples with a small computer program.

\ee

We make this more general.

\bp

$\hspace{0.01em}$


\label{Proposition NoWay-2}

We cannot make both $x+$ and x- contradictory.

\ep

The proof goes over several steps.
We will construct a sequence like ABcDef  \Xl., dscribing the valuation
$A,B,c,D,e,f$  \Xl. which will be consistent, i.e. there is no
contradiction between A and $c,$ etc., graphically no line $A$-c,
and any attempt to make
it inconsistent will result in a cycle of odd length, contradicting
Fact \ref{Fact Value2} (page \pageref{Fact Value2}).

\bd

$\hspace{0.01em}$


\label{Definition Connected}

 \xEh
 \xDH
$(A,a)$ and $(B,b)$ are directly connected iff $\{A,B\},$ or $\{A,b\},$ or
$\{a,B\},$ or
$\{a,b\}$ are contradictory, i.e., in our graphical notation, iff there is
a line from A to $B,$ or  \Xl..

(Recall than $\{A,a\}$ are contradictory for $x+.)$
 \xDH
$(A_{0},a_{0})$ and $(A_{n},a_{n})$ are connected iff there is a sequence
of directly
connected pairs $(A_{0},a_{0}),$ $(A_{1},a_{1}),$  \Xl, $(A_{n},a_{n}).$
 \xDH
A set of pairs $ \xda =\{(A_{0},a_{0}), \Xl,(A_{n},a_{n})\}$ is connected
iff all $(A_{0},a_{0})$ and $(A_{i},a_{i})$
are connected. (Instead of $(A_{0},a_{0})$ any other pair will do.)
 \xDH
A maximal connected set is called a bloc.
 \xEj

\ed

We assume now that all blocs are consistent (no odd loops of
contradictions), see
Fact \ref{Fact Value2} (page \pageref{Fact Value2}).

\bfa

$\hspace{0.01em}$


\label{Fact NoWay}

 \xEh
 \xDH
Let $ \xdb $ be a bloc. Fix $(A_{0},a_{0}) \xbe \xdb $ arbitrarily. Fix
$A_{0}$ (or $a_{0})$
arbitrarily. If $ \xdb =\{(A_{0},a_{0})\},$ we are done with this bloc.

 \xDH
Let $(A_{0},a_{0}) \xEd (X,x) \xbe \xdb.$ Then there is a path of
contradictions of
even length from $A_{0}$ to $X$ or $x.$

 \xDH
Example:

Recall that $ \xdb $ is connected. Let e.g.
$A_{0}-a_{0}-a_{1}-A_{1}-A_{2}-x.$ This has length 5,
but adding $x-X$ results in a path of even length 6.

We do this for all $(X,x) \xbe \xdb $ different from $(A_{0},a_{0}).$

 \xDH
This results in a set e.g. of $ \xdc =\{A_{0},X,x',x'',X''', \Xl \},$
where all $X,$ $x' $ etc.
have a path of contradictions of even length from $A_{0}$ - but also from
each other, we just have to go back via $A_{0}.$

 \xDH
We call this set $ \xdc $ an even choice set for $ \xdb,$ and write it
$\{ \xbb_{0},$ $ \xbb_{1},$  \Xl  $etc.\}.$
Seen as a (partial) valuation, it is consistent, i.e. there are no direct
lines linking e.g. $ \xbb_{0}- \xbb_{1},$ so $x+$ and x- cannot both be
contradictory.

 \xDH
Note that adding any new contradiction $ \xbb_{i}- \xbb_{j}$ results in an
odd loop
of contradictions, which cannot be, see
Fact \ref{Fact Value2} (page \pageref{Fact Value2}).

 \xDH
Suppose we have a set $\{(A_{i},a_{i}):i \xbe I\},$ composed of blocs $
\xdb_{j}:j \xbe J.$
Take now a valuation as above for each bloc $ \xdb_{j}.$ Then

 \xEh
 \xDH
The union of all even choice sets is free from contradictions.
 \xDH
We cannot add any contradiction within any one bloc - see above, (6).
 \xDH
Thus, if we have just one bloc, this valuation cannot be made
contradictory in a consistent way.
 \xDH
We can make the union of choice sets contradictory in a consistent way
by adding an inter-bloc contradiction.

 \xEj

 \xDH
Thus, we have created a full valuation which is consistent, and any
attempt
to make it inconsistent will result in a cycle of odd length.

 \xEj

\efa

$ \xcz $ Proposition 
\ref{Proposition NoWay-2} (page 
\pageref{Proposition NoWay-2})
\\[3ex]

$ \xCO $

$ \xCO $

$ \xCO $

\clearpage

\newsavebox{\immer}

\savebox{\immer}(10,10){

\put(10,80){$A$}
\put(10,0){$a$}
\put(90,80){$B$}
\put(90,0){$b$}
\put(170,80){$C$}
\put(170,0){$c$}
\put(250,80){$D$}
\put(250,0){$d$}

\put(12,10){\line(0,1){63}}
\put(92,10){\line(0,1){63}}
\put(172,10){\line(0,1){63}}
\put(252,10){\line(0,1){63}}

\put(20,83){\line(1,0){68}}

}

\clearpage

\begin{diagram}

\label{Diagram 4-case}
\index{Diagram 4-case}

\begin{picture}(250,300)(0,0)

\put(10,250){Example 4-case, Case (1)}

\put(10,165){\usebox{\immer}}

\put(120,210){\line(1,0){68}}

\put(10,100){Example 4-case, Case (2)}

\put(10,30){\usebox{\immer}}

\put(200,75){\line(1,0){68}}

\end{picture}

\end{diagram}

\vspace{10mm}

$ \xCO $

$ \xCO $

$ \xCO $
\clearpage
\section{
A Generalization of Yablo's Construction to $\xcO \xcU$
}

\label{Section Or-And}

We discuss here a general strategy and a number of examples
(in Example \ref{Example Or-And} (page \pageref{Example Or-And})).

They have in common that the formulas are of the type $ \xcO \xcU,$ i.e.
in
disjunctive normal form. The limiting cases are pure conjunctions
(as in Yablo's original approach) or pure disjunctions.

The examples are straightforward generalizations of Yablo's construction,
as we have here several columns, in Yablo's construction just one
column, and our choice functions $g$ (in all columns) correspond to the
choice
of one element in Yablo's construction.

Consider the first example below.

More precisely, as Yablo works with $ \xcU x_{i},$ there is one uniform
set of $x_{i}.$
We work with $ \xcO \xcU x_{i,j},$ so we have to distinguish the elements
in the $ \xcU $
from the sets in $ \xcO.$ We define for this purpose columns, whose
elements
are the elements in the $ \xcU,$ and the set of columns are the sets in $
\xcO.$
Negation is now slightly more complicated, not just OR of negated
elements,
but
OR of choice functions of negated elements in the columns.
We also need some enumeration of $ \xbo \xCK \xbo,$ or of a suitable set.

It is perhaps easiest to see the following example geometrically.
We have a vertical column $C_{i_{0}}$ where a certain property holds,
and a horizontal line (the choice function $g),$ where the opposite
holds, and column and horizontal line meet.

More precisely, the property will not necessarily hold in all of
$C_{i_{0}},$
but only from a certain height onward. As the choice functions $g$ will
chose even higher in the columns, the clash is assured.

For more details of the general strategy, see
Definition \ref{Definition Or-And} (page \pageref{Definition Or-And}),
Case \ref{Case DOA} (page \pageref{Case DOA}).
and
Example \ref{Example Or-And} (page \pageref{Example Or-And}),
in particular
Case \ref{Case OA6} (page \pageref{Case OA6}).

We first introduce some notation and definitions

\bd

$\hspace{0.01em}$


\label{Definition Or-And}

 \xEh
 \xDH
The examples will differ in the size of the $ \xcU $ and $ \xcO $ -
where the case of both finite
is trivial, no contradictions possible - and the relation in the
graph, i.e. which formulas are ``visible'' from a given formula.
(In Yablo's construction, the $x_{j},$ $j>i,$ are visible from $x_{i}.)$
We will write the relation by $ \xcP,$ $x \xcP y,$ but for simplicity
sometimes $x<y,$ too,
reminding us that variables inside the
formulas will be negated.
We consider linear and ranked orders, leaving general partial orders
aside.
All relations will be transitive and acyclic.

For the intuition (and beyond) we order the formulas in sets of
columns. Inside a column, the formulas are connected by $ \xcU,$ the
columns themselves are connected by $ \xcO.$ This is the logical
ordering, it is different from above order relation in the graph.

 \xDH The basic structure.

 \xEh
 \xDH
So, we have columns $C_{i},$ and inside the columns variables $x_{i,j}.$
As we might have finitely or countably infinitely many columns, of finite
or countably infinite size, we
write the set of columns $C:=\{C_{i}:i< \xba \},$ where $ \xba < \xbo +1,$
and
$C_{i}:=\{x_{i,j}:j< \xbb_{i}\},$ where $ \xbb_{i}< \xbo +1$ again. By
abuse of language, $C$ will
also denote the whole construction, $C:=\{x_{i,j}:i,j< \xbo \}.$
 \xDH
Given $C$ and $x_{i,j},$ $C \xex x_{i,j}:=\{x_{i',j' }:x_{i,j} \xcP x_{i'
,j' }\},$ the part of $C$ visible from
$x_{i,j}.$
 \xDH
Likewise, $C_{i' } \xex x_{i,j}:=\{x_{i',j' } \xbe C_{i' }:x_{i,j} \xcP
x_{i',j' }\}.$
 \xEj

 \xDH

Let $I(x_{i,j})$ $:=$ $\{i' < \xba:$ $C_{i' } \xex x_{i,j} \xEd \xCQ \}.$
(In some $i',$ there might be no $x_{i',j' }$ s.t. $x_{i,j} \xcP x_{i'
,j' }.)$

Given $i' \xbe I(x_{i,j}),$ let
$J(i',x_{i,j})$ $:=$ $\{j' < \xbb_{i' }:x_{i,j} \xcP x_{i',j' }\}.$
By $i' \xbe I(x_{i,j}),$ $J(i',x_{i,j}) \xEd \xCQ.$

 \xDH
Back to logic. Let $d(x_{i,j}):= \xcO \{ \xcU \{ \xCN x_{i',j' }:j' \xbe
J(i',x_{i,j})\}:i' \xbe I(x_{i,j})\},$
$(\xcu $ inside columns, $ \xco $ between columns).
We will sometimes abbreviate $d(x_{i,j})$ by $x_{i,j}.$
(Note that the $x_{i',j' }$ are exactly the elements visible from
$x_{i,j}.)$

 \xEh
 \xDH
If $x_{i,j}$ is true (written $x_{i,j}+),$ then all elements in one of the
$C_{i' },$
$i' \xbe I(x_{i,j}),$
and visible from $x_{i,j},$ must all be false - but we do not know in
which $C_{i' }.$
We denote this $C_{i' }$ by $C_{i(x_{i,j})},$ and define
$C[x_{i,j}]:=C_{i(x_{i,j})} \xex x_{i,j}.$
 \xDH
Conversely, suppose $x_{i,j}$ is false, $x_{i,j}-.$
Again, we consider only elements in $C \xex x_{i,j},$ i.e. visible from
$x_{i,j}.$
By distributivity, $x_{i,j}-$ $=$ $ \xcO \{ran(g):g \xbe \xbP \{C_{i' }
\xex x_{i,j}:i' \xbe I(x_{i,j})\}\}.$

(g is a choice function chosing in all columns $C_{i' }$ the ``sufficiently
big''
elements $x_{k,m},$ i.e. above $x_{i,j},$
$ran(g)$ its range or image.)

Note that the elements of $ran(g)$ are now positive!

The $ \xcO $ in above formula choses some such function $g,$ but we do not
know which.
Let $g[x_{i,j}]$ denote the chosen one.
 \xDH
Note that both $C[x_{i,j}]$ and $g[x_{i,j}]$ are undefined if there are no
$x_{i',j' }>x_{i,j}.$
 \xEj

 \xDH

\label{Case DOA}

The conflicts will be between the ``vertical'' (negative) columns, and
``horizontal'' (positive) lines of the $g.$ It is a very graphical
construction. If we start with a positive point, the
negative columns correspond to the $ \xcA $ in Yablo's
construction, the $g$ to the $ \xcE,$ if we start with a negative one, it
is
the other way round.

More precisely:

 \xEh
 \xDH Let $x_{i,j}+,$ consider $C[x_{i,j}] \xcc C_{i(x_{i,j})}$ -
recall all elements of $C[x_{i,j}]$ are negative.

If there is $x_{i',j' } \xbe C[x_{i,j}]$ which is not maximal in
$C[x_{i,j}],$ then
$g[x_{i',j' }]$ intersects $C[x_{i,j}],$ a contradiction, as all elements
of $ran(g[x_{i',j' }])$
are positive.

Of course, such non-maximal $x_{i',j' }$ need not exist. In that case, we
have to
try again with the negative element $x_{i',j' }$ and Case (5.2).

 \xDH
Let $x_{i,j}-,$ consider $g[x_{i,j}]$ - recall, all elements of
$ran(g[x_{i,j}])$ are
positive.

If there is $x_{i' j' } \xbe ran(g[x_{i,j}])$ such that $C[x_{i',j' }]
\xcs ran(g[x_{i,j}]) \xEd \xCQ,$
we have a contradiction. (This case is impossible in Yablo's original
construction.)

Otherwise, we have to try to work with the elements in $ran(g[x_{i,j}])$
and Case (5.1) above, etc.

Note that we can chose a suitable $x_{i',j' } \xbe ran(g[x_{i,j}]),$ in
particular
one which is not maximal (if such exist), but we have no control over
the choice of $C[x_{i',j' }],$ so we might have to exhaust all finite
columns, until the choice is only from a set of infinite columns.

 \xEj

See Diagram \ref{Diagram Or-And-1} (page \pageref{Diagram Or-And-1})  and
Example \ref{Example Or-And} (page \pageref{Example Or-And}),
Case \ref{Case OA6} (page \pageref{Case OA6}).

 \xEj
\clearpage

\ed

\be

$\hspace{0.01em}$


\label{Example Or-And}

 \xEh

 \xDH Case 1

\label{Case OA1}

Consider the structure $C:=\{x_{i,j}:i< \xba,j< \xbo \},$ with columns
$C_{i}:=\{x_{i,j}:j< \xbo \}.$

Take a standard enumeration $f$ of $C,$ e.g. $f(0):=x_{0,0},$ then
enumerate the $x_{i,j}$
s.t. $max\{i,j\}=1,$ then $max\{i,j\}=2,$ etc. As $f$ is bijective,
$f^{-1}(x_{i,j})$ is
defined.

(More precisely, let $m:=max\{i,j\},$ we go first horizontally from left
to
right over the columns up to column $m-1,$ then in column $m$ upwards,
i.e.
$x_{0,m}, \Xl,x_{m-1,m},x_{m,0}, \Xl,x_{m,m}$.)

Define the relation $x_{i,j} \xcP x_{i',j' }$ iff
$f^{-1}(x_{i,j})<f^{-1}(x_{i',j' }).$ Obviously,
$ \xcP $ is transitive and free from cycles.
$C \xex k:=\{x_{i,j} \xbe C:f^{-1}(x_{i,j}) \xCe k\}$ etc. are defined.

We now show that the structure has no truth values.

Suppose $x_{i,j}+.$

Consider $C[x_{i,j}],$ let $i':=i(x_{i,j})$ and chose $x_{i',j' } \xbe
C[x_{i,j}].$
$x_{i',j' }$ is false, $ran(g[x_{i',j' }])$ intersects $C_{i' }$ above
$x_{i',j' },$ so we have a contradiction.

In particular, $x_{0,0}+$ is impossible.

Suppose $x_{0,0}-,$ then $ran(g[x_{0,0}]) \xEd \xCQ,$ chose $x_{i,j} \xbe
ran(g[x_{0,0}]),$ so $x_{i,j}+,$
but we saw that this is impossible.

Note: for $ \xba =1,$ we have Yablo's construction.

 \xDH Case 2

\label{Case OA2}

We now show that the same construction with columns of height 2
does not work, it has an escape path.

Set $x_{i,0}+,$ $x_{i,1}-$ for all $i.$ $C_{i(x_{i,0})}$ might be $C_{i},$
so $C[x_{i,0}]=\{x_{i,1}\},$ which
is possible. $ran(g[x_{i,1}])$ might be $\{x_{j,0}:j>i\},$ which is
possible again.

 \xDH
Case 3

\label{Case OA3}

We change the order in
Case \ref{Case OA2} (page \pageref{Case OA2}),
to a (horizontally) ranked order:

$x_{i,j}<x_{i',j' }$ iff $i<i' $ (thus, between columns), we now have a
contradiction:

Consider $x_{i,0}-$ and $g[x_{i,0}].$ $g[x_{i,0}](i)$ is undefined. Let
$x_{i+1,j' }:=g[x_{i,0}](i+1),$ thus $x_{i+1,j' }+.$
Consider $C[x_{i+1,j' }].$ This must be some $C_{i' }$ for $i' >i+1,$
so $C[x_{i+1,j' }] \xcs ran(g[x_{i,0}]) \xEd \xCQ,$ and we have a
contradiction.

For $x_{0,0}+,$ consider $C[x_{0,0}],$ this must be some $C_{i},$ $i>0,$
take $x_{i,0} \xbe C_{i},$ $x_{i,0}-,$ and
continue as above.

 \xDH
Case 4

\label{Case OA4}

Modify
Case \ref{Case OA1} (page \pageref{Case OA1}),
to a ranked order, but this time horizontally:
$x_{i,j}<x_{i',j' }$ iff $j<j'.$

Take $x_{i,j}+,$ consider $C[x_{i,j}],$ this may be (part of) any $C_{i'
}$ (beginning
at $j+1).$ Take e.g. $x_{i',j+1}- \xbe C[x_{i,j}],$ consider $g[x_{i'
,j+1}],$
a choice function in $\{C_{k} \xex x_{i',j+1}:k, \xbo \}$
this will intersect $C[x_{i,j}].$
In particular, $x_{0,0}+$ is impossible.

Suppose $x_{0,0}-,$ take $x_{i,j}+ \xbe ran(g[x_{0,0}]),$ and continue as
above.

Note that the case with just one $C_{i}$ is the original Yablo
construction.

 \xDH
Case 5

\label{Case OA5}

We modify
Case \ref{Case OA1} (page \pageref{Case OA1})
again:
Consider the ranked order by $x_{i,j} \xcP x_{i',j' }$ iff
$max\{i,j\}<max\{i',j' \}.$

This is left to the reader as an exercise.

 \xDH Case 6

\label{Case OA6}

The general argument is as follows (and applies to general partial
orders, too):
 \xEh
 \xDH
We show that $x_{i,j}+$ leads to a contradiction.
(In our terminology, $x_{i,j}$ is the head.)
 \xEh
 \xDH
we find $C[x_{i,j}]$ (negative elements)
above $x_{i,j}$ - but we have no control over the choice of $C[x_{i,j}],$
 \xDH we chose $x_{i',j' } \xbe C[x_{i,j}]$ - if there is a minimal such,
chose this one,
it must not be a maximal element in $C[x_{i,j}]$ $(x_{i',j' }$ is the
knee),
 \xDH we find $g[x_{i',j' }]$ (positive elements)
above $x_{i',j' }$ - again we have no control over the
choice of this $g.$ But, as $C[x_{i,j}] \xex x_{i',j' }$ is not empty,
$ran(g[x_{i',j' }]) \xcs C[x_{i,j}] \xEd \xCQ,$ so we have a
contradiction
$(x_{i'',j'' } \xbe ran(g[x_{i',j' }]) \xcs C[x_{i,j}]$ is the foot).
 \xDH
We apply the reasoning to $x_{0,0}.$
 \xEj
 \xDH
We show that $x_{0,0}-$ leads to a contradiction.
 \xEh
 \xDH We have $g[x_{0,0}]$ (positive elements) -
but no control over the choice of $g.$ In particular,
it may be arbitrarily high up.
 \xDH we chose $x_{i,j} \xbe ran(g[x_{0,0}])$ with enough room above it
for the
argument about $x_{i,j}+$ in (6.1).
 \xEj

 \xDH Case 
\ref{Case OA3} (page 
\pageref{Case OA3})  is similar, except that we work
horizontally, not vertically.

 \xEj

 \xEj

\ee

$ \xCO $

$ \xCO $

$ \xCO $

\clearpage

\begin{diagram}

\label{Diagram Or-And-1}
\index{Diagram Or-And-1}

\unitlength0.5mm
\begin{picture}(150,180)(0,0)

\put(0,175){{\rm\bf Diagram for Definition
    \ref{Definition Or-And} Case \ref{Case DOA}}}

\put(0,150){Case \ref{Case DOA}, (4.1)}

\put(60,90){\line(0,1){60}}
\put(10,130){\line(1,0){90}}
\put(60,110){\circle*{1}}
\put(62,110){$x_{i',j'}$}
\put(55,85){$C[x_{i,j}]$, all -}
\put(0,110){$x_{i,j}+$}
\put(105,130){$g[x_{i',j'}]$, all +}

\put(0,60){Case \ref{Case DOA}, (4.2)}

\put(80,15){\circle*{1}}
\put(80,10){$x_{i',j'}$}
\put(80,15){\line(-2,1){60}}
\put(80,15){\line(2,1){40}}
\put(0,30){$x_{i,j}-$}
\put(125,35){$g[x_{i,j}]$, all +}
\put(50,30){\circle*{1}}
\put(52,31){$x_{i'',j''}$}

\end{picture}

\end{diagram}

\vspace{4mm}

\clearpage

\clearpage

\begin{diagram}

\label{Diagram Or-And}
\index{Diagram Or-And}

\unitlength0.5mm
\begin{picture}(150,180)(0,0)

\put(0,175){{\rm\bf Diagram for Example
    \ref{Example Or-And} Case \ref{Case OA1}}}

\put(5,10){\line(0,1){150}}
\put(35,10){\line(0,1){150}}
\put(65,10){\line(0,1){150}}
\put(95,10){\line(0,1){150}}
\put(125,10){\line(0,1){150}}

\put(5,10){\circle*{1}}
\put(7,15){$C_0$}
\put(5,5){$x_{0,0}$}
\put(65,5){$C_{i(x_{i,j})}$}
\put(36,30){$x_{i,j}$}
\put(35,30){\circle*{1}}

\put(66,60){$x_{i(x_{i,j}),j'}$}
\put(65,60){\circle*{1}}
\put(57,70){$C[x_{i,j}]$, all -}
\put(62,33){\line(0,1){127}}
\put(61,33){\line(1,0){2}}

\put(135,63){\line(0,1){97}}
\multiput(135,63)(-5,0){26}{\circle*{0.2}}
\put(133,63){\line(1,0){4}}
\put(138,90){$C \xex x_{i(x_{i,j}),j'}$}

\put(5,130){\line(1,-1){30}}
\put(65,130){\line(1,-1){30}}
\put(35,100){\line(1,1){30}}
\put(95,100){\line(1,1){30}}
\put(110,110){$g[x_{i(x_{i,j}),j'}]$, all +}

\end{picture}

\end{diagram}

\vspace{4mm}

\clearpage

\clearpage
\section{
Saw Blades
}

\label{Section Saw-Blades-All}
\subsection{
Introduction
}

Yablo works with contradictions in the form of $x_{0}= \xCN x_{1} \xcu
\xCN x_{2},$ $x_{1}= \xCN x_{2},$
graphically $x_{0} \xcP x_{1} \xcP x_{2},$ $x_{0} \xcP x_{2}.$ They are
combined in a formally simple total
order, which, however, blurs conceptual differences.

We discuss here different, conceptually very clear and simple, examples
of a Yablo-Like construction.

In particular, we emphasize the difference between $x_{1}$ and $x_{2}$ in
the
Yablo contradictions. The contradiction is
finished in $x_{2},$ but not in $x_{1},$ requiring the barring of escape
routes in $x_{1}.$
We repair the possible escape routes by constructing new contradictions
for the SAME origin. This is equivalent to closing under transitivity in
the
individual ``saw blades'' - see below.

So we use the same ``cells'' as Yablo does for the contradictions, but
analyse
the way they are put together.

We use the full strength of the conceptual difference between $x_{1}$ and
$x_{2}$
(in above notation)
only in Section 
\ref{Section Simplifications} (page 
\pageref{Section Simplifications}), where we show that
preventing
$x_{2}$ from being TRUE is sufficient, whereas we need $x_{1}$ to be
contradictory, see also
Fact \ref{Fact Infin-Branching} (page \pageref{Fact Infin-Branching}),
and
Remark \ref{Remark Contradictory} (page \pageref{Remark Contradictory}).
Thus, we obtain a minimal, i.e. necessary and sufficient, construction
for combining Yablo cells in this way.
\subsection{
Saw Blades
}

\label{Section Saw-Blades}
%
%

First, we show the escape route problem.

\be

$\hspace{0.01em}$


\label{Example Escape}

Consider Construction 
\ref{Construction Saw-Blade} (page 
\pageref{Construction Saw-Blade})
without closing under transitivity, i.e. the only arrows originating
in $x_{ \xbs,0}$ will be $x_{ \xbs,0} \xcP x_{ \xbs,1}$ and $x_{ \xbs
,0} \xcP y_{ \xbs,0},$ etc.

Let $x_{ \xbs,0}=TRUE,$ then $x_{ \xbs,1}$ is an $ \xco,$ and we
pursue the path $x_{ \xbs,0} \xcP x_{ \xbs,1} \xcP x_{ \xbs,2},$ this
has no contradiction so far, and we
continue with $x_{ \xbs,2}=TRUE,$ $x_{ \xbs,3}$ is $ \xco $ again,
we continue and have $x_{ \xbs,0} \xcP x_{ \xbs,1} \xcP x_{ \xbs,2}
\xcP x_{ \xbs,3} \xcP x_{ \xbs,4},$ etc.,
never meeting a contradiction, so we have an escape path.

\ee

$ \xCO $
%
%

\bcs

$\hspace{0.01em}$


\label{Construction Saw-Blade}

We construct a saw blade $ \xbs,$ $SB_{ \xbs }.$

 \xEh

 \xDH
``Saw Blades''
 \xEh
 \xDH
Let $x_{ \xbs,0} \xcP x_{ \xbs,1} \xcP x_{ \xbs,2} \xcP x_{ \xbs,3}
\xcP x_{ \xbs,4},$  \Xl.

$x_{ \xbs,0} \xcP y_{ \xbs,0},$ $x_{ \xbs,1} \xcP y_{ \xbs,0},$ $x_{
\xbs,1} \xcP y_{ \xbs,1},$ $x_{ \xbs,2} \xcP y_{ \xbs,1},$
$x_{ \xbs,2} \xcP y_{ \xbs,2},$ $x_{ \xbs,3} \xcP y_{ \xbs,2},$ $x_{
\xbs,3} \xcP y_{ \xbs,3},$ $x_{ \xbs,4} \xcP y_{ \xbs,3},$  \Xl.

we call the construction
a ``saw blade'', with ``teeth'' $y_{ \xbs,0},$ $y_{ \xbs,1},$ $y_{ \xbs
,2},$  \Xl.
and ``back'' $x_{ \xbs,0},$ $x_{ \xbs,1},$ $x_{ \xbs,2},$  \Xl.

We call $x_{ \xbs,0}$ the start of the blade.

See Diagram \ref{Diagram Sawblade-2} (page \pageref{Diagram Sawblade-2}).

 \xDH
Add (against escape), e.g. first $x_{ \xbs,0} \xcP x_{ \xbs,2},$ $x_{
\xbs,0} \xcP y_{ \xbs,1},$
then $x_{ \xbs,1} \xcP x_{ \xbs,3},$ $x_{ \xbs,1} \xcP y_{ \xbs,2},$
now
we have to add $x_{ \xbs,0} \xcP x_{ \xbs,3},$ $x_{ \xbs,0} \xcP y_{
\xbs,2},$ etc, recursively.
This is equivalent to
closing the saw blade under transitivity with negative arrows $ \xcP.$
This is easily seen.

 \xDH
We define the valuation by $d(x_{ \xbs,i}):= \xcU \xCN z_{ \xbs,j},$ for
all $x_{ \xbs,j}$
such that $x_{ \xbs,i} \xcP z_{ \xbs,j},$
as in the original Yablo construction.

 \xEj

 \xDH
Composition of saw blades
 \xEh
 \xDH
Add for the teeth $y_{ \xbs,0},$ $y_{ \xbs,1},$ $y_{ \xbs,2}$  \Xl.
their own saw blades, i.e.
start at $y_{ \xbs,0}$ a new saw blade $SB_{ \xbs,0}$ with $y_{ \xbs
,0}=x_{ \xbs,0,0},$ at
$y_{ \xbs,1}$ a new saw blade $SB_{ \xbs,1}$ with $y_{ \xbs,1}=x_{ \xbs
,1,0},$ etc.

 \xDH
Do this recursively.

I.e., at every tooth of every saw blade start a new saw blade.
See Diagram \ref{Diagram Sawblade-3} (page \pageref{Diagram Sawblade-3}).
 \xEj
 \xEj

Note:

It is NOT necessary to close the whole structure (the individual saw
blades
together) under transitivity.

\ecs

\bfa

$\hspace{0.01em}$


\label{Fact Saw-Blade}

All $z_{ \xbs,i}$ in all saw blades so constructed are contradictory,
i.e.
assigning them a truth value leads to a contradiction.

\efa

\subparagraph{
Proof
}

$\hspace{0.01em}$


Fix some saw blade $SB_{ \xbs }$ in the construction.

 \xEh

 \xDH
Take any $z_{ \xbs,i}$ with $z_{ \xbs,i}+,$ i.e. $z_{ \xbs,i}=TRUE.$ We
show that this is contradictory.

 \xEh

 \xDH
Case 1: $z_{ \xbs,i}$ is one of the $x_{ \xbs,i}$ i.e. it is in the back
of the blade.

Take any $x_{ \xbs,i' }$ in the back such that there is an arrow $x_{
\xbs,i} \xcP x_{ \xbs,i' }$
$(i':=i+1$ suffices).
Then $x_{ \xbs,i' }=FALSE,$ and we have an $ \xco $ at $x_{ \xbs,i' }.$
Take any $z_{ \xbs,j}$ such that $x_{ \xbs,i' } \xcP z_{ \xbs,j},$ by
transitivity, $x_{ \xbs,i} \xcP z_{ \xbs,j},$ so $z_{ \xbs,j}=FALSE,$
but as $x_{ \xbs,i' }=FALSE,$ $z_{ \xbs,j}=TRUE,$
contradiction.

 \xDH
Case 2: $z_{ \xbs,i}$ is one of the $y_{ \xbs,i},$ i.e. a tooth of the
blade.

Then $y_{ \xbs,i}$ is the start of the new blade starting at $y_{ \xbs
,i},$ and we
argue as above in Case 1.

 \xEj

 \xDH
Take any $z_{ \xbs,i}$ with $z_{ \xbs,i}-,$ i.e. $z_{ \xbs,i}=FALSE,$
and we have an $ \xco $ at $z_{ \xbs,i},$ and
one of the successors of $z_{ \xbs,i},$ say $z_{ \xbs,j},$ has to be
TRUE.
We just saw that this is impossible.

(For the intuition:
If $z_{ \xbs,i}$ is in the back of the blade, all of its successors are
in the
same blade.
If $z_{ \xbs,i}$ is one of the teeth of the blade, all of its successors
are
in the new blade, starting at $z_{ \xbs,i}.$
In both cases, $z_{ \xbs,j}=TRUE$ leads to a contradiction, as we saw
above.)

 \xEj

\br

$\hspace{0.01em}$


\label{Remark Saw-Blade}

Note that all $y_{ \xbs,i}$ are contradictory, too, not only the $x_{
\xbs,i}.$
We will see in
Section \ref{Section Simplifications} (page \pageref{Section Simplifications})
that we can achieve this by simpler means, as we need to consider here
the case $x_{ \xbs,i} \xco $ only, the contradiction for the case $x_{
\xbs,i} \xcu $ is already
treated.

Thus, we seemingly did not fully use here the conceptual clarity of
difference
between $x_{1}$ and $x_{2}$ alluded to in the beginning of
Section \ref{Section Saw-Blades} (page \pageref{Section Saw-Blades}).
See, however, the discussion in
Section \ref{Section InfBranch} (page \pageref{Section InfBranch}).

\er

$ \xCO $

$ \xCO $

\clearpage

\begin{diagram}

\label{Diagram Sawblade-2}
\index{Diagram Sawblade-2}

\unitlength0.6mm
\begin{picture}(150,100)(0,0)

\put(0,82){{\rm\bf Diagram Single Saw Blade }}
\put(0,75){Start of the saw blade $\sigma$ beginning at $x_{\sigma,0}$,}
\put(0,68){before closing under transitivity }

\put(5,50){$SB_{\sigma}$}


\put(20,60){$x_{_{\sigma,0}}$}
\put(20,40){$y_{_{\sigma,0}}$}
\put(26,61){\line(1,0){12}}
\put(32,60){\line(0,1){2}}
\put(22,43){\line(0,1){14}}
\put(21,50){\line(1,0){2}}
\put(24,43){\line(1,1){16}}
\put(31,51.5){\line(1,-1){1.5}}


\put(40,60){$x_{_{\sigma,1}}$}
\put(40,40){$y_{_{\sigma,1}}$}
\put(46,61){\line(1,0){12}}
\put(52,60){\line(0,1){2}}
\put(42,43){\line(0,1){14}}
\put(41,50){\line(1,0){2}}
\put(44,43){\line(1,1){16}}
\put(51,51.5){\line(1,-1){1.5}}


\put(60,60){$x_{_{\sigma,2}}$}
\put(60,40){$y_{_{\sigma,2}}$}
\put(66,61){\line(1,0){12}}
\put(72,60){\line(0,1){2}}
\put(62,43){\line(0,1){14}}
\put(61,50){\line(1,0){2}}
\put(64,43){\line(1,1){16}}
\put(71,51.5){\line(1,-1){1.5}}


\put(80,60){$x_{_{\sigma,3}}$}
\put(80,40){$y_{_{\sigma,3}}$}
\put(86,61){\line(1,0){12}}
\put(92,60){\line(0,1){2}}
\put(82,43){\line(0,1){14}}
\put(81,50){\line(1,0){2}}
\put(84,43){\line(1,1){16}}
\put(91,51.5){\line(1,-1){1.5}}


\put(100,60){$x_{_{\sigma,4}}$}
\put(100,40){$y_{_{\sigma,4}}$}
\put(106,61){\line(1,0){12}}
\put(112,60){\line(0,1){2}}
\put(102,43){\line(0,1){14}}
\put(101,50){\line(1,0){2}}
\put(104,43){\line(1,1){16}}
\put(111,51.5){\line(1,-1){1.5}}


\put(120,60){$x_{_{\sigma,5}}$}
\put(120,40){$y_{_{\sigma,5}}$}
\put(126,61){\line(1,0){12}}
\put(132,60){\line(0,1){2}}
\put(122,43){\line(0,1){14}}
\put(121,50){\line(1,0){2}}
\put(124,43){\line(1,1){16}}
\put(131,51.5){\line(1,-1){1.5}}


\put(140,60){$x_{_{\sigma,6}}$}
\put(140,40){$y_{_{\sigma,6}}$}
\multiput(146,61)(2,0){5}{\circle*{0.2}}
\put(142,43){\line(0,1){14}}
\put(141,50){\line(1,0){2}}
\multiput(144,43)(1.5,1.5){5}{\circle*{0.2}}

\end{picture}

\end{diagram}

\vspace{10mm}

Read $x_{ \xbs,0} \xcP x_{ \xbs,1} \xcP y_{ \xbs,0},$ $x_{ \xbs,0}
\xcP y_{ \xbs,0},$ etc, more precisely
$x_{ \xbs,0}= \xCN x_{ \xbs,1} \xcu \xCN y_{ \xbs,0},$ $x_{ \xbs,1}=
\xCN y_{ \xbs,0} \xcu \xCN x_{ \xbs,2} \xcu \xCN y_{ \xbs,1},$ etc.
\clearpage

\clearpage

\begin{diagram}

\label{Diagram Sawblade-3}
\index{Diagram Sawblade-3}

\unitlength0.6mm
\begin{picture}(150,190)(0,0)

\put(0,182){{\rm\bf Diagram Composition of Saw Blades }}
\put(0,175){Composition of saw blades (without additional arrows) }
\put(0,168){The fat dots indicate identity, e.g. $y_{0,0}=x_{0,0,0}$ }

\put(5,150){$SB_{0}$}


\put(20,160){$x_{_{0,0}}$}
\put(20,140){$y_{_{0,0}}$}
\put(26,161){\line(1,0){12}}
\put(32,160){\line(0,1){2}}
\put(22,143){\line(0,1){14}}
\put(21,150){\line(1,0){2}}
\put(24,143){\line(1,1){16}}
\put(31,151.5){\line(1,-1){1.5}}


\put(40,160){$x_{_{0,1}}$}
\put(40,140){$y_{_{0,1}}$}
\put(46,161){\line(1,0){12}}
\put(52,160){\line(0,1){2}}
\put(42,143){\line(0,1){14}}
\put(41,150){\line(1,0){2}}
\put(44,143){\line(1,1){16}}
\put(51,151.5){\line(1,-1){1.5}}


\put(60,160){$x_{_{0,2}}$}
\put(60,140){$y_{_{0,2}}$}
\put(66,161){\line(1,0){12}}
\put(72,160){\line(0,1){2}}
\put(62,143){\line(0,1){14}}
\put(61,150){\line(1,0){2}}
\put(64,143){\line(1,1){16}}
\put(71,151.5){\line(1,-1){1.5}}


\put(80,160){$x_{_{0,3}}$}
\put(80,140){$y_{_{0,3}}$}
\put(86,161){\line(1,0){12}}
\put(92,160){\line(0,1){2}}
\put(82,143){\line(0,1){14}}
\put(81,150){\line(1,0){2}}
\put(84,143){\line(1,1){16}}
\put(91,151.5){\line(1,-1){1.5}}


\put(100,160){$x_{_{0,4}}$}
\put(100,140){$y_{_{0,4}}$}
\put(106,161){\line(1,0){12}}
\put(112,160){\line(0,1){2}}
\put(102,143){\line(0,1){14}}
\put(101,150){\line(1,0){2}}
\put(104,143){\line(1,1){16}}
\put(111,151.5){\line(1,-1){1.5}}


\put(120,160){$x_{_{0,5}}$}
\put(120,140){$y_{_{0,5}}$}
\put(126,161){\line(1,0){12}}
\put(132,160){\line(0,1){2}}
\put(122,143){\line(0,1){14}}
\put(121,150){\line(1,0){2}}
\put(124,143){\line(1,1){16}}
\put(131,151.5){\line(1,-1){1.5}}


\put(140,160){$x_{_{0,6}}$}
\put(140,140){$y_{_{0,6}}$}
\multiput(146,161)(2,0){5}{\circle*{0.2}}
\put(142,143){\line(0,1){14}}
\put(141,150){\line(1,0){2}}
\multiput(144,143)(1.5,1.5){5}{\circle*{0.2}}

\put(45,120){$SB_{0,2}$}

\multiput(62,133)(0,2){3}{\circle*{0.5}}


\put(60,130){$x_{_{0,2,0}}$}
\put(60,110){$y_{_{0,2,0}}$}
\put(66,131){\line(1,0){12}}
\put(72,130){\line(0,1){2}}
\put(62,113){\line(0,1){14}}
\put(61,120){\line(1,0){2}}
\put(64,113){\line(1,1){16}}
\put(71,121.5){\line(1,-1){1.5}}


\put(80,130){$x_{_{0,2,1}}$}
\put(80,110){$y_{_{0,2,1}}$}
\put(86,131){\line(1,0){12}}
\put(92,130){\line(0,1){2}}
\put(82,113){\line(0,1){14}}
\put(81,120){\line(1,0){2}}
\put(84,113){\line(1,1){16}}
\put(91,121.5){\line(1,-1){1.5}}


\put(100,130){$x_{_{0,2,2}}$}
\put(100,110){$y_{_{0,2,2}}$}
\put(106,131){\line(1,0){12}}
\put(112,130){\line(0,1){2}}
\put(102,113){\line(0,1){14}}
\put(101,120){\line(1,0){2}}
\put(104,113){\line(1,1){16}}
\put(111,121.5){\line(1,-1){1.5}}


\put(120,130){$x_{_{0,2,3}}$}
\put(120,110){$y_{_{0,2,3}}$}
\put(126,131){\line(1,0){12}}
\put(132,130){\line(0,1){2}}
\put(122,113){\line(0,1){14}}
\put(121,120){\line(1,0){2}}
\put(124,113){\line(1,1){16}}
\put(131,121.5){\line(1,-1){1.5}}


\put(140,130){$x_{_{0,2,4}}$}
\put(140,110){$y_{_{0,2,4}}$}
\multiput(146,131)(2,0){5}{\circle*{0.2}}
\put(142,113){\line(0,1){14}}
\put(141,120){\line(1,0){2}}
\multiput(144,113)(1.5,1.5){5}{\circle*{0.2}}

\put(25,90){$SB_{0,1}$}

\multiput(42,103)(0,2){18}{\circle*{0.5}}


\put(40,100){$x_{_{0,1,0}}$}
\put(40,80){$y_{_{0,1,0}}$}
\put(46,101){\line(1,0){12}}
\put(52,100){\line(0,1){2}}
\put(42,83){\line(0,1){14}}
\put(41,90){\line(1,0){2}}
\put(44,83){\line(1,1){16}}
\put(51,91.5){\line(1,-1){1.5}}


\put(60,100){$x_{_{0,1,1}}$}
\put(60,80){$y_{_{0,1,1}}$}
\put(66,101){\line(1,0){12}}
\put(72,100){\line(0,1){2}}
\put(62,83){\line(0,1){14}}
\put(61,90){\line(1,0){2}}
\put(64,83){\line(1,1){16}}
\put(71,91.5){\line(1,-1){1.5}}


\put(80,100){$x_{_{0,1,2}}$}
\put(80,80){$y_{_{0,1,2}}$}
\put(86,101){\line(1,0){12}}
\put(92,100){\line(0,1){2}}
\put(82,83){\line(0,1){14}}
\put(81,90){\line(1,0){2}}
\put(84,83){\line(1,1){16}}
\put(91,91.5){\line(1,-1){1.5}}


\put(100,100){$x_{_{0,1,3}}$}
\put(100,80){$y_{_{0,1,3}}$}
\put(106,101){\line(1,0){12}}
\put(112,100){\line(0,1){2}}
\put(102,83){\line(0,1){14}}
\put(101,90){\line(1,0){2}}
\put(104,83){\line(1,1){16}}
\put(111,91.5){\line(1,-1){1.5}}


\put(120,100){$x_{_{0,1,4}}$}
\put(120,80){$y_{_{0,1,4}}$}
\put(126,101){\line(1,0){12}}
\put(132,100){\line(0,1){2}}
\put(122,83){\line(0,1){14}}
\put(121,90){\line(1,0){2}}
\put(124,83){\line(1,1){16}}
\put(131,91.5){\line(1,-1){1.5}}


\put(140,100){$x_{_{0,1,5}}$}
\put(140,80){$y_{_{0,1,5}}$}
\multiput(146,101)(2,0){5}{\circle*{0.2}}
\put(142,83){\line(0,1){14}}
\put(141,90){\line(1,0){2}}
\multiput(144,83)(1.5,1.5){5}{\circle*{0.2}}

\put(5,60){$SB_{0,0}$}

\multiput(22,73)(0,2){33}{\circle*{0.5}}


\put(20,70){$x_{_{0,0,0}}$}
\put(20,50){$y_{_{0,0,0}}$}
\put(26,71){\line(1,0){12}}
\put(32,70){\line(0,1){2}}
\put(22,53){\line(0,1){14}}
\put(21,60){\line(1,0){2}}
\put(24,53){\line(1,1){16}}
\put(31,61.5){\line(1,-1){1.5}}


\put(40,70){$x_{_{0,0,1}}$}
\put(40,50){$y_{_{0,0,1}}$}
\put(46,71){\line(1,0){12}}
\put(52,70){\line(0,1){2}}
\put(42,53){\line(0,1){14}}
\put(41,60){\line(1,0){2}}
\put(44,53){\line(1,1){16}}
\put(51,61.5){\line(1,-1){1.5}}


\put(60,70){$x_{_{0,0,2}}$}
\put(60,50){$y_{_{0,0,2}}$}
\put(66,71){\line(1,0){12}}
\put(72,70){\line(0,1){2}}
\put(62,53){\line(0,1){14}}
\put(61,60){\line(1,0){2}}
\put(64,53){\line(1,1){16}}
\put(71,61.5){\line(1,-1){1.5}}


\put(80,70){$x_{_{0,0,3}}$}
\put(80,50){$y_{_{0,0,3}}$}
\put(86,71){\line(1,0){12}}
\put(92,70){\line(0,1){2}}
\put(82,53){\line(0,1){14}}
\put(81,60){\line(1,0){2}}
\put(84,53){\line(1,1){16}}
\put(91,61.5){\line(1,-1){1.5}}


\put(100,70){$x_{_{0,0,4}}$}
\put(100,50){$y_{_{0,0,4}}$}
\put(106,71){\line(1,0){12}}
\put(112,70){\line(0,1){2}}
\put(102,53){\line(0,1){14}}
\put(101,60){\line(1,0){2}}
\put(104,53){\line(1,1){16}}
\put(111,61.5){\line(1,-1){1.5}}


\put(120,70){$x_{_{0,0,5}}$}
\put(120,50){$y_{_{0,0,5}}$}
\put(126,71){\line(1,0){12}}
\put(132,70){\line(0,1){2}}
\put(122,53){\line(0,1){14}}
\put(121,60){\line(1,0){2}}
\put(124,53){\line(1,1){16}}
\put(131,61.5){\line(1,-1){1.5}}


\put(140,70){$x_{_{0,0,6}}$}
\put(140,50){$y_{_{0,0,6}}$}
\multiput(146,71)(2,0){5}{\circle*{0.2}}
\put(142,53){\line(0,1){14}}
\put(141,60){\line(1,0){2}}
\multiput(144,53)(1.5,1.5){5}{\circle*{0.2}}

\put(5,30){$SB_{0,0,0}$}

\multiput(22,43)(0,2){3}{\circle*{0.5}}


\put(20,40){$x_{_{0,0,0,0}}$}
\put(20,20){$y_{_{0,0,0,0}}$}
\put(26,41){\line(1,0){12}}
\put(32,40){\line(0,1){2}}
\put(22,23){\line(0,1){14}}
\put(21,30){\line(1,0){2}}
\put(24,23){\line(1,1){16}}
\put(31,31.5){\line(1,-1){1.5}}


\put(40,40){$x_{_{0,0,0,1}}$}
\put(40,20){$y_{_{0,0,0,1}}$}
\put(46,41){\line(1,0){12}}
\put(52,40){\line(0,1){2}}
\put(42,23){\line(0,1){14}}
\put(41,30){\line(1,0){2}}
\put(44,23){\line(1,1){16}}
\put(51,31.5){\line(1,-1){1.5}}


\put(60,40){$x_{_{0,0,0,2}}$}
\put(60,20){$y_{_{0,0,0,2}}$}
\put(66,41){\line(1,0){12}}
\put(72,40){\line(0,1){2}}
\put(62,23){\line(0,1){14}}
\put(61,30){\line(1,0){2}}
\put(64,23){\line(1,1){16}}
\put(71,31.5){\line(1,-1){1.5}}


\put(80,40){$x_{_{0,0,0,3}}$}
\put(80,20){$y_{_{0,0,0,3}}$}
\put(86,41){\line(1,0){12}}
\put(92,40){\line(0,1){2}}
\put(82,23){\line(0,1){14}}
\put(81,30){\line(1,0){2}}
\put(84,23){\line(1,1){16}}
\put(91,31.5){\line(1,-1){1.5}}


\put(100,40){$x_{_{0,0,0,4}}$}
\put(100,20){$y_{_{0,0,0,4}}$}
\put(106,41){\line(1,0){12}}
\put(112,40){\line(0,1){2}}
\put(102,23){\line(0,1){14}}
\put(101,30){\line(1,0){2}}
\put(104,23){\line(1,1){16}}
\put(111,31.5){\line(1,-1){1.5}}


\put(120,40){$x_{_{0,0,0,5}}$}
\put(120,20){$y_{_{0,0,0,5}}$}
\put(126,41){\line(1,0){12}}
\put(132,40){\line(0,1){2}}
\put(122,23){\line(0,1){14}}
\put(121,30){\line(1,0){2}}
\put(124,23){\line(1,1){16}}
\put(131,31.5){\line(1,-1){1.5}}


\put(140,40){$x_{_{0,0,0,6}}$}
\put(140,20){$y_{_{0,0,0,6}}$}
\multiput(146,41)(2,0){5}{\circle*{0.2}}
\put(142,23){\line(0,1){14}}
\put(141,30){\line(1,0){2}}
\multiput(144,23)(1.5,1.5){5}{\circle*{0.2}}

\end{picture}

\end{diagram}

\vspace{4mm}

\clearpage

\clearpage
\subsection{
Discussion of Saw Blades
}
\subsubsection{
Simplifications
}

In general, and this does not only concern Saw Blade like constructions:

 \xEh

 \xDH
If, say, $x$ can be given the value TRUE (it has no successors, $x=y \xco
\xCN y,$
etc.), then we can simplify a variable $z$ where $x$ occurs. If $z=x \xcu
x',$ then
we may set $z=x',$ if $z=z \xco x',$ we may set $z=TRUE,$ etc.

 \xDH
If the structure below $x$ is a tree (no branches meet again), then we
have
no contradictions.

 \xDH
We may contract finitely many branching points to one branching point,
with equivalent structures (trivial), but not necessarily infinitely
many branching points, see
Example \ref{Example Procrastination} (page \pageref{Example Procrastination}).

 \xEj
\paragraph{
Simplifications that will not work
}

\label{Section NotWork}

We try to simplify here the Saw Blade construction.
Throughout, we consider formulas of pure conjunctions.

We start with a Yablo Cell, but try to continue otherwise.

So we have $x_{0} \xcP x_{1} \xcP x_{2},$ $x_{1} \xcP x_{2}.$ So $x_{0}+$
is impossible. We now try to treat $x_{0}-.$
We see in Section 
\ref{Section Simplifications} (page 
\pageref{Section Simplifications})
that appending $x_{2} \xch_{ \xCL }y_{2}$ may take care of the necessary
contradiction at $x_{2},$
see Diagram \ref{Diagram Sawblade-4} (page \pageref{Diagram Sawblade-4}).
When we try to do the same at $x_{1},$ i.e. some $x_{1} \xch_{ \xCL
}x_{3},$ we solve again
the necessary contradiction at $x_{1},$ but run into a problem with
$x_{0}+,$
as $x_{1}$ is an $ \xco.$ So $x_{3}$ has to be contradictory. If we
continue
$x_{3} \xch_{ \xCL }x_{4} \xch_{ \xCL }x_{5}$ etc., this will not work, as
we may set all such $x_{i}-,$
and have a model. In abstract terms, we only procrastinate the same
problem
without solving anything.
Of course, we could append after some time new
Yablo Cells, as in the saw blade construction, but this is cheating,
as the ``true'' construction begins only later.

Suppose we add not only $x_{1} \xch_{ \xCL }x_{3},$ but also $x_{0} \xch_{
\xCL }x_{3},$ then we solve
$x_{0}+,$ but $x_{0}-$ is not solved.

Working with cells of the type (2.1) in
Example \ref{Example Simple-Cells} (page \pageref{Example Simple-Cells})
will lead to similar problems.

Consequently, any attempt to use a ``pipeline'', avoiding infinite
branching, is doomed:
%
%

Instead of
$x_{0} \xcP x_{1},$ $x_{0} \xcP x_{2},$  \Xl. etc. we construct a
``pipeline'' of $x_{i}',$ with
$x_{0} \xcP x'_{1},$ $x_{1} \xcP x'_{2},$ etc, $x'_{1} \xcp x'_{2} \xcp
x'_{3} \Xl.,$ and $x'_{1} \xcp x_{1},$ $x'_{2} \xcp x_{2},$ etc.
or similarly, to have infinitely many contradictions for paths
from $x_{0}.$

As this is a set of classical formulas, this cannot achieve inconsistency,
see Fact \ref{Fact NotClass} (page \pageref{Fact NotClass}).
\subsubsection{
Infinite Branching and Recursive Contradictions are Necessary
}

\label{Section InfBranch}

\bfa

$\hspace{0.01em}$


\label{Fact Infin-Branching}

We need infinite branching in the saw blade construction at all $x_{i}.$

(We always use basic contradictions of the type $x_{i} \xcP x_{i+1} \xcP
y_{i},$ $x_{i} \xcP y_{i}$ -
which we abbreviate $ \xeA.)$

\efa

$ \xCO $

$ \xCO $

$ \xCO $

$ \xCO $
%
%

\subparagraph{
Proof
}

$\hspace{0.01em}$


 \xEh
 \xDH The argument

 \xEh
 \xDH

This is needed for $x_{0}$ only.

If $x_{0}-$ is impossible, then all arrows $x_{0} \xcP x_{i}$ have to lead
to
$ \xeA $ attached at $x_{i},$ so $x_{i}+$ must be impossible (recall
$x_{0}-$ means $ \xco $ at $x_{0}).$

(Suppose we try to stop at the first $ \xeA,$ then $x_{1} \xcp y_{0}$
leads in hindsight
to a contradiction, as it is again the $x_{0}$ of a new saw blade,
but as it has value -, we run into a circularity, we still have
to show that this is contradictory. For $x_{0}+,$ this is different,
we have to consider just one saw blade to see that this is contradictory.)

Thus, if $x_{0} \xcP x_{1},$ $x_{0} \xcP x_{i}$ exist, we attach at
$x_{1}$ and $x_{i}$ an $ \xeA,$
e.g. $x_{i} \xcP y_{i},$ $x_{i} \xcP x_{i+1} \xcP y_{i},$ etc.

This is important (and possible at this level of analysis) only for
$x_{0}.$

 \xDH

The following holds for all $i.$

$x_{i}+$ impossible $ \xch $ (by $x_{i} \xcP x_{i+1})$
all $x_{i+1} \xcP x_{j}$ and $x_{i+1} \xcP y_{j}$ must have a
contradiction
$x_{i} \xcP x_{j}$ viz. $x_{i} \xcP y_{j},$
so we have new arrows originating at $x_{i}.$

We need a branching at $x_{i+1},$ see
Example \ref{Example Paths} (page \pageref{Example Paths}),
Case (2.1).

This holds for all $i.$

 \xEj

 \xDH
So we have the following construction:

 \xEh
 \xDH
start with $ \xeA $ $x_{0} \xcP y_{0},$ $x_{0} \xcP x_{1} \xcP y_{0}.$
 \xEI
 \xDH
So $x_{1}$ is the knee in the cell $x_{0} \xcP y_{0},$ $x_{0} \xcP x_{1}
\xcP y_{0}.$
 \xEJ
 \xDH $x_{1}$
 \xEh
 \xDH
as $x_{0}-$ should be impossible, append a new $ \xeA $ to $x_{1},$ so
$x_{1}+$ is impossible:

$x_{0} \xcP x_{1} \xcP x_{2},$ $x_{0} \xcP y_{0},$ $x_{1} \xcP y_{0},$
$x_{1} \xcP y_{1},$ $x_{2} \xcP y_{1}.$

From now on, we will not mention all $y_{i},$ only all $x_{i}.$
 \xDH
as $x_{0}+$ should be impossible,
add $x_{0} \xcP x_{2}$ because of $x_{1} \xcP x_{2},$ and $x_{0} \xcP
y_{1}$ because of $x_{1} \xcP y_{1},$

so we have
$x_{0} \xcP x_{1} \xcP x_{2},$ $x_{0} \xcP x_{2},$
and the new arrow $x_{0} \xcP y_{1}$
 \xEI
 \xDH
So $x_{1}$ is the knee for $x_{0} \xcP x_{1} \xcP x_{2},$ $x_{0} \xcP
x_{2},$ too, and $x_{2}$ its foot.
 \xEJ
 \xEj
 \xDH $x_{2}$
 \xEh
 \xDH
as $x_{0}-$ should be impossible, append $ \xeA $ to $x_{2},$ because of
the new arrow $x_{0} \xcP x_{2},$
and $x_{2}+$ is impossible,

so we have
$x_{0} \xcP x_{1} \xcP x_{2} \xcP x_{3},$ $x_{0} \xcP x_{2}$
and the new arrows $ \xcP y_{j}$
 \xDH
as $x_{1}+$ should be impossible,
add $x_{1} \xcP x_{3}$ because of $x_{2} \xcP x_{3},$ and $x_{1} \xcP
y_{2}$ because of $x_{2} \xcP y_{2},$

so we have
$x_{0} \xcP x_{1} \xcP x_{2} \xcP x_{3},$ $x_{0} \xcP x_{2},$ $x_{1} \xcP
x_{3}$
and the new arrows $ \xcP y_{j}$
 \xEI
 \xDH So, for $x_{1} \xcP x_{2} \xcP x_{3},$ $x_{1} \xcP x_{3}$ $x_{1}$ is
the head, $x_{2}$ the knee, $x_{3}$ the foot.
 \xDH But, also, by $x_{0} \xcP x_{2} \xcP x_{3},$ $x_{0} \xcP x_{3},$
here $x_{0}$ is the head and $x_{2}$ the knee,
whereas in (2.2.1), $x_{0}$ was the head, and $x_{2}$ the foot.
 \xEJ
 \xDH
as $x_{0}+$ should be impossible,
add $x_{0} \xcP x_{3}$ because of $x_{1} \xcP x_{3},$ and $x_{0} \xcP
y_{2}$ because of $x_{1} \xcP y_{2},$

so we have
$x_{0} \xcP x_{1} \xcP x_{2} \xcP x_{3},$ $x_{0} \xcP x_{2},$ $x_{1} \xcP
x_{3},$ $x_{0} \xcP x_{3},$
and the new arrows $ \xcP y_{j}$
 \xEI
 \xDH So, for $x_{0} \xcP x_{1} \xcP x_{3},$ $x_{0} \xcP x_{3},$ $x_{0}$
is the head, $x_{1}$ the knee, $x_{3}$ the foot.
Etc.
 \xEJ

 \xEj
 \xDH $x_{3}$

 \xEh
 \xDH
as $x_{0}-$ should be impossible, append $ \xeA $ to $x_{3},$ because of
the new arrow $x_{0} \xcP x_{3},$
and $x_{3}+$ is impossible,

so we have
$x_{0} \xcP x_{1} \xcP x_{2} \xcP x_{3} \xcP x_{4},$ $x_{0} \xcP x_{2},$
$x_{1} \xcP x_{3},$ $x_{0} \xcP x_{3}$
and the new arrows $ \xcP y_{j}$

 \xDH
as $x_{2}+$ should be impossible,
add $x_{2} \xcP x_{4}$ because of $x_{3} \xcP x_{4},$ and $x_{2} \xcP
y_{3}$ because of $x_{3} \xcP y_{3},$

so we have
$x_{0} \xcP x_{1} \xcP x_{2} \xcP x_{3} \xcP x_{4},$ $x_{0} \xcP x_{2},$
$x_{1} \xcP x_{3},$ $x_{0} \xcP x_{3},$ $x_{2} \xcP x_{4}$
and the new arrows $ \xcP y_{j}$
 \xDH
as $x_{1}+$ should be impossible,
add $x_{1} \xcP x_{4}$ because of $x_{2} \xcP x_{4},$ and $x_{1} \xcP
y_{3}$ because of $x_{2} \xcP y_{3},$

so we have
$x_{0} \xcP x_{1} \xcP x_{2} \xcP x_{3} \xcP x_{4},$ $x_{0} \xcP x_{2},$
$x_{1} \xcP x_{3},$ $x_{0} \xcP x_{3},$ $x_{2} \xcP x_{4},$ $x_{1} \xcP
x_{4}$
and the new arrows $ \xcP y_{j}$
 \xDH
as $x_{0}+$ should be impossible,
add $x_{0} \xcP x_{4}$ because of $x_{1} \xcP x_{4},$ and $x_{0} \xcP
y_{3}$ because of $x_{1} \xcP y_{3},$

so we have
$x_{0} \xcP x_{1} \xcP x_{2} \xcP x_{3} \xcP x_{4},$ $x_{0} \xcP x_{2},$
$x_{1} \xcP x_{3},$ $x_{0} \xcP x_{3},$ $x_{2} \xcP x_{4},$ $x_{1} \xcP
x_{4},$ $x_{0} \xcP x_{4}$
and the new arrows $ \xcP y_{j}$
 \xEj
 \xDH
so we have a new arrow $x_{0} \xcP x_{4},$ and apply again $x_{0}-$, etc.

 \xEj

 \xDH
We see that the roles in the back of the saw blade change. $x_{1}$ begins
as a
knee, $x_{2}$ as a foot, later $x_{1}$ is a head, $x_{2}$ a knee, $x_{3}$
is a foot, etc.
Whereas it is simple to treat feet
(see Diagram \ref{Diagram Sawblade-4} (page \pageref{Diagram Sawblade-4})),
treating knees is more complicated, see also
Section \ref{Section NotWork} (page \pageref{Section NotWork}).
As the same nodes change roles,
we cannot have a ``pure'' construction according to our analysis
(separate treatment for knees and feet).
It seems difficult to separate the roles in a more complicated
construction, e.g. by working with mixed $ \xco \xcu $-formulas, see
in particular cases (2.2.1) and (2.3.2) where $x_{0}$ is the head in both
cases,
$x_{2}$ the foot in one, the knee in another.
 \xDH
Abstractly, we add complications in the steps (2.i), and repair them
in the steps (2.i.j), so, in the limit, all damage done will be
repaired. This is different from procrastination, where the same
problem is just pushed to the future.

 \xEj

Thus, we have infinite branching for this construction.

$ \xcz.$
\\[3ex]
%
%

The following remark shows that the construction has
contradictory truth values recursively often.

\br

$\hspace{0.01em}$


\label{Remark Contradictory}

We have a descending sequence of contradictory $x_{i}$ - i.e. without
attributable
truth value - recursively often.

There are at least two arguments to show this:
 \xEI
 \xDH
From the outside: otherwise, we could fill in truth values from the
bottom.
 \xDH
From the inside: we have to construct ever deeper $ \xeA $'s for $x_{0}$
to prevent
escape paths, see steps (2.i.1) in
Fact \ref{Fact Infin-Branching} (page \pageref{Fact Infin-Branching}).
 \xEJ
\clearpage
\subsection{
Simplifications of the Saw Blade Construction
}

\label{Section Simplifications}

\er

We show here that it is not necessary to make the $y_{ \xbs,i}$
contradictory
in a recursive construction, as in
Construction 
\ref{Construction Saw-Blade} (page 
\pageref{Construction Saw-Blade}).
It suffices to prevent them to be true.

We discuss three, much simplified, Saw Blade constructions.

Thus, we fully use here the conceptual difference of $x_{1}$ and $x_{2},$
as alluded to at the beginning of
Section \ref{Section Saw-Blades} (page \pageref{Section Saw-Blades}).

Note, however, that the back of each saw blade ``hides'' a Yablo
construction. The separate treatment of the teeth illustrates the
conceptual difference, but it cannot escape blurring it again in
the back of the blade.

See also Example 
\ref{Example Simple-Cells} (page 
\pageref{Example Simple-Cells})  for examples of simple
contradictions.

\bcs

$\hspace{0.01em}$


\label{Construction Simple-1}

 \xEh

 \xDH

Take ONE saw blade $ \xbs,$ and attach (after closing under transitivity)
at all $y_{ \xbs,i}$ a SINGLE Yablo Cell $y_{ \xbs,i} \xcP u_{ \xbs,i}
\xcP v_{ \xbs,i},$ $y_{ \xbs,i} \xcP v_{ \xbs,i}.$
We call this the decoration, it is not involved in closure under
transitivity.

 \xEh

 \xDH

Any node $z$ in the saw blade (back or tooth) cannot have $z+,$ this leads
to a contradiction:

If $z=x_{i}$ (in the back):

Let $x_{i}+:$
Take $x_{i} \xcP x_{i+1}$ (any $x_{j},$ $i<j$ would do), if $x_{i+1} \xcP
r,$ then by transitivity,
$x_{i} \xcP r,$ so we have a contradiction.

If $z=y_{i}$ (a tooth):

$y_{i}+$ is contradictory by the ``decoration'' appended to $y_{i}.$

 \xDH

Any $x_{i}-$ $(x_{i}$ in the back, as a matter of fact, $x_{0}-$ would
suffice)
is impossible:

Consider any $x_{i} \xcP r,$ then $r+$ is impossible, as we just saw.

Note: there are no arrows from the back of the blade to the decoration.

 \xEj

 \xDH

We can simplify even further. The only thing we need about the $y_{i}$ is
that
they cannot be $+.$ Instead of decorating them with a Yablo Cell, any
contradiction will do, the simplest one is $y_{i}=y'_{i} \xcu \xCN
y'_{i}.$ Even
just one $y' $ s.t. $y_{i}=y' \xcu \xCN y' $ for all $y_{i}$ would do. (Or
a constant
FALSE.)

See Diagram \ref{Diagram Sawblade-4} (page \pageref{Diagram Sawblade-4}).

Formally, we set

$x_{0}:= \xcU \{ \xCN x_{i}:i>0\} \xcu \xcU \{ \xCN y_{i}:i \xcg 0\},$

for $j>0:$

$x_{j}:= \xcU \{ \xCN x_{i}:i>j\} \xcu \xcU \{ \xCN y_{i}:i \xcg j-1\},$

and

$y_{j}:=y'_{j} \xcu \xCN y'_{j}.$

 \xDH

In a further step, we see that the $y_{i}$ (and thus the $y'_{i})$ need
not
be different from each other, one $y$ and one $y' $ suffice.

Thus, we set $x_{j}:= \xcU \{ \xCN x_{i}:i>j\} \xcu \xCN y,$ $y:=y' \xcu
\xCN y'.$

(Intuitively, the cells are arranged in a circle, with $y$ at the center,
and
$y' $ ``sticking out''. We might call this a ``curled saw blade''.

 \xDH
When we throw away the $y_{j}$ altogether, we have Yablo's construction.
this works, as we have the essential part in the $x_{i}$'s, and used the
$y_{j}' s$
only as a sort of scaffolding.

 \xEj

\ecs

\br

$\hspace{0.01em}$


\label{Remark Simple-2}

It seems difficult to conceptually simplify even further, as
Fact \ref{Fact Infin-Branching} (page \pageref{Fact Infin-Branching})
shows basically the need for the
construction of the single Saw Blades.
We have to do something about the teeth, and above
Construction 
\ref{Construction Simple-1} (page 
\pageref{Construction Simple-1}), in particular cases (2) and (3) are
simple solutions.

\er

The construction is robust, as the following easy remarks show:

 \xEh
 \xDH
Suppose we have ``gaps'' in the closure under transitivity, so, e.g. not
all $x_{0} \xcP x_{i}$ exist, they always exist only for $i>n.$ (And all
other $x_{k} \xcP x_{l}$
exist.) Then $x_{0}$ is still contradictory. Proof: Suppose $x_{0}+,$ then
we have the contradiction $x_{0} \xcP x_{n} \xcP x_{n+1}$ and $x_{0} \xcP
x_{n+1}.$ Suppose $x_{0}-,$ let $x_{0} \xcP x_{i}.$
As $x_{i}$ is unaffected, $x_{i}+$ is impossible.

 \xDH
Not only $x_{0}$ has gaps, but other $x_{i},$ too. Let again $x_{n}$ be an
upper bound for
the gaps. As above, we see that $x_{0}+,$ but also all $x_{i}+$ are
impossible.
If $x_{0} \xcP x_{i},$ as $x_{i}+$ is impossible, $x_{0}-$ is impossible.

 \xDH
$x_{0}$ has unboundedly often gaps, the other $x_{i}$ are not affected.
Thus, for $i \xEd 0,$ $x_{i}+$ and $x_{i}-$ are impossible. Thus, $x_{0}-$
is impossible, as
all $x_{i}+$ are, and $x_{0}+$ is, as all $x_{i}-$ are.

 \xEj

Finally,
instead of showing that two paths $ \xbp:x \Xl y$ and $ \xbs:x \Xl y$
are contradictory,
we may also show that all continuations $ \xbr:y \Xl z,$ $ \xbr':y \Xl
z' $ have
contradictions, $ \xbt:x \Xl z$ and $ \xbt':x \Xl z' $ to $ \xbs \xbr $
and $ \xbs \xbr' $ respectively.
Consider the following situation: $x_{0} \xcP x_{1} \xcP x_{2} \xcP
x_{3},$ $x_{0} \xcP x_{3},$ $x_{1} \xcP x_{3},$ but
$x_{0} \xcP x_{2}$ is missing, so we have no contradiction with $x_{0}
\xcP x_{1} \xcP x_{2}.$
$x_{0} \xcP x_{1} \xcP x_{2} \xcP x_{3}$ and $x_{1} \xcP x_{3}$ form no
contradiction, as both legths are odd.
We have a contradiction $x_{0} \xcP x_{1} \xcP x_{3}$ and $x_{0} \xcP
x_{3},$ but, as $x_{1}$ is an ``$ \xco $'',
we have to make sure that all continuations from $x_{1}$ have a
contradiction
with suitable $x_{0} \xcP x_{i},$ even without $x_{0} \xcP x_{2},$ they
have to meet later on.

\clearpage

\begin{diagram}

\label{Diagram Sawblade-4}
\index{Diagram Sawblade-4}

\unitlength0.6mm
\begin{picture}(150,190)(0,0)

\put(0,182){{\rm\bf Diagram Simplified Saw Blade }}
\put(0,175){Start of the saw blade before closing}
\put(0,168){the blade (without ``decoration'') under transitivity }



\put(20,160){$x_0$}
\put(20,140){$y_0$}
\put(20,130){$y'_0$}
\put(26,161){\line(1,0){12}}
\put(32,160){\line(0,1){2}}
\put(22,143){\line(0,1){14}}
\put(21,134){\line(0,1){4}}
\put(23,134){\line(0,1){4}}
\put(22,136){\line(1,0){2}}
\put(21,150){\line(1,0){2}}
\put(24,143){\line(1,1){16}}
\put(31,151.5){\line(1,-1){1.5}}


\put(40,160){$x_1$}
\put(40,140){$y_1$}
\put(40,130){$y'_1$}
\put(46,161){\line(1,0){12}}
\put(52,160){\line(0,1){2}}
\put(42,143){\line(0,1){14}}
\put(41,134){\line(0,1){4}}
\put(43,134){\line(0,1){4}}
\put(42,136){\line(1,0){2}}
\put(41,150){\line(1,0){2}}
\put(44,143){\line(1,1){16}}
\put(51,151.5){\line(1,-1){1.5}}


\put(60,160){$x_2$}
\put(60,140){$y_2$}
\put(60,130){$y'_2$}
\put(66,161){\line(1,0){12}}
\put(72,160){\line(0,1){2}}
\put(62,143){\line(0,1){14}}
\put(61,134){\line(0,1){4}}
\put(63,134){\line(0,1){4}}
\put(62,136){\line(1,0){2}}
\put(61,150){\line(1,0){2}}
\put(64,143){\line(1,1){16}}
\put(71,151.5){\line(1,-1){1.5}}


\put(80,160){$x_3$}
\put(80,140){$y_3$}
\put(80,130){$y'_3$}
\put(86,161){\line(1,0){12}}
\put(92,160){\line(0,1){2}}
\put(82,143){\line(0,1){14}}
\put(81,134){\line(0,1){4}}
\put(83,134){\line(0,1){4}}
\put(82,136){\line(1,0){2}}
\put(81,150){\line(1,0){2}}
\put(84,143){\line(1,1){16}}
\put(91,151.5){\line(1,-1){1.5}}


\put(100,160){$x_4$}
\put(100,140){$y_4$}
\put(100,130){$y'_4$}
\put(106,161){\line(1,0){12}}
\put(112,160){\line(0,1){2}}
\put(102,143){\line(0,1){14}}
\put(101,134){\line(0,1){4}}
\put(103,134){\line(0,1){4}}
\put(102,136){\line(1,0){2}}
\put(101,150){\line(1,0){2}}
\put(104,143){\line(1,1){16}}
\put(111,151.5){\line(1,-1){1.5}}


\put(120,160){$x_5$}
\put(120,140){$y_5$}
\put(120,130){$y'_5$}
\put(126,161){\line(1,0){12}}
\put(132,160){\line(0,1){2}}
\put(122,143){\line(0,1){14}}
\put(121,134){\line(0,1){4}}
\put(123,134){\line(0,1){4}}
\put(122,136){\line(1,0){2}}
\put(121,150){\line(1,0){2}}
\put(124,143){\line(1,1){16}}
\put(131,151.5){\line(1,-1){1.5}}


\put(140,160){$x_6$}
\put(140,140){$y_6$}
\put(140,130){$y'_6$}
\multiput(146,161)(2,0){5}{\circle*{0.2}}
\put(142,143){\line(0,1){14}}
\put(141,134){\line(0,1){4}}
\put(143,134){\line(0,1){4}}
\put(142,136){\line(1,0){2}}
\put(141,150){\line(1,0){2}}
\multiput(144,143)(1.5,1.5){5}{\circle*{0.2}}

\end{picture}

\end{diagram}

\vspace{10mm}

Read $y_{0} \xch_{ \xCL }y_{0}',$ $y_{0}=y_{0}' \xcu \xCN y_{0}',$ etc.
\clearpage
\clearpage
\subsection{
Paths Instead of Arrows
}

This section is not very systematic, and formulates some considerations
for more complicated situations.

Recall Section \ref{Section More} (page \pageref{Section More}),
discussing additional branchings.

Most of the basic reasoning here is analogous to the considerations in
Example \ref{Example Simple-Cells} (page \pageref{Example Simple-Cells}).
\subsubsection{
A Generalization: Paths Instead of Arrows from $x_0$ to $y_0$
}

We now consider paths instead of arrows from $x_{0}$ to $y_{0}.$
%
%

 \xEh

 \xDH

\be

$\hspace{0.01em}$


\label{Example Paths}

See also Example 
\ref{Example Simple-Cells} (page 
\pageref{Example Simple-Cells}).

So we have contradictory paths $ \xbs,$ $ \xbt $ from $x_{0}$ to $y_{0}.$
Suppose $ \xbs $ is the positive path (corresponding to $x_{0} \xcP x_{1}
\xcP y_{0}),$ and
$ \xbt $ the negative one.

 \xEh
 \xDH
Suppose
$ \xbs $ does not branch. Suppose now $x_{0}-,$ so we have $ \xco $ at
$x_{0},$ $y_{0}$ will be - and
$ \xco,$ too, so stacking such contradictions does not help, we have an
escape path downwards.
 \xDH
Suppose $ \xbs $ branches, say at $z,$ $ \xbs:x_{0}$  \Xl  $z$  \Xl
$y_{0}.$

Let $x_{0}+$
 \xEh
 \xDH
if $z+,$ too: We have the same situation as in Case (1) (constructed an
escape
route for the case $x_{0}-),$ so this is not interesting.
 \xDH
if $z-:$ This situation is similar to the Yablo construction, and
offers essentially nothing new.

The same considerations (infinity of branching, values $ \xCL)$ as for
the original construction apply.

 \xEj
 \xEj

\ee

 \xDH

We generalize now
Fact \ref{Fact Infin-Branching} (page \pageref{Fact Infin-Branching})
to paths.

See also Example 
\ref{Example Simple-Cells} (page 
\pageref{Example Simple-Cells}).

 \xEh
 \xDH The contradiction at $x_{0}$ cannot be $ \xbs:x_{0} \Xl y_{0},$ $
\ol{ \xbs }:x_{0} \Xl y_{0}$ without
branching by the argument against $x_{0} \xch_{ \xCL }y_{0}$ in
Example \ref{Example Simple-Cells} (page \pageref{Example Simple-Cells}).

 \xDH Assume by preprocessing that the first branching, say in $ \xbs,$
say in $x_{1},$
is equivalent
to an $ \xcO $ for $x_{0}+$ (i.e. $x_{1}$ is an $ \xcO,$ and $ \xbs $ up
to $x_{1}$ is positive, or
$x_{1}$ is an $ \xcU,$ and $ \xbs $ up to $x_{1}$ is negative)
(otherwise contract the branchings to $x_{0}).$

 \xDH As we have $ \xcO $ at $x_{1},$ every path leaving $x_{1}$ needs a
contradiction
with some other path from $x_{0}$ (and the latter does not go through
$x_{1},$
as a contradiction through $x_{1}$ would be invisible under the $ \xcO $
at $x_{1}),$
no matter what the choices in other $ \xcO $ are.

 \xDH For $x_{0}-,$ every path from $x_{0}$ has to lead to a
contradiction, see again
Fact \ref{Fact Infin-Branching} (page \pageref{Fact Infin-Branching}).

 \xDH
Apply the iterated reasoning in
Fact \ref{Fact Infin-Branching} (page \pageref{Fact Infin-Branching}).

 \xEj
%
%
%
%
%

 \xEj
\subsubsection{
Infinite Depth for Paths
}

\label{Section Inf-Depth}

We know that we need infinite depth, otherwise, we could fill in truth
values from the bottom.

We give here a constructive argument.

See Section 
\ref{Section InfBranch} (page 
\pageref{Section InfBranch})  for details of the saw blade (and
Yablo)
construction. We go here into more detail.

Consider the construction $x_{0} \xcP x_{n} \xcP x_{n+1},$ $x_{0} \xcP
x_{n+1}.$ If $x_{0}-,$ $x_{0} \xcP x_{n+1}$ has to
lead to a contradiction (recall, then $x_{0} \xco),$ so we cannot stop at
$x_{n+1},$
say we continue to $x_{m},$ then we have $x_{0} \xcP x_{m},$ etc.

The case where the contradiction to $x_{0} \xcP x_{n} \xcP x_{n+1}$ is not
a simple arrow
$x_{0} \xcP x_{n+1},$ but a path $ \xbs:x_{0} \Xl x_{n+1}$ is more
complicated, as
there might be a contradiction already on $ \xbs.$ We now show that this
does
not work.

Say $ \xbs $ has the form $x_{0} \xcP z \xcP w \xcP z' \xcP x_{n+1},$ with
detour $z \xcP z',$
$x_{0}+,$ $z-,$ with an OR at $z.$

So we have a contradiction on $ \xbs,$ but not on the detour.
But $x_{0} \xcP x_{n} \xcP x_{n+1}$ has to be contradicted in all cases of
the OR at $z,$
so this will not work.

See Diagram \ref{Diagram InfPaths} (page \pageref{Diagram InfPaths}).

There is no easy way out. Suppose, by some construction between $z' $ and
$x_{n+1}$
we could have a contradiction at $x_{n+1},$ i.e. $x_{n+1}-.$ Then,
irrespective
of the value of $z',$ we would have $x_{n+1}-.$ But then we could append
$x_{n+1} \xcp TRUE$ to $x_{n+1},$ and would have shown that $z' $ cannot
have a truth
value by a finite construction. We know that this is impossible.

Of course, the same argument applies if we try to contradict $x_{0} \xcP
x_{n} \xcP x_{n+1}$
further down the road at some $s,$ $s' $ etc. $x_{0} \xcP x_{n} \xcP
x_{n+1} \xcP  \Xl.s$ etc.

(By transitivity, we may take shortcuts, but never have only finitely
many steps.)

\clearpage

\begin{diagram}

\label{Diagram InfPaths}
\index{Diagram InfPaths}

\unitlength1.0mm
\begin{picture}(150,100)(0,0)

\put(0,80){{\rm\bf Infinite paths }}

\put(10,30){\circle*{1}}
\put(30,30){\circle*{1}}
\put(50,30){\circle*{1}}

\put(10,20){\circle*{1}}
\put(30,20){\circle*{1}}
\put(50,20){\circle*{1}}

\put(12,30){\line(1,0){16}}
\put(32,30){\line(1,0){16}}
\put(20,31){\line(0,-1){2}}
\put(40,31){\line(0,-1){2}}

\put(12,20){\line(1,0){16}}
\put(32,20){\line(1,0){16}}
\put(20,21){\line(0,-1){2}}
\put(40,21){\line(0,-1){2}}

\put(10,28){\line(0,-1){7}}
\put(50,28){\line(0,-1){7}}
\put(9,24){\line(1,0){2}}
\put(49,24){\line(1,0){2}}

\put(10,18){\line(0,-1){8}}
\put(50,18){\line(0,-1){8}}
\put(10,10){\line(1,0){40}}
\put(30,11){\line(0,-1){2}}

\put(8,33){$x_0$}
\put(28,33){$x_n$}
\put(48,33){$x_{n+1}$}

\put(6,19){$z$}
\put(28,22){$w$}
\put(52,19){$z'$}

\end{picture}

\end{diagram}

\bfa

$\hspace{0.01em}$


\label{Fact More-Infin-2}

We have infinite branching in Example 
\ref{Example Paths} (page 
\pageref{Example Paths}), too.

\efa

\subparagraph{
Proof
}

$\hspace{0.01em}$


Recall the argument in
Fact \ref{Fact Infin-Branching} (page \pageref{Fact Infin-Branching}).

 \xEh
 \xDH
(Case $x_{0}-$ in
Fact \ref{Fact Infin-Branching} (page \pageref{Fact Infin-Branching}).)

We need an $ \xeA $ at $z' $ in $ \xbs':x \Xl z \xcp z' $ because of
$x_{0}-.$
 \xDH
(Case $x_{i}+$ in
Fact \ref{Fact Infin-Branching} (page \pageref{Fact Infin-Branching}).)

 \xEh

 \xDH

We may have a contradiction to $x \xcP z \xcp z' $ by $x \xcp z'.$

This works fine for $x+,$ but not for $x-,$ as we have then an arrow $x
\xcp z',$
which does not lead to $ \xeA $ (as $z' -$), so we would have an escape
path
from $ \xco $ to $ \xco.$

 \xDH

As in
Fact \ref{Fact Infin-Branching} (page \pageref{Fact Infin-Branching}),
we have to introduce arrows
$x \xcP z'' $ und $x \xcP z'''.$

We now have new arrows originating in $x,$ we have to re-consider the case
$x_{0}-$ etc., as in
Fact \ref{Fact Infin-Branching} (page \pageref{Fact Infin-Branching}).

 \xEj

 \xEj

$ \xcz $
\\[3ex]

$ \xCO $

$ \xCO $

$ \xCO $

$ \xCO $

$ \xCO $

$ \xCO $
\clearpage
\subsection{
Nested Contradictory Cells
}

Diagram \ref{Diagram Nest} (page \pageref{Diagram Nest})
illustrates combinations of contradictory cells.

We have a contradictory cell $ \xBc b,c,d \xBe $ in the left hand diagram,
and may add new lines,
forming additional contradictory cells, like
the line $c-e$ in the central diagram, forming the cell $ \xBc c,d,e \xBe,$ or
the line $a-c$ in the right hand diagram, forming the cell $ \xBc a,b,c \xBe,$
(These two possibilities are equivalent.)

There is a mutitude of possibilities, e.g. $a-d$, $b$-e, etc.,
we have not investigated,
but we think they might not be very interesting - unless they
form a nested construction like in or similar to Yablo's construction.

See, however, the discussion in
Section \ref{Section Inf-Depth} (page \pageref{Section Inf-Depth})  and
Diagram \ref{Diagram InfPaths} (page \pageref{Diagram InfPaths}).
%
%

$ \xCO $

$ \xCO $

$ \xCO $

$ \xCO $

\begin{diagram}

\label{Diagram Nest}
\index{Diagram Nest}

\unitlength0.7mm
\begin{picture}(150,180)(0,0)

\put(0,175){{\rm\bf Diagram Nested Cells }}

\put(10,165){\circle*{1}}
\put(10,145){\circle*{1}}
\put(30,125){\circle*{1}}
\put(10,105){\circle*{1}}
\put(10,85){\circle*{1}}

\put(11,164){a}
\put(11,144){b}
\put(31,124){c}
\put(11,104){d}
\put(11,84){e}

\put(10,87){\line(0,1){16}}
\put(10,107){\line(0,1){36}}
\put(11,107){\line(1,1){16}}
\put(29,127){\line(-1,1){16}}
\put(10,147){\line(0,1){16}}


\put(50,165){\circle*{1}}
\put(50,145){\circle*{1}}
\put(70,125){\circle*{1}}
\put(50,105){\circle*{1}}
\put(50,85){\circle*{1}}

\put(51,164){a}
\put(51,144){b}
\put(71,124){c}
\put(51,104){d}
\put(51,84){e}

\put(50,87){\line(0,1){16}}
\put(50,107){\line(0,1){36}}
\put(51,107){\line(1,1){16}}
\put(69,127){\line(-1,1){16}}
\put(50,147){\line(0,1){16}}

\multiput(51,87)(1,2){20}{\circle*{0.2}}

\put(90,165){\circle*{1}}
\put(90,145){\circle*{1}}
\put(110,125){\circle*{1}}
\put(90,105){\circle*{1}}
\put(90,85){\circle*{1}}

\put(91,164){a}
\put(91,144){b}
\put(111,124){c}
\put(91,104){d}
\put(91,84){e}

\put(90,87){\line(0,1){16}}
\put(90,107){\line(0,1){36}}
\put(91,107){\line(1,1){16}}
\put(109,127){\line(-1,1){16}}
\put(90,147){\line(0,1){16}}

\multiput(109,127)(-1,2){20}{\circle*{0.2}}

\end{picture}

\end{diagram}

\clearpage

$ \xCO $

$ \xCO $


\clearpage



\begin{thebibliography}{xxxxxx}

\addcontentsline{toc}{section}{References}


\bibitem[AA11]{AA11}
M. Anderson, S. L. Anderson eds., ``Machine Ethics'',
Cambridge Univ. Press, 2011

\bibitem[AGM85]{AGM85}
C. Alchourron, P. Gardenfors, D. Makinson,
``On the logic of theory change: partial meet contraction and
revision functions'', Journal of Symbolic Logic, Vol. 50,
pp. 510--530, 1985

\bibitem[AIZ16]{AIZ16}
A. Azulay, E. Itskovits, A. Zaslaver,
``The $C.$ elegans connectome consists of homogenous circuits with
defined functional roles'',
PLoS Comput Biol $12(9),$ 2016

\bibitem[Auf17]{Auf17}
``Aufmerksamkeit'',
www.spektrum.de/lexikon/neurowissenschaft/
aufmerksamkeit/1072, 2017

\bibitem[Avr14]{Avr14}
A. Avron, ``What is relevance logic?'', Annals of Pure and Applied
Logic, 165 (2014) 26-48

\bibitem[BB11]{BB11}
A. G. Burgess, J. P. Burgess, ``Truth'', Princeton University Press,
Princeton, 2011

\bibitem[BBHLL10]{BBHLL10}
J. Ben$-Naim, J$-F. Bonnefon et al., ``Computer-mediated trust in
self-interested expert recommendations'',
AI and society 25 (4): 413-422, 2010

\bibitem[BP12]{BP12}
J. Ben-Naim, H. Prade, ``Evaluating trustworthiness from past
performances: interval-based approaches'',
Annals of Math. and AI, Vol. 64, 2-3, pp 247-268, 2012

\bibitem[BS17]{BS17}
T. Beringer, T. Schindler,
``A Graph-Theoretical Analysis of the Semantic Paradoxes'',
The Bulletin of Symbolic Logic, Vol. 23, No. 4, Dec. 2017

\bibitem[CCOM08]{CCOM08}
R. Cabeza, E. Ciaramelli, I. R. Olson, M. Moscovitch,
``Parietal cortex and episodic memory: An attentional account'',
Nat. Rev. Neurosci. 2008 Aug; $9(8):613-625$

\bibitem[Chu07]{Chu07}
P. M. Churchland, ``Neurophilosophy at Work'',
Cambridge University Press, 2007

\bibitem[Chu86]{Chu86}
P. S. Churchland, ``Neurophilosophy'',
MIT Press, Cambrige, Mass., 1986

\bibitem[Chu89]{Chu89}
P. M. Churchland, ``A Neurocomputational Perspective'', MIT Press,
1989

\bibitem[DR15]{DR15}
J. P. Delgrande, B. Renne, ``The logic of qualitative probability'',
IJCAI 2015, pp. 2904-2910

\bibitem[Dev65]{Dev65}
P. Devlin, ``The Enforcement of Morals'', Oxford, 1965

\bibitem[Dun95]{Dun95}
P. M. Dung, ``On the acceptability of arguments and its
fundamental role in nonmonotonic reasoning, logic programming and
$n$-person games'',
Artificial Intelligence 77 (1995), pp. 321--357

\bibitem[Dwo82]{Dwo82}
R. Dworkin, ``'Natural' Law Revisited'', University of Florida
Law Review, vol. 34, no. 2, pp. 165-188, 1982

\bibitem[Dwo86]{Dwo86}
R. Dworkin, ``Law's Empire'', Cambridge, USA, 1986

\bibitem[Ede04]{Ede04}
Gerald $M.$ Edelman, ``Wider than the sky'', Yale University Press, New
Haven 2004,
(German edition ``Das Licht des Geistes'', Rowohlt, 2007)

\bibitem[Ede89]{Ede89}
Gerald $M.$ Edelman, ``The remembered present'', Basic Books, New York,
1989

\bibitem[GLP17]{GLP17}
``Gehirn und Lernen - Plastizitaet'',
www.gehirnlernen.de/gehirn/plastizitaet, 2017

\bibitem[GR17]{GR17}
D. Gabbay, G. Rozenberg, et al.,
``Temporal aspects of many lives'', Paper 588

\bibitem[GS08f]{GS08f}
D. Gabbay, K. Schlechta, ``Logical tools for handling change in
agent-based systems''
Springer, Berlin, 2009, ISBN 978-3-642-04406-9.

\bibitem[GS10]{GS10}
D. Gabbay, K. Schlechta,
``Conditionals and modularity in general logics'',
Springer, Heidelberg, August 2011,
ISBN 978-3-642-19067-4,

\bibitem[GS16]{GS16}
D. Gabbay, K. Schlechta,
``A New Perspective on Nonmonotonic Logics'',
Springer, Heidelberg, Nov. 2016,
ISBN 978-3-319-46815-0,

\bibitem[Geg11]{Geg11}
K. R. Gegenfurtner,
``Gehirn und Wahrnehmung'', Fischer, Frankfurt 2011

\bibitem[HM17]{HM17}
D. Hassabis, E. A. Maguire,
``Deconstructing episodic memory with construction'',
Trends in Cognitive Sciences Vol.11 No.7

\bibitem[Haa14]{Haa14}
S. Haack, ``Evidence matters'', Cambridge University Press, 2014

\bibitem[Hab01]{Hab01}
J. Habermas, ``On the pragmatics of social interaction'',
MIT Press, 2001

\bibitem[Hab03]{Hab03}
J. Habermas, ``Truth and justification'',
MIT Press, 2003

\bibitem[Hab73]{Hab73}
J. Habermas, ``Wahrheitstheorien'', in
Fahrenbach (ed.), ``Wirklichkeit und Reflexion'', Pfuellingen, 1973

\bibitem[Hab90]{Hab90}
J. Habermas, ``Moral consciousness and communicative action'',
MIT Press, 1990

\bibitem[Hab96]{Hab96}
J. Habermas, ``Between facts and norms: contributions to a
discourse theory of law and democracy'', MIT Press, 1996

\bibitem[Han69]{Han69}
B. Hansson, ``An analysis of some deontic logics'', Nous 3, 373--398.
Reprinted in R. Hilpinen, ed. ``Deontic Logic: Introductory and Systematic
Readings'', Reidel, pp. 121--147, Dordrecht 1971

\bibitem[Heb49]{Heb49}
D. Hebb, ``The organization of behavior'', New York, Wiley, 1949

\bibitem[Hem35]{Hem35}
C. G. Hempel, ``On the logical positivists' theory of truth'',
Analysis, 2:49-59, 1935

\bibitem[IEP16]{IEP16}
``Philosophy of Law'', Internet Encyclopedia of Philosophy, 2016

\bibitem[KLM90]{KLM90}
S. Kraus, D. Lehmann, M. Magidor, ``Nonmonotonic reasoning, preferential
models and cumulative logics'', Artificial Intelligence, 44 (1--2),
pp. 167--207, July 1990.

\bibitem[KPP07]{KPP07}
B. Konikov, G. Petkov, N. Petrova,
``Context-sensitivy of human memory: Episode connectivity and its
influence on memory reconstruction'',
Context 2007: 317-329

\bibitem[Key21]{Key21}
J. M. Keynes, ``A treatise on probability'', London, 1921

\bibitem[Kri75]{Kri75}
S. Kripke, ``Outline of a Theory of Truth'',
The Journal of Philosophy, Vol. 72, No. 19, 1975, pp. 690-716

\bibitem[LMS01]{LMS01}
D. Lehmann, M. Magidor, K. Schlechta, ``Distance semantics for belief
revision'', Journal of Symbolic Logic, Vol. 66, No. 1,
pp. 295--317, March 2001

\bibitem[Leh96]{Leh96}
D. Lehmann, ``Generalized qualitative probability: Savage revisited'',
Proceedings $UAI' 96,$ pp. 381-388, Portland, Or, Aug. 1, 1996

\bibitem[Lew73]{Lew73}
D. Lewis, ``Counterfactuals'', Blackwell, Oxford, 1973

\bibitem[MP13]{MP13}
S. Modgil, H. Prakken, ``A general account of argumentation with
preferences'', Artificial Intelligence 195 (2013) 361-397

\bibitem[Mak19]{Mak19}
D. Makinson, ``Relevance via decomposition: a project, some results,
an open question'', Australasian Journal of Logic 14:3 2017, see also
``Sets, Logic and Maths for Computing'' (third edition), Springer 2020

\bibitem[Mil06]{Mil06}
J. S. Mill, ``On Liberty'', New York, 1906

\bibitem[Neu83]{Neu83}
O. Neurath, ``Philosophical papers 1913-46'', R. S. Cohen and M. Neurath
(eds.), Dordrecht and Boston, D. Reidel, 1983

\bibitem[OL09]{OL09}
M. Okun, I. Lampl, ``Balance of excitation and inhibition'',
Scholarpedia, $4(8):7467,$ 2009

\bibitem[Pul13]{Pul13}
F. Pulvermueller, ``How neurons make meaning: Brain mechanisms for
embodied and abstract-symbolic semantics'',
Trends in Cognitive Sciences, 17 (9), 458-470, 2013

\bibitem[RRM13]{RRM13}
L. Rabern, B. Rabern, M. Macauley,
``Dangerous reference graphs and semantic paradoxes'', in:
J. Philos. Logic (2013) 42:727-765

\bibitem[Rau21]{Rau21}
J. Rauch, ``The Constitution of Knowledge. A Defense of Truth'',
Brookings Press, Washington, 2021

\bibitem[Rot96]{Rot96}
G. Roth, ``Das Gehirn und seine Wirklichkeit'', Suhrkamp STW 1275,
Frankfurt 1996

\bibitem[Rus07]{Rus07}
B. Russell, ``On the nature of truth'', Proceedings of the
Aristotelian Society, 7:228-49, 1907

\bibitem[SEP13]{SEP13}
``Analogy and analogical reasoning'', Fall 13 edition,
Stanford Encyclopedia of Philosophy, 2013

\bibitem[SEP19c]{SEP19c}
``Analogy and analogical reasoning'',
Stanford Encyclopedia of Philosophy, 2019

\bibitem[SS05]{SS05}
J. Sabater, C. Sierra, ``Review on computational trust and reputation
models'', Artificial Intelligence Review, 2005

\bibitem[Sab14]{Sab14}
K. J. Sabo, ``Anankastic conditionals: If you want to go to Harlem \Xl'',
Draft for Semantics Companion, 2014

\bibitem[Sch04]{Sch04}
K. Schlechta, ``Coherent systems'', Elsevier, Amsterdam, 2004.

\bibitem[Sch18]{Sch18}
K. Schlechta, ``Formal Methods for Nonmonotonic and Related
Logics'',
Vol. 1: ``Preference and Size'',
Vol. 2: ``Theory Revision, Inheritance, and Various Abstract
Properties''
Springer, 2018

\bibitem[Sch18a]{Sch18a}
K. Schlechta, ``Formal Methods for Nonmonotonic and Related
Logics'',
Vol. 1: ``Preference and Size''
Springer, 2018

\bibitem[Sch18b]{Sch18b}
K. Schlechta, ``Formal Methods for Nonmonotonic and Related
Logics'',
Vol. 2: ``Theory Revision, Inheritance, and Various Abstract
Properties''
Springer, 2018

\bibitem[Sch18e]{Sch18e}
K. Schlechta, ``Operations on partial orders'',
arXiv 1809.10620

\bibitem[Sch95-3]{Sch95-3}
K. Schlechta, ``Preferential choice representation
theorems for branching time structures''
Journal of Logic and Computation, Oxford,
Vol.5, pp. 783--800, 1995

\bibitem[Sch97-2]{Sch97-2}
K. Schlechta, ``Nonmonotonic logics - basic concepts,
results, and techniques''
Springer Lecture Notes series, LNAI 1187, Jan. 1997.

\bibitem[Sha13]{Sha13}
R. Shafer-Landau ed., ``Ethical Theory'',
J. Wiley and Sons, 2013

\bibitem[Sim63]{Sim63}
G. G. Simpson, ``Historical science'', in
C. C. Albritton, ``Fabric of geology'', Stanford, 1963, pp.24-48

\bibitem[Sta17a]{Sta17a}
Stanford Encyclopedia of Philosophy,
``The coherence theory of truth'',
https://plato.stanford.edu/archives/fall2018/entries/truth-coherence
(accessed 2017)

\bibitem[Sta17b]{Sta17b}
Stanford Encyclopedia of Philosophy,
``The correspondence theory of truth''
https://plato.stanford.edu/archives/win2020/
entries/truth-correspondence
(accessed 2017)

\bibitem[Sta17c]{Sta17c}
Stanford Encyclopedia of Philosophy,
``Epistemology''
https://plato.stanford.edu/archives/fall2020/entries/epistemology
(accessed 2017)

\bibitem[Sta17d]{Sta17d}
Stanford Encyclopedia of Philosophy,
``The Philosophy of Neuroscience''
https://plato.stanford.edu/archives/fall2019/entries/neuroscience
(accessed 2017)

\bibitem[Sta18a]{Sta18a}
Stanford Encyclopedia of Philosophy,
``Metaethics''
https://plato.stanford.edu/archives/sum2014/entries/metaethics
(accessed 2018)

\bibitem[Sta18b]{Sta18b}
Stanford Encyclopedia of Philosophy,
``Juergen Habermas''
https://plato.stanford.edu/archives/fall2017/entries/habermas
(accessed 2018)

\bibitem[Sta18c]{Sta18c}
Stanford Encyclopedia of Philosophy,
``The legal concept of evidence''
https://plato.stanford.edu/archives/win2021/entries/evidence-legal
(accessed 2018)

\bibitem[Sta68]{Sta68}
R. Stalnaker, ``A theory of conditionals'', N. Rescher (ed.), ``Studies in
logical theory'', Blackwell, Oxford, pp. 98--112

\bibitem[Tha07]{Tha07}
P. Thagard, ``Coherence, truth and the development of scientific
knowledge'', Philosophy of Science, 74:26-47, 2007

\bibitem[WSFR02]{WSFR02}
P. Winkielman, N. Schwarz, T. A. Fazendeiro, R. Reber,
``The hedonic marking of processing fluency: implications for
evaluative judgement'', in:
J. Musch, K. C. Klauer eds., ``The psychology of evaluation: affective
processes in cognition and emotion'', 2002, Lawrence Erlbaum, Mahwah, NJ

\bibitem[Wik16a]{Wik16a}
Wikipedia, ``Rechtsphilosophie'',
https://de.wikipedia.org/wiki/Rechtsphilosophie
(accessed 2016)

\bibitem[Wik17a]{Wik17a}
Wikipedia, ``Memory'',
https://en.wikipedia.org/wiki/Memory
(accessed 2017)

\bibitem[Wik17b]{Wik17b}
Wikipedia, ``Semantic memory'',
https://en.wikipedia.org/wiki/Semantic-memory
(accessed 2017)

\bibitem[Wik17c]{Wik17c}
Wikipedia, ``Episodic memory'',
https://en.wikipedia.org/wiki/Episodic-memory
(accessed 2017)

\bibitem[Wik17d]{Wik17d}
Wikipedia, ``Recognition memory'',
https://en.wikipedia.org/wiki/Recognition-memory
(accessed 2017)

\bibitem[Wik17e]{Wik17e}
Wikipedia, ``Visual cortex'',
https://en.wikipedia.org/wiki/Visual-cortex
(accessed 2017)

\bibitem[Wik18a]{Wik18a}
Wikipedia, ``Empathy'',
https://en.wikipedia.org/wiki/Empathy
(accessed 2018)

\bibitem[Wik18b]{Wik18b}
Wikipedia, ``Diskursethik'',
https://de.wikipedia.org/wiki/Diskursethik
(accessed 2018)

\bibitem[Wik18c]{Wik18c}
Wikipedia, ``Philosophy of science'',
https://en.wikipedia.org/wiki/Philosophy-of-science
(accessed 2018)

\bibitem[Yab82]{Yab82}
S. Yablo, ``Grounding, dependence, and paradox'',
Journal Philosophical Logic, Vol. 11, No. 1, pp. 117-137, 1982

\bibitem[ZMM15]{ZMM15}
P. Zeidman, S. L. Mullally, E. A. Maguire,
``Constructing, perceiving, and maintainig scenes: Hippocampal activity
and connectivity'',
Cerebral Cortex, Oct. 2015, 25:3836-3855

\end{thebibliography}
\end{document}